\documentclass{aa}
\usepackage[varg]{txfonts}
\usepackage{natbib}
\usepackage{graphicx}
\usepackage{subcaption}
\usepackage{wasysym}
\usepackage[usenames,dvips]{color}
\usepackage{verbatim} 
\usepackage{rotating} 
\usepackage{multirow} 
\usepackage{soul} 
\usepackage{xcolor}
\usepackage{framed}
\usepackage{centernot}
\usepackage{amsmath}
\usepackage[percent]{overpic}
\usepackage{array}

\def\apjl{ApJ}%
\def\apjs{ApJS}%
\def\ao{Appl.~Opt.}%
\def\aap{A\&A}%
\title{An updated Type II supernova Hubble diagram\thanks{Tables A.1, A.3, A.5, A.7, A.9, A.11, A.13, A.15 and A.17 are also available in electronic format at the CDS via anonymous ftp to cdsarc.u-strasbg.fr (130.79.128.5) or via http://cdsweb.u-strasbg.fr/cgi-bin/qcat?J/A+A/.}}

\author{E.E.E. Gall \inst{\ref{inst1},\ref{inst2}\thanks{E-mail: egall01@qub.ac.uk}}
\and R. Kotak \inst{\ref{inst1}}  
\and B. Leibundgut \inst{\ref{inst3},\ref{inst4}}  
\and S. Taubenberger \inst{\ref{inst2},\ref{inst3}}
\and W. Hillebrandt \inst{\ref{inst2}} 
\and M. Kromer \inst{\ref{inst5},\ref{inst6},\ref{inst7}}  
\and W.S. Burgett \inst{\ref{inst8}}
\and K. Chambers \inst{\ref{inst9}}
\and H. Flewelling \inst{\ref{inst9}}
\and M. E. Huber \inst{\ref{inst9}}
\and N. Kaiser \inst{\ref{inst9}}
\and R.P. Kudritzki \inst{\ref{inst9}}
\and E.A. Magnier \inst{\ref{inst9}}
\and N. Metcalfe \inst{\ref{inst10}}
\and K. Smith \inst{\ref{inst1}} 
\and J.L. Tonry \inst{\ref{inst9}}
\and R.J. Wainscoat \inst{\ref{inst9}}
\and C. Waters \inst{\ref{inst9}}
}

\institute{Astrophysics Research Centre, School of Mathematics and Physics, Queen's University Belfast, Belfast BT7 1NN, UK\label{inst1}
\and Max-Planck-Institut f\"{u}r Astrophysik, Karl-Schwarzschild-Str. 1, DE-85748 Garching-bei-M\"{u}nchen, Germany\label{inst2}
\and ESO, Karl-Schwarzschild-Strasse 2, 85748 Garching, Germany\label{inst3}
\and Excellence Cluster Universe, Technische Universit\"{a}t M\"{u}nchen, Boltzmannstrasse 2, DE-85748 Garching-bei-M\"{u}nchen, Germany\label{inst4}
\and Heidelberger Institut f\"{u}r Theoretische Studien, Schloss-Wolfsbrunnenweg 35, D-69118 Heidelberg, Germany \label{inst5}
\and Zentrum für Astronomie der Universit\"{a}t Heidelberg, Institut f\"{u}r Theoretische Astrophysik, Philosophenweg 12, D-69120 Heidelberg, Germany \label{inst6}
\and The Oskar Klein Centre \& Department of Astronomy, Stockholm University, AlbaNova, SE-106 91 Stockholm, Sweden\label{inst7}
\and GMTO Corporation, 465 N. Halstead St., Suite 250 Pasadena, California 91107, USA\label{inst8}
\and Institute for Astronomy, University of Hawaii at Manoa, 2680 Woodlawn Drive, Honolulu, Hawaii 96822, USA\label{inst9}
\and Centre for Extragalactic Astronomy, Department of Physics, Durham University, South Road, Durham DH1 3LE, UK\label{inst10}
}

\date{Received 30 May 2017 / Accepted 10 October 2017 }

\abstract{We present photometry and spectroscopy of nine Type II-P/L supernovae (SNe) with redshifts in the $0.045 \lesssim z \lesssim 0.335$ range, with a view to re-examining their utility as distance indicators. Specifically, we apply the expanding photosphere method (EPM) and the standardized candle method (SCM) to each target, and find that both methods yield distances that are in reasonable agreement with each other.
The current record-holder for the highest-redshift spectroscopically confirmed SN II-P is PS1-13bni ($z$ = $0.335^{+0.009}_{-0.012}$), and illustrates the promise of Type II SNe as cosmological tools. We updated existing EPM and SCM Hubble diagrams by adding our sample to those previously published. 
Within the context of Type II SN distance measuring techniques, we investigated two related questions. 
First, we explored the possibility of utilising spectral lines other than the traditionally used Fe\,{\sc ii}\,$\lambda$5169 to infer the photospheric velocity of SN ejecta. Using local well-observed 
objects, we derive an epoch-dependent relation between the strong Balmer line and Fe\,{\sc ii}\,$\lambda$5169 velocities that is applicable 30 to 40\,days post-explosion. Motivated in part by the continuum of key observables such as rise time and decline rates exhibited from II-P to II-L SNe, we assessed the possibility of using Hubble-flow Type II-L SNe as distance indicators. These yield similar distances as the Type II-P SNe. Although these initial results are encouraging, a significantly larger sample of SNe II-L would be required to draw definitive conclusions. }

\keywords{Stars: supernovae -- distance scale -- supernovae: individual: SN~2013eq, SN~2013ca, LSQ13cuw, PS1-13wr, PS1-14vk, PS1-12bku, PS1-13abg, PS1-13baf, PS1-13bmf, PS1-13atm, PS1-13bni}

\begin{document}

\maketitle

\section{Introduction}
\label{section:intro}

The past decades have been marked by an ongoing revolution in cosmology and distance estimation techniques. Following the astounding discovery that the Universe was expanding at an accelerating rate \citep{Riess1998a,Perlmutter1999a}, the quest to determine the precise value of the Hubble constant, $H_0$, has inspired a host of new distance determination techniques, systematic improvements to old approaches, as well as large-scale focussed projects dedicated to improving the precision in the measurements of $H_0$ in order to shed light on the constituents of the energy density of the Universe. 

Already in the 1980s i.e., before the launch of the Hubble Space Telescope (HST), the measurement of the Hubble constant with an uncertainty of $\lesssim$10\% was chosen as one of three HST ``Key Projects''.
\citet{Freedman2001} presented the final results of this endeavour. Using an array of secondary distance indicators, calibrated using Cepheid distances, they reported a value for $H_0$ of 72\,$\pm$\,8\,km\,s$^{-1}$\,Mpc$^{-1}$. 

Galaxies for which distance estimates from multiple sources are available are crucial as anchors for the extragalactic distance scale. 
One example is NGC~4258, which has estimates for a geometric distance through mega masers \citep{Humphreys2013}, a Cepheid distance \citep[e.g.][]{Fiorentino2013}, and a Type II-P SN distance \citep{Polshaw2015}. NGC~4258 may therefore be a more suitable anchor galaxy than the Large Magellanic Cloud, which was used as the first rung on the distance ladder for a variety of distance estimation techniques \citep[][and references therein]{Riess2011b,Riess2011a}.

Recently, the \citet{PlanckCollaboration2016} presented the outcomes of observations of the cosmic microwave background, which enabled them to restrict the Hubble constant to $H_{0}$ = 67.8\,$\pm$\,0.9\,km\,s$^{-1}$\,Mpc$^{-1}$, corresponding to an uncertainty of only 1.3\%. 

In the latest update, \citet{Riess2016} combined both newly calibrated distance measurements of Cepheid stars and Type Ia SNe, to constrain the Hubble constant to only 2.4\% as $H_{0}$ = 73.0\,$\pm$\,1.8\,km\,s$^{-1}$\,Mpc$^{-1}$.   

While the uncertainties in $H_0$ have decreased significantly over the years, a disagreement at the $2.0-2.5\,\sigma$ level between the values derived via SN Ia cosmology and measurements of the cosmic microwave background has emerged, reflecting a tension between local and global measurements of $H_0$. Such a discrepancy could either imply unknown systematic uncertainties in the measurements of $H_0$ or that current cosmological models have to be revised \citep[e.g.][]{Bennett2014}. Nevertheless, \citet{Bennett2014} also maintain that the various estimates of $H_0$ are not inconsistent with each other.

The ``Expanding Photosphere Method'' \citep[EPM;][]{Kirshner1974} and the ``Standardized Candle Method'' \citep[SCM;][]{Hamuy2002a} and their more recent derivatives provide independent routes to $H_0$.
On the most fundamental level, the EPM relies on the comparison of the angular size of an object with the ratio between its observed and theoretical flux, whereas the SCM uses the relation between the expansion velocities of SNe II-P during the plateau phase and its plateau luminosity to determine the distance. The EPM and SCM
are subject to different systematics than the previously mentioned techniques and they have both undergone significant improvements from their original description (e.g. \citealt{Wagoner1981,Schmidt1994b,Eastman1996,Hamuy2001,Dessart2005,Gall2016} for the EPM or \citealt{Nugent2006,Poznanski2009} for the SCM), aiming to reduce their systematic uncertainties. In particular, the EPM is independent of the cosmic distance ladder. Additionally, SNe II-P are more common albeit less luminous than SNe Ia, and bear the potential to be observed in statistically  significant numbers. 

Several other methods have been proposed to either improve on EPM or reduce the observational effort, mostly spectroscopy, for distance determination with Type II SNe. The application of a single dilution factor for the black body radiation has been shown to be a clear limitation of EPM. The Spectral Expanding Atmosphere Method \citep[SEAM;][]{Mitchell2002, Baron2004, Dessart2006} attempts to establish the physical conditions in the supernova through detailed spectral fitting. In this case, the emergent flux is determined through detailed model fitting for each spectral epoch. This method requires excellent spectral data, and has so far been applied only to the brightest and nearest supernovae (SNe~1987A and 1999em). Variants of the SCM include the photospheric magnitude method \cite[][]{Rodriguez2014} and the photometric candle method \citep[][]{deJaeger2015, deJaeger2017a}. These rely almost exclusively on photometry, and have been mostly developed to exploit the large SN samples that are expected to become available in future surveys. A separate approach has been taken by \citet{Pejcha2015}, who combine all available photometric and spectroscopic information to generate best fit synthetic light curves, yielding parameters such as reddening and nickel mass in addition to a relative distance. This method requires exquisite observational data to function, and in particular, it depends on a good sampling of the light curve and the velocity evolution for many supernovae. Although this method does not provide an independent distance zeropoint, it obtains very good relative distances. Thus it is clear that no single approach is equally applicable at all redshifts.

\citet{Schmidt1994a} considered Type II SNe out to $cz \sim 5500$\,km\,s$^{-1}$, and developed much of the EPM framework.
In this study, we build upon our previous work \citep{Gall2015,Gall2016} where we re-examined relativistic effects -- specifically, the difference between ``angular size'' and ``luminosity'' distance -- that come into play when applying the EPM to SNe at non-negligible redshifts, a subtlety that was neglected in previous studies. We re-derived the basic equations of the EPM, and showed that for distant SNe, the angular size, $\theta$, should be corrected by a factor of $(1+z)^2$, and that the observed flux has to be transformed into the SN rest frame. These findings were applied to SN~2013eq ($z$ = 0.041\,$\pm$\,0.001) in \citet{Gall2016}, implementing both the EPM and SCM, and demonstrated that the two techniques give consistent results.

However, even in the local Universe the division between SNe II-P -- which are typically used for the EPM and SCM -- and SNe II-L is ambiguous and has given rise to extensive discussions on whether SNe II-P/L should \citep{Anderson2014b,Sanders2014}, or should not \citep{Arcavi2012,Faran2014b} be viewed as members of the same class with a continuum of properties. The rise times of a sample of 20 Type II-P and II-L SNe were analyzed in \citet{Gall2015}, amongst them LSQ13cuw, a Type II-L SN with excellent constraints on its explosion epoch ($<$\,1\,d). We found some evidence for SNe II-L having longer rise times and higher luminosities than SNe II-P, but also indications that a clear separation of SNe II-P/L into two distinct classes can be challenging \citep[see also][]{Gonzalez-Gaitan2015, Rubin16}. Given that this distinction is marginal, we investigate the possibility of including SNe II-L into the Hubble diagram.

We push the EPM and SCM techniques further, by applying both techniques to a set of Type II-P/L SNe, with redshifts in the following range: $0.04 < z < 0.34$. Three SNe in our sample exhibited post-peak decline rates that would be consistent with a Type II-L classification. This has little bearing on our results. Our sample also includes PS1-13bni with a redshift of $z$ = $0.335^{+0.009}_{-0.012}$ -- to the best of our knowledge -- the highest-redshift Type II-P SN discovered yet. We prepare the path for future surveys that will find statistically significant numbers of SNe II-P/L in the Hubble flow, and provide an independent channel with which to gain insights into the expansion history of the Universe, its geometry, the nature of dark energy, and other cosmological parameters. Similar analyses on different data sets have been performed by \citet{Rodriguez2014} and \citet{deJaeger2015}.

The paper is divided into the following parts: observations of the SNe in our sample are presented in \S\ref{section:Observations_and_data_reduction}; distance measurements using the Expanding Photosphere Method as well as the Standard Candle Method are performed in \S\ref{section:Results_and_discussion}; our main conclusions are given in \S\ref{section:Conclusions}.

\section{Observations and data reduction}
\label{section:Observations_and_data_reduction}

We acquired a sample of eleven Type II SNe ranging in redshifts from $z \sim 0.04$ to $z \sim 0.34$ (see Table \ref{table:whole_sample}). 
The majority of these objects were discovered by the Panoramic Survey Telescope \& Rapid Response System 1 \citep[Pan-STARRS1;][]{Kaiser2010}. Observations were obtained in the Pan-STARRS1 (PS1) filter system  $g_\mathrm{PS1}$, $r_\mathrm{PS1}$, $i_\mathrm{PS1}$, $z_\mathrm{PS1}$, and $y_\mathrm{PS1}$ \citep{Tonry2012b}. Aside from lying in the desired redshift range, targets were selected for their young age to allow for photometric and spectroscopic follow-up within the first 50 days after explosion.

Additionally, we obtained follow-up observations of SN 2013ca (also known as LSQ13aco) after announcements both in the Astronomers' Telegrams\footnote{http://www.astronomerstelegram.org.} \citep[ATels;][]{2013ATel5056} and the Central Bureau Electronic Telegrams\footnote{http://www.cbat.eps.harvard.edu/index.html.} \citep[CBET;][]{2013CBET3508}. We also include LSQ13cuw \citep{Gall2015} and SN 2013eq \citep{Gall2016} in our sample. 
All three objects fall in the redshift range of interest and fulfill the our selection criteria.
In the following we present the data reduction procedures as well as an overview of the photometric and spectroscopic observations.

\begin{table*}
 \caption{Sample of intermediate redshift Type II-P/L SNe}
 \label{table:whole_sample}
 \centering
 \resizebox{1.03\textwidth}{!}{%
 \begin{tabular}{c   l                       c                       r		                 c                              c c l c }
 \hline                 
 \multirow{2}{*}{SN} & \multicolumn{1}{c}{\multirow{2}{*}{Type}} & \multirow{2}{*}{R.A.} & \multicolumn{1}{c}{\multirow{2}{*}{Dec.}} & \multirow{2}{*}{Host galaxy} & \multicolumn{2}{c}{$E(B-V)$ [mag]} & \multicolumn{1}{c}{\multirow{2}{*}{Redshift}} &  \\ 
                     & & & &   & galactic & host & &  \\
\hline                 
 SN 2013eq & II-P & 17:33:15.73 & +36:28:35.2  & -                        & 0.034 & 0.062\,$\pm$\,0.028 & 0.041\,$\pm$\,0.001       & a     \\
 SN 2013ca & II-P & 11:58:43.25 & +19:08:56.2  & PGC 37698                & 0.029 & $<$\,0.04           & 0.045\,$\pm$\,0.001       & b,c   \\
 LSQ13cuw  & II-L & 02:39:57.35 & -08:31:24.2  & SDSS J023957.37-083123.8 & 0.023 & $<$\,0.16           & 0.045\,$\pm$\,0.003       & d     \\
 PS1-13wr  & II-P & 14:15:57.97 & +14:15:58.0 & SDSS J141558.95+520302.9 & 0.012 & 0.110\,$\pm$\,0.049 & 0.076\,$\pm$\,0.001       & b,c,e \\
 PS1-14vk  & II-L & 12:10:45.47 & +48:41:46.7  & SDSS J121045.27+484143.9 & 0.020 & $<$\,0.09           & 0.080\,$\pm$\,0.001       & b,c,e \\
 PS1-12bku & II-P & 22:16:40.20 & -01:08:24.4  & SDSS J221640.32-010824.7 & 0.029 & $<$\,0.27           & $0.088^{+0.006}_{-0.002}$ & b,c   \\
 PS1-13abg & II-P & 10:00:51.10 & +02:11:40.0  & zCOSMOS 824314           & 0.014 & $<$\,0.07           & 0.123\,$\pm$\,0.001       & b,c,f \\
 PS1-13baf & II-P & 16:16:01.61 & +54:58:31.3  & SDSS J161601.54+545833.0 & 0.007 & $<$\,0.36           & 0.144\,$\pm$\,0.001       & b,c   \\
 PS1-13bmf & II-L & 12:23:10.93 & +47:44:24.9  & SDSS J122311.46+474426.7 & 0.011 & -                   & 0.163\,$\pm$\,0.001       & b,c,e \\
 PS1-13atm & II-L? & 09:59:16.60 & +03:18:28.3  & SDSS J095916.19+031848.5 & 0.020 & -                   & 0.220\,$\pm$\,0.001       & b,c   \\
 PS1-13bni & II-P & 16:12:37.31 & +56:18:36.8  & SDSS J161236.57+561832.4 & 0.007 & -                   & $0.335^{+0.009}_{-0.012}$ & b,c   \\
 \hline  
 \end{tabular}}
 \\[1.5ex]
 \flushleft
 a) \citet{Gall2016};   
 b) this study;             
 c) NASA/IPAC Extragalactic Database;   
 d) \citet{Gall2015};        
 e) Sloan Digital Sky Survey; 
 f) \citet{Lilly2007}     
\end{table*}

\begin{figure*}
   \centering
   \begin{subfigure}[t]{0.295\textwidth}
       \includegraphics[width=\columnwidth]{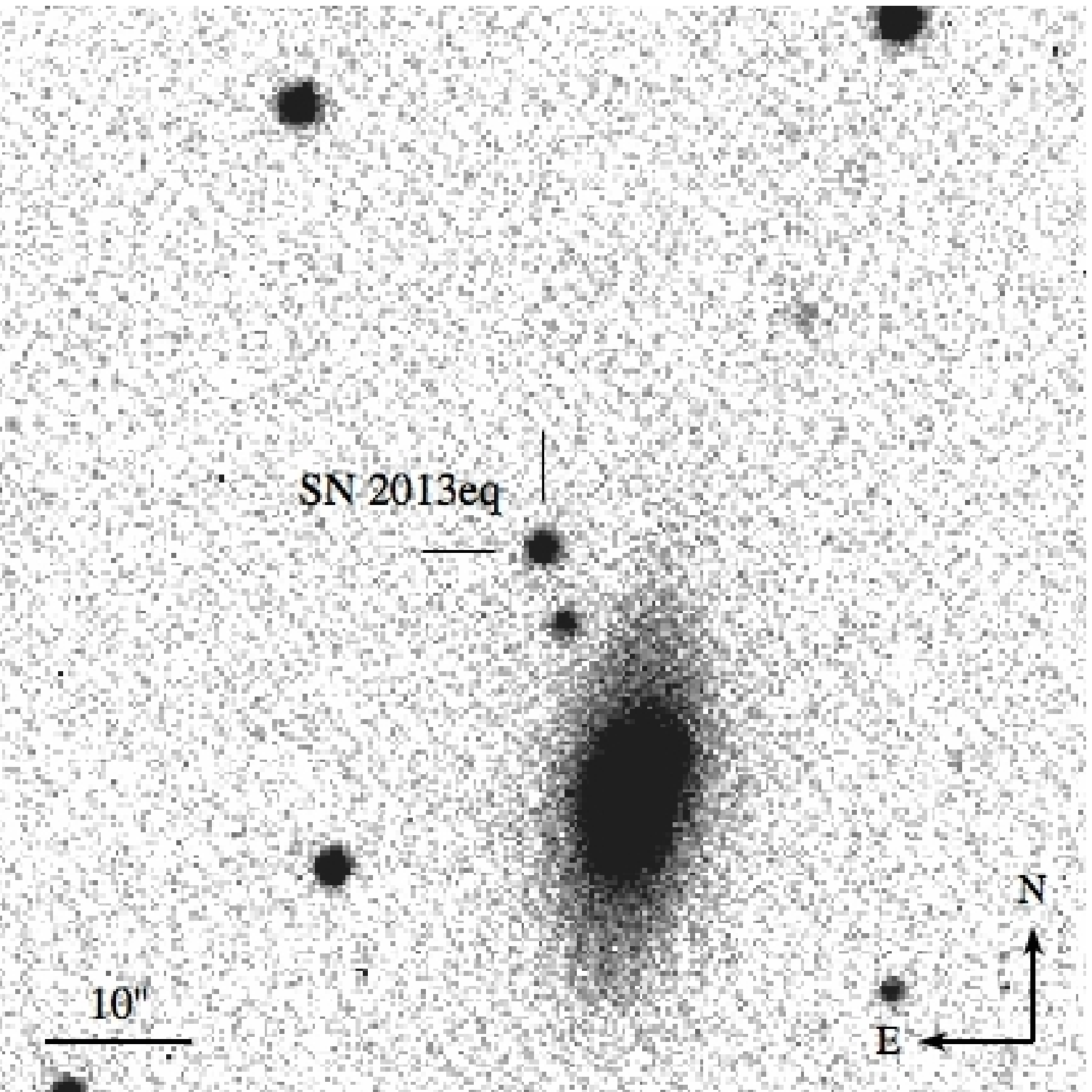} %
   \end{subfigure}%
   \hspace{2.0mm}
   \begin{subfigure}[t]{0.295\textwidth}
       \includegraphics[width=\columnwidth]{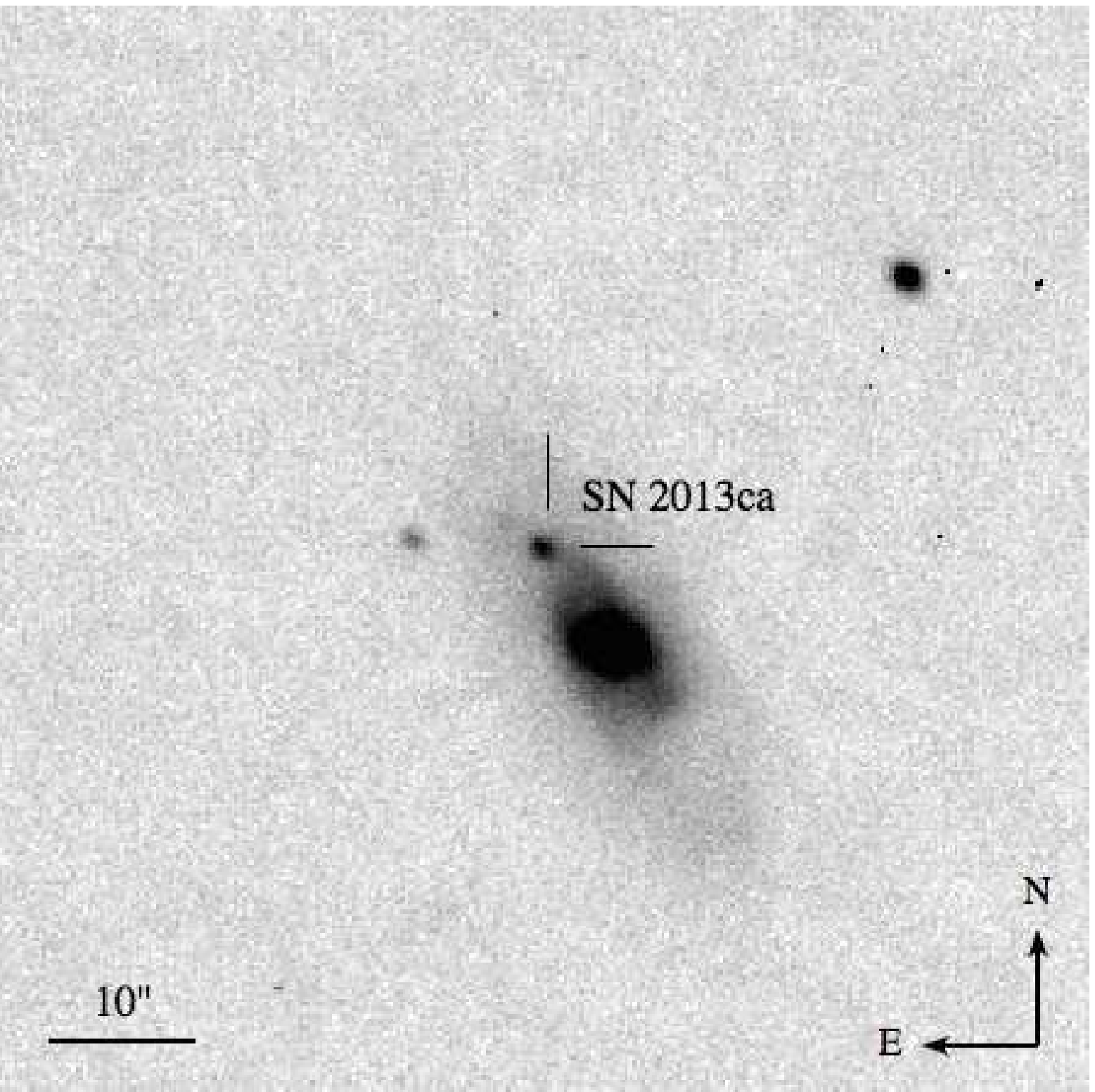} %
   \end{subfigure}%
   \hspace{2.0mm}
   \begin{subfigure}[t]{0.295\textwidth}
       \includegraphics[width=\columnwidth]{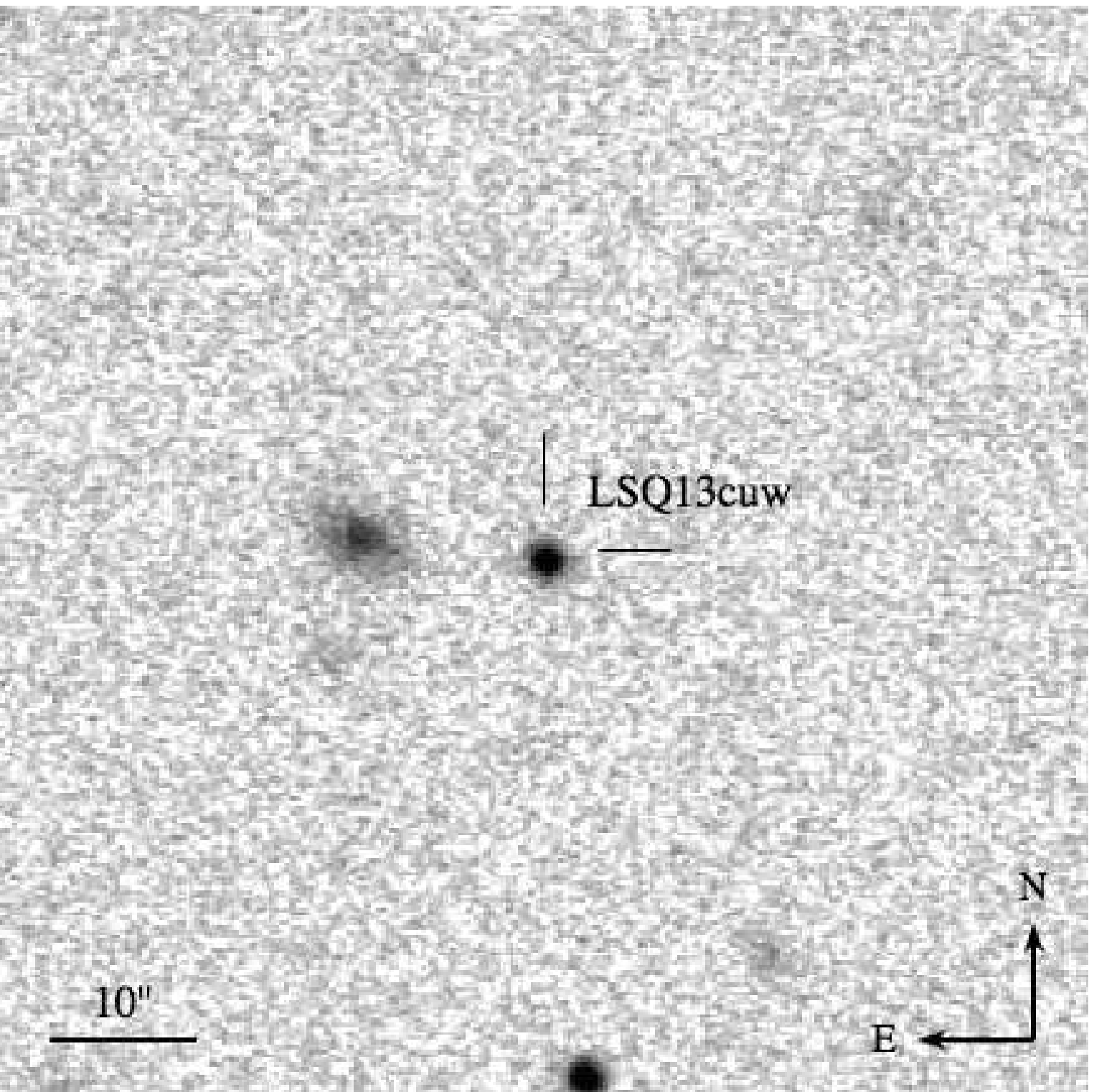} %
   \end{subfigure}%
   
   \vspace{2.0mm}
   \begin{subfigure}[t]{0.295\textwidth}
       \includegraphics[width=\columnwidth]{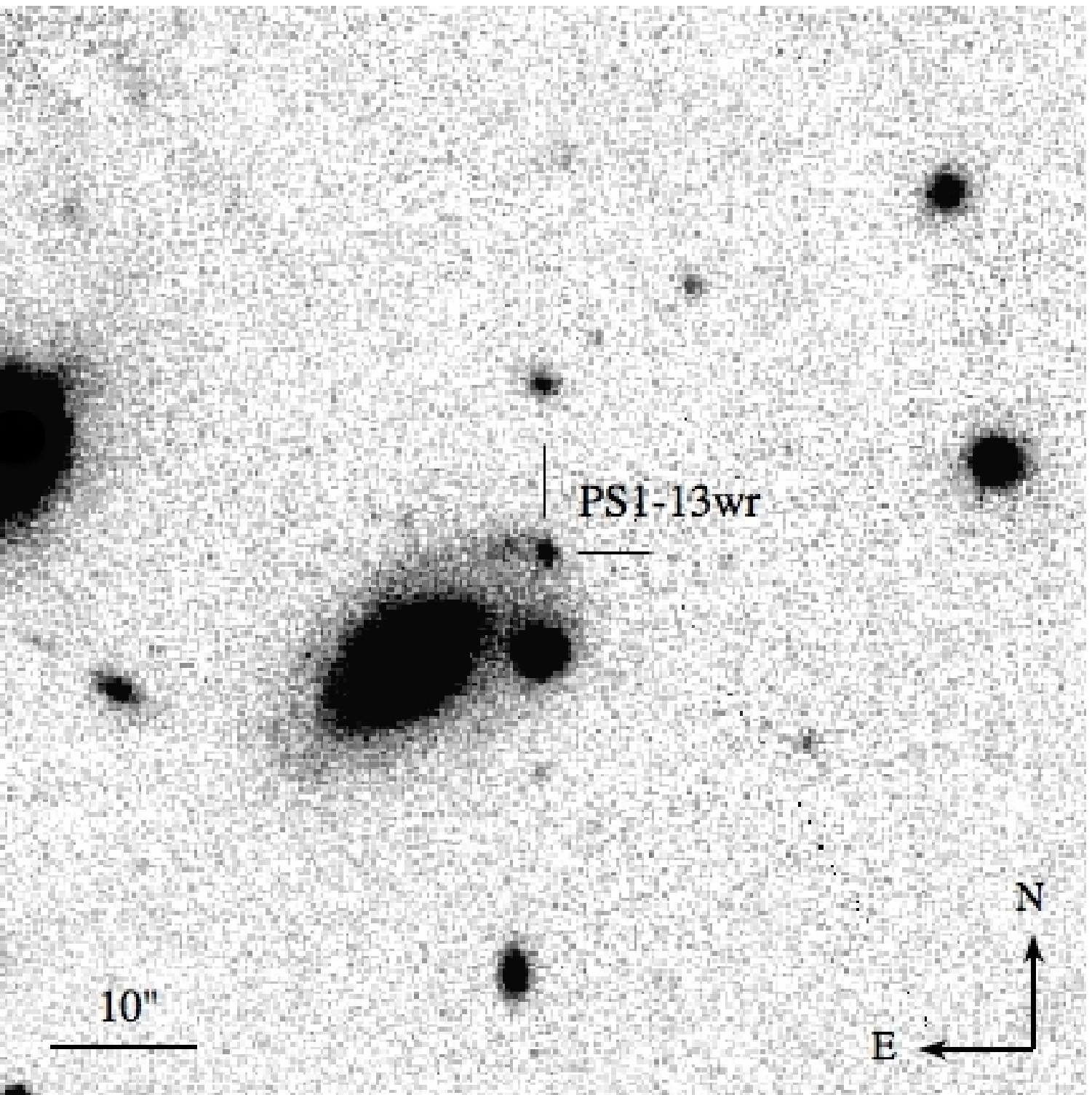} %
   \end{subfigure}%
   \hspace{2.0mm}
   \begin{subfigure}[t]{0.295\textwidth}
       \includegraphics[width=\columnwidth]{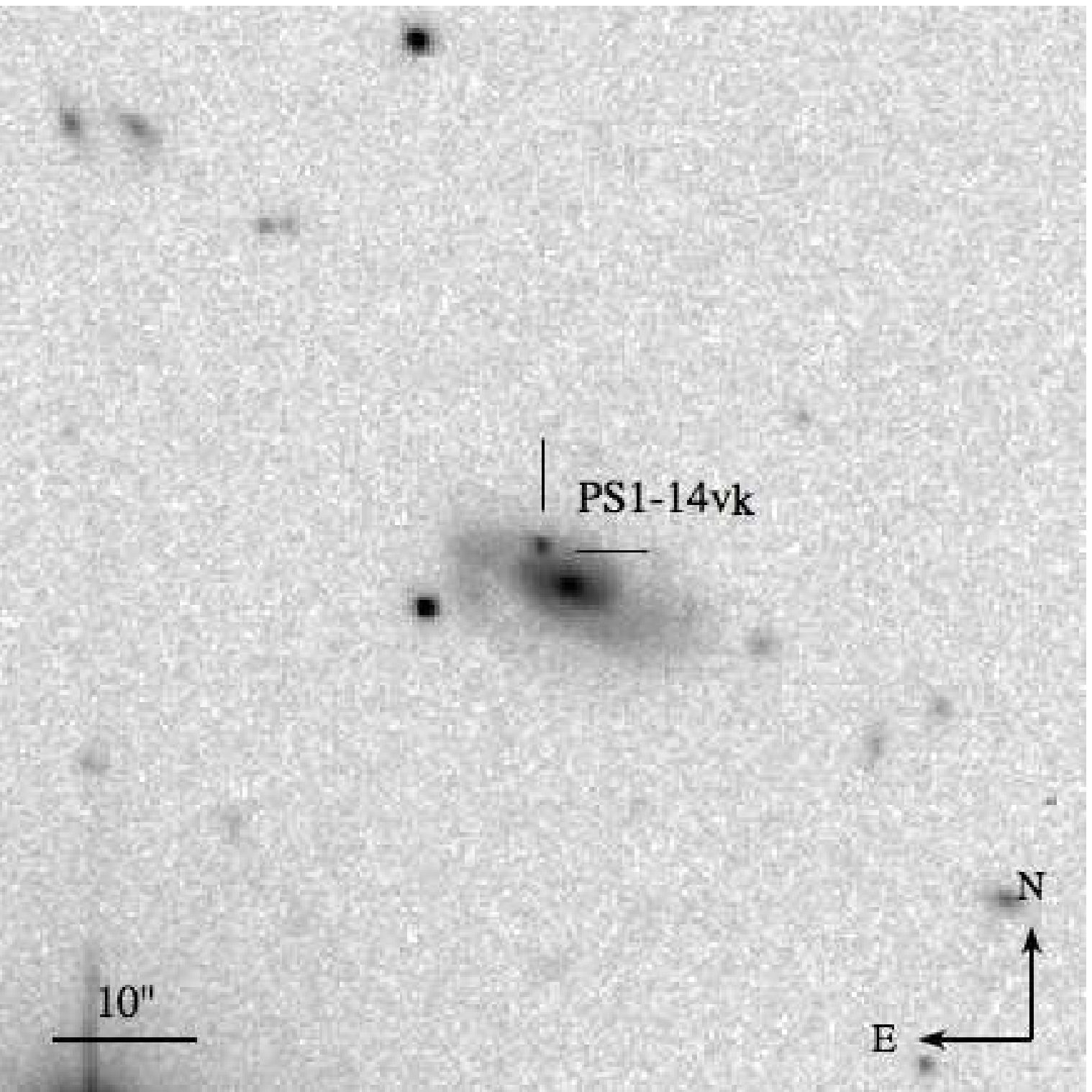} %
   \end{subfigure}%
   \hspace{2.0mm}
   \begin{subfigure}[t]{0.295\textwidth}
       \includegraphics[width=\columnwidth]{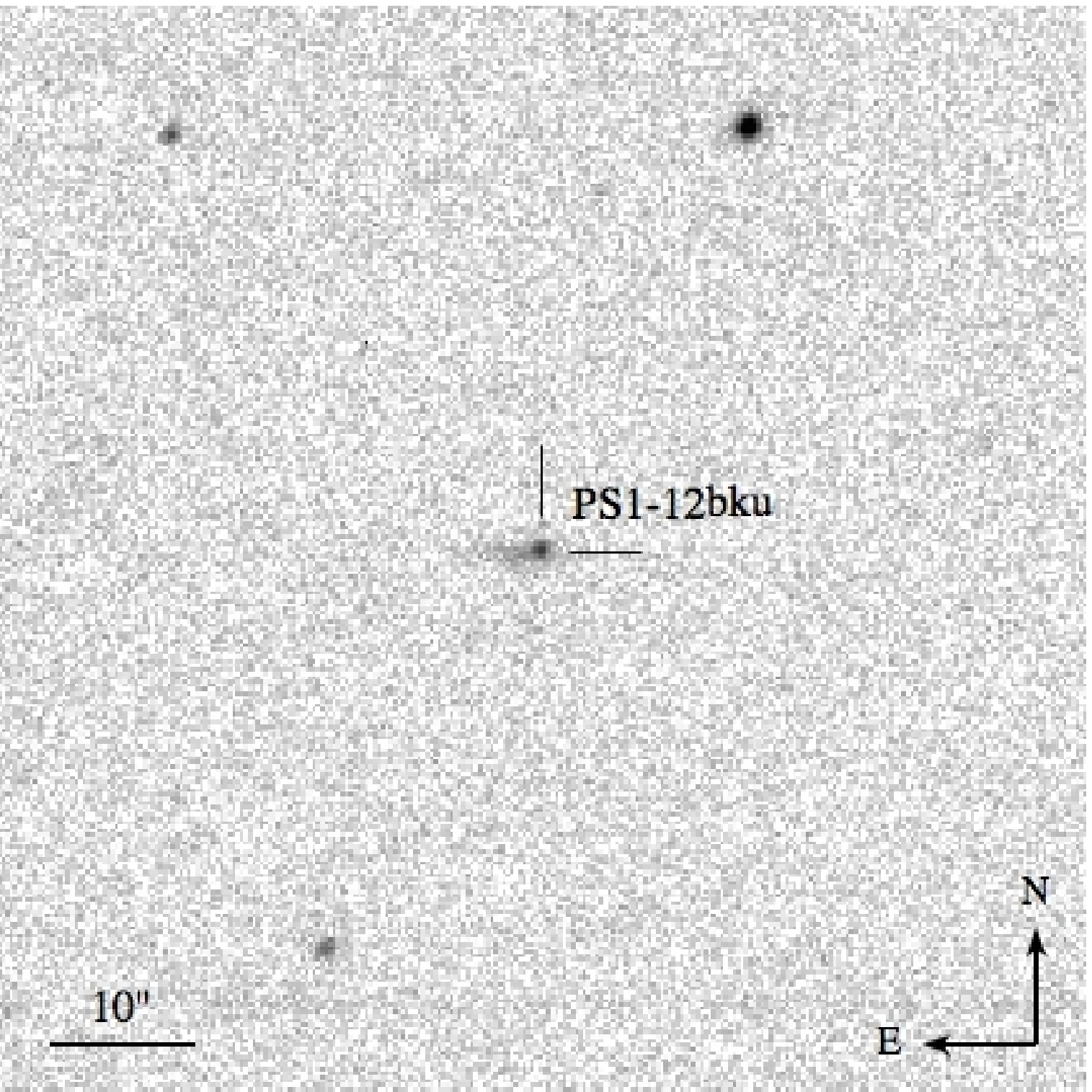} %
   \end{subfigure}%

   \vspace{2.0mm}
   \begin{subfigure}[t]{0.295\textwidth}
       \includegraphics[width=\columnwidth]{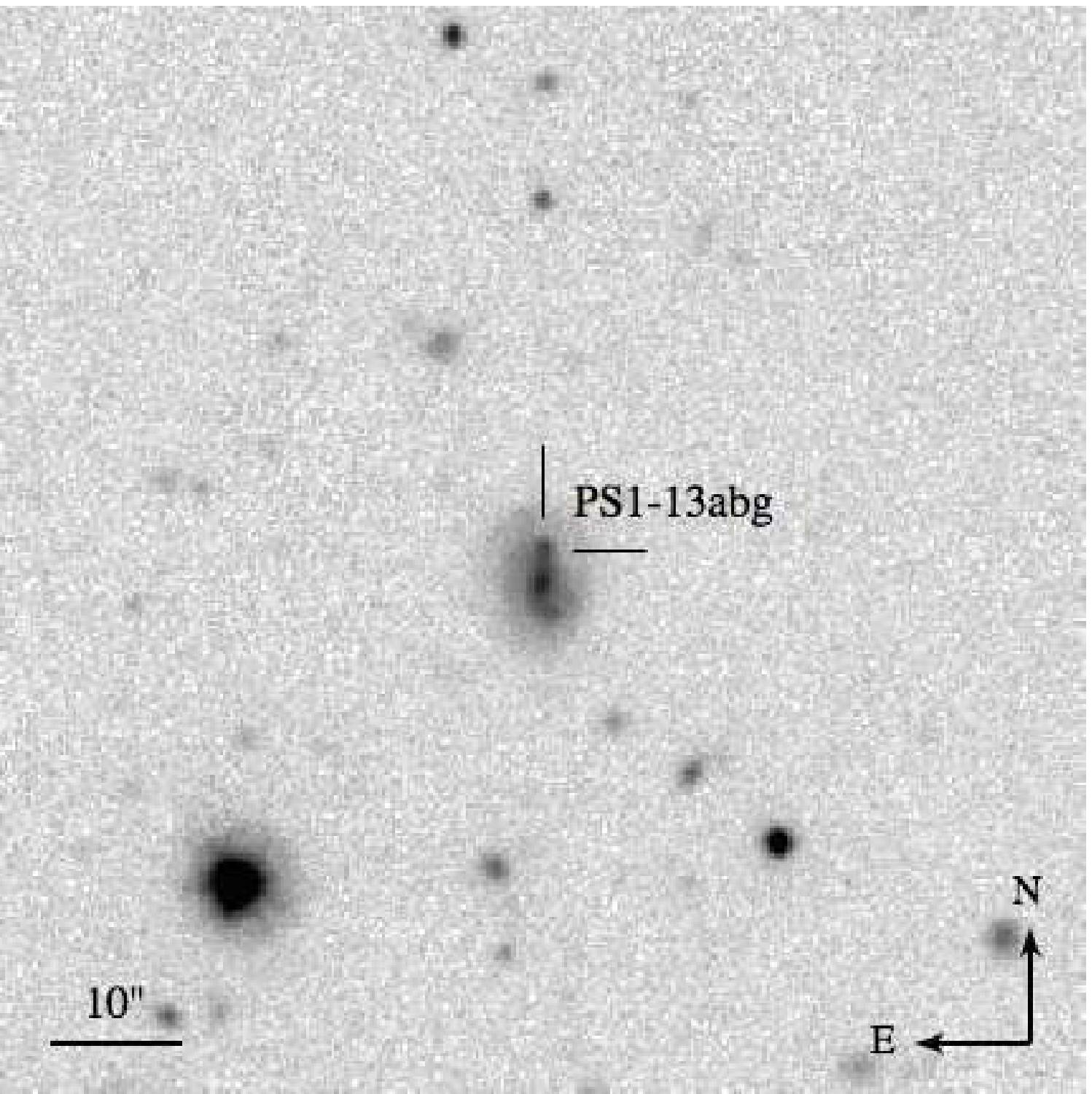} %
   \end{subfigure}%
   \hspace{2.0mm}
   \begin{subfigure}[t]{0.295\textwidth}
       \includegraphics[width=\columnwidth]{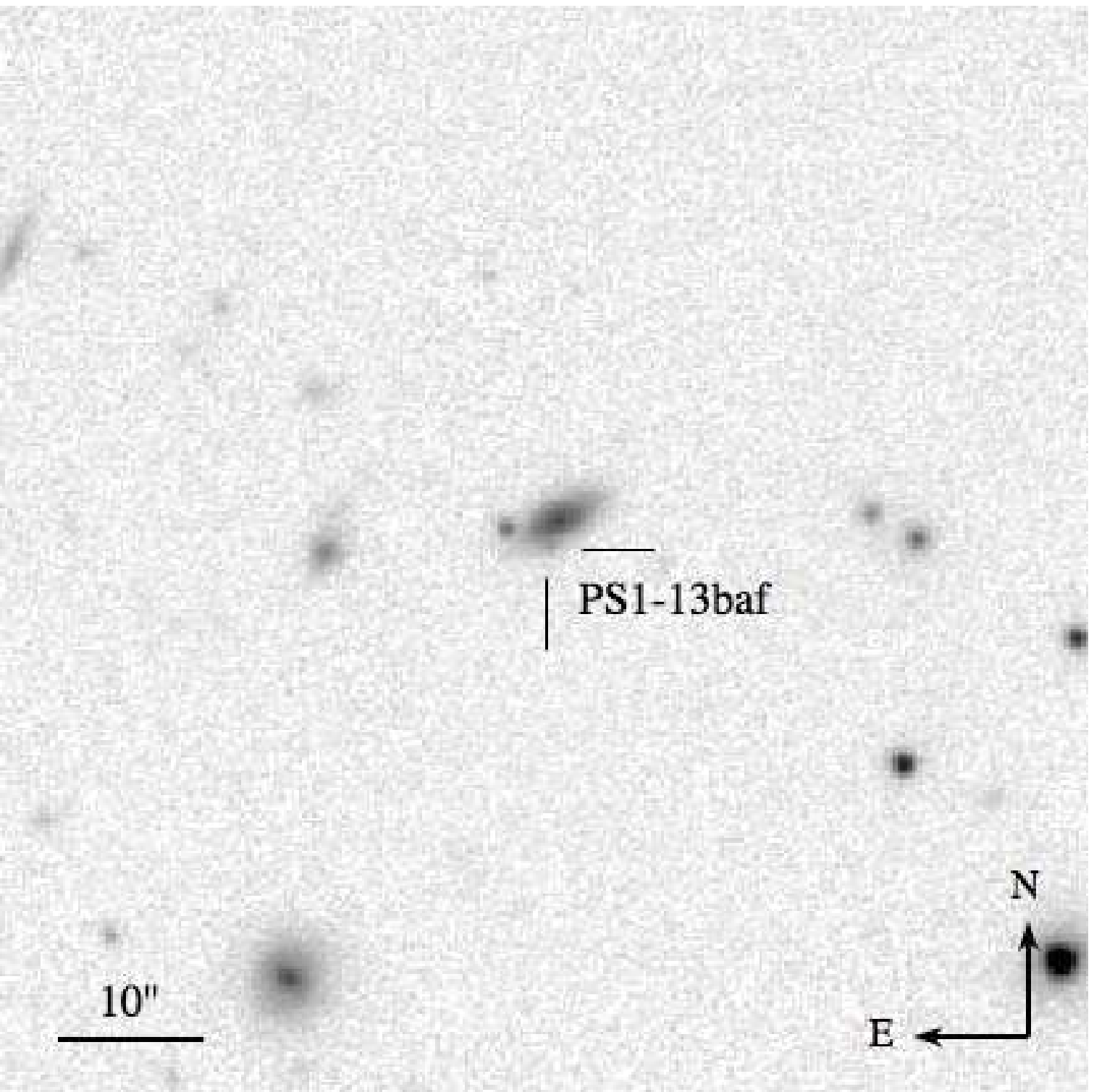} %
   \end{subfigure}%
   \hspace{2.0mm}
   \begin{subfigure}[t]{0.295\textwidth}
       \includegraphics[width=\columnwidth]{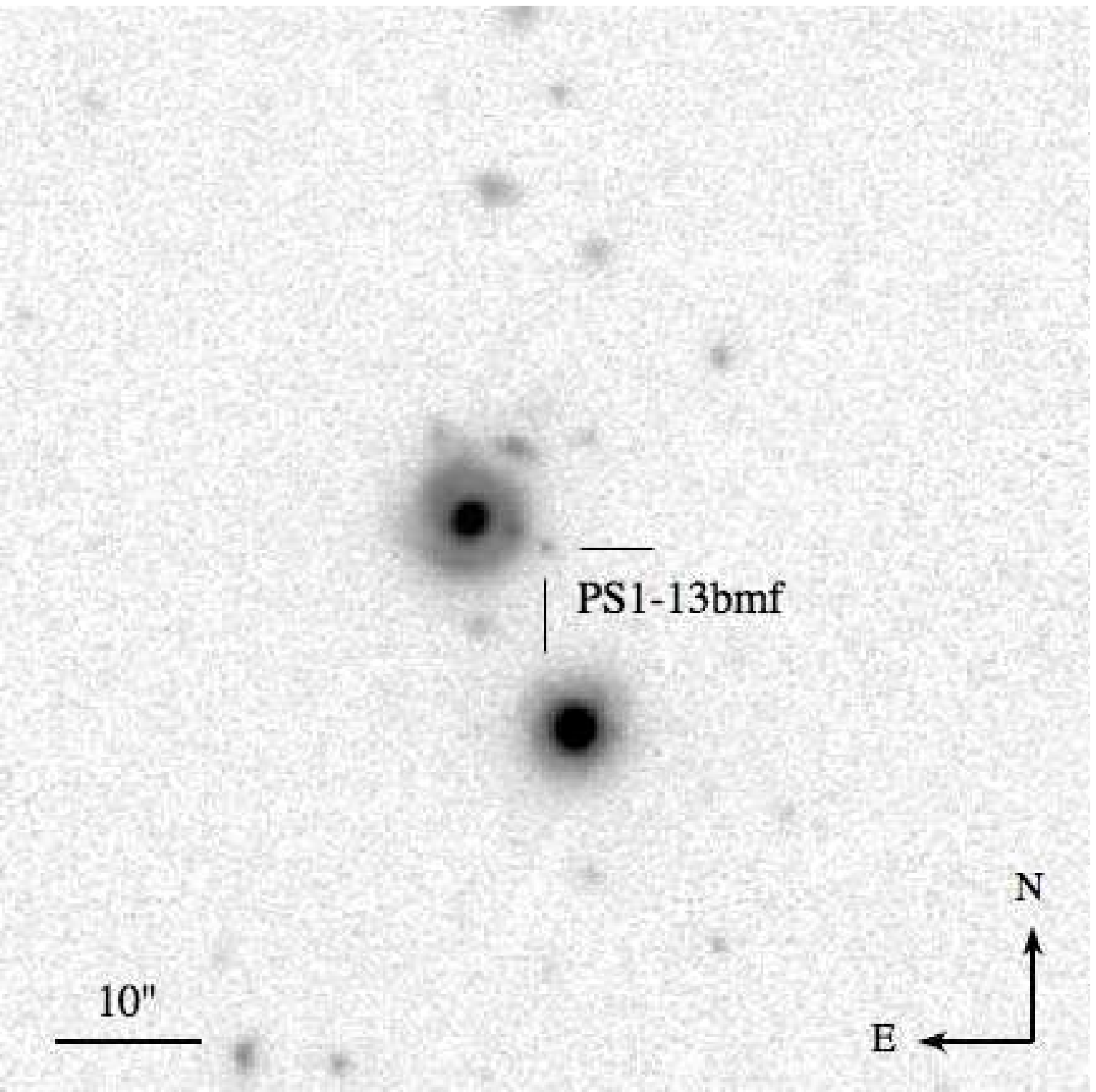} %
   \end{subfigure}%

   \vspace{2.0mm}
   \begin{subfigure}[t]{0.295\textwidth}
   \includegraphics[width=\columnwidth]{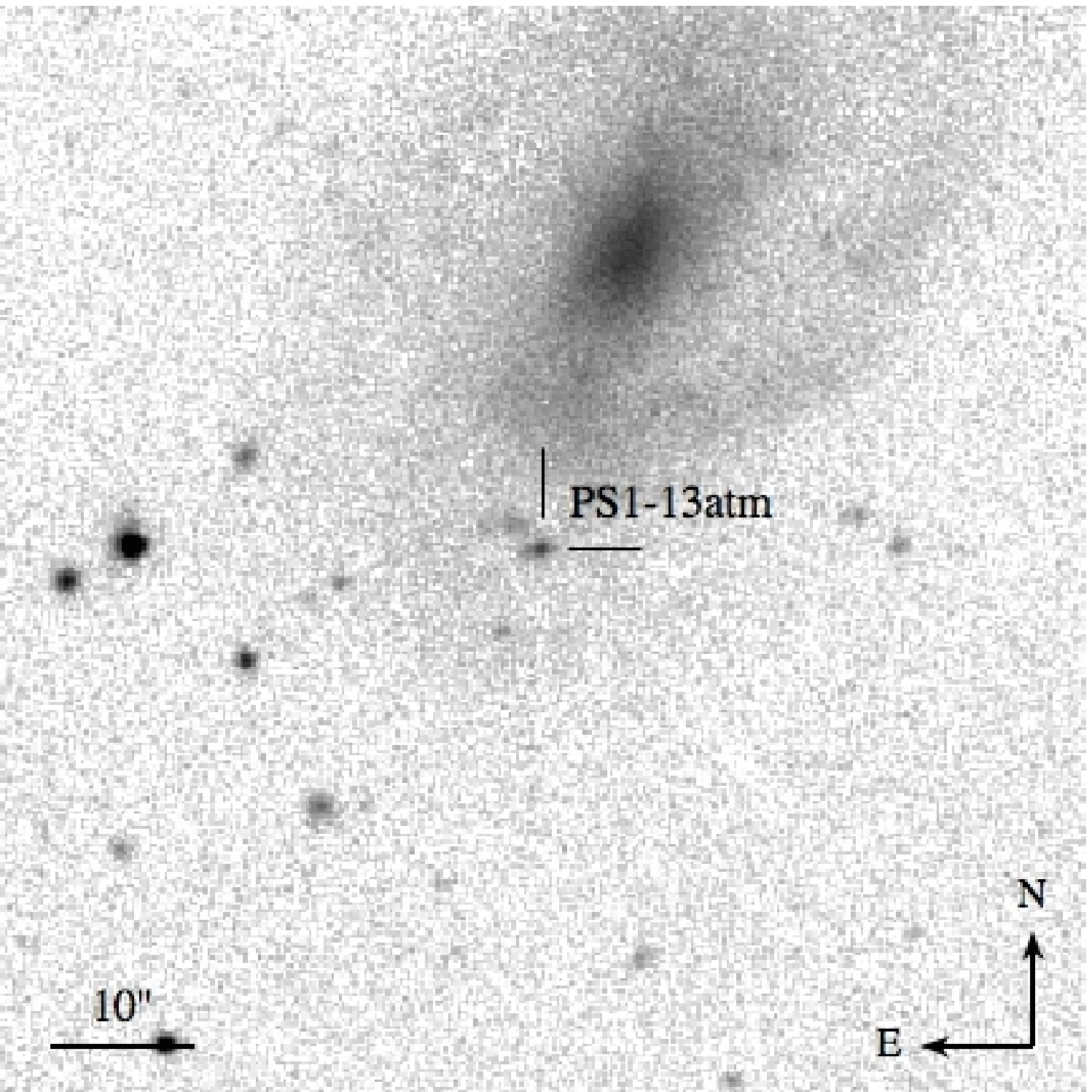} %
   \end{subfigure}%
   \hspace{2.0mm}
   \begin{subfigure}[t]{0.295\textwidth}
       \includegraphics[width=\columnwidth]{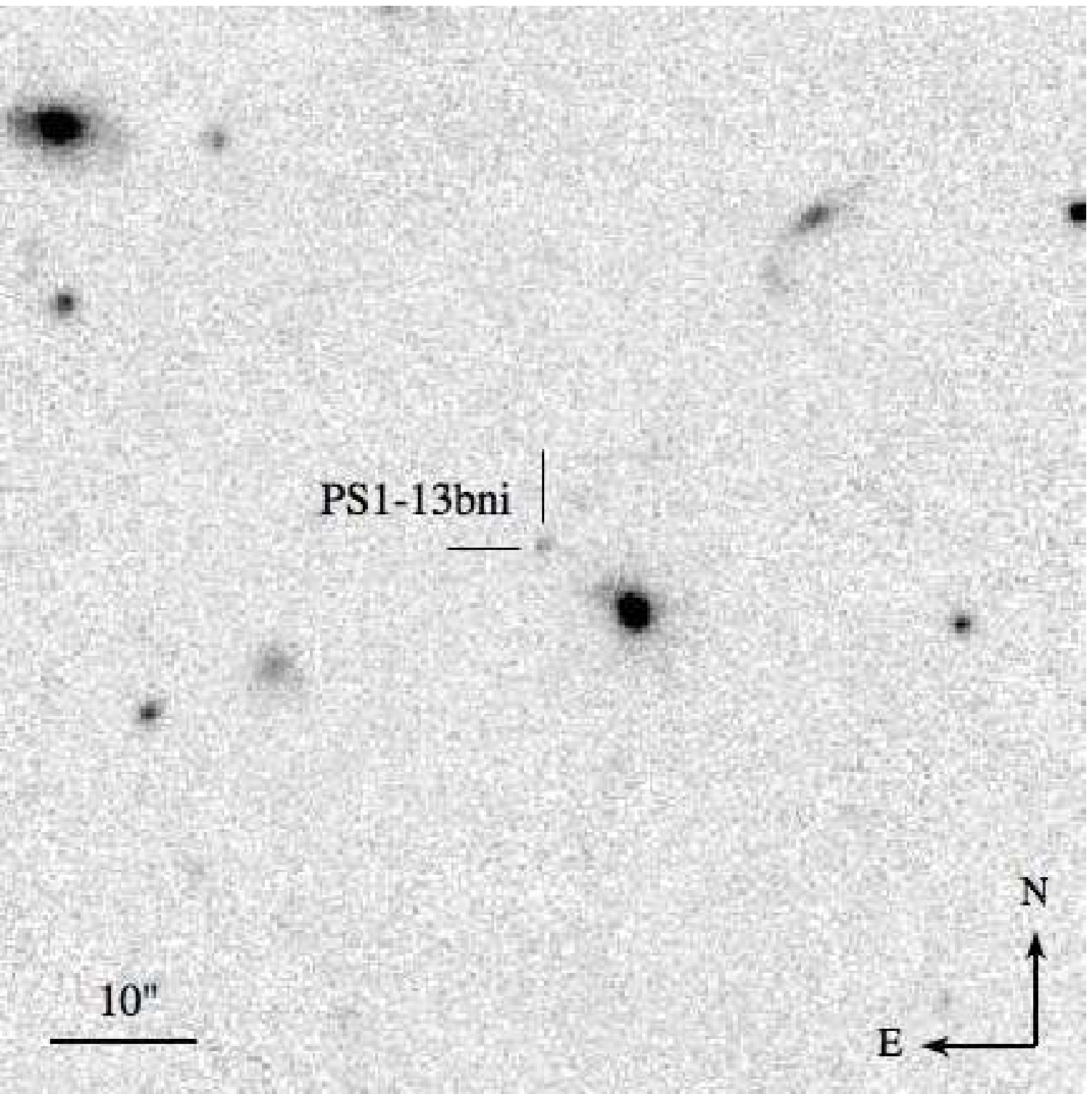} %
   \end{subfigure}%
   \caption{The SNe in our sample and their environments. Short dashes mark the location of the respective supernova (see Table \ref{table:whole_sample} for the exact coordinates). The images were taken in 
the SDSS $i'$-band on MJD 56520.89 for SN 2013eq, 
the $i'$-band on MJD 56462.90 for SN 2013ca, 
the $r'$-band on MJD 56625.03 for LSQ13cuw, 
the $i_{\mathrm{PS1}}$-band on MJD 56462.31 for PS1-13wr, 
the $i'$-band on MJD 56768.92 for PS1-14vk, 
the $i'$-band on MJD 56205.88 for PS1-12bku, 
the $i_{\mathrm{PS1}}$-band on MJD 56422.29 for PS1-13abg, 
the $i_{\mathrm{PS1}}$-band on MJD 56414.52 for PS1-13baf, 
the $i_{\mathrm{PS1}}$-band on MJD 56422.34 for PS1-13bmf, 
the $i_{\mathrm{PS1}}$-band on MJD 56422.29 for PS1-13atm, and 
the $i_{\mathrm{PS1}}$-band on MJD 56432.50 for PS1-13bni.
}
   \label{figure:Host_Galaxies}
\end{figure*}

\subsection{Data reduction}
\label{section:Data_reduction}

The PS1 data were processed by the Image Processing Pipeline \citep{Magnier2006}, which carries out a number of steps including bias correction, flat-fielding, bad pixel masking and artefact location. Difference imaging was performed by subtracting high quality reference images from the new observations. The subtracted images then formed the basis for point-spread function (PSF) fitting photometry.

In cases where the Pan-STARRS1 photometry did not provide sufficient coverage, ancillary imaging was obtained with either RATCam or the Optical Wide Field Camera, IO:O, mounted on the 2m Liverpool Telescope (LT; $g'r'i'$ filters) or the Andalucia Faint Object Spectrograph and Camera, ALFOSC mounted on the Nordic Optical Telescope (NOT; $u'g'r'i'z'$ filters). 
All data were reduced in the standard fashion using either the LT pipelines or {\sc iraf}\footnote{{\sc iraf} (Image Reduction and Analysis Facility) is distributed by the National Optical Astronomy Observatories, which are operated by the Association of Universities for Research in Astronomy, Inc., under cooperative agreement with the National Science Foundation.}. This includes trimming, bias subtraction, and flat-fielding.
For PS1-14vk we additionally performed a template subtraction due to its location within the host galaxy (it has a projected distance of $\sim$ 4.5\,kpc from the center of the host galaxy). Template images were obtained from the SDSS catalogue in the filters $g'$, $r'$, and $i'$. Point-spread function (PSF) fitting photometry was carried out on all LT and NOT images using the custom built {\sc SNOoPY}\footnote{SuperNOva PhotometrY, a package for SN photometry implemented in IRAF by E. Cappellaro; http://sngroup.oapd.inaf.it/snoopy.html.} package within {\sc iraf}. Photometric zero points and colour terms were derived using observations of Landolt standard star fields \citep{Landolt1992a} in photometric nights and their averaged values were then used to calibrate the magnitudes of a set of local sequence stars that were in turn used to calibrate the photometry of the SNe in the remainder of nights. 

We estimated the uncertainties of the PSF-fitting via artificial star experiments. An artificial star of the same magnitude as the SN was placed close to the position of the SN. The magnitude was measured, and the process was repeated for several positions around the SN. The standard deviation of the magnitudes of the artificial star were combined in quadrature with the uncertainty of the PSF-fit and the uncertainty of the photometric zeropoint to give the final uncertainty of the magnitude of the SN. 

Within the restrictions imposed by instrument availability and weather conditions, we obtained a series of three to six optical spectra per SN with the Optical System for Imaging and low-Intermediate-Resolution Integrated Spectroscopy (OSIRIS, grating IDs R300R or R300B) mounted on the Gran Telescopio CANARIAS (GTC).
The spectra were reduced using {\sc iraf} following standard procedures. These included trimming, bias subtraction, flat-fielding, optimal extraction, wavelength calibration via arc lamps, flux calibration via spectrophotometric standard stars, and re-calibration of the spectral fluxes to match the photometry. The spectra were also corrected for telluric absorption using a model spectrum of the telluric bands, which was created using the standard star spectrum.

\subsection{The Type II SN sample}

Table~\ref{table:whole_sample} presents an overview of the SN properties. A more detailed description of each SN is given in Appendix~\ref{section:appendix:observations_of_individual_SNe} alongside Figures~\ref{figure:SN2013ca} to \ref{figure:PS1-13bni} and Tables \ref{table:SN2013ca_photometry}-\ref{table:PS1-13bni_journal_spectra} that present the photometric and spectroscopic observations. Photometry and spectroscopy for LSQ13cuw and SN 2013eq are adopted from \citet{Gall2015} and \citet{Gall2016}, respectively. 

The photometric coverage ranges from rather poor (with just one photometric epoch in $g_\mathrm{PS1}$ and $r_\mathrm{PS1}$ for PS1-13atm) to very good (e.g. for PS1-13baf, which has an average 5-8 day cadence coverage in $g_{\mathrm{PS1}}r_{\mathrm{PS1}}i_{\mathrm{PS1}}z_{\mathrm{PS1}}$ up to $\sim$\,100\,d after discovery).
Most SNe show clear plateaus in their light curves identifying them as SNe II-P. The $r_{\mathrm{PS1}}i_{\mathrm{PS1}}z_{\mathrm{PS1}}$ decline rates of PS1-13bni have relatively large uncertainties, however they point towards a II-P classification. 

A few SNe, namely LSQ13cuw, PS1-14vk and PS1-13baf, display decline rates after maximum that are higher than  0.5\,mag/50\,d, which -- following the definition by \citet{Li2011a} -- places them in the SN II-L class. 
PS1-13atm shows only a very weak H$\alpha$ absorption component, which is a SN~II-L characteristic \citep[see][]{Gutierrez2014}. Furthermore the $i_{\mathrm{PS1}}$-band photometry is suggestive of a relatively long rise time ($\sim$\,$>$\,15\,d), which could also be an indication that PS1-13atm is a Type II-L SN. 

We are able to constrain the explosion epoch for LSQ13cuw, PS1-12bku, PS1-13abg, PS1-13abg, PS1-13bmf and PS1-13atm to a precision ranging from $\pm$\,0.1\,d to $\pm$\,5.0\,d using only photometric data, i.e. via pre-discovery non-detections, or a low-order polynomial fit to the rise time photometry.

All our SNe display the typical lines of H$\alpha$ and H$\beta$ in their spectra. Features stemming from weak lines of iron, in particular Fe\,{\sc ii}\,$\lambda$5169 are visible in at least one spectrum for most, albeit not all, of SNe.

Redshifts are derived either from the host galaxy or directly from the SN spectra. 
The redshifts are summarized in Table~\ref{table:whole_sample}, while details of the redshift determination for the individual objects are presented in Appendix~\ref{section:appendix:observations_of_individual_SNe}.

The galactic extinction was adopted from the NASA/IPAC Extragalactic Database and is based on the values published by \cite{Schlafly2011}.
A measure of the dust extinction within the SN host galaxy was obtainable only for selected SNe. For SN 2013eq a value of $E(B-V)^{\mathrm{host}}_{\mathrm{SN\ 2013eq}}$ = 0.062\,$\pm$\,0.028 was adopted from \citet{Gall2016}. Weak Na\,{\sc i}\,D absorptions, are visible in the PS1-13wr spectra, which we use to determine the host galaxy extinction: $E(B-V)^{\mathrm{host}}_{\mathrm{PS1-13wr}}$ = 0.110\,$\pm$\,0.049 \citep[applying Equation 9 from][]{Poznanski2012}. In most other cases we derived an upper limit to the equivalent width of the Na\,{\sc i}\,D absorption and the extinction within the host galaxy. This process is described in detail in Appendix \ref{section:Observations:SN_2013ca}. For our three highest-$z$ SNe PS1-13bmf, PS1-13atm and PS1-13bni, we are not able to obtain a meaningful upper limit for the EW of the Na\,{\sc i}\,D blend, due to the poor signal-to-noise ratio of their spectra.

\section{Results and discussion}
\label{section:Results_and_discussion}

The EPM has been applied to a variety of SNe II. Detailed discussions of the EPM as well as examples of applying this technique in practice can be found e.g. in \citet{Kirshner1974}, \citet{Schmidt1994a}, \citet{Hamuy2001}, \citet{Leonard2002a}, \citet{Dessart2005}, or \citet{Jones2009}. We will closely follow the approach presented in \citet{Gall2016}. We will use the dilution factors for the filter combination $\{BVI\}$ as presented by \citet{Hamuy2001} and \citet{Dessart2005}, respectively.

The SCM has been developed in \citet{Hamuy2002a}, \citet{Nugent2006}, \citet{Poznanski2009}, and \citet{DAndrea2010}. We follow the approach of \citet{Nugent2006}, who modified the technique to be applicable for SNe at cosmologically significant redshifts. 

In the following, we give a short overview of the use of SNe II-L as distance indicators up to this point (see Section \ref{Section:Type_IILSNe_as_distance_indicators}) and then present the preparatory steps required to apply either the EPM or the SCM. These are the application of $K$-corrections (Section \ref{section:Kcorrections}) and the determination of the temperatures (Section \ref{Section:Temperature_evolution}) and expansion velocities (Section \ref{Section:Velocities}). In particular, we will explore the possibility of using the H$\alpha$- or $H\beta$- velocities to estimate the photospheric velocity of Type II-P SNe (see Section \ref{section:velocities:HalphaHbeta_Fe5169_relation}). We then apply the EPM (Section \ref{section:EPM_distances}) and SCM (Section \ref{section:SCM_distances}) to our sample and compare the results in Section \ref{section:Comparison_EPM_SCM}. Finally we create an EPM and a SCM Hubble diagram (Section \ref{Section:The_Hubble_diagram}) and investigate the implications of applying the EPM and SCM also to Type II-L SNe (Section \ref{Section:Applying_EPM_SCM_to_TypeIILSNe}).

\subsection{Type II-L SNe as distance indicators}
\label{Section:Type_IILSNe_as_distance_indicators}

The Type II-L SN 1979C was used by \citet{Eastman1996} to derive an EPM distance of 15\,$\pm$\,4\,Mpc, which is consistent with the Cepheid distance of its host galaxy, NGC 4321 \citep[$\sim$\,17\,Mpc, e.g.][]{Freedman1994}. Another case is the Type II-L SN 1990K which was included in the SCM sample of \citet{Hamuy2002a}. Some SNe in the intermediate redshift SCM sample of \citet{Nugent2006} appear to have relatively steeply declining light curves (see e.g. SNLS-03D4cw in their Figure 7 or SNLS-04D1ln in their Figure 8).

\citet{Poznanski2009} select only objects with the lowest decline rates -- i.e. Type II-P SNe -- for their SCM sample, claiming that this reduces the scatter in the Hubble diagram. However, they also admit that only some of the steeper declining SNe II defy the velocity-luminosity correlation, while others do appear to be as close to the Hubble line as the rest of their sample. This was noted also by \citet{DAndrea2010}, who find that ``none of the five most deviant SNe in [their] sample would be removed using the decline rate method.''

From a physical point of view we expect no fundamental difference between the progenitors of SNe II-P and II-L, i.e. we assume a one-parameter continuum depending mainly on the mass of the hydrogen envelope of the progenitor star at the time of explosion. This is corroborated by the work of \citet{Pejcha2015}, who find that the main parameter determining the light curve shape is the photospheric temperature.
Of course, other factors (such as metallicity) may influence the details of the explosion and its observational characteristics, thereby affecting the precision of distance measurements.  

As was done by \citet{Rodriguez2014} and \citet{deJaeger2017a}, we include supernovae with relatively
fast decline rates in what follows, but we will discuss their application to distance measurements separately where appropriate.

\subsection{K-corrections}
\label{section:Kcorrections}

In order to correctly apply the EPM in combination with the $\{BVI\}$ dilution factors the photometry needs to be converted to the Johnson-Cousins filter system. EPM and SCM require the photometry to be transformed into the rest frame of each SN \citep[see Equation 13 in][]{Gall2016}. 

We perform the transformations from the observed photometry to the rest frame Johnson-Cousins $BVI$ filters in one step by calculating $K$-corrections with the {\sc snake} code (SuperNova Algorithm for $K$-correction Evaluation) within the S3 package \citep{Inserra2016}. 

SNe with $z < 0.13$: 
The spectral coverage is sufficient to estimate valid $K$-corrections. 

SNe PS1-13baf, PS1-13bmf and PS1-13atm: due to their redshifts above $z \sim 0.14$ the spectra of these SNe only partly cover the rest frame $I$-band. For this reason we combined the spectra of these three SNe with spectra from SNe 2013eq and 2013ca at similar epochs to obtain more accurate estimates for the respective $K$-corrections. 

The $K$-corrections for PS1-13bni ($z$ = $0.335^{+0.009}_{-0.012}$) to the rest frame $B$- and $V$-band are determined using the available spectroscopy. The rest frame $I$-band $K$-corrections for PS1-13bni are calculated by interpolating the $K$-corrections from SNe~2013eq and 2013ca spectra to the epochs of the PS1-13bni spectra. This should provide valid results considering that most SNe II-P/L are relatively homogeneous in their spectral evolution. A small additional error in the $I$-band $K$-corrections for PS1-13bni cannot be excluded due to the unknown explosion times and the exact relative epochs of either SNe~2013eq, 2013ca and PS1-13bni.

\subsection{Temperature evolution}
\label{Section:Temperature_evolution}

In preparation for a temperature determination, the uncorrected observed photometry is interpolated to the epochs of spectroscopic observations, dereddened and $K$-corrected. For PS1-12bku, where photometry from multiple sources is available we first interpolate the $K$-corrections for each instrument to the epochs of photometric observations and then correct the photometry for dust extinction. The dereddened and $K$-corrected $B$, $V$, and $I$-band light curves are then interpolated to the epochs of spectroscopic observations. 

Finally, the rest frame $BVI$ magnitudes are converted into physical fluxes. The temperature at each epoch is estimated via a blackbody fit to the $BVI$-fluxes.
To estimate the temperature uncertainties we performed additional blackbody fits spanning the extrema of the measured SN fluxes. 
The standard deviation of the resulting array of temperatures was taken as a conservative estimate of the temperature's uncertainty. 
 
The results are presented in Table \ref{table:EPM_photometry} in the appendix.

\subsection{Velocities}
\label{Section:Velocities}

\subsubsection{Fe\,{\sc ii}\,$\lambda$5169}

The EPM and the SCM require an estimate of the photospheric velocities. It has been argued that the Fe\,{\sc ii}\,$\lambda$5169 absorption minimum provides a reasonable estimate of the photospheric velocity \citep{Dessart2005}. While the use of other weak lines, such as Fe\,{\sc ii} $\lambda$5018, or Fe\,{\sc ii} $\lambda\lambda$4629,4670,5276,5318 has also been discussed \citep[e.g.][]{Leonard2002a,Leonard2002b}, here we will advance using the Fe\,{\sc ii}\,$\lambda$5169 velocities when available. The relatively low quality of the spectra for the more distant SNe renders the measurement of weak lines impossible. For this same reason we also explore the possibility to estimate the Fe\,{\sc ii}\,$\lambda$5169 velocity via the H$\alpha$ or H$\beta$ velocities.

\subsubsection{Using H$\alpha$ and H$\beta$ to estimate photospheric velocities}
\label{section:velocities:HalphaHbeta_Fe5169_relation}

\citet{Nugent2006} explored the ratio between the H$\beta$ and the Fe\,{\sc ii}\,$\lambda$5169 velocities and found a correlation. \citet{Poznanski2010} later improved this correlation and found that the Fe\,{\sc ii}\,$\lambda$5169 velocities stand in a linear relation with the H$\beta$ velocities\footnote{\citet{Poznanski2010} give the relation in the form: v$_{\mathrm{Fe\,5169}}$ = (0.84\,$\pm$\,0.05) v$_{\mathrm{H}\beta}$. We converted this into a v$_{\mathrm{H}\beta}$/v$_{\mathrm{Fe\,5169}}$ ratio for easier comparability with our results.}: v$_{\mathrm{H}\beta}$/v$_{\mathrm{Fe\,5169}}$ = 1.19$^{+0.08}_{-0.07}$. They included spectra from epochs between 5 and 40 days post-explosion. Our analysis is comparable to similar studies performed by \citep{Takats2014, Rodriguez2014, deJaeger2017a}.

In order to test this relation we collected five well observed Type II-P SNe from the literature: 
SN 1999em \citep{Leonard2002a,Leonard2002b}, 
SN 1999gi \citep{Leonard2002c}, 
SN 2004et \citep{Sahu2006}, 
SN 2005cs \citep{Pastorello2006,Pastorello2009a}, and 
SN 2006bp \citep{Quimby2007}. These objects were carefully selected on the basis of the quality and cadence of the available spectroscopy as well as for the good constraints on the explosion epochs. Note that none of the objects is of Type II-L, which might potentially exhibit different velocity evolutions than SNe II-P. A more detailed discussion follows in Section~\ref{section:IIL_velocities}.

We determined H$\beta$ and Fe\,{\sc ii}\,$\lambda$5169 velocities up to $\sim$\,70\,d post explosion using {\sc iraf} by fitting a Gaussian function to the minima of the respective lines and assumed an uncertainty of 5\,\% for all velocity measurements. \citet{Takats2012} provided a detailed comparison of different measurement techniques for expansion velocities. The absorption method employed here was found to be equivalent to cross-correlation methods. In particular, they found that the Fe\,{\sc ii}\,$\lambda$5169 tends to underestimate the photospheric velocity at phases less than 40 days, while H$\beta$ consistently overestimates the photospheric velocity. They further showed significant divergence of the H$\beta$ velocities from the models at phases larger than 40 days. We determined exponential fits to the velocity evolution using:
\begin{equation}
\label{equation:velocity_evolution_fit}
v(t^\star) = a\,e^{-b\,t^\star} + c .
\end{equation}
Here $v(t^\star)$ is the velocity of a particular line at time $t^\star$ since explosion (rest frame)\footnote{Throughout this paper we adopt the notation of \citet{Gall2016} in that variables in the SN rest frame will be marked with ``$\star$'', while ``$\Diamond$'' denotes variables in the observer frame.} and $a$, $b$, and $c$ are fit parameters. The uncertainty of each fit was estimated by calculating the root mean square of the deviation between the data and the fit. 
The explosion times for the five SNe were assumed to be as given in \citet[][and references therein]{Gall2015}. The fit parameters are given in Table~\ref{table:velocity_fit_parameters}.

\begin{figure}
  \centering
    \includegraphics[trim = 1mm 1.5mm 1.5mm 1mm, clip, width=\columnwidth]{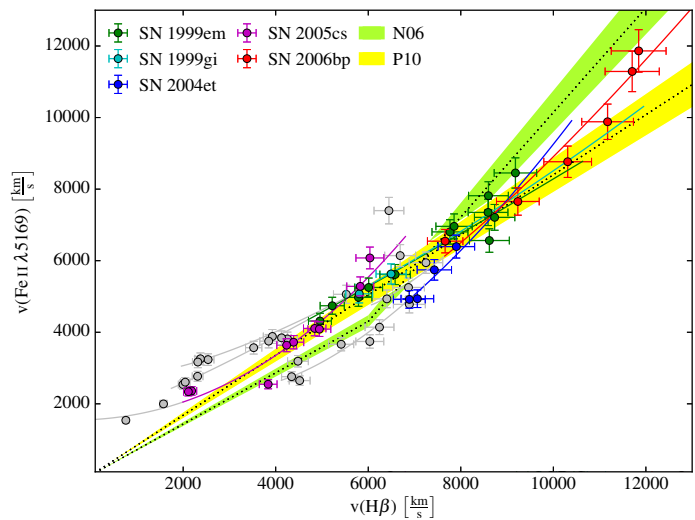}
    \caption{Comparison of H$\beta$ and Fe\,{\sc ii}\,$\lambda$5169 velocities. The coloured points correspond to velocities measured between 5 and 40 days from explosion, while the gray points represent epochs either within 5 days of the explosion or after 40 days. The solid lines correspond to the fitted evolution of the H$\beta$ and Fe\,{\sc ii}\,$\lambda$5169 velocities for each SN. The green and yellow shaded regions reflect the v$_{\mathrm{H}\beta}$/v$_{\mathrm{Fe\,5169}}$ ratios as presented by \citet{Nugent2006} and \citet{Poznanski2010}.}
    \label{figure:Hbeta_Fe5169_direct}
\end{figure}

In Figure~\ref{figure:Hbeta_Fe5169_direct} we show the resulting relation between the H$\beta$ and Fe\,{\sc ii}\,$\lambda$5169 velocities in comparison with the corresponding relation as given by \citet{Nugent2006} and \citet{Poznanski2010} (areas shaded in green and yellow). The coloured points mark velocities measured between 5 and 40 days after explosion, while the gray points depict velocities measured either before day 5 or after day 40. While the relations from \citet{Nugent2006} and \citet{Poznanski2010} follow the general trend of the data, it is also clear that in particular at later epochs the velocities of the individual SNe evolve ``away'' from the linear relations. The fitted velocity evolution for each SN -- depicted as solid lines in Figure \ref{figure:Hbeta_Fe5169_direct} -- is not linear (as the \citet{Poznanski2010} relation) but rather a curve, only part of which lies in the linear regime. 
 
We take a slightly different approach than \citet{Nugent2006} or \citet{Poznanski2010} in that we investigate the ratio between the H$\beta$ and the Fe\,{\sc ii}\,$\lambda$5169 velocities not as a function of velocity but rather as a function of time (i.e. epoch since explosion). At the same time, we also explore the viability of a relation between the H$\alpha$ and the Fe\,{\sc ii}\,$\lambda$5169 velocities, which might be a potential asset when dealing with spectra from high-$z$ SNe II (such as those that will be routinely discovered by the LSST\footnote{Large Synoptic Survey Telescope: www.lsst.org.})
in which neither Fe\,{\sc ii}\,$\lambda$5169 nor H$\beta$ can be detected. 

\begin{figure*}[t!]
   \centering
   \begin{subfigure}[t]{0.49\textwidth}
       \includegraphics[trim = 1mm 2mm 1.8mm 2mm, clip, width=\columnwidth]{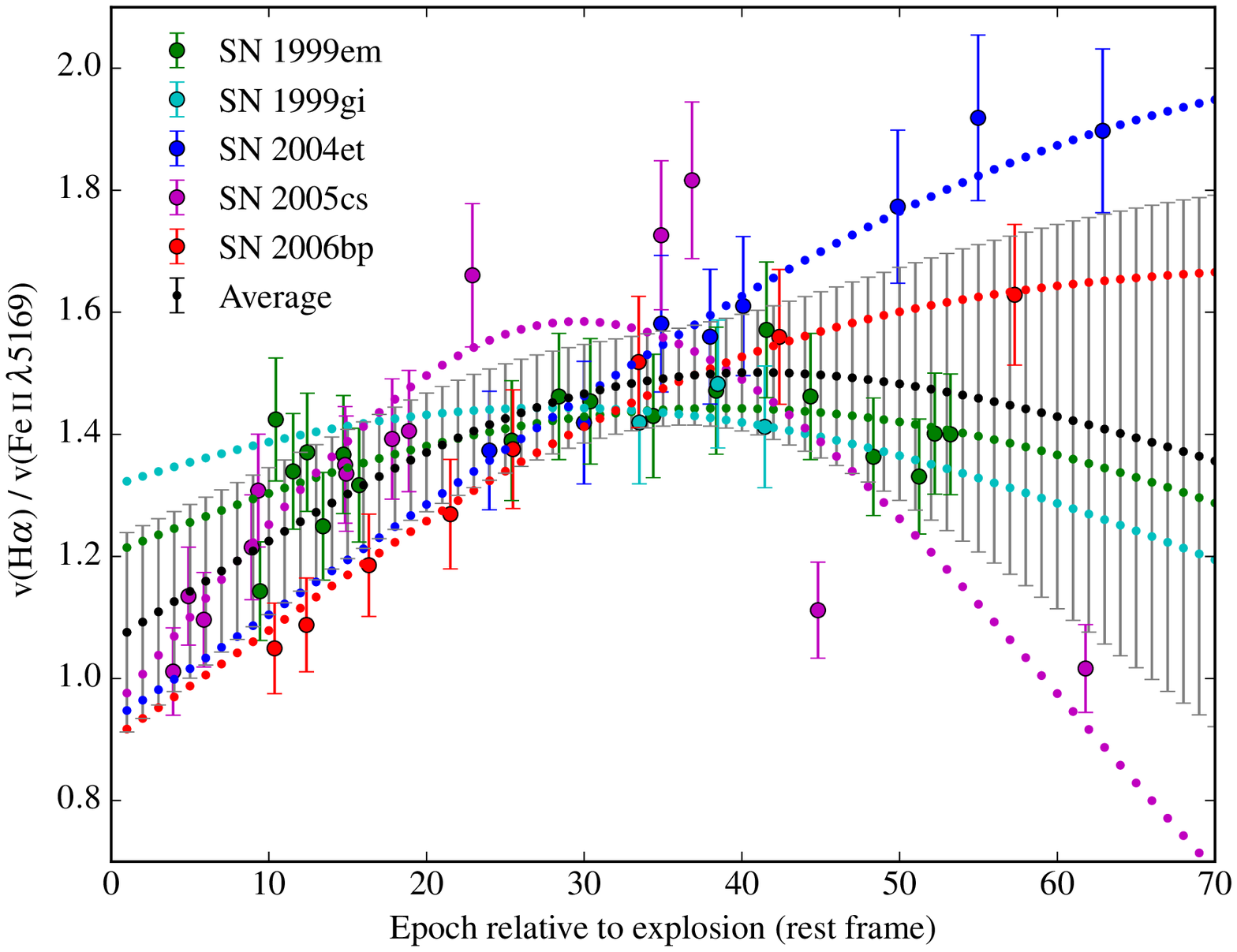} 
   \end{subfigure}%
   \begin{subfigure}[t]{0.49\textwidth}
       \includegraphics[trim = 0.8mm 1.3mm 1.7mm 1mm, clip, width=\columnwidth]{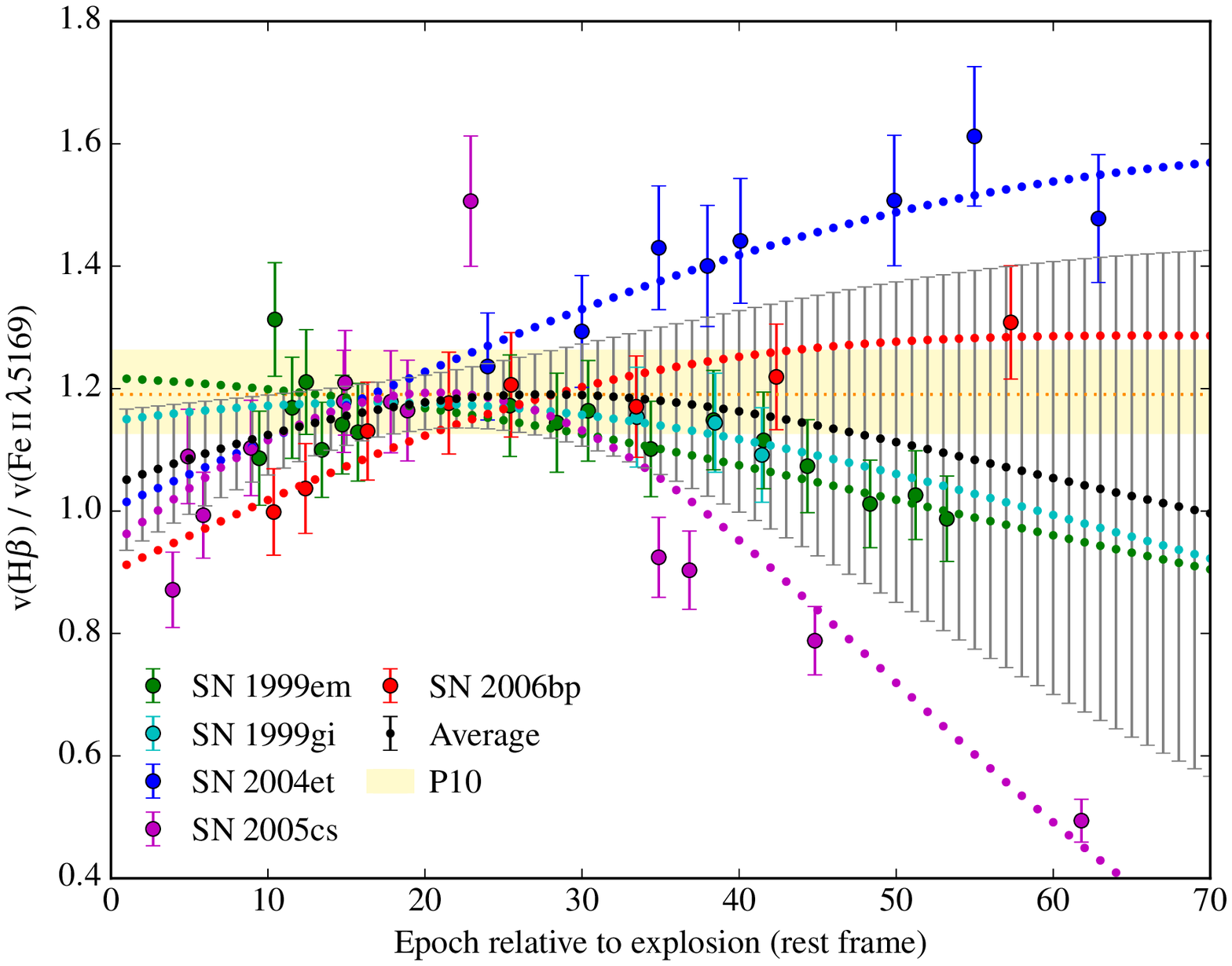} 
   \end{subfigure}%
   \caption{v$_{\mathrm{H}\alpha}$/v$_{\mathrm{Fe\,5169}}$ ratio (left panel) and v$_{\mathrm{H}\beta}$/v$_{\mathrm{Fe\,5169}}$ ratio (right panel) compared to the rest frame epoch (from explosion). The thick dotted lines represent the v$_{\mathrm{H}\alpha}$/v$_{\mathrm{Fe\,5169}}$ and v$_{\mathrm{H}\beta}$/v$_{\mathrm{Fe\,5169}}$ ratios for the individual SNe as well as the averaged ratios. For reasons of better visibility the uncertainty is depicted only for the averaged ratios and not for the fitted velocity ratios of the individual SN. The circle markers represent the measured v$_{\mathrm{H}\alpha}$/v$_{\mathrm{Fe\,5169}}$ and v$_{\mathrm{H}\beta}$/v$_{\mathrm{Fe\,5169}}$ ratios for those epochs and SNe where spectroscopic data was available and both H$\alpha$ (or H$\beta$) and the Fe\,{\sc ii}\,$\lambda$5169 velocity could be measured. The yellow shaded band in the right panel corresponds to the ratio of v$_{\mathrm{H}\beta}$/v$_{\mathrm{Fe\,5169}}$ = 1.19$^{+0.08}_{-0.07}$ as given by \citet{Poznanski2010}.}
   \label{figure:HalphaHbeta_Fe5169_ratios}
\end{figure*}

We interpolated the H$\alpha$, H$\beta$ and Fe\,{\sc ii}\,$\lambda$5169 velocities to all epochs up to $\sim$\,70 days after explosion and then calculated the 
v$_{\mathrm{H}\alpha}$/v$_{\mathrm{Fe\,5169}}$ and v$_{\mathrm{H}\beta}$/v$_{\mathrm{Fe\,5169}}$ ratios for each epoch and each SN, respectively (Figure \ref{figure:HalphaHbeta_Fe5169_ratios}). The uncertainties of the individual line fits were propagated to estimate the uncertainty of the ratios. 
We then averaged the v$_{\mathrm{H}\alpha}$/v$_{\mathrm{Fe\,5169}}$ and v$_{\mathrm{H}\beta}$/v$_{\mathrm{Fe\,5169}}$ ratios of the five SNe for each epoch (represented as black points in Figure \ref{figure:HalphaHbeta_Fe5169_ratios}). The uncertainties were estimated by adding the uncertainties of the individual SN line ratios and the standard deviation of the five values at each epoch in quadrature. The latter is the dominant contributor to the uncertainty, indicating that the velocity evolution is distinct for each SN. 
The results are presented in Figure~\ref{figure:HalphaHbeta_Fe5169_ratios} and listed in
Table~\ref{table:Halpha_Fe5169_ratio} for the v$_{\mathrm{H}\alpha}$/v$_{\mathrm{Fe\,5169}}$ ratio and
Table~\ref{table:Hbeta_Fe5169_ratio} for the v$_{\mathrm{H}\beta}$/v$_{\mathrm{Fe\,5169}}$ ratio.

At early epochs the individual v$_{\mathrm{H}\alpha}$/v$_{\mathrm{Fe\,5169}}$ or v$_{\mathrm{H}\beta}$/v$_{\mathrm{Fe\,5169}}$ ratios of the five SNe scatter less than 10\%. After about 40 days for H$\alpha$ and about 30 days for H$\beta$ the velocity ratios to Fe\,{\sc ii}\,$\lambda$5169 of the various SNe begin to diverge significantly. A possible explanation is that during the cooling phase of the SN after the shock breakout, Type II-P SNe display relatively homogeneous properties: the hydrogen envelope is still fully ionized and the ejecta are expanding and cooling. Hydrogen recombination sets in only after a few weeks and differences between the individual SNe in progenitor size, composition, or hydrogen envelope mass become apparent. This is reflected in the variety of luminosities and velocities observed for SNe II-P/L -- in fact the relation between luminosity and velocity builds the basis for the standardized candle method \citep{Hamuy2002a}. These differences are reflected also in the  v$_{\mathrm{H}\alpha}$/v$_{\mathrm{Fe\,5169}}$ or v$_{\mathrm{H}\beta}$/v$_{\mathrm{Fe\,5169}}$ ratios. A related pattern seems to be that the most extreme SNe in terms of deviation from the average ratios, SNe 2004et and 2005cs, display a relatively high ($\sim$\,$-17.2$\,mag for SN 2004et) or rather low ($\sim$\,$-15.1$\,mag for SN 2005cs) $R$/$r$-band maximum luminosity compared to the other objects (ranging between $-16.5$ and $-16.9$\,mag). 

The v$_{\mathrm{H}\beta}$/v$_{\mathrm{Fe\,5169}}$ ratio is flatter than the v$_{\mathrm{H}\alpha}$/v$_{\mathrm{Fe\,5169}}$ ratio, suggesting that it depends less on epoch. The v$_{\mathrm{H}\beta}$/v$_{\mathrm{Fe\,5169}}$ ratio as constant within the errors between 5 and 30 days after explosion, which is in agreement with the results from \citet[][shown in yellow in the right panel of Figure \ref{figure:HalphaHbeta_Fe5169_ratios}.]{Poznanski2010}. This result is also consistent with the findings of \citet{Takats2012} for early phases, where a nearly constant ratio of v$_{\mathrm{H}\beta}$/v$_{\mathrm{Fe\,5169}}$ can be inferred. The v$_{\mathrm{H}\alpha}$/v$_{\mathrm{Fe\,5169}}$ ratio displays a stronger evolution with time.

\begin{table}
  \caption{v$_{\mathrm{H}\alpha}$/v$_{\mathrm{Fe\,5169}}$ ratio}
  \label{table:Halpha_Fe5169_ratio}
  \centering
  \begin{tabular}{c c |    c c        }
  \hline 
  Epoch$^*$ & Ratio$^{**}$ & Epoch$^*$ & Ratio$^{**}$ \\
 \hline 
  5.0 & 1.14\,$\pm$\,0.14 & 23.0 & 1.41\,$\pm$\,0.09 \\ 
  6.0 & 1.16\,$\pm$\,0.14 & 24.0 & 1.42\,$\pm$\,0.09 \\ 
  7.0 & 1.18\,$\pm$\,0.13 & 25.0 & 1.43\,$\pm$\,0.09 \\ 
  8.0 & 1.19\,$\pm$\,0.13 & 26.0 & 1.44\,$\pm$\,0.09 \\ 
  9.0 & 1.21\,$\pm$\,0.12 & 27.0 & 1.44\,$\pm$\,0.09 \\  
 10.0 & 1.23\,$\pm$\,0.12 & 28.0 & 1.45\,$\pm$\,0.08 \\  
 11.0 & 1.24\,$\pm$\,0.12 & 29.0 & 1.46\,$\pm$\,0.08 \\
 12.0 & 1.26\,$\pm$\,0.11 & 30.0 & 1.47\,$\pm$\,0.08 \\
 13.0 & 1.27\,$\pm$\,0.11 & 31.0 & 1.47\,$\pm$\,0.08 \\ 
 14.0 & 1.29\,$\pm$\,0.11 & 32.0 & 1.48\,$\pm$\,0.08 \\ 
 15.0 & 1.30\,$\pm$\,0.11 & 33.0 & 1.48\,$\pm$\,0.08 \\
 16.0 & 1.32\,$\pm$\,0.10 & 34.0 & 1.49\,$\pm$\,0.08 \\
 17.0 & 1.33\,$\pm$\,0.10 & 35.0 & 1.49\,$\pm$\,0.08 \\
 18.0 & 1.34\,$\pm$\,0.10 & 36.0 & 1.49\,$\pm$\,0.08 \\
 19.0 & 1.36\,$\pm$\,0.10 & 37.0 & 1.50\,$\pm$\,0.08 \\
 20.0 & 1.37\,$\pm$\,0.10 & 38.0 & 1.50\,$\pm$\,0.09 \\
 21.0 & 1.38\,$\pm$\,0.10 & 39.0 & 1.50\,$\pm$\,0.09 \\
 22.0 & 1.39\,$\pm$\,0.09 & 40.0 & 1.50\,$\pm$\,0.10 \\ 
  \hline  
  \end{tabular}
  \\[1.5ex]
  \flushleft
  $^*$Epoch from explosion in SN rest frame. \\
  $^{**}$Averaged ratio between H$\alpha$ and Fe\,{\sc ii}\,$\lambda$5169 velocities.
\end{table}

\begin{table}
  \caption{v$_{\mathrm{H}\beta}$/v$_{\mathrm{Fe\,5169}}$ ratio}
  \label{table:Hbeta_Fe5169_ratio}
  \centering
  \begin{tabular}{  c c  |  c c  }
  \hline 
  Epoch$^*$ & Ratio$^{**}$ & Epoch$^*$ & Ratio$^{**}$ \\
\hline 
  5.0 & 1.09\,$\pm$\,0.09 & 18.0 & 1.17\,$\pm$\,0.04 \\
  6.0 & 1.09\,$\pm$\,0.09 & 19.0 & 1.17\,$\pm$\,0.04 \\
  7.0 & 1.10\,$\pm$\,0.08 & 20.0 & 1.18\,$\pm$\,0.04 \\
  8.0 & 1.11\,$\pm$\,0.07 & 21.0 & 1.18\,$\pm$\,0.05 \\
  9.0 & 1.12\,$\pm$\,0.07 & 22.0 & 1.18\,$\pm$\,0.05 \\
 10.0 & 1.12\,$\pm$\,0.07 & 23.0 & 1.19\,$\pm$\,0.05 \\
 11.0 & 1.13\,$\pm$\,0.06 & 24.0 & 1.19\,$\pm$\,0.05 \\
 12.0 & 1.14\,$\pm$\,0.06 & 25.0 & 1.19\,$\pm$\,0.06 \\
 13.0 & 1.14\,$\pm$\,0.05 & 26.0 & 1.19\,$\pm$\,0.06 \\
 14.0 & 1.15\,$\pm$\,0.05 & 27.0 & 1.19\,$\pm$\,0.07 \\
 15.0 & 1.16\,$\pm$\,0.05 & 28.0 & 1.19\,$\pm$\,0.07 \\
 16.0 & 1.16\,$\pm$\,0.05 & 29.0 & 1.19\,$\pm$\,0.08 \\
 17.0 & 1.17\,$\pm$\,0.04 & 30.0 & 1.19\,$\pm$\,0.08 \\
  \hline  
  \end{tabular}
  \\[1.5ex]
  \flushleft
  $^*$Epoch from explosion in SN rest frame. \\
  $^{**}$Averaged ratio between H$\beta$ and Fe\,{\sc ii}\,$\lambda$5169 velocities.
\end{table}

\subsubsection{Velocities in Type II-L SNe}
\label{section:IIL_velocities}

The v$_{\mathrm{H}\alpha}$-v$_{\mathrm{Fe\,5169}}$ and v$_{\mathrm{H}\beta}$-v$_{\mathrm{Fe\,5169}}$ relations were derived using only Type II-P SNe. While we expect SNe II-L to display similar properties in their velocity evolution as SNe II-P, we caution that applying the relations also for SNe II-L might introduce some systematic biases, e.g. due to the fact that SNe II-L are generally believed to produce more energetic ejecta than SNe II-P. 
It is also unclear whether the v$_{\mathrm{H}\alpha}$/v$_{\mathrm{Fe\,5169}}$ and v$_{\mathrm{H}\beta}$/v$_{\mathrm{Fe\,5169}}$ ratios of individual SNe II-L would diverge at the same epoch as for SNe II-P. 

Additionally, SNe II-L typically show smaller absorption components in their H$\alpha$ features compared to SNe II-P \citep{Gutierrez2014}. The extreme LSQ13cuw shows  this can make a velocity determination challenging (see Section \ref{section:EPM:LSQ13cuw}).

\subsection{EPM distances}
\label{section:EPM_distances}

To measure the EPM distances for the SNe in our sample we apply Equations 13 and 14 from \citet{Gall2016}:
\begin{equation}
\theta^\dag = 2 \sqrt{\frac{f^\mathrm{dered}_{\star,F} }{\zeta_{\star,BVI}^2(T_{\star,BVI}) \pi B(\lambda_{\star,\mathrm{eff}_F},T_{\star,BVI})}} ,
\end{equation}
and
\begin{equation}
\label{equation:chi_t}
\chi = \frac{\theta^\dag}{2v} = \frac{t^\star-t^\star_0}{D_L} ,
\end{equation}
where $\theta^\dag = \theta/(1+z)^2$ is the modified ``angular size'' of the SN,
$f^\mathrm{dered}_{\star,F}$ is the rest frame flux for the filter $F$,
$\zeta_{\star,BVI}(T_{\star,BVI})$ is the rest frame dilution factor for the filter combination $\{BVI\}$ at the rest frame $BVI$-temperature $T_{\star,BVI}$, 
$\lambda_{\star,\mathrm{eff}_F}$ is the effective wavelength of the corresponding filter $F$, in the rest frame, 
$B(\lambda_{\star,\mathrm{eff}_F},T_{\star,BVI})$ is the black body function, 
$v$ is the photospheric velocity,
$t^\star$ the time after explosion at $t^\star_0$ in the SN frame, and
$D_L$ is the luminosity distance of the SN. 

As in \citet{Gall2016} we use the temperature dependent \{$BVI$\} dilution factors presented by \citet[][based on the dilution factors calculated by \citealt{Eastman1996}]{Hamuy2001} and \citet{Dessart2005}. 
The Fe\,{\sc ii}\,$\lambda$5169 velocities were used for the distance determinations if measureable. For the SNe where the Fe\,{\sc ii}\,$\lambda$5169 feature could not be securely identified, we applied the v$_{\mathrm{H}\alpha}$/v$_{\mathrm{Fe\,5169}}$ and/or v$_{\mathrm{H}\beta}$/v$_{\mathrm{Fe\,5169}}$ ratio, to estimate the photospheric velocities and the SN distance. 

As pointed out by \citet{Schmidt1992} EPM suffers less from extinction than other methods. An error in the extinction correction will lead to a compensating effect between temperature and luminosity of the supernova. These lead to a reduction of the influence of extinction on the distance determination through EPM.

The distances were calculated by determining $\chi$ for each filter ($B$, $V$, and $I$) and epoch. 
To estimate the uncertainties of both the distance $D_L$ and the time of explosion $t_0^\star$, we performed additional fits to all combinations of adding or subtracting the uncertainties of $\chi$. 
The standard deviation of the resulting arrays of distances and explosion times was taken as a conservative estimate of their respective uncertainties.

\subsubsection{Commonalities}
\label{section:EPM_Commonalities}

\begin{figure*}[t!]
   \centering
   \begin{subfigure}[t]{0.47\textwidth}
      \includegraphics[width=\columnwidth]{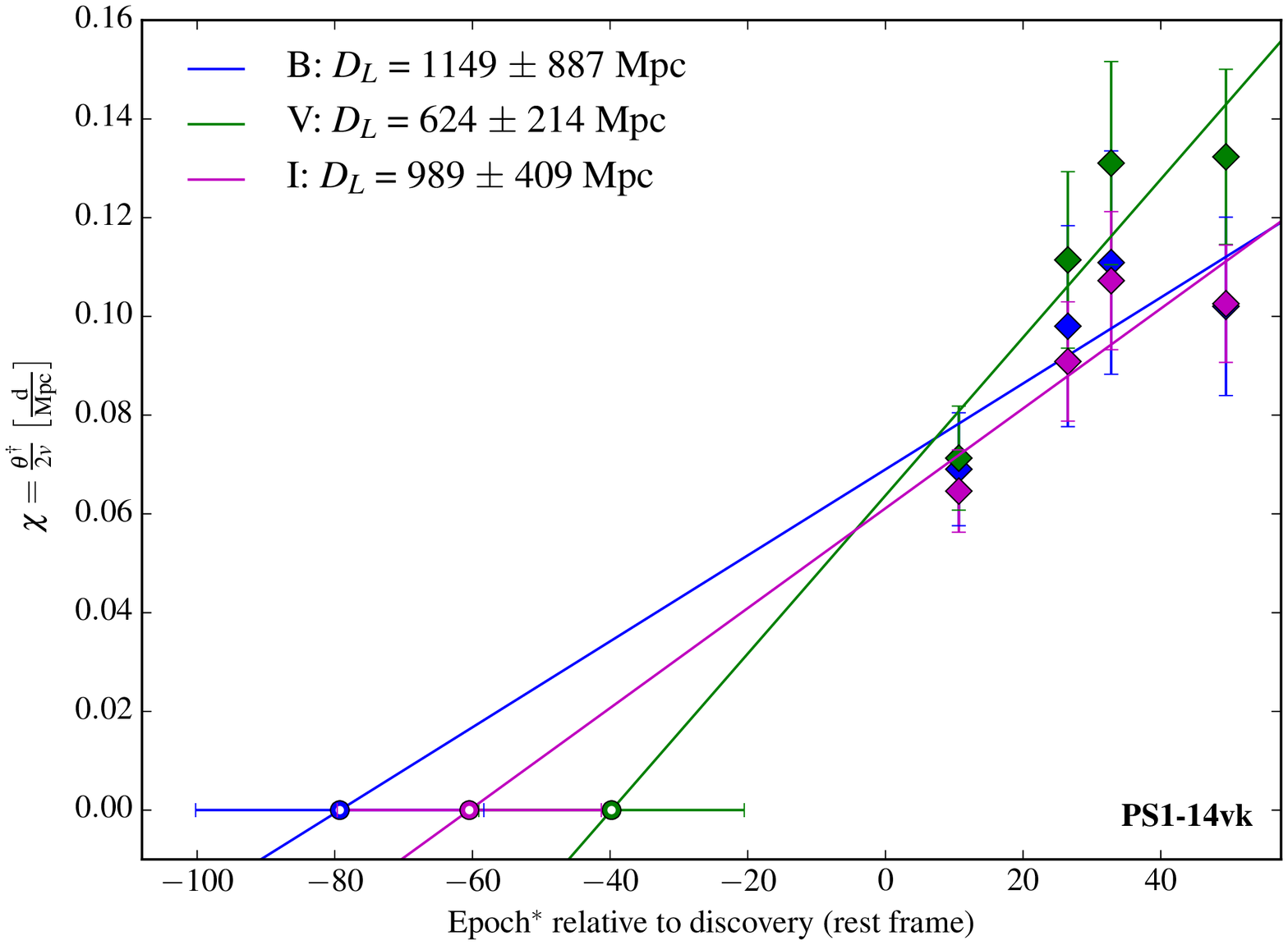}
   \end{subfigure}%
   \begin{subfigure}[t]{0.47\textwidth}
      \includegraphics[width=\columnwidth]{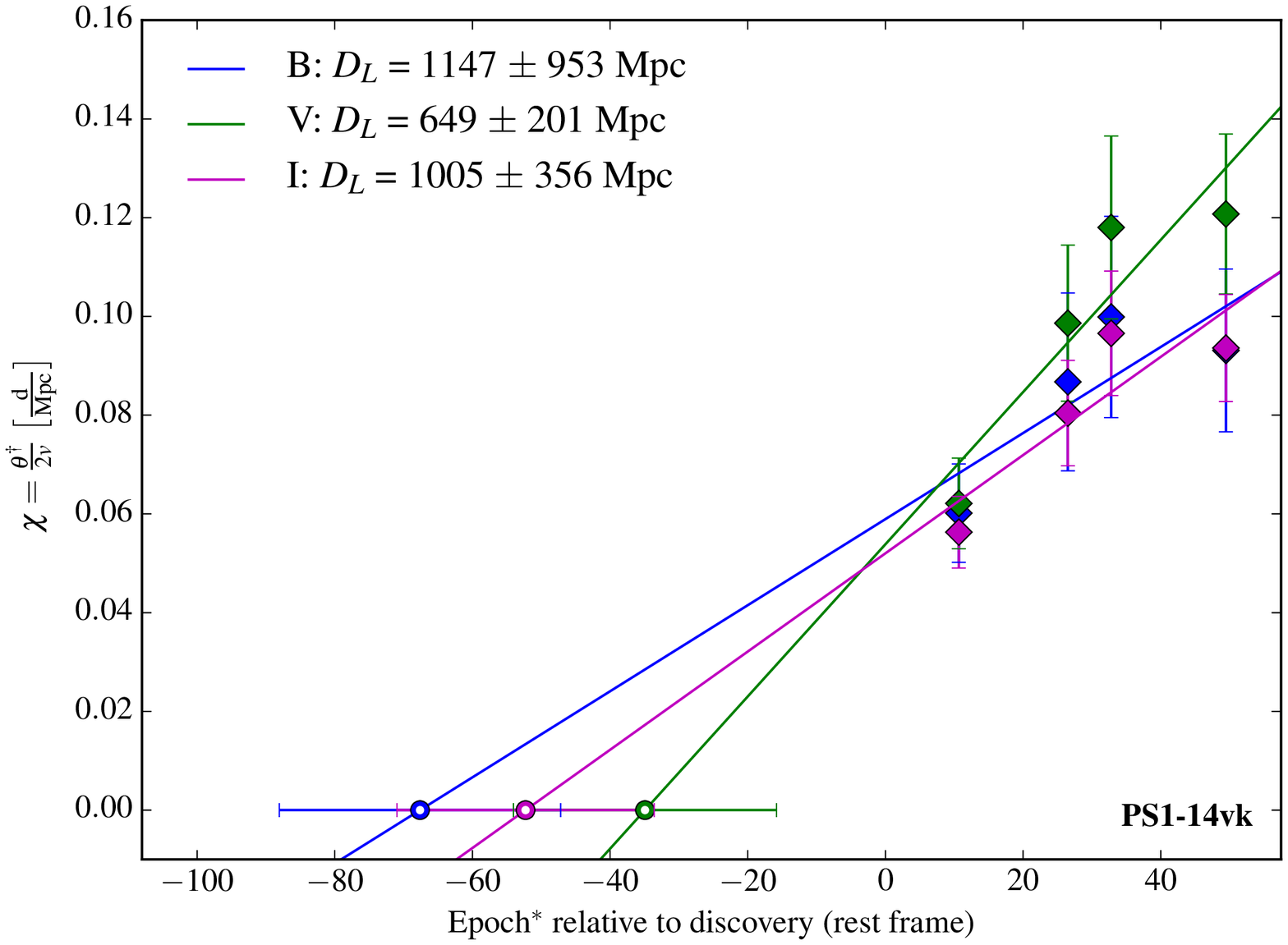}
   \end{subfigure}%
   \caption{Distance fits for PS1-14vk using all available epochs and $\zeta_{BVI}$ as given in \citet{Hamuy2001} (left panel) and \citet{Dessart2005} (right panel). The diamond markers denote values of $\chi$ through which the fit is made.}
   \label{figure:PS1-14vk_EPM_distances_all}
\end{figure*}
\begin{figure*}[t!]
   \centering
   \begin{subfigure}[t]{0.47\textwidth}
      \includegraphics[width=\columnwidth]{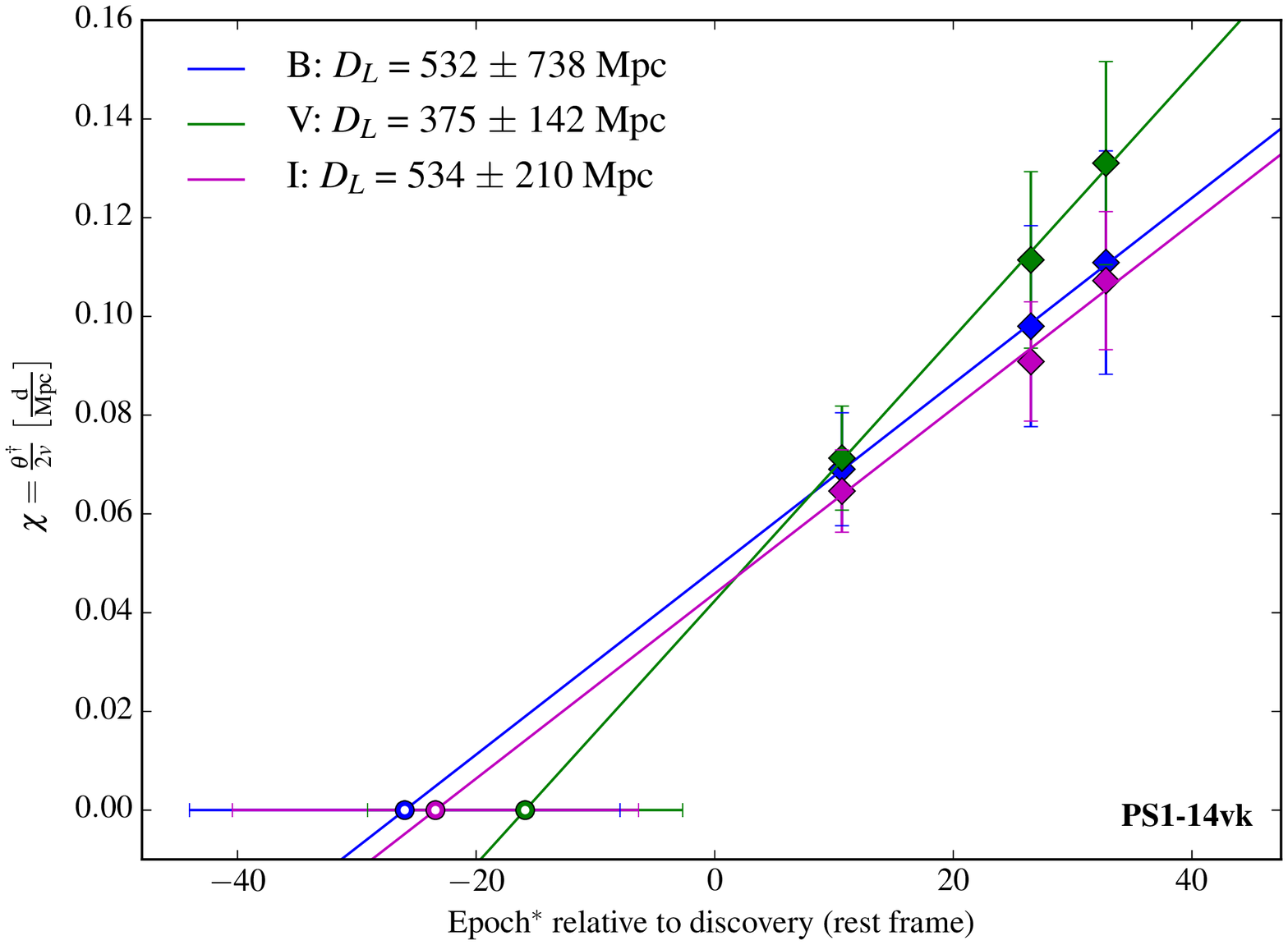}
   \end{subfigure}%
   \begin{subfigure}[t]{0.47\textwidth}
      \includegraphics[width=\columnwidth]{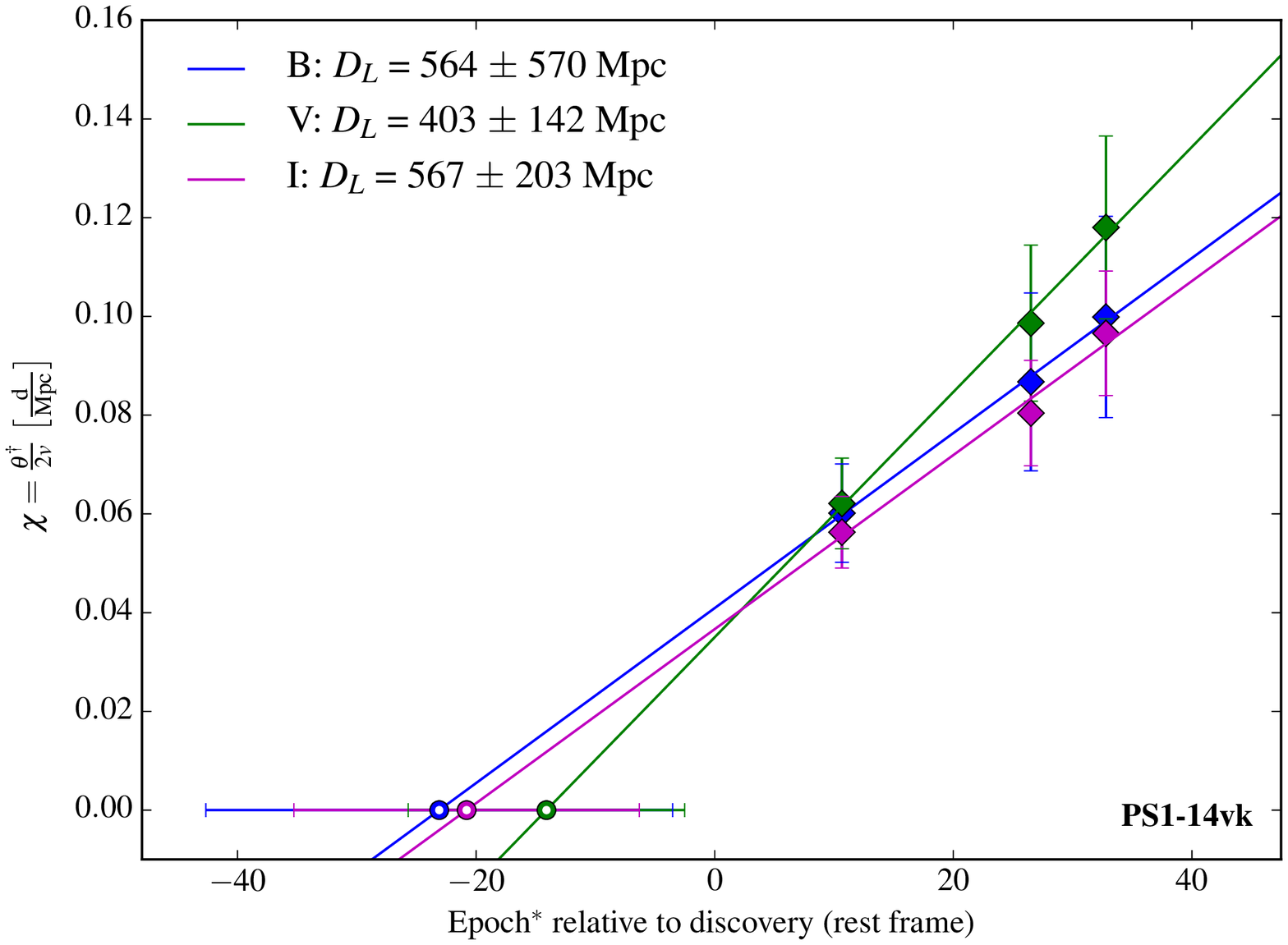}
   \end{subfigure}%
   \caption{Distance fits for PS1-14vk using only epochs that follow a linear relation and $\zeta_{BVI}$ as given in \citet{Hamuy2001} (left panel) and \citet{Dessart2005} (right panel). The diamond markers denote values of $\chi$ through which the fit is made.}
   \label{figure:PS1-14vk_EPM_distances_lin}
\end{figure*}

Figures~\ref{figure:PS1-14vk_EPM_distances_all} and \ref{figure:PS1-14vk_EPM_distances_lin} show the distance fits for PS1-14vk and are representative for the $\chi$-$t^\star$-fits made for all SNe (see Figures \ref{figure:SN2013ca_EPM_distances}-\ref{figure:PS1-13bni_EPM_distances} in Appendix C). We use PS1-14vk as an example to outline a number of commonalities.
\begin{itemize}

\item The results using either the dilution factors by \citet{Hamuy2001} or \citet{Dessart2005} are very similar for all SNe. The dilution factors by \citet{Dessart2005} systematically yield slightly larger distances.

\item The $V$-band distance is systematically smaller and the explosion epoch later than for the $BI$-bands.
The exception of the rule is PS1-13baf, where the $V$-band distances are larger than derived from $B$- and $I$-band. Discrepancies in the EPM distances when using different filter combinations have previously been observed \citep[e.g.][]{Hamuy2001,Jones2009}.

\item For some SNe the photometry provides an independent constraint on the time of explosion. In theses cases we use the ``observed'' explosion epoch as an additional data point in the fit and are able to significantly reduce the error in the distance determination. 

\item The v$_{\mathrm{H}\beta}$/v$_{\mathrm{Fe\,5169}}$ ratio was applied only up to $\sim$\,30 days from explosion.

\item \citet{Jones2009} argue that after around 40 days from explosion the linearity of the $\theta/v$ versus $t$ relation in Type II-P SNe deteriorates. Considering the scarcity of data points for our SNe, we use data up to $\sim$\,60 days from explosion for the distance fits, whenever viable. The $\chi$-$t^\star$ relation seems to be linear also in this extended regime see~\ref{section:appendix:Temperature_evolution}. For PS1-14vk, a Type II-L SN, we observe a breakdown in the linearity of the $\chi$-$t^\star$ relation after 30\,days after discovery cf. Fig.~\ref{figure:PS1-14vk_EPM_distances_lin}. Taking into account the epoch of explosion $\sim$\,20 days before discovery, this corresponds to a breakdown in the linearity of the $\chi$-$t^\star$ relation sometime between $\sim$\,50 and $\sim$\,70 days post explosion. When performing a $\chi$-$t^\star$ fit beyond the linear regime the distances are overestimated significantly and the estimated epoch of explosion is considerably earlier than when using only the linear regime. 

\item Our errors on the distances (averaged over the $BVI$ filters) span a wide range between $\sim$\,3\,\% and $\sim$\,54\,\%, essentially depending on the quality of the available data for each SN. A strong constraint on the epoch of explosion reduces the uncertainty of the distance fit significantly. The errors account for the uncertainties from the photometry, the SN redshift, the $K$-corrections, the photospheric velocities and -- for SNe~2013eq and PS1-13wr -- the dust extinction in the host galaxy. 

\end{itemize}

The velocities and dilution factors that were used to derive various values of $\theta^{\dag}$ are presented in Table \ref{table:EPM_quantities} in the appendix, while our final distance results are summarized in Table \ref{table:EPM_distances}. In the following we outline the particularities for each individual SN. 

\begin{table*}
  \caption{EPM distances and explosion times for the SNe in our sample}
  \label{table:EPM_distances}
  \centering
  \resizebox{0.93\textwidth}{!}{%
  \begin{tabular}{c   c          c                         r@{\,$\pm$\,}l       c                c            c                c        c          }
  \hline 
  \multirow{2}{*}{SN} & Dilution & \multirow{2}{*}{Filter} & \multicolumn{2}{c}{$D_L$} & Averaged $D_L$ & $t^\star_0$      & Averaged $t^\star_0$ & $t^\Diamond_0$ & Estimate   \\
                      & factor   &                         & \multicolumn{2}{c}{Mpc}   & Mpc             & days$^{*}$ & days$^{*}$     & MJD   & of $t_0$ via \\
\hline 
\multirow{6}{*}{SN 2013ca} 
  & \multirow{3}{*}{H01} & $B$ & 237&129 & \multirow{3}{*}{216\,$\pm$\,51} & 13.5\,$\pm$\,21.5 & \multirow{3}{*}{9.3$^{+9.7}_{-9.3}$} & \multirow{3}{*}{56382.3$^{+9.7}_{-10.1}$} & \multirow{6}{*}{EPM} \\
  &                      & $V$ & 170&43  &                                 &  \ 0.9\,$\pm$\,10.6 &                                      &                                           &                      \\
  &                      & $I$ & 242&72  &                                 & 13.6\,$\pm$\,16.4 &                                      &                                           &                      \\
  & \multirow{3}{*}{D05} & $B$ & 247&115 & \multirow{3}{*}{227\,$\pm$\,47} &  \ 9.3\,$\pm$\,21.3 & \multirow{3}{*}{5.7$^{+9.0}_{-5.7}$} & \multirow{3}{*}{56386.1$^{+5.9}_{-9.4}$}  &                      \\
  &                      & $V$ & 180&43  &                                 & -1.7\,$\pm$\,9.2  &                                      &                                           &                      \\
  &                      & $I$ & 253&68  &                                 &  \ 9.4\,$\pm$\,13.6 &                                      &                                           &                      \\
\hline 
\multirow{6}{*}{LSQ13cuw} 
  & \multirow{3}{*}{H01 -- H$\beta$ } & $B$ & 267&49 & \multirow{3}{*}{267\,$\pm$\,24} & \multirow{6}{*}{-} & \multirow{6}{*}{-} & \multirow{6}{*}{56593.4\,$\pm$\,0.7} & \multirow{6}{*}{G15} \\
  &                                   & $V$ & 270&42 &                                 &                     &                     &                                       &                       \\
  &                                   & $I$ & 263&34 &                                 &                     &                     &                                       &                       \\
  & \multirow{3}{*}{D05 -- H$\beta$ } & $B$ & 306&57 & \multirow{3}{*}{306\,$\pm$\,28} &                     &                     &                                       &                       \\
  &                                   & $V$ & 310&48 &                                 &                     &                     &                                       &                       \\
  &                                   & $I$ & 303&39 &                                 &                     &                     &                                       &                       \\
\hline 
  \multirow{6}{*}{PS1-13wr} 
  & \multirow{3}{*}{H01} & $B$ & 325&200 & \multirow{3}{*}{309\,$\pm$\,93} & 19.2\,$\pm$\,22.7 & \multirow{3}{*}{17.7\,$\pm$\,11.6} & \multirow{3}{*}{56330.5\,$\pm$\,12.5} & \multirow{6}{*}{EPM} \\
  &                      & $V$ & 282&161 &                                 & 15.9\,$\pm$\,22.1 &                                    &                                       &                      \\
  &                      & $I$ & 319&113 &                                 & 18.1\,$\pm$\,14.4 &                                    &                                       &                      \\
  & \multirow{3}{*}{D05} & $B$ & 347&200 & \multirow{3}{*}{330\,$\pm$\,94} & 17.5\,$\pm$\,24.1 & \multirow{3}{*}{16.2\,$\pm$\,11.3} & \multirow{3}{*}{56332.2\,$\pm$\,12.1} &                      \\
  &                      & $V$ & 302&161 &                                 & 14.5\,$\pm$\,19.7 &                                    &                                       &                      \\
  &                      & $I$ & 341&115 &                                 & 16.5\,$\pm$\,13.1 &                                    &                                       &                      \\
\hline 
\multirow{12}{*}{PS1-14vk} 
  & \multirow{3}{*}{H01 -- all} & $B$ & 1149&887 & \multirow{3}{*}{921\,$\pm$\,333} & 79.3\,$\pm$\,20.9 & \multirow{3}{*}{59.8\,$\pm$\,11.4} & \multirow{3}{*}{56675.9\,$\pm$\,12.4} & \multirow{12}{*}{EPM} \\
  &                             & $V$ &  624&214 &                                  & 39.8\,$\pm$\,19.3 &                                    &                                       &                       \\
  &                             & $I$ &  989&409 &                                  & 60.5\,$\pm$\,19.2 &                                    &                                       &                       \\
  & \multirow{3}{*}{H01 -- lin} & $B$ &  532&738 & \multirow{3}{*}{480\,$\pm$\,260} & 26.0\,$\pm$\,18.0 & \multirow{3}{*}{21.8\,$\pm$\,9.4}  & \multirow{3}{*}{56717.0\,$\pm$\,10.1} &                       \\
  &                             & $V$ &  375&142 &                                  & 15.9\,$\pm$\,13.2 &                                    &                                       &                       \\
  &                             & $I$ &  534&210 &                                  & 23.4\,$\pm$\,17.0 &                                    &                                       &                       \\
  & \multirow{3}{*}{D05 -- all} & $B$ & 1147&953 & \multirow{3}{*}{934\,$\pm$\,346} & 67.6\,$\pm$\,20.4 & \multirow{3}{*}{51.6\,$\pm$\,11.2} & \multirow{3}{*}{56684.8\,$\pm$\,12.1} &                       \\
  &                             & $V$ &  649&201 &                                  & 34.9\,$\pm$\,19.1 &                                    &                                       &                       \\
  &                             & $I$ & 1005&356 &                                  & 52.3\,$\pm$\,18.7 &                                    &                                       &                       \\
  & \multirow{3}{*}{D05 -- lin} & $B$ &  564&570 & \multirow{3}{*}{511\,$\pm$\,207} & 23.1\,$\pm$\,19.6 & \multirow{3}{*}{19.3\,$\pm$\,9.0}  & \multirow{3}{*}{56719.6\,$\pm$\,9.7}  &                       \\
  &                             & $V$ &  403&142 &                                  & 14.1\,$\pm$\,11.6 &                                    &                                       &                       \\
  &                             & $I$ &  567&203 &                                  & 20.8\,$\pm$\,14.5 &                                    &                                       &                       \\
\hline 
  \multirow{6}{*}{PS1-12bku} 
  & \multirow{3}{*}{H01} & $B$ & 357&43 & \multirow{3}{*}{357\,$\pm$\,21} & \multirow{6}{*}{-} & \multirow{6}{*}{-} & \multirow{6}{*}{56160.9\,$\pm$\,0.4} & \multirow{6}{*}{PS1} \\
  &                      & $V$ & 349&34 &                                 &                    &                    &                                       &                      \\
  &                      & $I$ & 366&29 &                                 &                    &                    &                                       &                      \\
  & \multirow{3}{*}{D05} & $B$ & 409&49 & \multirow{3}{*}{409\,$\pm$\,24} &                    &                    &                                       &                      \\
  &                      & $V$ & 399&39 &                                 &                    &                    &                                       &                      \\
  &                      & $I$ & 419&33 &                                 &                    &                    &                                       &                      \\
\hline 
\multirow{6}{*}{PS1-13abg} 
  & \multirow{3}{*}{H01} & $B$ & 474&54 & \multirow{3}{*}{459\,$\pm$\,29} & \multirow{6}{*}{-} & \multirow{6}{*}{-} & \multirow{6}{*}{56375.4\,$\pm$\,5.0} & \multirow{6}{*}{PS1} \\
  &                      & $V$ & 440&47 &                                 &                    &                    &                                      &                      \\
  &                      & $I$ & 463&47 &                                 &                    &                    &                                      &                      \\
  & \multirow{3}{*}{D05} & $B$ & 505&58 & \multirow{3}{*}{489\,$\pm$\,30} &                    &                    &                                      &                      \\
  &                      & $V$ & 469&50 &                                 &                    &                    &                                      &                      \\
  &                      & $I$ & 494&50 &                                 &                    &                    &                                      &                      \\
\hline 
\multirow{6}{*}{PS1-13baf} 
  & \multirow{3}{*}{H01 -- H$\beta$ } & $B$ & 689&740 & \multirow{3}{*}{690\,$\pm$\,294} & \multirow{6}{*}{-} & \multirow{6}{*}{-} & \multirow{6}{*}{56408.0\,$\pm$\,1.5} & \multirow{6}{*}{PS1} \\
  &                                   & $V$ & 717&426 &                                  &                     &                     &                                       &                       \\
  &                                   & $I$ & 666&221 &                                  &                     &                     &                                       &                       \\
  & \multirow{3}{*}{D05 -- H$\beta$ } & $B$ & 790&854 & \multirow{3}{*}{792\,$\pm$\,339} &                     &                     &                                       &                       \\
  &                                   & $V$ & 821&490 &                                  &                     &                     &                                       &                       \\
  &                                   & $I$ & 763&254 &                                  &                     &                     &                                       &                       \\
\hline 
\multirow{6}{*}{PS1-13bmf} 
  & \multirow{3}{*}{H01} & $B$ & 620&40 & \multirow{3}{*}{622\,$\pm$\,21} & \multirow{6}{*}{-} & \multirow{6}{*}{-} & \multirow{6}{*}{56420.0\,$\pm$\,0.1} & \multirow{6}{*}{PS1} \\
  &                      & $V$ & 606&34 &                                 &                     &                     &                                      &                      \\
  &                      & $I$ & 639&35 &                                 &                     &                     &                                      &                      \\
  & \multirow{3}{*}{D05} & $B$ & 705&45 & \multirow{3}{*}{707\,$\pm$\,24} &                     &                     &                                      &                      \\
  &                      & $V$ & 689&39 &                                 &                     &                     &                                      &                      \\
  &                      & $I$ & 727&41 &                                 &                     &                     &                                      &                      \\
\hline 
\multirow{6}{*}{PS1-13bni} 
  & \multirow{3}{*}{H01 -- H$\beta$} & $B$ & 1699&451  & \multirow{3}{*}{1772\,$\pm$\,538} &  7.3\,$\pm$\,12.4  & \multirow{3}{*}{8.1\,$\pm$\,5.9} & \multirow{3}{*}{56401.3\,$\pm$\,7.9} & \multirow{6}{*}{EPM} \\
  &                      & $V$ & 1538&1109 &                                   &  5.4\,$\pm$\,8.3  &                                  &                                      &                      \\
  &                      & $I$ & 2078&1082 &                                   &  11.7\,$\pm$\,9.7  &                                  &                                      &                      \\
  & \multirow{3}{*}{D05 -- H$\beta$} & $B$ & 2019&542  & \multirow{3}{*}{2110\,$\pm$\,658} &  8.6\,$\pm$\,13.6 & \multirow{3}{*}{9.5\,$\pm$\,6.4} & \multirow{3}{*}{56400.0\,$\pm$\,8.6} &                      \\
  &                      & $V$ & 1823&1349 &                                   &  6.6\,$\pm$\,8.8  &                                  &                                      &                      \\
  &                      & $I$ & 2488&1336 &                                   & 13.4\,$\pm$\,10.5 &                                  &                                      &                      \\
\hline  
  \end{tabular}}
  \\[1.5ex]
  \flushleft
  $^*$Days in SN rest frame before discovery. H01 = \citet{Hamuy2001}; D05 = \citet{Dessart2005}; G15 = \citet{Gall2015}. PS1 = PanSTARRS1 photometry and/or non-detections. 
\end{table*}

\subsubsection{SN 2013eq}

SN 2013eq is presented in detail in \citet{Gall2016}. Using the dilution factors from \citet{Hamuy2001}, \citet{Gall2016} found a luminosity distance of $D_L$ = 151\,$\pm$\,18\,Mpc and an explosion time of 4.1\,$\pm$\,4.4 days before discovery (rest frame), corresponding to a $t^\Diamond_0$ of MJD 56499.6\,$\pm$\,4.6 (observer frame). Applying the dilution factors from \citet{Dessart2005} $D_L$ = 164\,$\pm$\,20\,Mpc and an explosion time of 3.1\,$\pm$\,4.1 days before discovery (rest frame) corresponding to a $t^\Diamond_0$ of MJD 56500.7\,$\pm$\,4.3 (observer frame) were measured.

\subsubsection{SN 2013ca}

Good quality spectroscopy is available for SN 2013ca allowing us to determine the photospheric velocity directly from the Fe\,{\sc ii}\,$\lambda$5169 line. The photometric coverage is poor, with only 2 points observed in each of the LT $g'r'i'$ filters. The flux at the epochs of spectroscopic observations was linearly interpolated through those two data points. This should be a reasonable assumption for a Type II-P SN during the plateau phase.

\subsubsection{LSQ13cuw (SN II-L)}
\label{section:EPM:LSQ13cuw}

LSQ13cuw photometry {and spectroscopy was adopted from \citet{Gall2015}. They constrain the epoch of explosion to MJD 56593.4\,$\pm$\,0.7. This was used as a further data point in the distance fit and provides a more accurate result. Unfortunately, lines of Fe\,{\sc ii}\,$\lambda$5169 are not visible in any of the LSQ13cuw spectra and we applied the v$_{\mathrm{H}\beta}$/v$_{\mathrm{Fe\,5169}}$ ratio to estimate the photospheric velocities.

However, the spectra of LSQ13cuw are characterised by H$\alpha$ and H$\beta$ features that show almost no absorption component, which makes a velocity determination difficult. We were only able to confidently measure the H$\beta$ velocity after +32\,days. However, as outlined in Section \ref{section:velocities:HalphaHbeta_Fe5169_relation} the v$_{\mathrm{H}\beta}$ - v$_{\mathrm{Fe\,5169}}$ relation is not reliable after $\sim$\,30 days from explosion. This leaves us with only one estimate for the photospheric velocity from the +32\,d spectrum.

\subsubsection{PS1-13wr}

The reasonably high signal-to-noise spectra of PS1-13wr allow us to determine the photospheric velocity directly from the Fe\,{\sc ii}\,$\lambda$5169 line. The errors in the distance determination of PS1-13wr are larger than for other SNe due to the uncertainty of the dust extinction within the host galaxy (Table~\ref{table:whole_sample}).

\subsubsection{PS1-14vk (SN II-L)}
\label{section:EPM:PS1_14vk}

The photospheric velocity of PS1-14vk was measured from the Fe\,{\sc ii}\,$\lambda$5169 line for all epochs, except $\sim$\,2\,d and $\sim$\,14\,d post-discovery, where the line was either not visible or blended.

The values of $\chi$ in PS1-14vk for the epochs +11, +27, and +33\,d follow a clearly linear relation in contrast to the last epoch at +49\,d after discovery (see discussion in Section~\ref{section:EPM_Commonalities}). This is independent whether we apply the dilution factors given by \citet{Hamuy2001} or \citet{Dessart2005}. 
Consequently, we performed fits only through $\chi$ values at +11, +27, and +33\,days.

\subsubsection{PS1-12bku}

The Fe\,{\sc ii}\,$\lambda$5169 line in the PS1-12bku spectra was used to estimate the photospheric velocity. We can additionally constrain the fits by using the estimate for the explosion epoch available for PS1-12bku. This reduces the distance uncertainties significantly.

\subsubsection{PS1-13abg}

Only poor quality spectra are available for PS1-13abg and we can identify the Fe\,{\sc ii}\,$\lambda$5169 line only in the +44\,d spectrum. We use this measurement and the loose constraint for the explosion epoch to estimate the distance to PS1-13abg.

\subsubsection{PS1-13baf}

The spectra of PS1-13baf have a poor signal-to-noise ratio and we are not able to identify Fe\,{\sc ii}\,$\lambda$5169 line in any of the spectra. We apply the v$_{\mathrm{H}\beta}$/v$_{\mathrm{Fe\,5169}}$ relation to the first two epochs (i.e. before $\sim$\,30\,days), to estimate the photospheric velocities. The estimate for the explosion epoch reduces the error in the distance.

\subsubsection{PS1-13bmf (SN II-L)}

The spectra of PS1-13bmf at the first two epochs (+5\,d and +8\,d) are basically featureless. The Fe\,{\sc ii}\,$\lambda$5169 line could be identified at +29 and +47\,days. We additionally constrain the fits by using the estimate for the explosion epoch available for PS1-13bmf (Section \ref{appendix:section:PS1_13bmf_observations} in the appendix).

\subsubsection{PS1-13atm (SN II-L)}

The photometric coverage of PS1-13atm, our second-highest redshift SN, is insufficient to perform reliable fits for an estimate of the $BVI$ magnitudes at the epochs of spectroscopic observations. The $g_{\mathrm{PS1}}$- and $r_{\mathrm{PS1}}$-band have only a single data point each. We therefore forgo an attempt to calculate an EPM distance to PS1-13atm.

\subsubsection{PS1-13bni}
\label{section:EPM:PS1_13bni}

The Fe\,{\sc ii}\,$\lambda$5169 line could not be clearly identified at any epoch of PS1-13bni. We relied entirely on estimates of the photospheric velocities using the H$\beta$ line (at +15 and +24\,days) and the v$_{\mathrm{H}\beta}$/v$_{\mathrm{Fe\,5169}}$ relation. We note that the relation is prone to large uncertainties at epochs $\gtrsim$\,50 days post explosion.

Since the v$_{\mathrm{H}\beta}$/v$_{\mathrm{Fe\,5169}}$ relation is epoch dependent, but no estimate of an explosion epoch from other sources is available for PS1-13bni, we pursue an iterative approach to derive the time of explosion. For the initial iteration we assume the first detection to be the time of explosion and evaluate the v$_{\mathrm{H}\beta}$/v$_{\mathrm{Fe\,5169}}$ relation at the resulting epochs. Then we fit for   the distance and the explosion epoch as would be done for any other SN. The derived explosion epoch is then used as a base for the second iteration, and the spectral epochs as well as v$_{\mathrm{H}\beta}$/v$_{\mathrm{Fe\,5169}}$ ratios are adjusted accordingly. We repeat this process until the explosion time converges.

\subsection{SCM distances}
\label{section:SCM_distances}

The $V$- and $I$-band corrected photometry was also used for SCM. The light curves are interpolated to 50\,days after explosion. 

The expansion velocity is determined using the relation published by \citet[][Equation 2]{Nugent2006}: 
\begin{equation}
v_{50} = v(t^\star) \left( \frac{t^\star}{50} \right)^{0.464\,\pm\,0.017} ,
\end{equation}
where $v(t^\star)$ is the Fe\,{\sc ii}\,$\lambda$5169 velocity at time $t^\star$ after explosion (rest frame). For the SNe where we could not identify the Fe\,{\sc ii}\,$\lambda$5169 feature in their spectra we first applied the v$_{\mathrm{H}\beta}$/v$_{\mathrm{Fe\,5169}}$ relation and then utilized the derived Fe\,{\sc ii}\,$\lambda$5169 velocities to estimate the expansion velocity at day 50. This procedure was carried out twice for those SNe where estimates of the explosion epoch where available via the EPM (depending on dilution factors).

Finally, we use Equation 1 in \citet{Nugent2006} to derive the distance modulus:
\begin{equation}
M_{I_{50}} = -\alpha\, \mathrm{log}_{10}\left(\frac{ v_{50,\mathrm{Fe\,\textsc {ii} }} }{5000}\right) - 1.36 \left[ (V - I)_{50} - (V - I)_0 \right] + M_{I_0} ,
\end{equation}
where $M_I$ is the rest frame $I$-band magnitude, $(V - I)$ the colour, and $v$ is the expansion velocity each evaluated at 50 days after explosion. The parameters are set as follows: $\alpha = 5.81$, $M_{I_0} = -17.52$ (for an $H_0$ of 70\,km\,s$^{-1}$\,Mpc$^{-1}$) and $(V - I)_0 = 0.53$, following \citet{Nugent2006}.  

Our results are shown in Table~\ref{table:SCM_distance}. Additionally, we adopt the SCM distances derived by \citet{Gall2016} for SN 203eq: $D_L$ = 160\,$\pm$\,32\,Mpc and $D_L$ = 157\,$\pm$\,31\,Mpc, using the explosion epochs calculated via the EPM and utilizing the dilution factors either from \citet{Hamuy2001} or \citet{Dessart2005}. 

The final errors on individual distances span a range between 11 and 35\,\% depending mainly on whether the explosion epoch is well constrained or not. While the $I$-band magnitude and the $(V - I)$ colour will not change significantly during the plateau phase of Type II-P SNe and are therefore relatively robust; however, this is not true for the expansion velocity. Any uncertainty in the explosion epoch directly translates into an uncertainty in the 50\,d velocity and thereby affects the precision of the distance measurement. In our sample this is borne out in the fact that the SCM distances derived using estimates for the explosion epoch from photometry, have significantly smaller relative uncertainties, than those derived using estimates via the EPM.

\begin{table*}
  \small
  \caption{SCM quantities and distances}
  \label{table:SCM_distance}
  \centering
  \begin{tabular}{c   c              l       c              c               l               c         c        r@{\,$\pm$\,}l                   }
  \hline                   
  \multirow{2}{*}{SN} & Estimate     & \multicolumn{1}{c}{$t^\Diamond_0$} & $V^{*}_{50}$ & $I^{*}_{50}$ & \multicolumn{1}{c}{$v_{50}$}      & Estimate of & $\mu$ & \multicolumn{2}{c}{$D_L$}              \\
                      & of $t_0$ via & \multicolumn{1}{c}{mjd}   & mag          & mag           & \multicolumn{1}{c}{km\,s$^{-1}$} & velocity via   & mag    &    \multicolumn{2}{c}{Mpc}             \\
  \hline 
 \vspace{0.5mm}
\multirow{2}{*}{SN 2013ca} 
  & EPM -- H01 & 56382.3$^{+9.7}_{-10.1}$ & 19.08\,$\pm$\,0.10 & 18.56\,$\pm$\,0.08 & 5427\,$\pm$\,798 & \multirow{2}{*}{Fe\,{\sc ii}\,$\lambda$5169} & 36.28\,$\pm$\,0.43 & 180&36 \\
 \vspace{0.5mm}
  & EPM -- D05 & 56386.1$^{+5.9}_{-9.4}$  & 19.12\,$\pm$\,0.10 & 18.59\,$\pm$\,0.08 & 5228\,$\pm$\,758 &                                              & 36.22\,$\pm$\,0.42 & 176&34 \\
LSQ13cuw & G15 & 56593.4\,$\pm$\,0.7 & 20.61\,$\pm$\,0.10 & 20.00\,$\pm$\,0.09 & 5616\,$\pm$\,655 & H$\beta$  & 37.93\,$\pm$\,0.38 & 385&67 \\
\multirow{2}{*}{PS1-13wr} 
  & EPM -- H01 & 56330.5\,$\pm$\,12.5 & 21.38\,$\pm$\,0.06 & 20.49\,$\pm$\,0.07 & 4458\,$\pm$\,963 & \multirow{2}{*}{Fe\,{\sc ii}\,$\lambda$5169} & 38.21\,$\pm$\,0.57 & 438&115 \\
  & EPM -- D05 & 56332.2\,$\pm$\,12.1 & 21.39\,$\pm$\,0.06 & 20.49\,$\pm$\,0.07 & 4368\,$\pm$\,959 &                                              & 38.17\,$\pm$\,0.58 & 430&114 \\
\multirow{2}{*}{PS1-14vk} 
  & EPM -- H01 & 56717.0\,$\pm$\,10.1 & 20.95\,$\pm$\,0.29 & 20.63\,$\pm$\,0.24 & 5228\,$\pm$\,818 & \multirow{2}{*}{Fe\,{\sc ii}\,$\lambda$5169} & 37.98\,$\pm$\,0.75 & 394&136 \\
  & EPM -- D05 & 56719.6\,$\pm$\,9.7  & 21.02\,$\pm$\,0.27 & 20.68\,$\pm$\,0.23 & 5093\,$\pm$\,816 &                                              & 37.99\,$\pm$\,0.73 & 396&133 \\
PS1-12bku & PS1 & 56160.9\,$\pm$\,0.4 & 20.70\,$\pm$\,0.07 & 20.18\,$\pm$\,0.06 & 4258\,$\pm$\,291 & Fe\,{\sc ii}\,$\lambda$5169 & 37.28\,$\pm$\,0.23 & 286&31 \\
PS1-13abg & PS1 & 56375.4\,$\pm$\,5.0 & 22.27\,$\pm$\,0.08 & 21.12\,$\pm$\,0.09 & 4672\,$\pm$\,438 & Fe\,{\sc ii}\,$\lambda$5169 & 39.32\,$\pm$\,0.32 & 730&107 \\
PS1-13baf & PS1 & 56408.0\,$\pm$\,1.5 & 23.06\,$\pm$\,0.22 & 22.38\,$\pm$\,0.12 & 4093\,$\pm$\,473 & H$\beta$  & 39.60\,$\pm$\,0.51 & 832&194 \\
PS1-13bmf & PS1 & 56420.0\,$\pm$\,0.1 & 22.51\,$\pm$\,0.06 & 21.92\,$\pm$\,0.11 & 4363\,$\pm$\,256 & Fe\,{\sc ii}\,$\lambda$5169 & 39.18\,$\pm$\,0.28 & 684&87 \\
\multirow{2}{*}{PS1-13bni} 
  & EPM -- H01 & 56401.3\,$\pm$\,7.9 & 23.39\,$\pm$\,0.26 & 23.18\,$\pm$\,0.20 & 5814\,$\pm$\,1175 & \multirow{2}{*}{H$\beta$} & 40.65\,$\pm$\,0.76 & 1348&470 \\
  & EPM -- D05 & 56400.0\,$\pm$\,8.6 & 23.39\,$\pm$\,0.26 & 23.19\,$\pm$\,0.20 & 5913\,$\pm$\,1237 &                           & 40.68\,$\pm$\,0.77 & 1368&487 \\
\hline  
  \end{tabular}
  \\[1.5ex]
  \flushleft
  $^{*}K$-corrected magnitudes in the Johnson-Cousins Filter System. H01: \citet{Hamuy2001}; D05: \citet{Dessart2005}. 
\end{table*}

\subsection{Comparison of EPM and SCM distances}
\label{section:Comparison_EPM_SCM}

\begin{figure}
  \centering
    \includegraphics[width=\columnwidth]{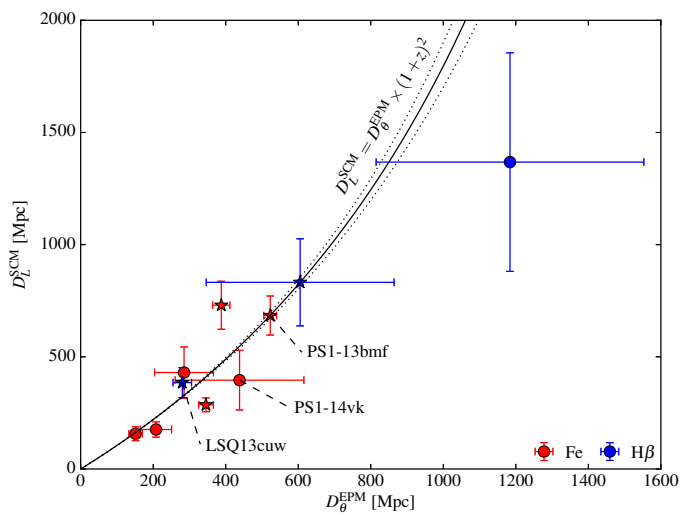}
    \caption{Comparison of EPM and SCM distances, using the dilution factors by \citet{Dessart2005}. Filled circles denote SNe with the explosion epoch obtained via EPM, while stars mark SNe with the explosion epoch estimated from pre-discovery photometry. The three Type II-L SNe are labelled specifically. Different colours denote the line that was used to estimate the photospheric velocities: red corresponds to Fe\,{\sc ii}\,$\lambda$5169, and dark blue to H$\beta$. The solid line shows the $(1+z)^2$ relation between luminosity and angular-size distances: $D_L^\mathrm{SCM} = D_\theta^\mathrm{EPM} \times (1+z)^2$. $\Lambda$CDM cosmology with $H_0$ = 70\,$\pm$\,5\,km\,s$^{-1}$\,Mpc$^{-1}$, $\Omega_m = 0.3$ and $\Omega_\Lambda = 0.7$ is assumed throughout. }
    \label{figure:EPM_vs_SCM}
\end{figure}

Fig.~\ref{figure:EPM_vs_SCM} compares the EPM and SCM distances for the entire sample. There appears no obvious trend for one technique to systematically result in longer or shorter distances; a shift would indicate $H_0$ $\ne$ 70\,km\,s$^{-1}$\,Mpc$^{-1}$. No obvious systematic shift can be discerned amongst the SNe using the Fe\,{\sc ii}\,$\lambda$5169 line for the EPM and SCM distances as an estimator for the photospheric velocity.

\subsection{The Hubble diagram}
\label{Section:The_Hubble_diagram}

Figure \ref{figure:Hubble_diagram} shows the Hubble diagrams using EPM (left panel) and SCM distances (right panel), respectively. The red and blue points represent SNe from our sample for which either Fe\,{\sc ii}\,$\lambda$5169 or H$\beta$ was used to estimate the photospheric velocities. For reasons of better visibility we only depict our distance results using the \citet{Dessart2005} dilution factors, which give somewhat larger distances than the \citet{Hamuy2001} dilution factors. Our conclusions are the same regardless of which set of dilution factors is used. The grey points depict SNe from other samples. 
The solid line in both panels represents a $\Lambda$CDM cosmology ($H_0$ = 70\,km\,s$^{-1}$\,Mpc$^{-1}$, $\Omega_m$ = 0.3 and $\Omega_\Lambda$ = 0.7.\footnote{The choice of $H_0$ = 70\,km\,s$^{-1}$\,Mpc$^{-1}$ is arbitrary and adopted mainly for consistency with the SCM parameters suggested by \citet{Nugent2006}. The general principles and our conclusions are the same independent of the exact choice of $H_0$.}). The three Type II-L SNe~LSQ13cuw, PS1-14vk and PS1-13bmf are labelled in both the EPM and the SCM Hubble diagrams.

\begin{figure*}[t!]
   \centering
   \begin{subfigure}[t]{0.49\textwidth}
      \includegraphics[width=\columnwidth]{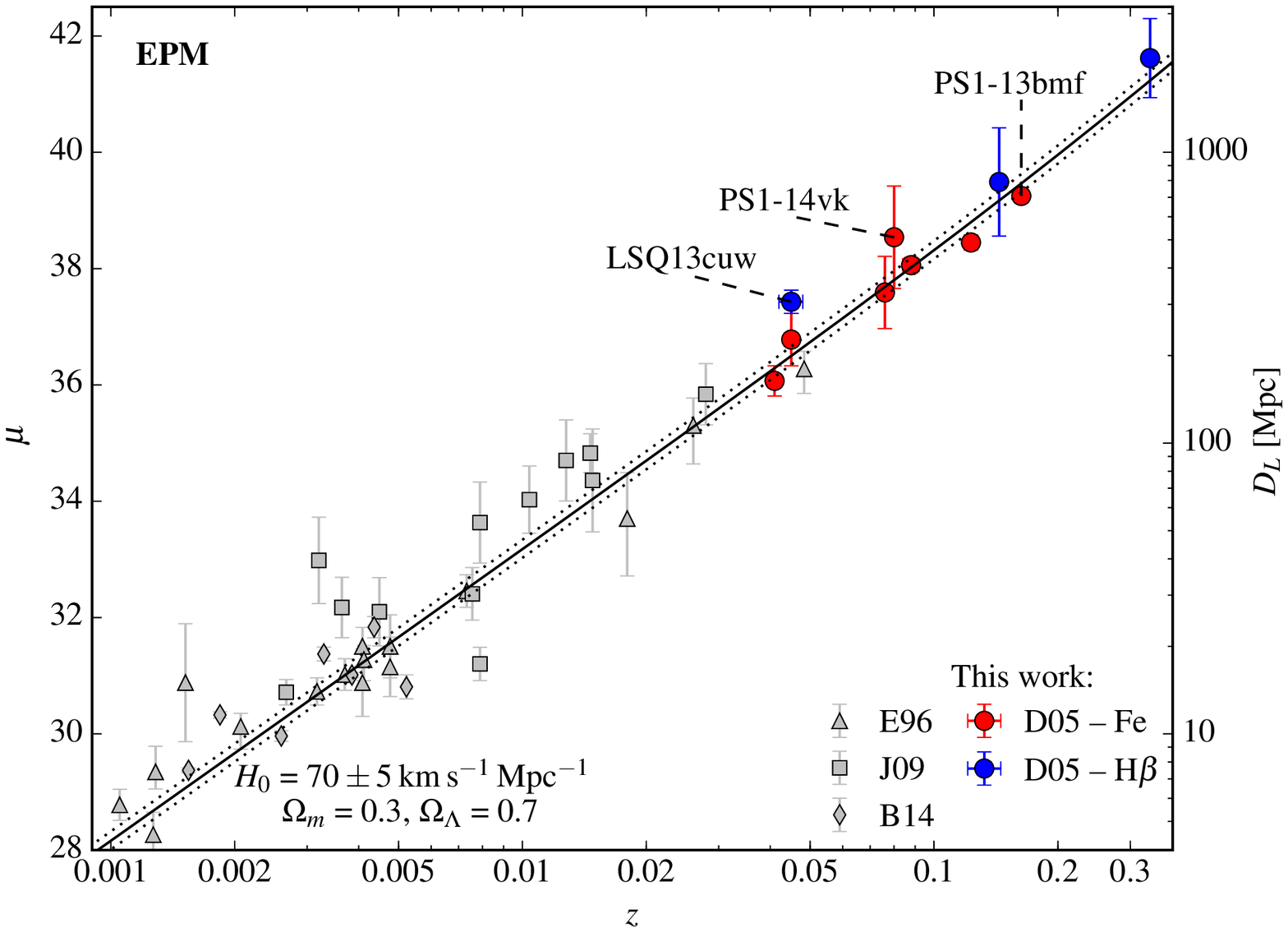}
   \end{subfigure}%
   \begin{subfigure}[t]{0.49\textwidth}
      \includegraphics[width=\columnwidth]{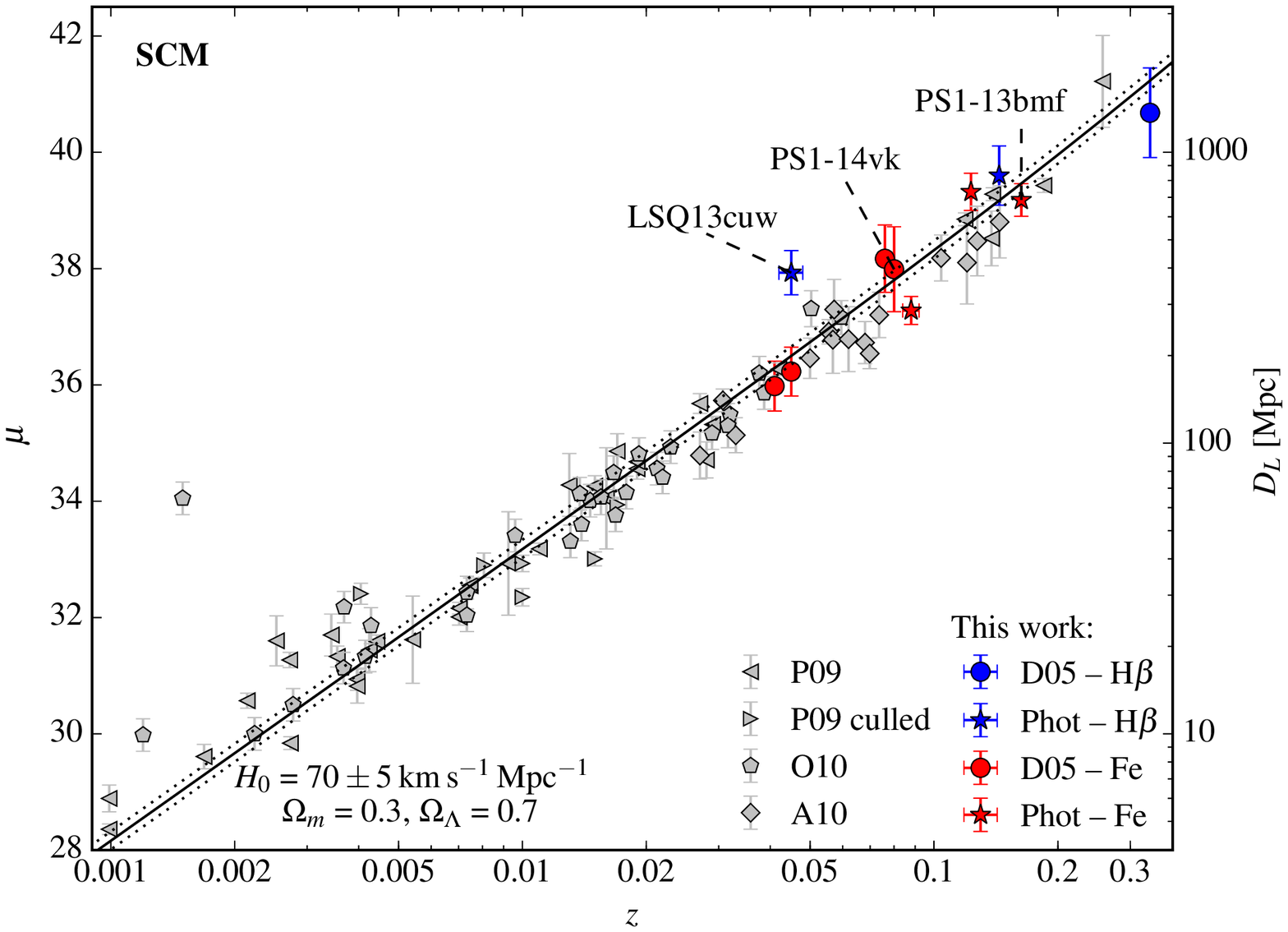}
   \end{subfigure}%
   \caption{SN II Hubble diagrams using the distances determined via EPM (left panel) and SCM (right panel). EPM Hubble diagram (left): the distances derived for our sample (circles) use the dilution factors from \citet{Dessart2005}. The different colours denote the absorption line that was used to estimate the photospheric velocities (Fe\,{\sc ii}\,$\lambda$5169 -- red; H$\beta$ -- blue). 
We also included EPM measurements from \citet[][E96]{Eastman1996}, \citet[][J09]{Jones2009} and \citet[][B14]{Bose2014}.
The solid line corresponds to a $\Lambda$CDM cosmology with $H_0$ = 70\,km\,s$^{-1}$\,Mpc$^{-1}$, $\Omega_m$ = 0.3 and $\Omega_\Lambda$ = 0.7, and the dotted lines to the range covered by an uncertainty on $H_0$ of 5\,km\,s$^{-1}$\,Mpc$^{-1}$. 
   SCM Hubble diagram (right): circle markers depict the SCM distances derived for our sample using the explosion epochs previously derived via EPM and the dilution factors from \citet{Dessart2005}. The star shaped markers depict SNe for which an independent estimate of the explosion time was available via photometry. The colours are coded in the same way as for the EPM Hubble diagram. Similarly, the solid and dotted lines portray the same relation between redshift and distance modulus as in the left panel. 
   We also included SCM distances from \citet[][P09]{Poznanski2009}, \citet[][O10]{Olivares2010} and \citet[][A10]{DAndrea2010}. We separated the objects ``culled'' by \citet{Poznanski2009} from the rest of the sample by using a different symbol.  
   The three Type II-L SNe~LSQ13cuw, PS1-14vk and PS1-13bmf are identified in both the EPM and the SCM Hubble diagram.
}
   \label{figure:Hubble_diagram}
\end{figure*}

\subsubsection{EPM Hubble diagram}

Our EPM measurements are complemented by EPM distances from the samples of \citet[][Table~6]{Eastman1996}, \citet[][Table~5]{Jones2009} and \citet[][Table~3]{Bose2014} in the EPM Hubble diagram (left panel of Figure~\ref{figure:Hubble_diagram}). In the cases of \citet{Jones2009} and \citet{Bose2014} we selected the distances given using the \citet{Dessart2005} dilution factors. In addition, \citet{Bose2014} give alternate results for the SNe 2004et, 2005cs, and 2012aw, for which constraints for the explosion epoch are available. We chose those values rather than the less constrained distance measurements. Note that SN~1992ba appears in \citet{Jones2009} and \citet{Eastman1996}, while SN~1999gi was published in \citet{Jones2009} and \citet{Bose2014}. 

The SN distances trace the slope of the Hubble line within the uncertainties. This is an indication that the relative distances are measured to a rather high accuracy.

\subsubsection{SCM Hubble diagram}

The SCM Hubble diagram shows our sample alongside SCM distances from \citet[][Table 2]{Poznanski2009}, \citet[][Table 8]{Olivares2010} and \citet[][Table 3]{DAndrea2010}. The \citet{Poznanski2009} sample contains all objects from \citet{Nugent2006}. We also included those objects that \citet{Poznanski2009} rejected due to their higher decline rates. \citet{DAndrea2010} do not give the distance measurements directly but rather their derived values for the $I$-band magnitude, the $(V-I)$-colour and the velocity 50 days after explosion (rest frame). We used these to apply the same equation and parameters from \citet{Nugent2006} as for our own sample, to find the distances to these objects. 
Note that the \citet{Poznanski2009} and \citet{Olivares2010} have a number of SNe in common: SNe 1991al, 1992af, 1992ba, 1999br, 1999cr, 1999em, 1999gi, 2003hl, 2003iq, and 2004et. 

Our SCM distances scatter around the $H_0$ = 70\,km\,s$^{-1}$\,Mpc$^{-1}$ as SCM is based on a previously chosen value of $H_0$ \citep[following][]{Nugent2006}.

There seems to be no obvious difference in the scatter for SNe with an estimate of the explosion epoch based on SN photometry or those relying on an EPM estimate for the time of explosion. This implies either that the epochs of explosion derived via the EPM are fairly accurate, or that constraints on the explosion epoch of only a few days, are not relevant for precise SCM measurements.

\subsubsection{PS1-13bni}

Finally, we would like to point out that our distance measurements to PS1-13bni (at a redshift of $z$ = $0.335^{+0.009}_{-0.012}$) demonstrates that both the EPM and the SCM bear great potential for cosmology. This statement is, however, tainted by the large uncertainties in the decline rates of PS1-13bni, and the implication that PS1-13bni could be a Type II-L SN and the open question whether these can be used as distance indicators.

\subsection{Applying the EPM and SCM to Type II-L SNe}
\label{Section:Applying_EPM_SCM_to_TypeIILSNe}

When considering the use of SNe II-L for cosmology a few peculiarities must be considered. As discussed in Section \ref{section:IIL_velocities} it is unclear whether SNe II-L display the same velocity evolution as SNe II-P and whether the v$_{\mathrm{H}\alpha}$/v$_{\mathrm{Fe\,5169}}$ and v$_{\mathrm{H}\beta}$/v$_{\mathrm{Fe\,5169}}$ ratios evolve similarly for SNe~II-L and SNe~II-P. \citet{Pejcha2015} find a physical answer to this question by showing that the light curve shape is mostly determined by temperature changes in the photosphere. For EPM with SNe~II-L the additional question arises whether the relation between $\chi$ and $t^\star$ is linear for SNe II-L (see Equation \ref{equation:chi_t}). The case of PS1-14vk indicates that this linear relation might be valid for a similar period of time as for SNe II-P (see Section \ref{section:EPM_Commonalities}). 

Although SNe~II-L fade faster than SNe II-P, they are on average brighter. \citet{Gall2015} suggest that SNe II-L might have slightly longer rise times than SNe II-P. This somewhat facilitates obtaining pre-maximum data, resulting in a better constraint on the explosion epoch. The uncertainty in the explosion epoch contributes significantly to the total error budget \citep[c.f. also][]{Pejcha2015}.

The three Type II-L SNe LSQ13cuw, PS1-14vk, and PS1-13bmf, are indistinguishable from the rest of the sample, in the EPM and the SCM Hubble diagrams. We find once again (c.f. \S\ref{section:intro}) that there is no clear observational distinction between the SNe II-P and II-L classes nor is one expected on theoretical grounds. Testing the validity of Type II-P SN relations, like the $\chi$-$t^\star$ relation in the EPM, or the velocity-luminosity correlation for the SCM for SNe with a range of decline rates, will help to better understand the sources of systematic uncertainties for both techniques.

\section{Conclusions}
\label{section:Conclusions}

Optical light curves and spectra of nine Type\,II-P/L SNe with redshifts between $z$ = 0.045 and $z$ = 0.335: SN 2013ca, PS1-13wr, PS1-14vk, PS1-12bku, PS1-13abg, PS1-13baf, PS1-13bmf, PS1-13atm and PS1-13bni were presented. To this sample, we added the Type II-P SN 2013eq \citep[][$z$ = 0.041\,$\pm$\,0.001]{Gall2016} and the Type II-L SN LSQ13cuw \citep[][$z$ = 0.045\,$\pm$\,0.003]{Gall2015} to derive distances utilizing the expanding photosphere method and the standardized candle method. The procedures presented in \citet{Gall2016} for EPM at cosmologically significant redshifts and the approach of \citet{Nugent2006} for the SCM were adopted. 

The possibility of estimating the photospheric velocities in Type II-P/L SNe H$\alpha$ and H$\beta$ was explored. We used five well observed SNe (SNe~1999gi, 1999em, 2004et, 2005cs, and 2006bp) to calculate an epoch-dependent relation between the H$\alpha$ and H$\beta$ velocities relative to the Fe\,{\sc ii}\,$\lambda$5169 velocity. 
The v$_{\mathrm{H}\alpha}$/v$_{\mathrm{Fe\,5169}}$ and v$_{\mathrm{H}\beta}$/v$_{\mathrm{Fe\,5169}}$ ratios are in good agreement at early epochs ($\le$\,40 days after explosion) for H$\alpha$ and $\le$\,30 days for H$\beta$. The scatter in these velocity ratios is between 4 and 14\%. At later epochs the v$_{\mathrm{H}\alpha}$/v$_{\mathrm{Fe\,5169}}$ and v$_{\mathrm{H}\beta}$/v$_{\mathrm{Fe\,5169}}$ ratios diverge significantly between individual SNe. 

For the utilization in the EPM and the SCM we used the Fe\,{\sc ii}\,$\lambda$5169 line as an estimator of the photospheric velocity where ever possible. In cases where no Fe\,{\sc ii}\,$\lambda$5169 feature could be identified we applied the v$_{\mathrm{H}\beta}$/v$_{\mathrm{Fe\,5169}}$ relation to obtain an estimate for the velocity. 

Our EPM and SCM distances are in good agreement with the expectations from standard $\Lambda$CDM cosmology. This is especially encouraging given the varied quality of the available data for local versus intermediate- or high-redshift objects. Comparable precision to other distance indicators can feasibly be achieved via larger samples and higher quality data of Type II SNe. 

The case of PS1-13bni at a spectroscopically derived redshift of $z$ = $0.335^{+0.009}_{-0.012}$ is the highest-$z$ SN II for which a distance measurement was ever attempted, further  demonstrating the potential of the EPM and SCM for cosmological applications.

Finally, we evaluated the implications of using Type II-L SNe for cosmology. Our sample contains three Type II-L SNe that yield similar distances compared to the SNe II-P. Including SNe II-L would allow for larger samples to be considered. However, we note that that future investigations on the differences and similarities between SNe II-L and II-P are necessary to robustly assess any systematics. Although currently only demonstrated for local SNe \citep{Maguire2010a}, further improvements may also be possible by incorporating infrared datasets for SNe within the Hubble flow.

\textit{Note added in proof:} During the refereeing process \citet{deJaeger2017b} was published. It reports the observations of a more distant Type II supernova, SN 2016jhj at $z = 0.34$, although with a single spectrum and hence the application of the SCM only.

\section*{Acknowledgements} 

Based in part on observations made with the Gran Telescopio Canarias (GTC2007-12ESO, ESO program ID: 189.D-2007, PI:RK), installed in the Spanish Observatorio del Roque de los Muchachos of the Instituto de Astrofísica de Canarias, on the island of La Palma. 

EEEG, BL, ST, and WH acknowledge partial support by the Deutsche Forschungsgemeinschaft through the TransRegio project TRR33 ‘The Dark Universe’.
R.K. acknowledges funding from STFC (ST/L000709/1).
N.M. acknowledges support from the Science and Technology Facilities Council [ST/L00075X/1].

Based in part on observations obtained at the Liverpool Telescope (program ID: XIL10B01, PI:RK), operated on the island of La Palma by Liverpool John Moores University in the Spanish Observatorio del Roque de los Muchachos of the Instituto de Astrofísica de Canarias with financial support from the UK Science and Technology Facilities Council.

Based in part on observations obtained at the Nordic Optical Telescope, operated by the Nordic Optical Telescope Scientific Association at the Observatorio del Roque de los Muchachos, La Palma, Spain, of the Instituto de Astrofísica de Canarias.

This research has made use of the NASA/IPAC Extragalactic Database (NED) which is operated by the Jet Propulsion Laboratory, California Institute of Technology, under contract with the National Aeronautics and Space Administration.

The Pan-STARRS1 Surveys (PS1) have been made possible through contributions of the Institute for Astronomy, the University of Hawaii, the Pan-STARRS Project Office, the Max-Planck Society and its participating institutes, the Max Planck Institute for Astronomy, Heidelberg and the Max Planck Institute for Extraterrestrial Physics, Garching, The Johns Hopkins University, Durham University, the University of Edinburgh, Queen's University Belfast, the Harvard-Smithsonian Center for Astrophysics, the Las Cumbres Observatory Global Telescope Network Incorporated, the National Central University of Taiwan, the Space Telescope Science Institute, the National Aeronautics and Space Administration under Grant No. NNX08AR22G issued through the Planetary Science Division of the NASA Science Mission Directorate, the National Science Foundation under Grant No. AST-1238877, the University of Maryland, and Eotvos Lorand University (ELTE) and the Los Alamos National Laboratory.

Funding for the Sloan Digital Sky Survey (SDSS) and SDSS-II has been provided by the Alfred P. Sloan Foundation, the Participating Institutions, the National Science Foundation, the U.S. Department of Energy, the National Aeronautics and Space Administration, the Japanese Monbukagakusho, the Max Planck Society, and the Higher Education Funding Council for England. The SDSS Web Site is http://www.sdss.org/.

The SDSS is managed by the Astrophysical Research Consortium for the Participating Institutions. The Participating Institutions are the American Museum of Natural History, Astrophysical Institute Potsdam, University of Basel, University of Cambridge, Case Western Reserve University, University of Chicago, Drexel University, Fermilab, the Institute for Advanced Study, the Japan Participation Group, Johns Hopkins University, the Joint Institute for Nuclear Astrophysics, the Kavli Institute for Particle Astrophysics and Cosmology, the Korean Scientist Group, the Chinese Academy of Sciences (LAMOST), Los Alamos National Laboratory, the Max-Planck-Institute for Astronomy (MPIA), the Max-Planck-Institute for Astrophysics (MPA), New Mexico State University, Ohio State University, University of Pittsburgh, University of Portsmouth, Princeton University, the United States Naval Observatory, and the University of Washington.

\bibliography{EPMSample_paper} \bibliographystyle{aa}

\appendix

\section{Photometric and spectroscopic observations of the individual SNe} 
\label{section:appendix:observations_of_individual_SNe}

An overview of SN sample is given in Section~\ref{section:Observations_and_data_reduction}, Table~\ref{table:whole_sample}. In the following we present the data for each individual SN. All epochs are given in the SN rest frame.

\subsection{SN 2013eq}
\label{section:Observations:SN2013eq}
Photometric data up to 76 days after discovery and a series of five spectra ranging from 7 to 65 days after discovery for SN 2013eq are presented in \citet{Gall2016}. They find a redshift of $z = 0.041\,\pm\,0.001$, a host galaxy extinction of $E(B-V)_{\mathrm{host}} = 0.062\,\pm\,0.028$\,mag and a Galactic extinction of $E(B -V)_\mathrm{{Gal}} = 0.034$\,mag from \citet{Schlafly2011}. We adopt these values also for this study.

\subsection{SN 2013ca}
\label{section:Observations:SN_2013ca}

SN 2013ca was discovered on 2013 April 21 as ``LSQ13aco'' by the La Silla Quest Supernova Survey \citep[LSQ;][]{Baltay2013}, and spectroscopically classified as a Type II-P SN \citep{2013ATel5056}.
It was independently discovered on 2013 May 1 and likewise classified as a Type II-P SN by \citet{2013CBET3508}, who also report further detections on 2013 April 10 and May 3. 

We obtained LT $g'r'i'$ photometry 41 and 68 days after the first detection and a series of four GTC spectra ranging between +30 and +60\,days after the first detection. 

In order to determine the redshift of SN 2013ca we extracted host galaxy spectra alongside the SN spectra and measured the redshift of the narrow H$\alpha$ emission line. Averaging over all epochs results in $z=0.045\,\pm\,0.001$.

Features attributable to the Na\,{\sc i} doublet from interstellar gas either in the Milky Way or the host galaxy are not apparent in the spectra of SN 2013ca. In order to derive an upper limit to the equivalent width (EW) of a Na\,{\sc i}\,D absorption, we constructed a weighted stack of all SN 2013ca spectra, where the weights reflect the signal-to-noise ratio. We then created a series of artificial Gaussian profiles centered at 5893\,{\AA} and subtracted these from the stacked spectrum. The FWHM of the Gaussian profiles was set to 17.0\,{\AA}, which corresponds to our lowest resolution spectrum. We increased the EW of the artificial profile in steps of 0.1\,{\AA} in order to determine the limiting EW at which an artificial line profile as described above would be detectable in the stacked spectrum. Using this method we derive an upper limit for the EW of the Na\,{\sc i}\,D $\lambda\lambda$\,5890,5896 blend to be $<$\,0.4\,{\AA}. Applying the relation in Equation 9 of \citet{Poznanski2012}, we find $E(B-V)_{\mathrm{host}} <$\,0.04\,mag. 

The light curves of SN 2013eq are presented in the left panel of Figure~\ref{figure:SN2013ca}. Table~\ref{table:SN2013ca_photometry} shows the log of photometry and the calibrated magnitudes. The photometric coverage of SN 2013ca is sparse. The two available data points in  $r'$ Yield a decline rate of 0.34\,$\pm$\,0.18\,mag/50\,d, which is below the threshold of 0.5\,mag/50\,d set by \citet{Li2011a} to distinguish SNe II-P from SNe II-L. 

\begin{figure*}[t!]
   \centering
   \begin{subfigure}[t]{0.49\textwidth}
      \includegraphics[width=\columnwidth]{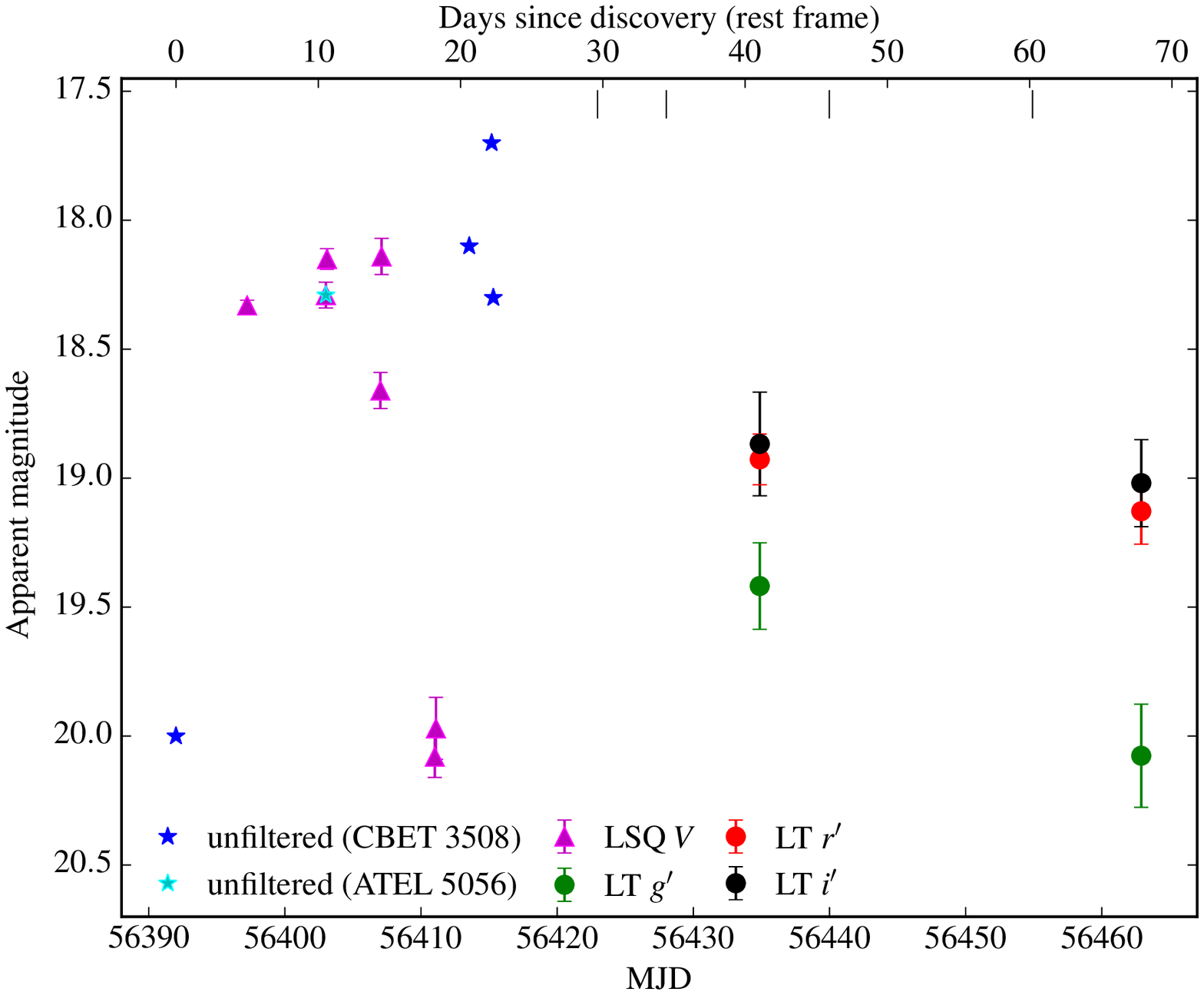}
   \end{subfigure}%
   \begin{subfigure}[t]{0.49\textwidth}
      \includegraphics[width=\columnwidth]{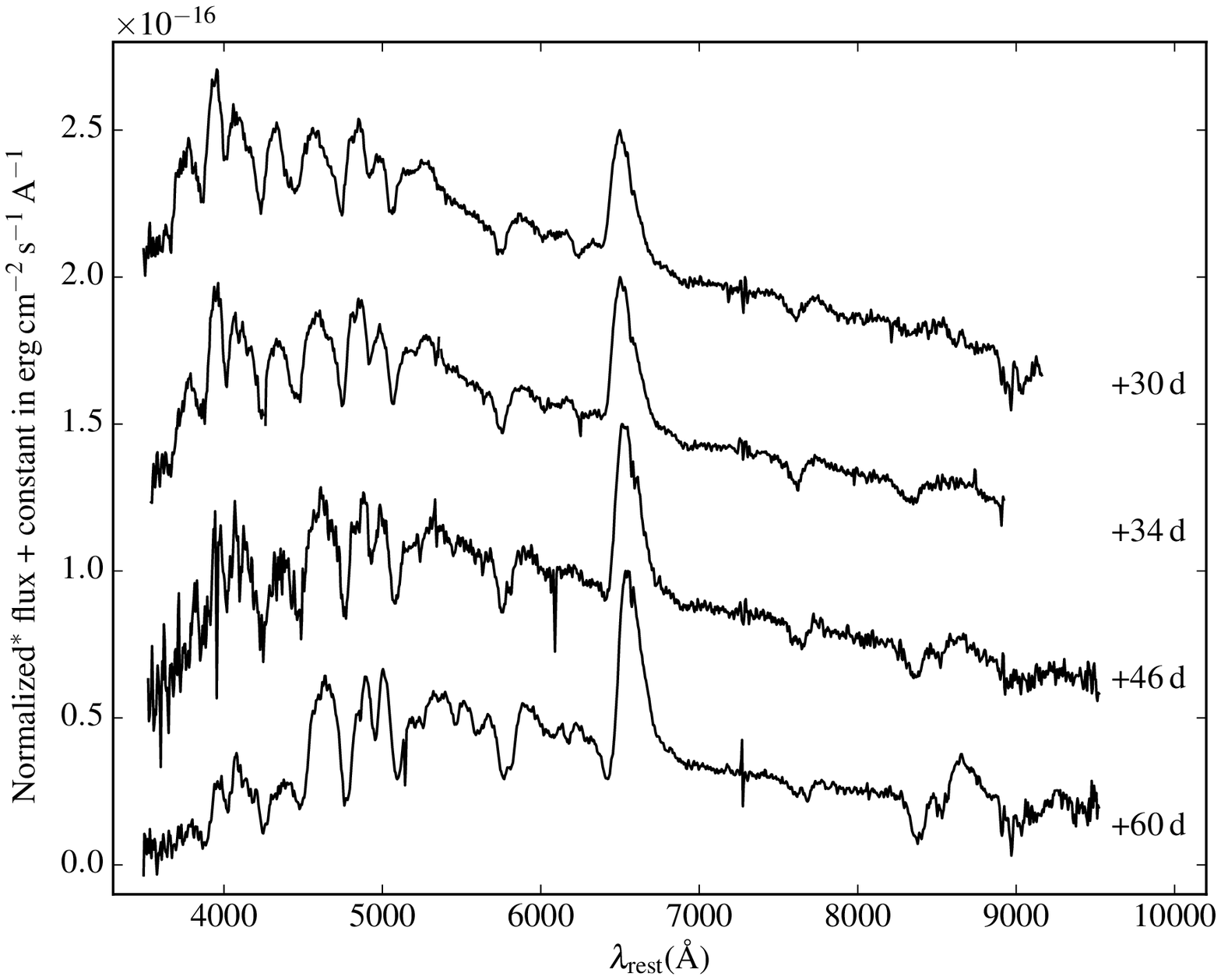}
   \end{subfigure}%
   \caption{SN 2013ca. Left panel: LT $g'r'i'$ and LSQ $V$ light curves of SN 2013ca. The vertical ticks on the top mark the epochs of the observed spectra. ATel 5056: \citet{2013ATel5056}; CBET 3508: \citet{2013CBET3508}. Right panel: SN 2013ca spectroscopy. $^*$Flux normalized to the maximum H$\alpha$ flux for better visibility of the features. The exact normalizations are: 
flux/1.3 for the +30\,d spectrum;
flux/1.4 for +34\,d;
flux/1.5 for +46\,d;
flux/1.5 for +60\,d.
}
   \label{figure:SN2013ca}
\end{figure*}

\begin{table*}
  \small
  \caption{SN 2013ca: Photometric observations}
  \label{table:SN2013ca_photometry}
  \centering
  \begin{tabular}{c   c                      r            c                            c                        c           }
  \hline               
  \multirow{2}{*}{Date} & \multirow{2}{*}{MJD} & \multicolumn{1}{c}{Rest frame} & \multirow{2}{*}{Magnitude} & \multirow{2}{*}{Filter} & \multirow{2}{*}{Telescope} \\ 
                      &                      & \multicolumn{1}{c}{epoch$^*$}  &                            &                         &              \\
  \hline 
  2013-05-22 & 56434.88 & +41.03 & 19.42\,$\pm$\,0.17 & $g^\prime$ & LT \\
  2013-06-19 & 56462.90 & +67.84 & 20.08\,$\pm$\,0.20 & $g^\prime$ & LT \\
  \hline  
  2013-05-22 & 56434.88 & +41.03 & 18.93\,$\pm$\,0.10 & $r^\prime$ & LT \\
  2013-06-19 & 56462.90 & +67.85 & 19.13\,$\pm$\,0.13 & $r^\prime$ & LT \\
  \hline  
  2013-05-22 & 56434.88 & +41.04 & 18.87\,$\pm$\,0.20 & $i^\prime$ & LT \\
  2013-06-19 & 56462.90 & +67.85 & 19.02\,$\pm$\,0.17 & $i^\prime$ & LT \\
  \hline  
  \end{tabular}
  \\[1.5ex]
  \flushleft
  The tabulated magnitudes are given ``as observed'', i.e. neither corrected for dust extinction nor $K$-corrected. $^*$Rest frame epochs (assuming a redshift of 0.045) with respect to the first detection on 56392.0 (MJD). 
LT = Liverpool Telescope. 
\end{table*}

Table~\ref{table:SN2013ca_journal_spectra} gives the journal of spectroscopic observations. The calibrated spectra of SN 2013ca are presented in the right panel of Figure~\ref{figure:SN2013ca}. They are corrected for galactic reddening and redshift ($z$ = 0.045). 
The H$\alpha$ line shows only a small absorption at +30 and +34\,days, which becomes more prominent at +46\,d and +60\,days. Weak Fe\,{\sc ii}\,$\lambda$5169 is visible at all four epochs. 

\begin{table*}
  \caption{SN 2013ca: Journal of spectroscopic observations}
  \label{table:SN2013ca_journal_spectra}
  \centering
  \begin{tabular}{c c     c            c              c            c                               }
  \hline 
\multirow{2}{*}{Date} & \multirow{2}{*}{MJD} & Epoch$^*$  & Wavelength   & Resolution & \multirow{2}{*}{Telescope+Instrument} \\
             &           & rest frame & range in \AA &  \AA       &                               \\
\hline 
  2013-05-10 & 56422.97  & +29.63 & 3650 - 9595 &  17.0      & GTC+OSIRIS+R300B              \\
  2013-05-15 & 56428.02  & +34.46 & 3700 - 9345 &  16.9      & GTC+OSIRIS+R300B              \\
  2013-05-27 & 56439.99  & +45.92 & 3680 - 9970 &  16.9      & GTC+OSIRIS+R300B              \\
  2013-06-11 & 56454.91  & +60.20 & 3650 - 9970 &  16.8      & GTC+OSIRIS+R300B              \\
  \hline  
  \end{tabular}
  \\[1.5ex]
  \flushleft
  $^*$Rest frame epochs (assuming a redshift of 0.045) with respect to the first detection on 56392.0 (MJD). The resolution was determined from the FWHM of the O\,{\sc i} $\lambda$5577.34 sky line.
\end{table*}

\subsection{LSQ13cuw}
Photometric data up to 100 days after discovery and 5 spectra ranging between 25 and 84\,days after discovery for LSQ13cuw were presented by \citet{Gall2015}. Performing a fit to the pre-peak photometry \citet{Gall2015} are able to constrain the epoch of explosion to be MJD 56593.42\,$\pm$\,0.68. They find a redshift of $z$ = 0.045\,$\pm$\,0.003, a host galaxy extinction $E(B-V)_{\mathrm{host}}$ $<$ 0.16\,mag and a Galactic extinction of $E(B -V)_\mathrm{{Gal}}$ = 0.023\,mag from \citet{Schlafly2011}. We adopt these values here.

\subsection{PS1-13wr}
PS1-13wr was discovered on 2013 February 26 by the PS1 MD survey. $g_{\mathrm{PS1}}r_{\mathrm{PS1}}i_{\mathrm{PS1}}z_{\mathrm{PS1}}y_{\mathrm{PS1}}$ photometry was obtained in the course of the PS1 MD survey up to 150\,days after the first detection. We  obtained 4 GTC spectra ranging from +12 to +41\,days after discovery. 

PS1-13wr lies close to the galaxy SDSS J141558.95+520302.9 for which the SDSS reports a spectroscopic redshift of 0.076\,$\pm$\,0.001. This redshift was adopted also for PS1-13wr and is consistent with the redshift of the Na\,{\sc i}\,D doublet visible in the spectra of PS1-13wr.

In order to obtain an accurate measure of the Na\,{\sc i}\,D equivalent width we constructed a weighted stack of all PS1-13wr spectra, where the weights reflect the signal-to-noise ratio in the 5850-6000\,{\AA} region. We measure an equivalent width of EW$_{\mathrm{Na\,I\,D}} = 0.76\,\pm\,0.07$\,{\AA} in the stacked PS1-13wr spectrum. Applying the empirical relation between Na\,{\sc i}\,D absorption and dust extinction given in \citet[][Equation~9]{Poznanski2012}, this translates into an extinction within the host galaxy of $E(B-V)^{\mathrm{host}}_{\mathrm{PS1-13wr}} = 0.110\,\pm\,0.049$\,mag. 

The light curves of PS1-13wr are presented in the left panel of Figure~\ref{figure:PS1-13wr}. Table~\ref{table:PS1-13wr_photometry} gives the log of imaging observations and the calibrated magnitudes.
PS1-13wr displays a distinct plateau in the $r_{\mathrm{PS1}}$-, $i_{\mathrm{PS1}}$-, $z_{\mathrm{PS1}}$-, and $y_{\mathrm{PS1}}$-bands until about 80\,days after discovery, when it drops onto the radioactive tail. The $r_{\mathrm{PS1}}$-band decline rate is $\sim$0.20\,$\pm$\,0.07\,mag/50\,days characterising it as a II-P SN.

\begin{figure*}[t!]
   \centering
   \begin{subfigure}[t]{0.49\textwidth}
      \includegraphics[width=\columnwidth]{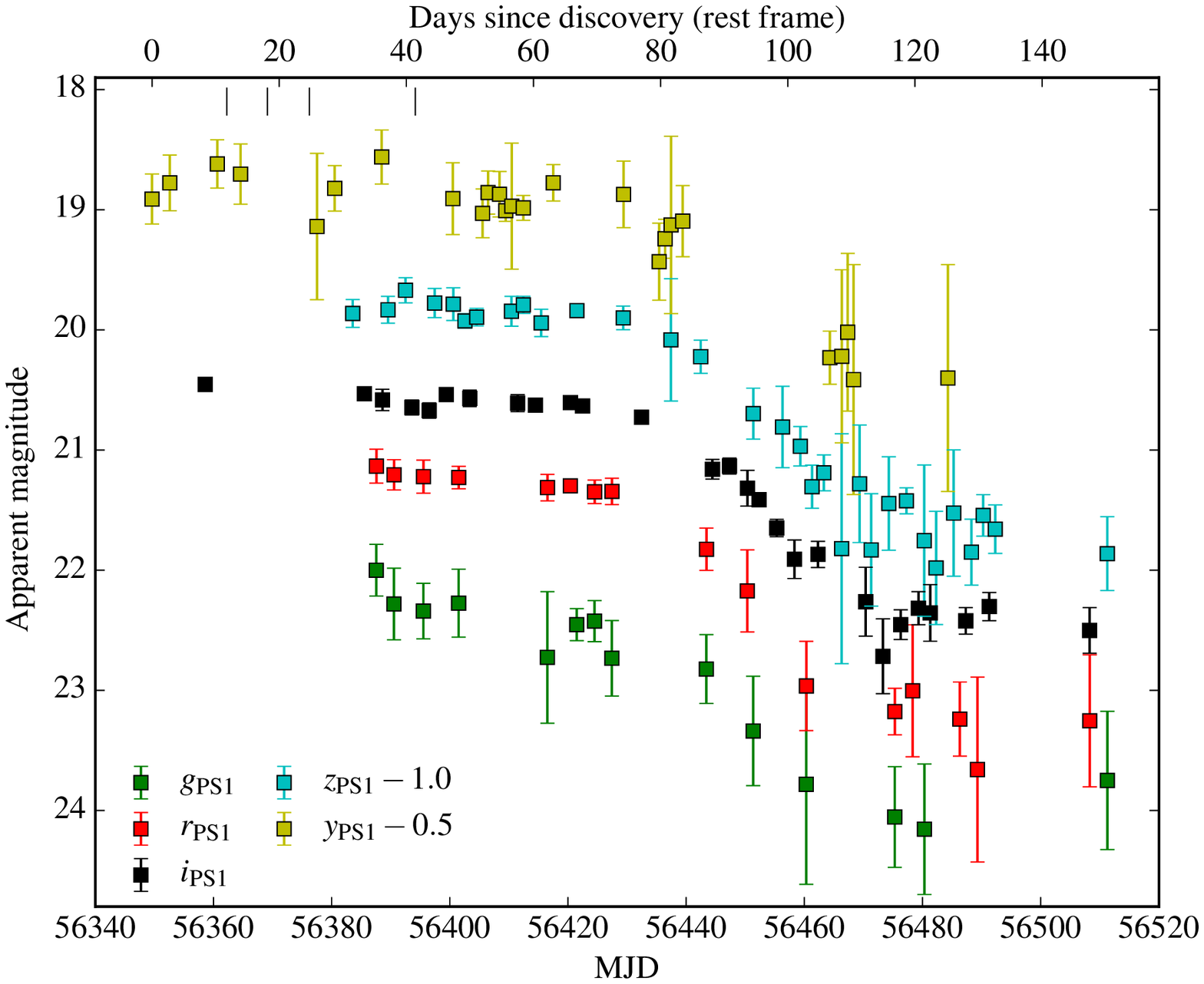}
   \end{subfigure}%
   \begin{subfigure}[t]{0.49\textwidth}
      \includegraphics[width=\columnwidth]{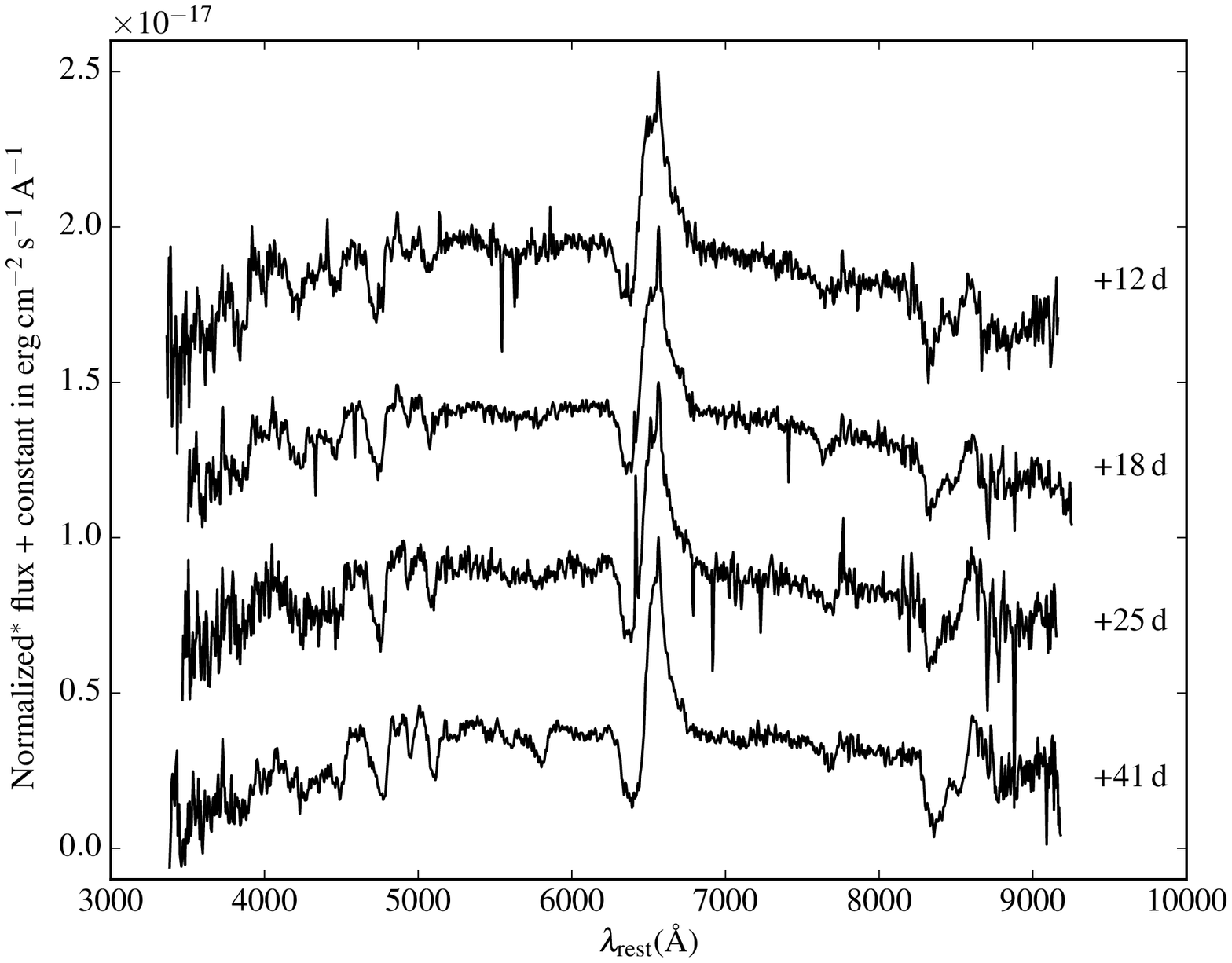}
\end{subfigure}%
   \caption{PS1-13wr. Left panel: $grizy_\mathrm{PS1}$ light curves of PS1-13wr. The vertical ticks on the top mark the epochs of the observed spectra. Right panel: PS1-13wr spectroscopy. $^*$Flux normalized to the maximum H$\alpha$ flux for better visibility of the features. The exact normalizations are: 
flux/3.4 for the +12\,d spectrum;
flux/3.5 for +18\,d;
flux/3.6 for +25\,d;
flux/3.6 for +41\,d.
}
   \label{figure:PS1-13wr}
\end{figure*}

\begin{table*}
  \small
  \caption{PS1-13wr: Photometric observations}
  \label{table:PS1-13wr_photometry}

  \centering
  \begin{tabular}{c   c                      r            c                            c                        c           }
  \hline               
  \multirow{2}{*}{Date} & \multirow{2}{*}{MJD} & \multicolumn{1}{c}{Rest frame} & \multirow{2}{*}{Magnitude} & \multirow{2}{*}{Filter} & \multirow{2}{*}{Telescope} \\ 
                      &                      & \multicolumn{1}{c}{epoch$^*$}  &                            &                         &              \\
  \hline 
  2013-04-05 & 56387.56 & +35.28  & 22.00\,$\pm$\,0.21 & $g_\mathrm{PS1}$ & PS1 \\
  2013-04-08 & 56390.55 & +38.05  & 22.28\,$\pm$\,0.30 & $g_\mathrm{PS1}$ & PS1 \\
  2013-04-13 & 56395.53 & +42.68  & 22.34\,$\pm$\,0.23 & $g_\mathrm{PS1}$ & PS1 \\
  2013-04-19 & 56401.49 & +48.22  & 22.27\,$\pm$\,0.28 & $g_\mathrm{PS1}$ & PS1 \\
  2013-05-04 & 56416.52 & +62.19  & 22.73\,$\pm$\,0.55 & $g_\mathrm{PS1}$ & PS1 \\
  2013-05-09 & 56421.49 & +66.81  & 22.45\,$\pm$\,0.13 & $g_\mathrm{PS1}$ & PS1 \\
  2013-05-12 & 56424.50 & +69.60  & 22.42\,$\pm$\,0.17 & $g_\mathrm{PS1}$ & PS1 \\
  2013-05-15 & 56427.43 & +72.33  & 22.73\,$\pm$\,0.31 & $g_\mathrm{PS1}$ & PS1 \\
  2013-05-31 & 56443.42 & +87.19  & 22.82\,$\pm$\,0.29 & $g_\mathrm{PS1}$ & PS1 \\
  2013-06-08 & 56451.34 & +94.55  & 23.34\,$\pm$\,0.46 & $g_\mathrm{PS1}$ & PS1 \\
  2013-06-17 & 56460.33 & +102.90 & 23.78\,$\pm$\,0.83 & $g_\mathrm{PS1}$ & PS1 \\
  2013-07-02 & 56475.30 & +116.82 & 24.05\,$\pm$\,0.42 & $g_\mathrm{PS1}$ & PS1 \\
  2013-07-07 & 56480.31 & +121.48 & 24.16\,$\pm$\,0.54 & $g_\mathrm{PS1}$ & PS1 \\
  2013-08-07 & 56511.29 & +150.27 & 23.75\,$\pm$\,0.58 & $g_\mathrm{PS1}$ & PS1 \\
  \hline
  2013-04-05 & 56387.58 & +35.29  & 21.13\,$\pm$\,0.14 & $r_\mathrm{PS1}$ & PS1 \\
  2013-04-08 & 56390.56 & +38.06  & 21.21\,$\pm$\,0.13 & $r_\mathrm{PS1}$ & PS1 \\
  2013-04-13 & 56395.54 & +42.69  & 21.22\,$\pm$\,0.14 & $r_\mathrm{PS1}$ & PS1 \\
  2013-04-19 & 56401.50 & +48.24  & 21.23\,$\pm$\,0.09 & $r_\mathrm{PS1}$ & PS1 \\
  2013-05-04 & 56416.53 & +62.20  & 21.31\,$\pm$\,0.11 & $r_\mathrm{PS1}$ & PS1 \\
  2013-05-08 & 56420.40 & +65.79  & 21.30\,$\pm$\,0.05 & $r_\mathrm{PS1}$ & PS1 \\
  2013-05-12 & 56424.51 & +69.62  & 21.35\,$\pm$\,0.10 & $r_\mathrm{PS1}$ & PS1 \\
  2013-05-15 & 56427.44 & +72.34  & 21.34\,$\pm$\,0.11 & $r_\mathrm{PS1}$ & PS1 \\
  2013-05-31 & 56443.43 & +87.20  & 21.83\,$\pm$\,0.18 & $r_\mathrm{PS1}$ & PS1 \\
  2013-06-07 & 56450.35 & +93.63  & 22.17\,$\pm$\,0.34 & $r_\mathrm{PS1}$ & PS1 \\
  2013-06-17 & 56460.34 & +102.91 & 22.96\,$\pm$\,0.37 & $r_\mathrm{PS1}$ & PS1 \\
  2013-07-02 & 56475.32 & +116.84 & 23.18\,$\pm$\,0.19 & $r_\mathrm{PS1}$ & PS1 \\
  2013-07-05 & 56478.32 & +119.63 & 23.00\,$\pm$\,0.55 & $r_\mathrm{PS1}$ & PS1 \\
  2013-07-13 & 56486.32 & +127.06 & 23.24\,$\pm$\,0.31 & $r_\mathrm{PS1}$ & PS1 \\
  2013-07-16 & 56489.30 & +129.83 & 23.66\,$\pm$\,0.77 & $r_\mathrm{PS1}$ & PS1 \\
  2013-08-04 & 56508.30 & +147.49 & 23.25\,$\pm$\,0.55 & $r_\mathrm{PS1}$ & PS1 \\
  \hline
  2013-03-07 & 56358.60 & +8.36   & 20.45\,$\pm$\,0.04 & $i_\mathrm{PS1}$ & PS1 \\
  2013-04-03 & 56385.54 & +33.40  & 20.53\,$\pm$\,0.05 & $i_\mathrm{PS1}$ & PS1 \\
  2013-04-06 & 56388.60 & +36.24  & 20.58\,$\pm$\,0.09 & $i_\mathrm{PS1}$ & PS1 \\
  2013-04-11 & 56393.58 & +40.87  & 20.65\,$\pm$\,0.06 & $i_\mathrm{PS1}$ & PS1 \\
  2013-04-14 & 56396.51 & +43.60  & 20.67\,$\pm$\,0.06 & $i_\mathrm{PS1}$ & PS1 \\
  2013-04-17 & 56399.40 & +46.28  & 20.54\,$\pm$\,0.04 & $i_\mathrm{PS1}$ & PS1 \\
  2013-04-21 & 56403.38 & +49.98  & 20.57\,$\pm$\,0.07 & $i_\mathrm{PS1}$ & PS1 \\
  2013-04-29 & 56411.47 & +57.50  & 20.61\,$\pm$\,0.07 & $i_\mathrm{PS1}$ & PS1 \\
  2013-05-02 & 56414.50 & +60.31  & 20.63\,$\pm$\,0.04 & $i_\mathrm{PS1}$ & PS1 \\
  2013-05-08 & 56420.41 & +65.81  & 20.61\,$\pm$\,0.03 & $i_\mathrm{PS1}$ & PS1 \\
  2013-05-10 & 56422.46 & +67.71  & 20.63\,$\pm$\,0.04 & $i_\mathrm{PS1}$ & PS1 \\
  2013-05-20 & 56432.44 & +76.99  & 20.73\,$\pm$\,0.06 & $i_\mathrm{PS1}$ & PS1 \\
  2013-06-01 & 56444.45 & +88.15  & 21.16\,$\pm$\,0.08 & $i_\mathrm{PS1}$ & PS1 \\
  2013-06-04 & 56447.33 & +90.83  & 21.13\,$\pm$\,0.07 & $i_\mathrm{PS1}$ & PS1 \\
  2013-06-07 & 56450.37 & +93.65  & 21.32\,$\pm$\,0.15 & $i_\mathrm{PS1}$ & PS1 \\
  2013-06-09 & 56452.33 & +95.47  & 21.41\,$\pm$\,0.05 & $i_\mathrm{PS1}$ & PS1 \\
  2013-06-12 & 56455.33 & +98.26  & 21.65\,$\pm$\,0.07 & $i_\mathrm{PS1}$ & PS1 \\
  2013-06-15 & 56458.32 & +101.04 & 21.91\,$\pm$\,0.16 & $i_\mathrm{PS1}$ & PS1 \\
  2013-06-19 & 56462.31 & +104.74 & 21.87\,$\pm$\,0.11 & $i_\mathrm{PS1}$ & PS1 \\
  2013-06-27 & 56470.39 & +112.25 & 22.26\,$\pm$\,0.29 & $i_\mathrm{PS1}$ & PS1 \\
  2013-06-30 & 56473.30 & +114.96 & 22.72\,$\pm$\,0.31 & $i_\mathrm{PS1}$ & PS1 \\
  2013-07-03 & 56476.31 & +117.76 & 22.45\,$\pm$\,0.12 & $i_\mathrm{PS1}$ & PS1 \\
  2013-07-06 & 56479.31 & +120.55 & 22.32\,$\pm$\,0.14 & $i_\mathrm{PS1}$ & PS1 \\
  2013-07-08 & 56481.30 & +122.40 & 22.36\,$\pm$\,0.24 & $i_\mathrm{PS1}$ & PS1 \\
  2013-07-14 & 56487.32 & +127.99 & 22.42\,$\pm$\,0.11 & $i_\mathrm{PS1}$ & PS1 \\
  2013-07-18 & 56491.30 & +131.69 & 22.30\,$\pm$\,0.12 & $i_\mathrm{PS1}$ & PS1 \\
  2013-08-04 & 56508.28 & +147.47 & 22.50\,$\pm$\,0.19 & $i_\mathrm{PS1}$ & PS1 \\
  \hline
  2013-04-01 & 56383.57 & +31.56  & 20.86\,$\pm$\,0.12 & $z_\mathrm{PS1}$ & PS1 \\
  2013-04-07 & 56389.54 & +37.11  & 20.83\,$\pm$\,0.11 & $z_\mathrm{PS1}$ & PS1 \\
  2013-04-10 & 56392.51 & +39.88  & 20.67\,$\pm$\,0.10 & $z_\mathrm{PS1}$ & PS1 \\
  2013-04-15 & 56397.43 & +44.45  & 20.78\,$\pm$\,0.12 & $z_\mathrm{PS1}$ & PS1 \\
  2013-04-18 & 56400.59 & +47.38  & 20.79\,$\pm$\,0.14 & $z_\mathrm{PS1}$ & PS1 \\
  2013-04-20 & 56402.56 & +49.22  & 20.93\,$\pm$\,0.06 & $z_\mathrm{PS1}$ & PS1 \\
  2013-04-22 & 56404.55 & +51.06  & 20.89\,$\pm$\,0.07 & $z_\mathrm{PS1}$ & PS1 \\
  2013-04-28 & 56410.43 & +56.54  & 20.85\,$\pm$\,0.12 & $z_\mathrm{PS1}$ & PS1 \\
  2013-04-30 & 56412.45 & +58.41  & 20.79\,$\pm$\,0.07 & $z_\mathrm{PS1}$ & PS1 \\
  \hline  
  \end{tabular}
\end{table*}

\begin{table*}
  \small
  \caption*{Table \ref{table:PS1-13wr_photometry}: (continued)}
  \centering
  \begin{tabular}{c   c                      r            c                            c                        c           }
  \hline               
  \multirow{2}{*}{Date} & \multirow{2}{*}{MJD} & \multicolumn{1}{c}{Rest frame} & \multirow{2}{*}{Magnitude} & \multirow{2}{*}{Filter} & \multirow{2}{*}{Telescope} \\ 
                      &                      & \multicolumn{1}{c}{epoch$^*$}  &                            &                         &              \\
  \hline
  2013-05-03 & 56415.47 & +61.21  & 20.94\,$\pm$\,0.11 & $z_\mathrm{PS1}$ & PS1 \\
  2013-05-09 & 56421.50 & +66.82  & 20.84\,$\pm$\,0.05 & $z_\mathrm{PS1}$ & PS1 \\
  2013-05-17 & 56429.34 & +74.11  & 20.90\,$\pm$\,0.10 & $z_\mathrm{PS1}$ & PS1 \\
  2013-05-25 & 56437.42 & +81.62  & 21.08\,$\pm$\,0.51 & $z_\mathrm{PS1}$ & PS1 \\
  2013-05-30 & 56442.46 & +86.30  & 21.22\,$\pm$\,0.14 & $z_\mathrm{PS1}$ & PS1 \\
  2013-06-08 & 56451.36 & +94.57  & 21.70\,$\pm$\,0.21 & $z_\mathrm{PS1}$ & PS1 \\
  2013-06-13 & 56456.31 & +99.17  & 21.81\,$\pm$\,0.34 & $z_\mathrm{PS1}$ & PS1 \\
  2013-06-16 & 56459.32 & +101.97 & 21.97\,$\pm$\,0.16 & $z_\mathrm{PS1}$ & PS1 \\
  2013-06-18 & 56461.30 & +103.81 & 22.31\,$\pm$\,0.18 & $z_\mathrm{PS1}$ & PS1 \\
  2013-06-20 & 56463.31 & +105.67 & 22.19\,$\pm$\,0.15 & $z_\mathrm{PS1}$ & PS1 \\
  2013-06-23 & 56466.32 & +108.47 & 22.82\,$\pm$\,0.96 & $z_\mathrm{PS1}$ & PS1 \\
  2013-06-26 & 56469.35 & +111.29 & 22.28\,$\pm$\,0.49 & $z_\mathrm{PS1}$ & PS1 \\
  2013-06-28 & 56471.32 & +113.12 & 22.83\,$\pm$\,0.47 & $z_\mathrm{PS1}$ & PS1 \\
  2013-07-01 & 56474.30 & +115.89 & 22.45\,$\pm$\,0.39 & $z_\mathrm{PS1}$ & PS1 \\
  2013-07-04 & 56477.29 & +118.67 & 22.42\,$\pm$\,0.11 & $z_\mathrm{PS1}$ & PS1 \\
  2013-07-07 & 56480.29 & +121.46 & 22.75\,$\pm$\,0.63 & $z_\mathrm{PS1}$ & PS1 \\
  2013-07-09 & 56482.29 & +123.32 & 22.98\,$\pm$\,0.47 & $z_\mathrm{PS1}$ & PS1 \\
  2013-07-12 & 56485.27 & +126.09 & 22.53\,$\pm$\,0.53 & $z_\mathrm{PS1}$ & PS1 \\
  2013-07-15 & 56488.28 & +128.88 & 22.85\,$\pm$\,0.27 & $z_\mathrm{PS1}$ & PS1 \\
  2013-07-17 & 56490.27 & +130.73 & 22.54\,$\pm$\,0.17 & $z_\mathrm{PS1}$ & PS1 \\
  2013-07-19 & 56492.31 & +132.63 & 22.66\,$\pm$\,0.20 & $z_\mathrm{PS1}$ & PS1 \\
  2013-08-07 & 56511.27 & +150.25 & 22.86\,$\pm$\,0.31 & $z_\mathrm{PS1}$ & PS1 \\
  \hline
  2013-02-26 & 56349.60 & +0.00   & 20.41\,$\pm$\,0.21 & $y_\mathrm{PS1}$ & PS1 \\
  2013-03-01 & 56352.60 & +2.79   & 20.28\,$\pm$\,0.23 & $y_\mathrm{PS1}$ & PS1 \\
  2013-03-09 & 56360.65 & +10.26  & 20.12\,$\pm$\,0.20 & $y_\mathrm{PS1}$ & PS1 \\
  2013-03-13 & 56364.59 & +13.93  & 20.20\,$\pm$\,0.25 & $y_\mathrm{PS1}$ & PS1 \\
  2013-03-26 & 56377.52 & +25.94  & 20.64\,$\pm$\,0.61 & $y_\mathrm{PS1}$ & PS1 \\
  2013-03-29 & 56380.52 & +28.74  & 20.32\,$\pm$\,0.19 & $y_\mathrm{PS1}$ & PS1 \\
  2013-04-06 & 56388.48 & +36.13  & 20.06\,$\pm$\,0.23 & $y_\mathrm{PS1}$ & PS1 \\
  2013-04-18 & 56400.50 & +47.31  & 20.41\,$\pm$\,0.30 & $y_\mathrm{PS1}$ & PS1 \\
  2013-04-23 & 56405.55 & +52.00  & 20.53\,$\pm$\,0.20 & $y_\mathrm{PS1}$ & PS1 \\
  2013-04-24 & 56406.49 & +52.87  & 20.36\,$\pm$\,0.18 & $y_\mathrm{PS1}$ & PS1 \\
  2013-04-26 & 56408.42 & +54.66  & 20.37\,$\pm$\,0.19 & $y_\mathrm{PS1}$ & PS1 \\
  2013-04-27 & 56409.47 & +55.64  & 20.51\,$\pm$\,0.09 & $y_\mathrm{PS1}$ & PS1 \\
  2013-04-28 & 56410.48 & +56.57  & 20.47\,$\pm$\,0.52 & $y_\mathrm{PS1}$ & PS1 \\
  2013-04-30 & 56412.43 & +58.39  & 20.48\,$\pm$\,0.10 & $y_\mathrm{PS1}$ & PS1 \\
  2013-05-05 & 56417.51 & +63.11  & 20.28\,$\pm$\,0.15 & $y_\mathrm{PS1}$ & PS1 \\
  2013-05-17 & 56429.40 & +74.16  & 20.37\,$\pm$\,0.28 & $y_\mathrm{PS1}$ & PS1 \\
  2013-05-23 & 56435.44 & +79.77  & 20.93\,$\pm$\,0.32 & $y_\mathrm{PS1}$ & PS1 \\
  2013-05-24 & 56436.44 & +80.70  & 20.74\,$\pm$\,0.16 & $y_\mathrm{PS1}$ & PS1 \\
  2013-05-25 & 56437.45 & +81.64  & 20.63\,$\pm$\,0.74 & $y_\mathrm{PS1}$ & PS1 \\
  2013-05-27 & 56439.42 & +83.47  & 20.59\,$\pm$\,0.30 & $y_\mathrm{PS1}$ & PS1 \\
  2013-06-21 & 56464.32 & +106.61 & 21.73\,$\pm$\,0.22 & $y_\mathrm{PS1}$ & PS1 \\
  2013-06-23 & 56466.34 & +108.49 & 21.72\,$\pm$\,0.72 & $y_\mathrm{PS1}$ & PS1 \\
  2013-06-24 & 56467.34 & +109.42 & 21.52\,$\pm$\,0.66 & $y_\mathrm{PS1}$ & PS1 \\
  2013-06-25 & 56468.34 & +110.35 & 21.91\,$\pm$\,0.96 & $y_\mathrm{PS1}$ & PS1 \\
  2013-07-11 & 56484.30 & +125.19 & 21.90\,$\pm$\,0.94 & $y_\mathrm{PS1}$ & PS1 \\
  \hline  
  \end{tabular}
  \\[1.5ex]
  \flushleft
  The tabulated magnitudes are given ``as observed'', i.e. neither corrected for dust extinction nor $K$-corrected. $^*$Rest frame epochs (assuming a redshift of 0.076) with respect to the first detection on 56349.60 (MJD). 
PS1 = Panoramic Survey Telescope \& Rapid Response System 1.
\end{table*} 

Table~\ref{table:PS1-13wr_journal_spectra} lists the journal of spectroscopic observations for PS1-13wr. The calibrated spectra of PS1-13wr are presented in the right panel of Figure~\ref{figure:PS1-13wr}. They are corrected for galactic and host galaxy reddening as well as redshift ($z$ = 0.076). 
The spectra display the typical features of H$\alpha$ and H$\beta$. The H$\alpha$ lines in all four spectra have a clear P-Cygni profile shape. It additionally displays a narrow emission component which is likely a contamination from the host galaxy. Weak lines of iron, in particular Fe\,{\sc ii}\,$\lambda$5169 are visible in all four spectra. 

\begin{table*}
  \caption{PS1-13wr: Journal of spectroscopic observations}
  \label{table:PS1-13wr_journal_spectra}
  \centering
  \begin{tabular}{c c     c            c              c            c                               }
  \hline 
\multirow{2}{*}{Date} & \multirow{2}{*}{MJD} & Epoch$^*$  & Wavelength   & Resolution & \multirow{2}{*}{Telescope+Instrument} \\
             &           & rest frame & range in \AA &  \AA       &                               \\
\hline 
  2013-03-11 & 56362.26  & +11.77 & 3600 - 9970 &  17.0      & GTC+OSIRIS+R300B              \\
  2013-03-17 & 56369.12  & +18.14 & 3770 - 9970 &  17.0      & GTC+OSIRIS+R300B              \\
  2013-03-25 & 56376.23  & +24.74 & 3730 - 9865 &  16.7      & GTC+OSIRIS+R300B              \\
  2013-04-11 & 56394.15  & +41.40 & 3640 - 9895 &  16.9      & GTC+OSIRIS+R300B              \\
  \hline  
  \end{tabular}
  \\[1.5ex]
  \flushleft
  $^*$Rest frame epochs (assuming a redshift of 0.076) with respect to the first detection on 56349.60 (MJD). The resolution was determined from the FWHM of the O\,{\sc i} $\lambda$5577.34 sky line.
\end{table*}

\subsection{PS1-14vk}
PS1-14vk was discovered on 2014 March 24 by the PS1 3$\pi$ survey. We obtained LT $g'r'i'$ photometry up to 68\,days after discovery and 6 GTC spectra ranging from +2 to +49\,days after first detection. 

PS1-14vk lies in the vicinity of the galaxy SDSS J121045.27+484143.9 for which the SDSS reports a spectroscopic redshift of 0.080\,$\pm$\,0.001, which we adopt also for PS1-14vk.

Features attributable to the Na\,{\sc i} doublet from interstellar gas either in the Milky Way or the host galaxy, are not apparent in the spectra of PS1-14vk. We derived an upper limit for the equivalent width of the Na\,{\sc i}\,D $\lambda\lambda$\,5890,5896 blend in the same manner as for SN 2013ca (Section~\ref{section:Observations:SN2013eq}) and we find EW$_\mathrm{Na\,I\,D}$ $<$\,0.7\,{\AA}, which translates into a host extinction of $E(B-V)_{\mathrm{host}} <$\,0.09\,mag \citep[Equation~9 of][]{Poznanski2012}.

The light curves of PS1-14vk are presented in the left panel of Figure \ref{figure:PS1-14vk}. Table \ref{table:PS1-14vk_photometry} shows the log of imaging observations and the calibrated magnitudes.
PS1-14vk declines linearly in the $g'$-, $r'$- and $i'$-band with an averaged decline rate of 1.16\,$\pm$\,0.23\,mag/50\,d in the $r'$-band. We classify PS1-14vk as a Type II-L SN, following \citet{Li2011a}.

\begin{figure*}[t!]
   \centering
   \begin{subfigure}[t]{0.49\textwidth}
      \includegraphics[width=\columnwidth]{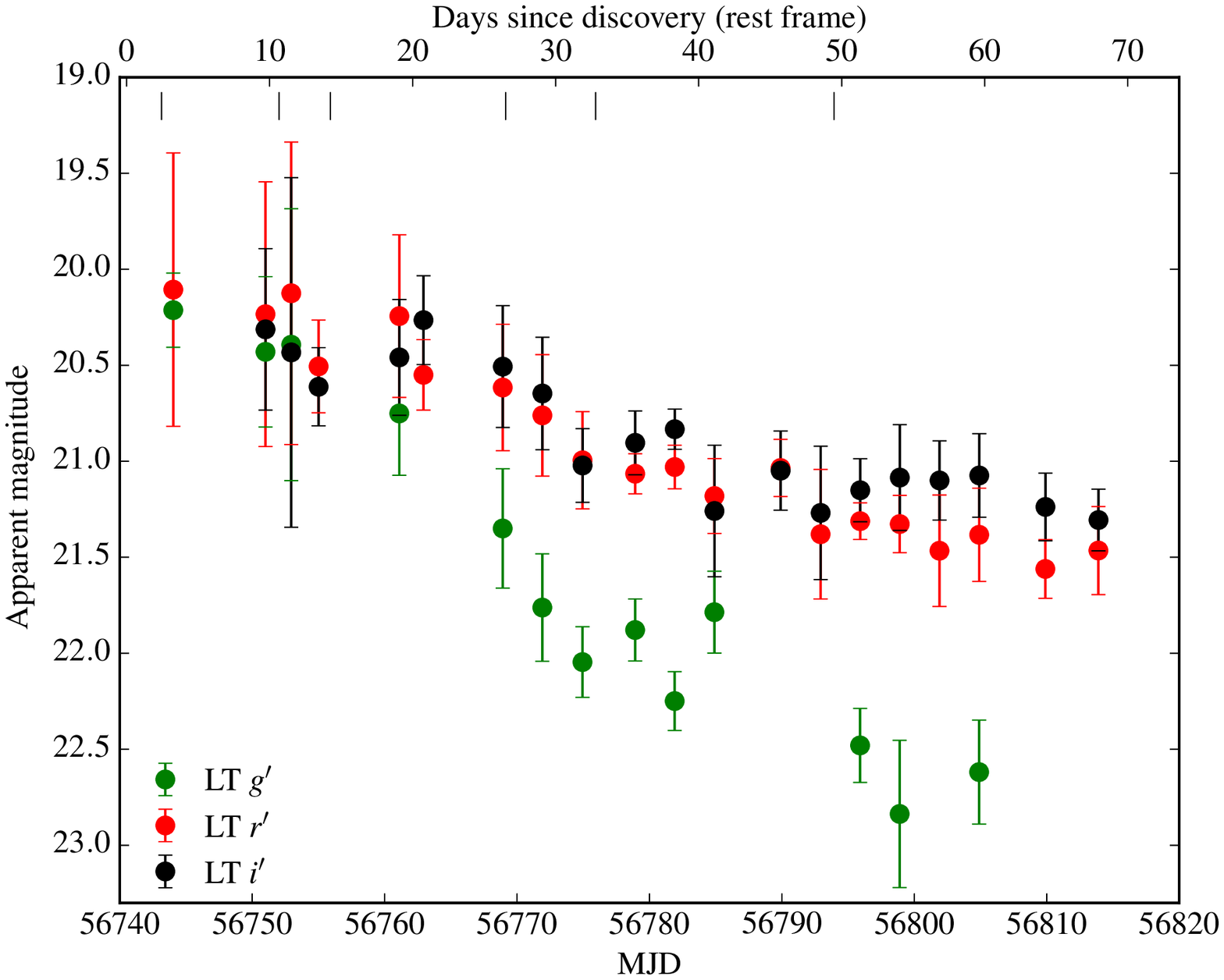}
   \end{subfigure}%
   \begin{subfigure}[t]{0.49\textwidth}
      \includegraphics[width=\columnwidth]{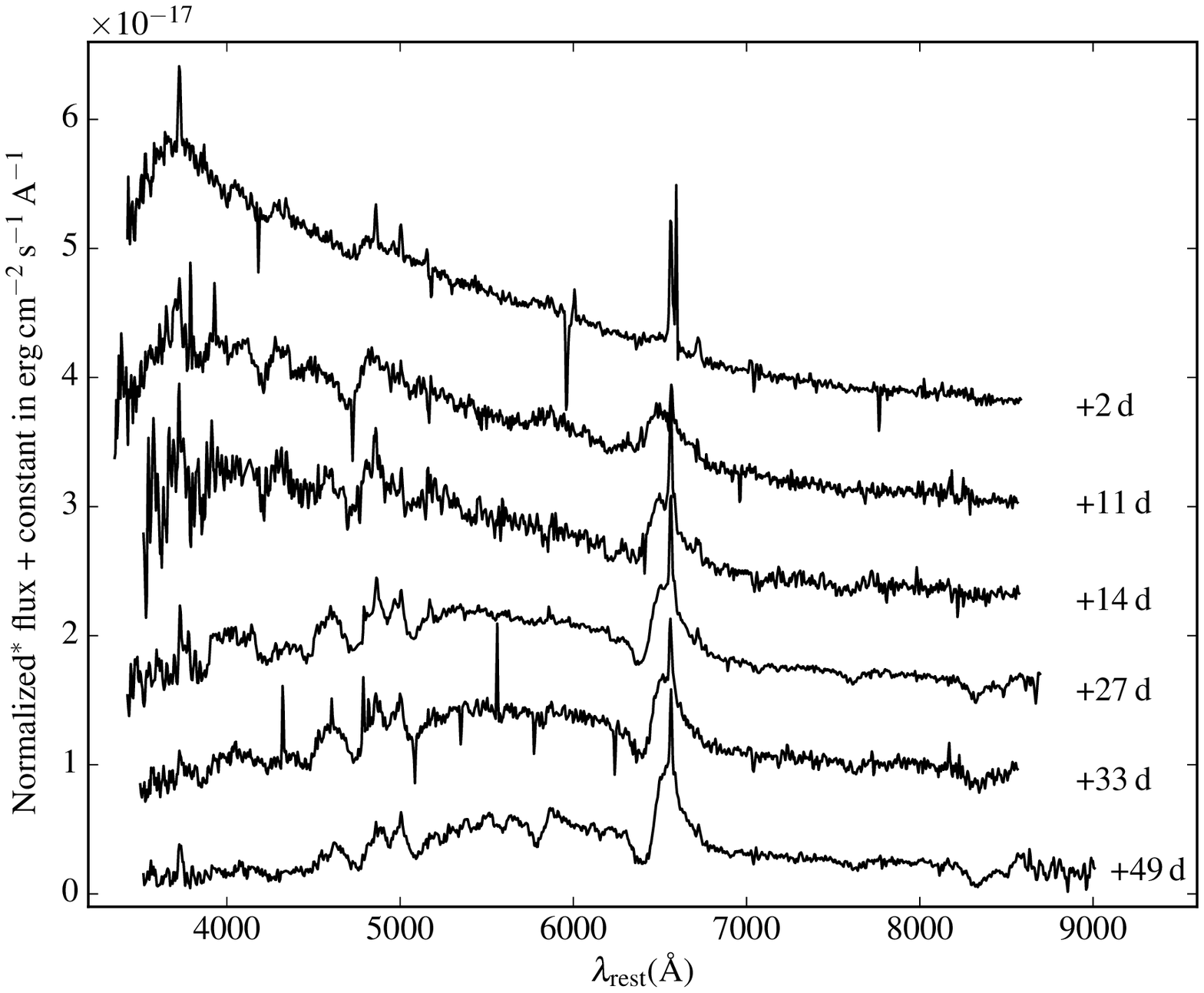}
\end{subfigure}%
   \caption{PS1-14vk. Left panel: LT $g'r'i'$ light curves of PS1-14vk. The vertical ticks on the top mark the epochs of the observed spectra. Right panel: PS1-14vk spectroscopy. $^*$Flux normalized to the maximum H$\alpha$ flux for better visibility of the features. The exact normalizations are: 
flux/2.6 for the +2\,d spectrum;
flux/2.7 for +11\,d;
flux/2.8 for +14\,d;
flux/2.3 for +27\,d;
flux/2.1 for +33\,d;
flux/1.9 for +49\,d.
}
   \label{figure:PS1-14vk}
\end{figure*}

\begin{table*}
  \small
  \caption{PS1-14vk: Photometric observations}
  \label{table:PS1-14vk_photometry}
  \centering
  \begin{tabular}{c   c                      r            c                            c                        c           }
  \hline               
  \multirow{2}{*}{Date} & \multirow{2}{*}{MJD} & \multicolumn{1}{c}{Rest frame} & \multirow{2}{*}{Magnitude} & \multirow{2}{*}{Filter} & \multirow{2}{*}{Telescope} \\ 
                      &                      & \multicolumn{1}{c}{epoch$^*$}  &                            &                         &              \\
  \hline 
2014-03-27 & 56744.04 &  +3.26 & 20.21\,$\pm$\,0.19 & $g^\prime$ & LT \\
2014-04-03 & 56751.00 &  +9.71 & 20.43\,$\pm$\,0.39 & $g^\prime$ & LT \\
2014-04-05 & 56752.95 & +11.51 & 20.39\,$\pm$\,0.71 & $g^\prime$ & LT \\
2014-04-13 & 56761.11 & +19.07 & 20.75\,$\pm$\,0.32 & $g^\prime$ & LT \\
2014-04-21 & 56768.91 & +26.30 & 21.35\,$\pm$\,0.31 & $g^\prime$ & LT \\
2014-04-24 & 56771.91 & +29.07 & 21.76\,$\pm$\,0.28 & $g^\prime$ & LT \\
2014-04-27 & 56774.94 & +31.88 & 22.05\,$\pm$\,0.18 & $g^\prime$ & LT \\
2014-05-01 & 56778.92 & +35.56 & 21.88\,$\pm$\,0.16 & $g^\prime$ & LT \\
2014-05-04 & 56781.90 & +38.32 & 22.25\,$\pm$\,0.15 & $g^\prime$ & LT \\
2014-05-07 & 56784.90 & +41.10 & 21.79\,$\pm$\,0.21 & $g^\prime$ & LT \\
2014-05-18 & 56795.90 & +51.29 & 22.48\,$\pm$\,0.19 & $g^\prime$ & LT \\
2014-05-21 & 56798.89 & +54.05 & 22.84\,$\pm$\,0.38 & $g^\prime$ & LT \\
2014-05-27 & 56804.89 & +59.61 & 22.62\,$\pm$\,0.27 & $g^\prime$ & LT \\
\hline
2014-03-27 & 56744.04 &  +3.27 & 20.11\,$\pm$\,0.71 & $r^\prime$ & LT \\
2014-04-03 & 56751.00 &  +9.71 & 20.23\,$\pm$\,0.69 & $r^\prime$ & LT \\
2014-04-05 & 56752.95 & +11.51 & 20.12\,$\pm$\,0.79 & $r^\prime$ & LT \\
2014-04-07 & 56755.01 & +13.43 & 20.51\,$\pm$\,0.24 & $r^\prime$ & LT \\
2014-04-13 & 56761.11 & +19.07 & 20.24\,$\pm$\,0.42 & $r^\prime$ & LT \\
2014-04-15 & 56762.93 & +20.76 & 20.55\,$\pm$\,0.18 & $r^\prime$ & LT \\
2014-04-21 & 56768.92 & +26.30 & 20.62\,$\pm$\,0.33 & $r^\prime$ & LT \\
2014-04-24 & 56771.91 & +29.07 & 20.76\,$\pm$\,0.32 & $r^\prime$ & LT \\
2014-04-27 & 56774.94 & +31.88 & 21.00\,$\pm$\,0.25 & $r^\prime$ & LT \\
2014-05-01 & 56778.92 & +35.56 & 21.07\,$\pm$\,0.10 & $r^\prime$ & LT \\
2014-05-04 & 56781.91 & +38.33 & 21.03\,$\pm$\,0.11 & $r^\prime$ & LT \\
2014-05-07 & 56784.90 & +41.10 & 21.18\,$\pm$\,0.19 & $r^\prime$ & LT \\
2014-05-12 & 56789.90 & +45.73 & 21.04\,$\pm$\,0.15 & $r^\prime$ & LT \\
2014-05-15 & 56792.92 & +48.53 & 21.38\,$\pm$\,0.34 & $r^\prime$ & LT \\
2014-05-18 & 56795.91 & +51.29 & 21.31\,$\pm$\,0.10 & $r^\prime$ & LT \\
2014-05-21 & 56798.90 & +54.06 & 21.33\,$\pm$\,0.15 & $r^\prime$ & LT \\
2014-05-24 & 56801.90 & +56.84 & 21.47\,$\pm$\,0.29 & $r^\prime$ & LT \\
2014-05-27 & 56804.90 & +59.62 & 21.38\,$\pm$\,0.24 & $r^\prime$ & LT \\
2014-06-01 & 56809.89 & +64.24 & 21.56\,$\pm$\,0.15 & $r^\prime$ & LT \\
2014-06-05 & 56813.90 & +67.95 & 21.46\,$\pm$\,0.23 & $r^\prime$ & LT \\
\hline
2014-04-03 & 56751.00 &  +9.71 & 20.31\,$\pm$\,0.42 & $i^\prime$ & LT \\
2014-04-05 & 56752.95 & +11.51 & 20.43\,$\pm$\,0.91 & $i^\prime$ & LT \\
2014-04-07 & 56755.01 & +13.43 & 20.61\,$\pm$\,0.20 & $i^\prime$ & LT \\
2014-04-13 & 56761.11 & +19.07 & 20.46\,$\pm$\,0.30 & $i^\prime$ & LT \\
2014-04-15 & 56762.93 & +20.76 & 20.27\,$\pm$\,0.23 & $i^\prime$ & LT \\
2014-04-21 & 56768.92 & +26.30 & 20.51\,$\pm$\,0.32 & $i^\prime$ & LT \\
2014-04-24 & 56771.91 & +29.07 & 20.65\,$\pm$\,0.29 & $i^\prime$ & LT \\
2014-04-27 & 56774.95 & +31.88 & 21.02\,$\pm$\,0.19 & $i^\prime$ & LT \\
2014-05-01 & 56778.92 & +35.56 & 20.90\,$\pm$\,0.17 & $i^\prime$ & LT \\
2014-05-04 & 56781.91 & +38.33 & 20.83\,$\pm$\,0.10 & $i^\prime$ & LT \\
2014-05-07 & 56784.91 & +41.11 & 21.26\,$\pm$\,0.34 & $i^\prime$ & LT \\
2014-05-12 & 56789.90 & +45.73 & 21.05\,$\pm$\,0.21 & $i^\prime$ & LT \\
2014-05-15 & 56792.93 & +48.53 & 21.27\,$\pm$\,0.35 & $i^\prime$ & LT \\
2014-05-18 & 56795.91 & +51.30 & 21.15\,$\pm$\,0.16 & $i^\prime$ & LT \\
2014-05-21 & 56798.90 & +54.06 & 21.09\,$\pm$\,0.28 & $i^\prime$ & LT \\
2014-05-24 & 56801.90 & +56.84 & 21.10\,$\pm$\,0.21 & $i^\prime$ & LT \\
2014-05-27 & 56804.90 & +59.62 & 21.07\,$\pm$\,0.22 & $i^\prime$ & LT \\
2014-06-01 & 56809.90 & +64.25 & 21.24\,$\pm$\,0.18 & $i^\prime$ & LT \\
2014-06-05 & 56813.91 & +67.96 & 21.31\,$\pm$\,0.16 & $i^\prime$ & LT \\
  \hline  
  \end{tabular}
  \\[1.5ex]
  \flushleft
  The tabulated magnitudes are given ``as observed'', i.e. neither corrected for dust extinction nor $K$-corrected. $^*$Rest frame epochs (assuming a redshift of 0.080) with respect to the first detection on 56740.51 (MJD). 
LT = Liverpool Telescope.
\end{table*}

Table~\ref{table:PS1-14vk_journal_spectra} gives the journal of spectroscopic observations for PS1-14vk. The fully reduced and calibrated spectra of PS1-14vk are presented in the right panel of Figure \ref{figure:PS1-14vk}. They are corrected for galactic reddening and redshift ($z$ = 0.080). 
The spectra at all epochs show features of H$\alpha$ and H$\beta$. Narrow H$\alpha$ emission is visible at all epochs. We attribute this to contamination by the host galaxy, due to the projected distance of the SN of only $\sim$ 4.5\,kpc from the center of the host galaxy. H$\alpha$ displays only a very weak absorption until +14\,d after discovery, as is also observed by \citet{Gutierrez2014} for other SNe II-L. At later epochs the H$\alpha$ absorption becomes more prominent. A faint Fe\,{\sc ii}\,$\lambda$5169 line can be seen in the spectra after +11\,days.

\begin{table*}
  \caption{PS1-14vk: Journal of spectroscopic observations}
  \label{table:PS1-14vk_journal_spectra}
  \centering
  \begin{tabular}{c c     c            c              c            c                               }
  \hline 
\multirow{2}{*}{Date} & \multirow{2}{*}{MJD} & Epoch$^*$  & Wavelength   & Resolution & \multirow{2}{*}{Telescope+Instrument} \\
             &           & rest frame & range in \AA &  \AA       &                               \\
\hline 
  2014-03-26 & 56743.15  &  +2.44 & 3700 - 9290 &  17.0      & GTC+OSIRIS+R300B              \\
  2014-04-04 & 56752.03  & +10.66 & 3620 - 9270 &  16.7      & GTC+OSIRIS+R300B              \\
  2014-04-08 & 56755.91  & +14.26 & 3800 - 9280 &  16.5      & GTC+OSIRIS+R300B              \\
  2014-04-21 & 56769.13  & +26.50 & 3700 - 9410 &  16.6      & GTC+OSIRIS+R300B              \\
  2014-04-28 & 56775.93  & +32.79 & 3780 - 9270 &  17.0      & GTC+OSIRIS+R300B              \\
  2014-05-16 & 56793.93  & +49.46 & 3800 - 9750 &  16.8      & GTC+OSIRIS+R300B              \\
  \hline  
  \end{tabular}
  \\[1.5ex]
  \flushleft
  $^*$Rest frame epochs (assuming a redshift of 0.080) with respect to the first detection on 56740.51 (MJD). The resolution was determined from the FWHM of the O\,{\sc i} $\lambda$5577.34 sky line.
\end{table*}

\subsection{PS1-12bku}
PS1-12bku was discovered on 2012 August 22 by the PS1 MD survey. $g_{\mathrm{PS1}}r_{\mathrm{PS1}}i_{\mathrm{PS1}}z_{\mathrm{PS1}}y_{\mathrm{PS1}}$ photometry was obtained in the course of the PS1 MD survey and NOT $u'g'r'i'z'$ photometry up to 114\,days after the first detection. We obtained four GTC spectra ranging from +23 to +81\,days. 

PS1-12bku displays neither narrow emissions from the host galaxy nor is a spectroscopic redshift for its host galaxy available in a public data base. With the objective of estimating the redshift of PS1-12bku, we performed a series of cross correlations using SNID \citep[Supernova Identification,][]{Blondin2007}. The redshifts range from 0.086 to 0.094 with a median of 0.088. We adopt this value as the redshift of the SN, and give the range in redshifts mentioned above as a conservative estimate of its uncertainty: $z = 0.088^{+0.006}_{-0.002}$.

No features attributable to the Na\,{\sc i} doublet from interstellar gas either in the Milky Way or the host galaxy, are apparent in the spectra of PS1-12bku. An upper limit for the equivalent width of the Na\,{\sc i}\,D $\lambda\lambda$\,5890,5896 blend was derived in the same manner as for SN~2013ca (see Section \ref{section:Observations:SN2013eq}). We find EW$_\mathrm{Na\,I\,D}$ $<$\,1.1\,{\AA}, which translates into a host extinction of \mbox{$E(B-V)_{\mathrm{host}}$}\,$<$\,0.27\,mag \citep[Equation 9 of][]{Poznanski2012}.

The light curves of PS1-12bku are presented in the left panel of Figure \ref{figure:PS1-12bku}. Table \ref{table:PS1-12bku_photometry} shows the log of imaging observations and the calibrated magnitudes.
PS1-12bku was discovered only 0.8\,d after the last non-detection in the $z_\mathrm{PS1}$-band on MJD 56160.51. This allows us to constrain the explosion epoch of PS1-12bku to MJD 56160.9\,$\pm$\,0.4. The $r_{\mathrm{PS1}}$-, $i_{\mathrm{PS1}}$-, $z_{\mathrm{PS1}}$-, $y_{\mathrm{PS1}}$ light curves display a plateau that to lasts until about 100 days after discovery where the $r$-band data suggests a drop in luminosity. 

\begin{figure*}[t!]
   \centering
   \begin{subfigure}[t]{0.49\textwidth}
      \includegraphics[width=\columnwidth]{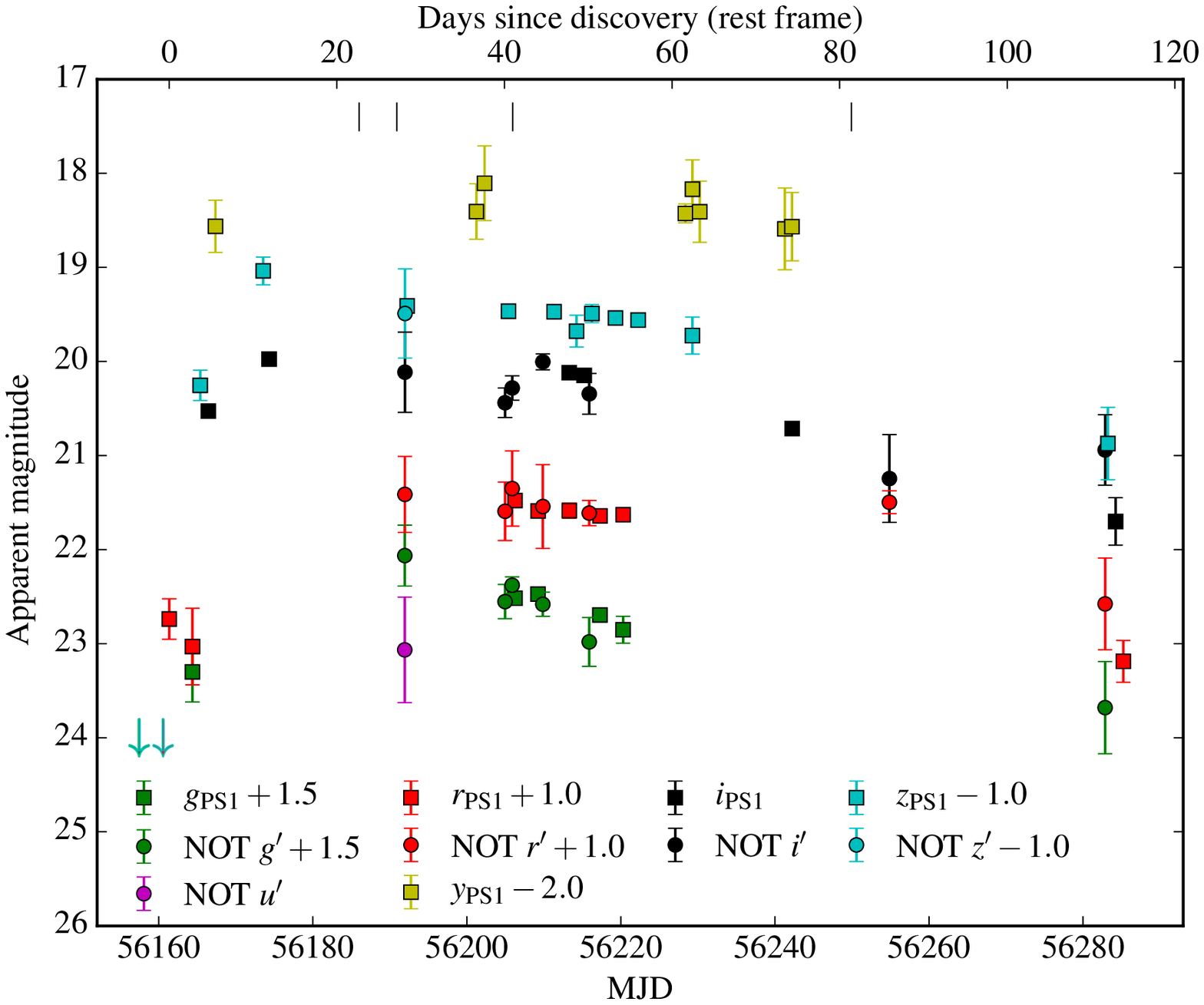}
   \end{subfigure}%
   \begin{subfigure}[t]{0.49\textwidth}
      \includegraphics[width=\columnwidth]{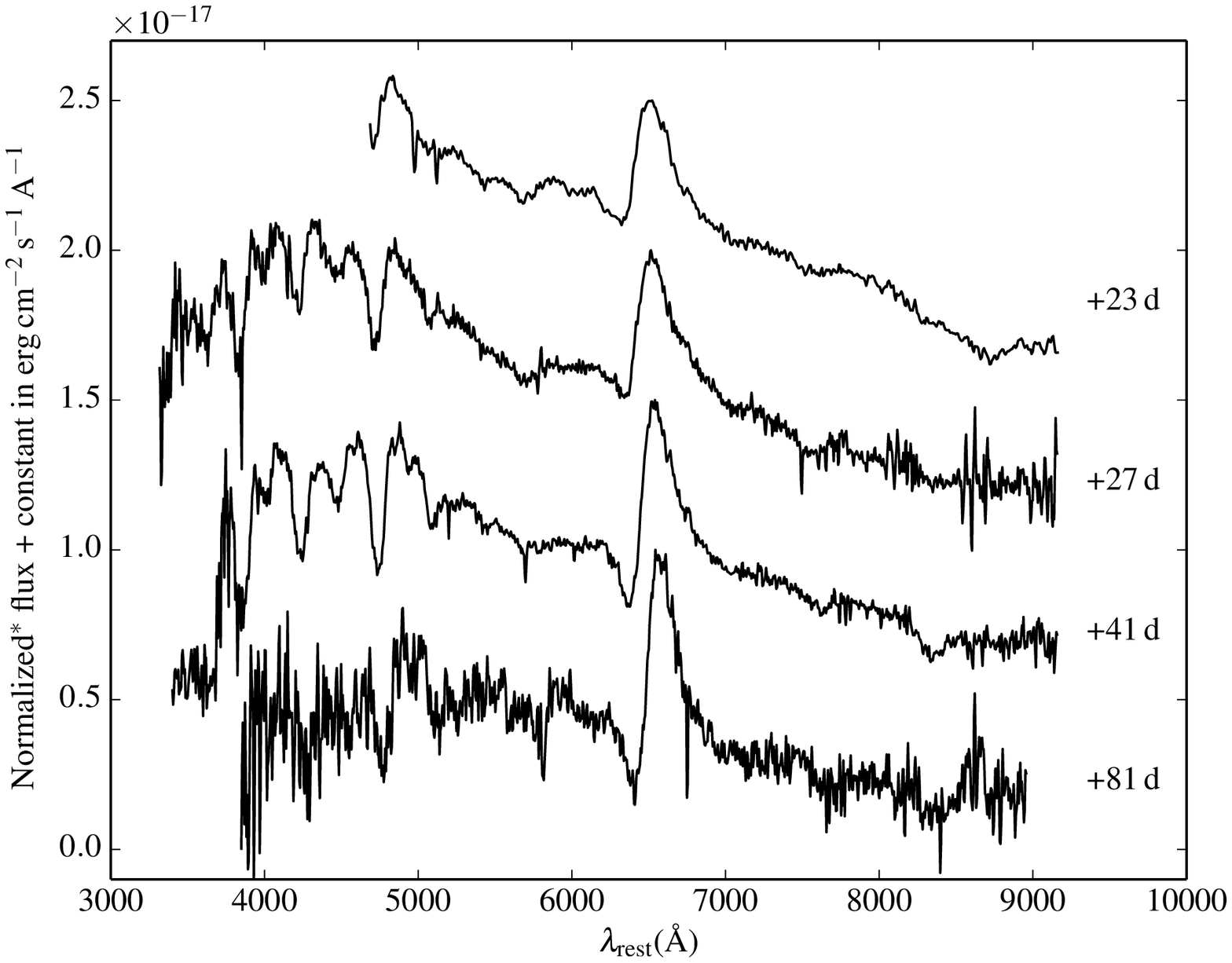}
\end{subfigure}%
   \caption{PS1-12bku. Left panel: $grizy_\mathrm{PS1}$ and NOT $u'g'r'i'z'$ light curves of PS1-12bku. The vertical ticks on the top mark the epochs of the observed spectra. Right panel: PS1-12bku spectroscopy. $^*$Flux normalized to the maximum H$\alpha$ flux for better visibility of the features. The exact normalizations are: 
flux/3.3 for the +23\,d spectrum;
flux/3.6 for +27\,d;
flux/3.5 for +41\,d;
flux/2.9 for +81\,d.
}
   \label{figure:PS1-12bku}
\end{figure*}

\begin{table*}
  \small
  \caption{PS1-12bku: Photometric observations}
  \label{table:PS1-12bku_photometry}
  \centering
  \begin{tabular}{c   c                      r            c                            c                        c           }
  \hline               
  \multirow{2}{*}{Date} & \multirow{2}{*}{MJD} & \multicolumn{1}{c}{Rest frame} & \multirow{2}{*}{Magnitude} & \multirow{2}{*}{Filter} & \multirow{2}{*}{Telescope} \\ 
                      &                      & \multicolumn{1}{c}{epoch$^*$}  &                            &                         &              \\
  \hline 
  2012-09-21 & 56191.93 &  +28.10 & 23.07\,$\pm$\,0.56 & $u$ & NOT \\
  \hline 
  2012-08-25 & 56164.36 &   +2.76 & 21.80\,$\pm$\,0.32 & $g_\mathrm{PS1}$ & PS1 \\
  2012-10-06 & 56206.24 &  +41.25 & 21.01\,$\pm$\,0.06 & $g_\mathrm{PS1}$ & PS1 \\
  2012-10-09 & 56209.24 &  +44.01 & 20.97\,$\pm$\,0.04 & $g_\mathrm{PS1}$ & PS1 \\
  2012-10-17 & 56217.28 &  +51.40 & 21.20\,$\pm$\,0.07 & $g_\mathrm{PS1}$ & PS1 \\
  2012-10-20 & 56220.28 &  +54.16 & 21.35\,$\pm$\,0.14 & $g_\mathrm{PS1}$ & PS1 \\
\hline 
  2012-09-21 & 56191.94 &  +28.11 & 20.56\,$\pm$\,0.32 & $g^\prime$ & NOT \\
  2012-10-04 & 56204.95 &  +40.07 & 21.05\,$\pm$\,0.18 & $g^\prime$ & NOT \\
  2012-10-05 & 56205.88 &  +40.92 & 20.88\,$\pm$\,0.09 & $g^\prime$ & NOT \\
  2012-10-09 & 56209.86 &  +44.58 & 21.08\,$\pm$\,0.13 & $g^\prime$ & NOT \\
  2012-10-15 & 56215.88 &  +50.12 & 21.48\,$\pm$\,0.26 & $g^\prime$ & NOT \\
  2012-12-21 & 56282.86 & +111.68 & 22.18\,$\pm$\,0.49 & $g^\prime$ & NOT \\
  \hline 
  2012-08-22 & 56161.36 &   +0.00 & 21.74\,$\pm$\,0.22 & $r_\mathrm{PS1}$ & PS1 \\
  2012-08-25 & 56164.37 &   +2.77 & 22.03\,$\pm$\,0.41 & $r_\mathrm{PS1}$ & PS1 \\
  2012-10-06 & 56206.25 &  +41.26 & 20.48\,$\pm$\,0.03 & $r_\mathrm{PS1}$ & PS1 \\
  2012-10-09 & 56209.25 &  +44.02 & 20.59\,$\pm$\,0.04 & $r_\mathrm{PS1}$ & PS1 \\
  2012-10-13 & 56213.31 &  +47.75 & 20.59\,$\pm$\,0.08 & $r_\mathrm{PS1}$ & PS1 \\
  2012-10-17 & 56217.29 &  +51.41 & 20.64\,$\pm$\,0.05 & $r_\mathrm{PS1}$ & PS1 \\
  2012-10-20 & 56220.30 &  +54.17 & 20.63\,$\pm$\,0.07 & $r_\mathrm{PS1}$ & PS1 \\
  2012-12-24 & 56285.22 & +113.85 & 22.19\,$\pm$\,0.22 & $r_\mathrm{PS1}$ & PS1 \\
  \hline 
  2012-09-21 & 56191.95 &  +28.12 & 20.41\,$\pm$\,0.40 & $r^\prime$ & NOT \\
  2012-10-04 & 56204.96 &  +40.07 & 20.59\,$\pm$\,0.31 & $r^\prime$ & NOT \\
  2012-10-05 & 56205.88 &  +40.92 & 20.35\,$\pm$\,0.40 & $r^\prime$ & NOT \\
  2012-10-09 & 56209.86 &  +44.58 & 20.54\,$\pm$\,0.45 & $r^\prime$ & NOT \\
  2012-10-15 & 56215.89 &  +50.12 & 20.61\,$\pm$\,0.13 & $r^\prime$ & NOT \\
  2012-11-23 & 56254.86 &  +85.94 & 20.50\,$\pm$\,0.12 & $r^\prime$ & NOT \\
  2012-12-21 & 56282.87 & +111.68 & 21.58\,$\pm$\,0.49 & $r^\prime$ & NOT \\
  \hline 
  2012-08-27 & 56166.43 &   +4.66 & 20.53\,$\pm$\,0.05 & $i_\mathrm{PS1}$ & PS1 \\
  2012-09-04 & 56174.33 &  +11.92 & 19.98\,$\pm$\,0.03 & $i_\mathrm{PS1}$ & PS1 \\
  2012-10-13 & 56213.29 &  +47.73 & 20.12\,$\pm$\,0.06 & $i_\mathrm{PS1}$ & PS1 \\
  2012-10-15 & 56215.23 &  +49.51 & 20.15\,$\pm$\,0.06 & $i_\mathrm{PS1}$ & PS1 \\
  2012-11-11 & 56242.23 &  +74.34 & 20.71\,$\pm$\,0.06 & $i_\mathrm{PS1}$ & PS1 \\
  2012-12-23 & 56284.22 & +112.93 & 21.70\,$\pm$\,0.25 & $i_\mathrm{PS1}$ & PS1 \\
  \hline 
  2012-09-21 & 56191.96 &  +28.13 & 20.11\,$\pm$\,0.43 & $i^\prime$ & NOT \\
  2012-10-04 & 56204.96 &  +40.08 & 20.44\,$\pm$\,0.16 & $i^\prime$ & NOT \\
  2012-10-05 & 56205.88 &  +40.92 & 20.28\,$\pm$\,0.13 & $i^\prime$ & NOT \\
  2012-10-09 & 56209.87 &  +44.59 & 20.00\,$\pm$\,0.08 & $i^\prime$ & NOT \\
  2012-10-15 & 56215.89 &  +50.12 & 20.34\,$\pm$\,0.22 & $i^\prime$ & NOT \\
  2012-11-23 & 56254.86 &  +85.94 & 21.25\,$\pm$\,0.47 & $i^\prime$ & NOT \\
  2012-12-21 & 56282.87 & +111.68 & 20.94\,$\pm$\,0.37 & $i^\prime$ & NOT \\
\hline 
  2012-08-26 & 56165.38 &   +3.70 & 21.25\,$\pm$\,0.16 & $z_\mathrm{PS1}$ & PS1 \\
  2012-09-03 & 56173.55 &  +11.21 & 20.04\,$\pm$\,0.18 & $z_\mathrm{PS1}$ & PS1 \\
  2012-09-22 & 56192.26 &  +28.40 & 20.41\,$\pm$\,0.06 & $z_\mathrm{PS1}$ & PS1 \\
  2012-10-05 & 56205.40 &  +40.48 & 20.47\,$\pm$\,0.02 & $z_\mathrm{PS1}$ & PS1 \\
  2012-10-11 & 56211.33 &  +45.93 & 20.47\,$\pm$\,0.04 & $z_\mathrm{PS1}$ & PS1 \\
  2012-10-14 & 56214.25 &  +48.61 & 20.68\,$\pm$\,0.17 & $z_\mathrm{PS1}$ & PS1 \\
  2012-10-16 & 56216.23 &  +50.43 & 20.49\,$\pm$\,0.10 & $z_\mathrm{PS1}$ & PS1 \\
  2012-10-19 & 56219.29 &  +53.25 & 20.54\,$\pm$\,0.04 & $z_\mathrm{PS1}$ & PS1 \\
  2012-10-22 & 56222.24 &  +55.96 & 20.56\,$\pm$\,0.04 & $z_\mathrm{PS1}$ & PS1 \\
  2012-10-29 & 56229.29 &  +62.43 & 20.73\,$\pm$\,0.20 & $z_\mathrm{PS1}$ & PS1 \\
  2012-12-22 & 56283.23 & +112.02 & 21.87\,$\pm$\,0.38 & $z_\mathrm{PS1}$ & PS1 \\
  \hline 
  2012-09-21 & 56191.97 &  +28.14 & 20.49\,$\pm$\,0.47 & $z^\prime$ & NOT \\
  \hline 
  2012-08-28 & 56167.36 &   +5.52 & 20.56\,$\pm$\,0.28 & $y_\mathrm{PS1}$ & PS1 \\
  2012-10-01 & 56201.24 &  +36.65 & 20.41\,$\pm$\,0.29 & $y_\mathrm{PS1}$ & PS1 \\
  2012-10-02 & 56202.30 &  +37.63 & 20.11\,$\pm$\,0.40 & $y_\mathrm{PS1}$ & PS1 \\
  2012-10-28 & 56228.36 &  +61.59 & 20.43\,$\pm$\,0.10 & $y_\mathrm{PS1}$ & PS1 \\
  2012-10-29 & 56229.29 &  +62.44 & 20.17\,$\pm$\,0.31 & $y_\mathrm{PS1}$ & PS1 \\
  2012-10-30 & 56230.23 &  +63.30 & 20.41\,$\pm$\,0.33 & $y_\mathrm{PS1}$ & PS1 \\
  2012-11-10 & 56241.29 &  +73.46 & 20.59\,$\pm$\,0.43 & $y_\mathrm{PS1}$ & PS1 \\
  2012-11-11 & 56242.20 &  +74.31 & 20.57\,$\pm$\,0.36 & $y_\mathrm{PS1}$ & PS1 \\
  \hline  
  \end{tabular}
  \\[1.5ex]
  \flushleft
  The tabulated magnitudes are given ``as observed'', i.e. neither corrected for dust extinction nor $K$-corrected. $^*$Rest frame epochs (assuming a redshift of 0.088) with respect to the discovery on 56161.36 (MJD). 
NOT = Nordic Optical Telescope; PS1 = Panoramic Survey Telescope \& Rapid Response System 1. 
\end{table*}

Table~\ref{table:PS1-12bku_journal_spectra} gives the journal of spectroscopic observations for PS1-12bku. The calibrated spectra of PS1-12bku are presented in the right panel of Figure \ref{figure:PS1-12bku}. They are corrected for galactic reddening and redshift ($z$ = 0.088). 
The spectra of PS1-12bku display strong H$\alpha$ and H$\beta$ lines. A weak Fe\,{\sc ii}\,$\lambda$5169 absorption can also be seen at all epochs. 

\begin{table*}
  \caption{PS1-12bku: Journal of spectroscopic observations}
  \label{table:PS1-12bku_journal_spectra}
  \centering
  \begin{tabular}{c c     c            c              c            c                               }
  \hline 
\multirow{2}{*}{Date} & \multirow{2}{*}{MJD} & Epoch$^*$  & Wavelength   & Resolution & \multirow{2}{*}{Telescope+Instrument} \\
             &           & rest frame & range in \AA &  \AA       &                               \\
\hline 
  2012-09-16 & 56186.01  & +22.66 & 5100 - 9999  &  31.2      & GTC+OSIRIS+R300R              \\
  2012-09-20 & 56190.90  & +27.15 & 3400 - 9965  &  21.3      & GTC+OSIRIS+R300B              \\
  2012-10-05 & 56205.94  & +40.98 & 3700 - 9965  &  21.6      & GTC+OSIRIS+R300B              \\
  2012-11-17 & 56249.92  & +81.40 & 3980 - 9755  &  21.4      & GTC+OSIRIS+R300B              \\
  \hline  
  \end{tabular}
  \\[1.5ex]
  \flushleft
  $^*$Rest frame epochs (assuming a redshift of 0.088) with respect to the first detection on 56161.36 (MJD). The resolution was determined from the FWHM of the O\,{\sc i} $\lambda$5577.34 sky line.
\end{table*}

\subsection{PS1-13abg}
PS1-13abg was discovered on 2013 March 29 by the PS1 MD survey. $g_{\mathrm{PS1}}r_{\mathrm{PS1}}i_{\mathrm{PS1}}z_{\mathrm{PS1}}y_{\mathrm{PS1}}$ photometry was obtained in the course of the PS1 MD survey up to 38\,days past the first detection. We  obtained three GTC spectra ranging from +7 to +44\,days after discovery. 

PS1-13abg lies close to the galaxy zCOSMOS 824314, for which \citet{Lilly2007} report a redshift of 0.123\,$\pm$\,0.001. This redshift was adopted for PS1-13abg.

No features attributable to the Na\,{\sc i} doublet from interstellar gas either in the Milky Way or the host galaxy, are apparent in the spectra of PS1-13abg. An upper limit for the equivalent width of the Na\,{\sc i}\,D $\lambda\lambda$\,5890,5896 blend was measured and yields EW$_\mathrm{Na\,I\,D}$ $<$\,0.6\,{\AA}, which corresponds to a host extinction of \mbox{$E(B-V)_{\mathrm{host}}$}\,$<$\,0.07\,mag \citep[Section~\ref{section:Observations:SN2013eq} and Equation 9 of][]{Poznanski2012}.

The light curves of PS1-13abg are presented in the left panel of Figure \ref{figure:PS1-13abg}. Table \ref{table:PS1-13abg_photometry} shows the log of imaging observations and the calibrated magnitudes.
PS1-13abg was discovered 10 days after the last non-detection in the $z_\mathrm{PS1}$-band on MJD 56370.34. This allows us to put a loose constraint on the explosion epoch of PS1-13abg to MJD 56375.4\,$\pm$\,5.0. The $r_{\mathrm{PS1}}$-, $i_{\mathrm{PS1}}$-, $z_{\mathrm{PS1}}$-, $y_{\mathrm{PS1}}$ light curves display a clear plateau. The drop onto the radioactive tail is not observed.

\begin{figure*}[t!]
   \centering
   \begin{subfigure}[t]{0.49\textwidth}
      \includegraphics[width=\columnwidth]{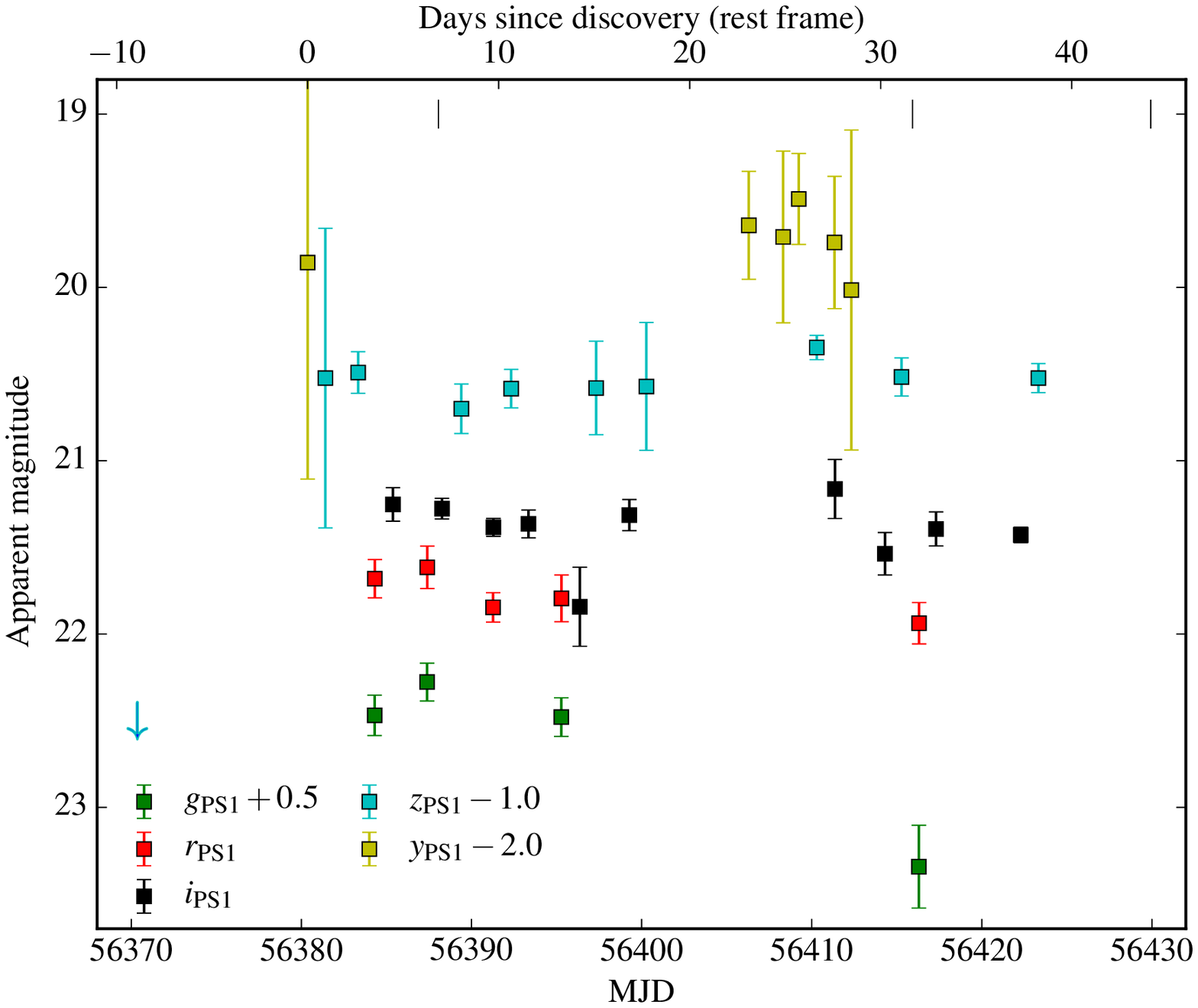}
   \end{subfigure}%
   \begin{subfigure}[t]{0.49\textwidth}
      \includegraphics[width=\columnwidth]{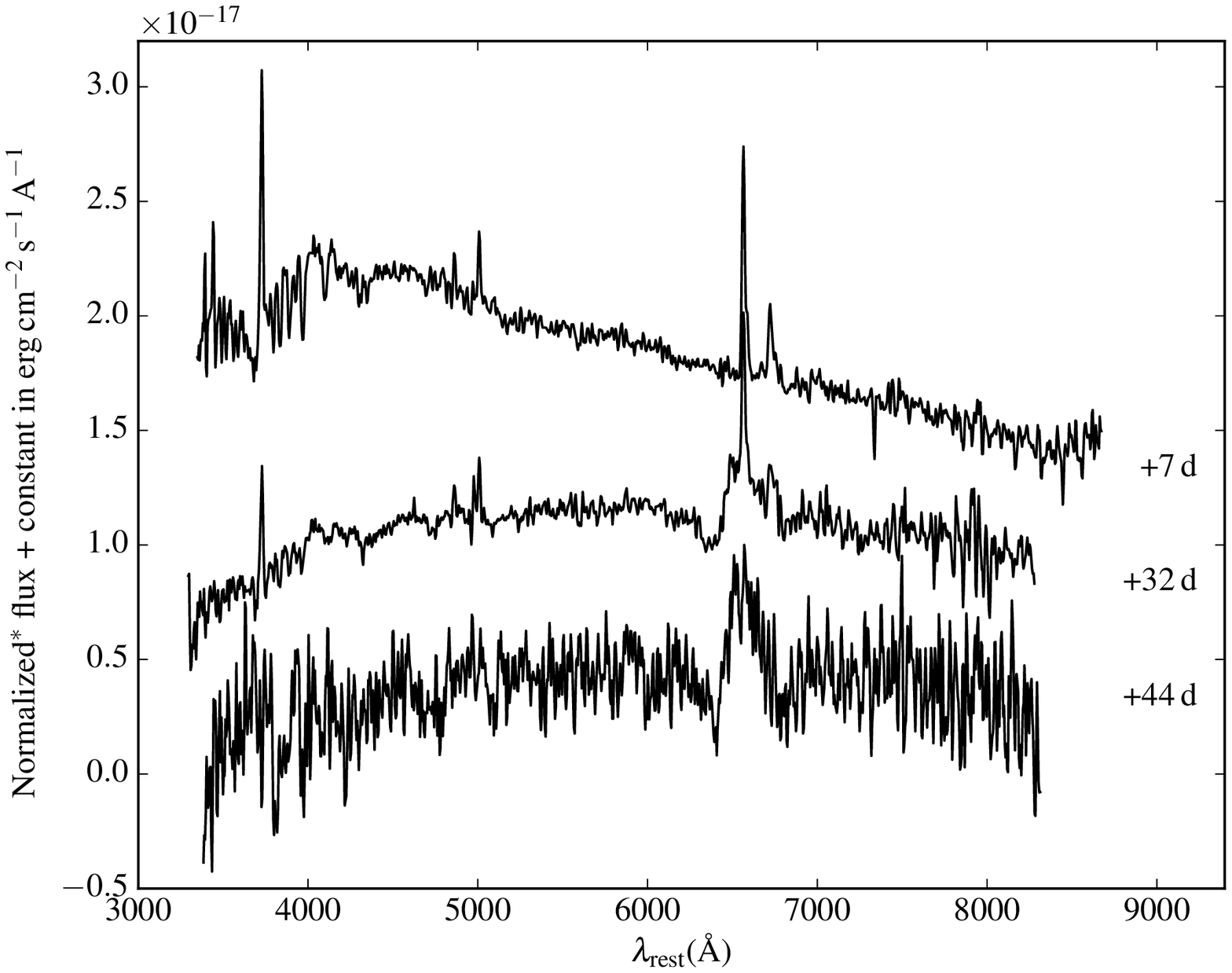}
\end{subfigure}%
   \caption{PS1-13abg. Left panel: $grizy_\mathrm{PS1}$ light curves of PS1-13abg. The vertical ticks on the top mark the epochs of the observed spectra. Right panel: PS1-13abg spectroscopy. $^*$Flux normalized to the maximum H$\alpha$ flux for better visibility of the features. The exact normalizations are: 
flux/0.7 for the +7\,d spectrum;
flux/0.8 for +32\,d;
flux/1.1 for +44\,d.
}
   \label{figure:PS1-13abg}
\end{figure*}

\begin{table*}
  \small
  \caption{PS1-13abg: Photometric observations}
  \label{table:PS1-13abg_photometry}
  \centering
  \begin{tabular}{c   c                      r            c                            c                        c           }
  \hline               
  \multirow{2}{*}{Date} & \multirow{2}{*}{MJD} & \multicolumn{1}{c}{Rest frame} & \multirow{2}{*}{Magnitude} & \multirow{2}{*}{Filter} & \multirow{2}{*}{Telescope} \\ 
                      &                      & \multicolumn{1}{c}{epoch$^*$}  &                            &                         &              \\
  \hline 
2013-04-02 & 56384.31 & +3.51  & 21.97\,$\pm$\,0.12 & $g_\mathrm{PS1}$ & PS1 \\
2013-04-05 & 56387.39 & +6.25  & 21.78\,$\pm$\,0.11 & $g_\mathrm{PS1}$ & PS1 \\
2013-04-13 & 56395.28 & +13.28 & 21.98\,$\pm$\,0.11 & $g_\mathrm{PS1}$ & PS1 \\
2013-05-04 & 56416.30 & +32.00 & 22.84\,$\pm$\,0.24 & $g_\mathrm{PS1}$ & PS1 \\
\hline
2013-04-02 & 56384.32 & +3.52  & 21.68\,$\pm$\,0.11 & $r_\mathrm{PS1}$ & PS1 \\
2013-04-05 & 56387.40 & +6.26  & 21.61\,$\pm$\,0.12 & $r_\mathrm{PS1}$ & PS1 \\
2013-04-09 & 56391.27 & +9.71  & 21.85\,$\pm$\,0.08 & $r_\mathrm{PS1}$ & PS1 \\
2013-04-13 & 56395.29 & +13.29 & 21.79\,$\pm$\,0.13 & $r_\mathrm{PS1}$ & PS1 \\
2013-05-04 & 56416.31 & +32.01 & 21.94\,$\pm$\,0.12 & $r_\mathrm{PS1}$ & PS1 \\
\hline
2013-04-03 & 56385.38 & +4.47  & 21.25\,$\pm$\,0.10 & $i_\mathrm{PS1}$ & PS1 \\
2013-04-06 & 56388.27 & +7.03  & 21.28\,$\pm$\,0.06 & $i_\mathrm{PS1}$ & PS1 \\
2013-04-09 & 56391.29 & +9.73  & 21.38\,$\pm$\,0.05 & $i_\mathrm{PS1}$ & PS1 \\
2013-04-11 & 56393.35 & +11.56 & 21.36\,$\pm$\,0.08 & $i_\mathrm{PS1}$ & PS1 \\
2013-04-14 & 56396.37 & +14.25 & 21.84\,$\pm$\,0.23 & $i_\mathrm{PS1}$ & PS1 \\
2013-04-17 & 56399.28 & +16.84 & 21.31\,$\pm$\,0.09 & $i_\mathrm{PS1}$ & PS1 \\
2013-04-29 & 56411.37 & +27.61 & 21.16\,$\pm$\,0.17 & $i_\mathrm{PS1}$ & PS1 \\
2013-05-02 & 56414.31 & +30.22 & 21.54\,$\pm$\,0.12 & $i_\mathrm{PS1}$ & PS1 \\
2013-05-05 & 56417.32 & +32.90 & 21.39\,$\pm$\,0.10 & $i_\mathrm{PS1}$ & PS1 \\
2013-05-10 & 56422.29 & +37.33 & 21.43\,$\pm$\,0.04 & $i_\mathrm{PS1}$ & PS1 \\
\hline
2013-03-30 & 56381.41 & +0.93  & 21.52\,$\pm$\,0.86 & $z_\mathrm{PS1}$ & PS1 \\
2013-04-01 & 56383.34 & +2.64  & 21.49\,$\pm$\,0.12 & $z_\mathrm{PS1}$ & PS1 \\
2013-04-07 & 56389.40 & +8.05  & 21.70\,$\pm$\,0.14 & $z_\mathrm{PS1}$ & PS1 \\
2013-04-10 & 56392.33 & +10.66 & 21.58\,$\pm$\,0.11 & $z_\mathrm{PS1}$ & PS1 \\
2013-04-15 & 56397.33 & +15.11 & 21.58\,$\pm$\,0.27 & $z_\mathrm{PS1}$ & PS1 \\
2013-04-18 & 56400.28 & +17.73 & 21.57\,$\pm$\,0.37 & $z_\mathrm{PS1}$ & PS1 \\
2013-04-28 & 56410.30 & +26.65 & 21.35\,$\pm$\,0.07 & $z_\mathrm{PS1}$ & PS1 \\
2013-05-03 & 56415.28 & +31.09 & 21.52\,$\pm$\,0.11 & $z_\mathrm{PS1}$ & PS1 \\
2013-05-11 & 56423.33 & +38.25 & 21.52\,$\pm$\,0.08 & $z_\mathrm{PS1}$ & PS1 \\
\hline
2013-03-29 & 56380.37 & +0.00 & 21.86\,$\pm$\,1.25 & $y_\mathrm{PS1}$ & PS1 \\
2013-04-24 & 56406.31 & +23.10 & 21.64\,$\pm$\,0.31 & $y_\mathrm{PS1}$ & PS1 \\
2013-04-26 & 56408.32 & +24.89 & 21.71\,$\pm$\,0.50 & $y_\mathrm{PS1}$ & PS1 \\
2013-04-27 & 56409.24 & +25.71 & 21.49\,$\pm$\,0.26 & $y_\mathrm{PS1}$ & PS1 \\
2013-04-29 & 56411.34 & +27.58 & 21.74\,$\pm$\,0.38 & $y_\mathrm{PS1}$ & PS1 \\
2013-04-30 & 56412.33 & +28.46 & 22.02\,$\pm$\,0.92 & $y_\mathrm{PS1}$ & PS1 \\
  \hline  
  \end{tabular}
  \\[1.5ex]
  \flushleft
  The tabulated magnitudes are given ``as observed'', i.e. neither corrected for dust extinction nor $K$-corrected. $^*$Rest frame epochs (assuming a redshift of 0.123) with respect to the first detection on 56380.37 (MJD). 
PS1 = Panoramic Survey Telescope \& Rapid Response System 1.
\end{table*}

Table \ref{table:PS1-13abg_journal_spectra} gives the journal of spectroscopic observations for PS1-13abg. The calibrated spectra of PS1-13abg are presented in the right panel of Figure \ref{figure:PS1-13abg}. They are corrected for galactic reddening and redshift ($z$ = 0.123). 
The +7\,d spectrum of PS1-13abg is an almost featureless continuum with the exception of two narrow lines in the regions of H$\alpha$ and H$\beta$, respectively. The +32 and +44\,d spectra exhibit H$\alpha$ and H$\beta$, the +32\,d H$\alpha$ line showing a strong narrow component. Narrow components are not observed in the +44\,d spectrum, this could, however, be due to the poor signal-to-noise ratio in the last spectrum. A weak Fe\,{\sc ii}\,$\lambda$5169 absorption can also be found in the +44\,d spectrum.

\begin{table*}
  \caption{PS1-13abg: Journal of spectroscopic observations}
  \label{table:PS1-13abg_journal_spectra}
  \centering
  \begin{tabular}{c c     c            c              c            c                               }
  \hline 
\multirow{2}{*}{Date} & \multirow{2}{*}{MJD} & Epoch$^*$  & Wavelength   & Resolution & \multirow{2}{*}{Telescope+Instrument} \\
             &           & rest frame & range in \AA &  \AA       &                               \\
\hline 
  2013-04-05 & 56388.06  &  +6.85 & 3760 - 9740 &  16.8      & GTC+OSIRIS+R300B              \\
  2013-05-03 & 56415.93  & +31.66 & 3700 - 9800 &  17.2      & GTC+OSIRIS+R300B              \\
  2013-05-17 & 56429.93  & +44.13 & 3800 - 9335 &  20.6      & GTC+OSIRIS+R300B              \\
  \hline  
  \end{tabular}
  \\[1.5ex]
  \flushleft
  $^*$Rest frame epochs (assuming a redshift of 0.123) with respect to the first detection on 56380.37 (MJD). The resolution was determined from the FWHM of the O\,{\sc i} $\lambda$5577.34 sky line.
\end{table*}

\subsection{PS1-13baf}
PS1-13baf was discovered on 2013 August 13 by the PS1 MD survey. $g_{\mathrm{PS1}}r_{\mathrm{PS1}}i_{\mathrm{PS1}}z_{\mathrm{PS1}}y_{\mathrm{PS1}}$ photometry was obtained in the course of the PS1 MD survey up to 113\,days after the first detection. We  obtained three GTC spectra ranging from +16 to +39\,days after discovery. 

Spectroscopic redshifts of the host galaxy were not available for PS1-13baf. Its redshift was therefore determined by measuring the narrow H$\alpha$ and H$\beta$ components (presumably from the host galaxy) in the +28 and +39\,d spectra and averaging the respective results: $z$ = 0.144\,$\pm$\,0.001.

No features attributable to the Na\,{\sc i} doublet from interstellar gas either in the Milky Way or the host galaxy, are apparent in the spectra of PS1-13baf. An upper limit for the equivalent width of the Na\,{\sc i}\,D $\lambda\lambda$\,5890,5896 blend was derived in the same manner as for SN 2013ca (Section~\ref{section:Observations:SN2013eq}). We find EW$_\mathrm{Na\,I\,D}$ $<$\,1.2\,{\AA}, which can be converted into a host extinction of $E(B-V)_{\mathrm{host}} <$\,0.36\,mag \citep[Equation 9 of][]{Poznanski2012}.

The light curves of PS1-13baf are presented in the left panel of Figure \ref{figure:PS1-13baf}. Table \ref{table:PS1-13baf_photometry} shows the log of imaging observations and the calibrated magnitudes.
PS1-13baf was discovered three days after the last non-detection in the $y_\mathrm{PS1}$-band on MJD 56406.59. This constrains the explosion epoch of PS1-13abg to MJD 56408.0\,$\pm$\,1.5. The light curves in all filters display a clear plateau until at least 100\,d (rest frame) after discovery.

\begin{figure*}[t!]
   \centering
   \begin{subfigure}[t]{0.49\textwidth}
      \includegraphics[width=\columnwidth]{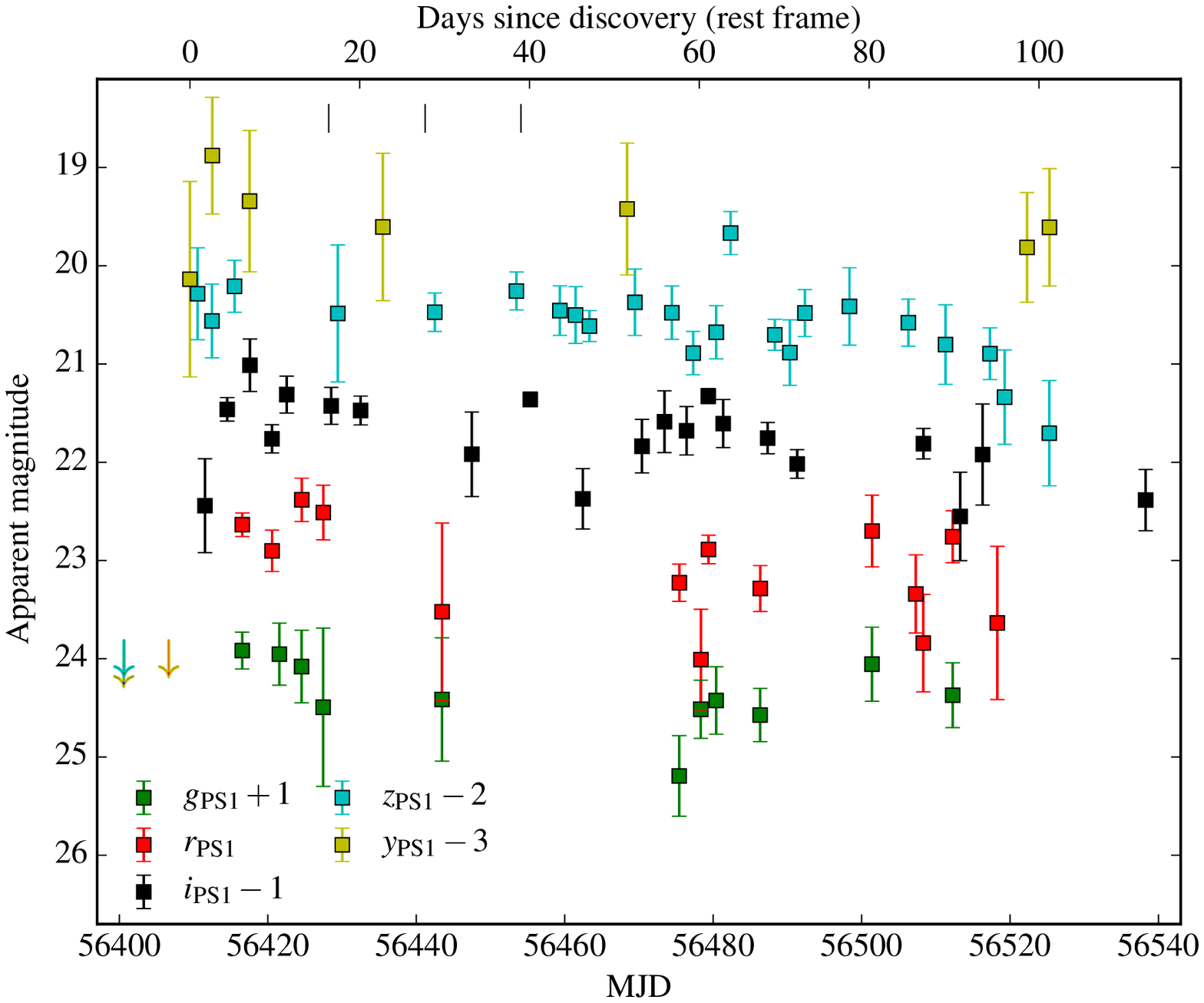}
   \end{subfigure}%
   \begin{subfigure}[t]{0.49\textwidth}
      \includegraphics[width=\columnwidth]{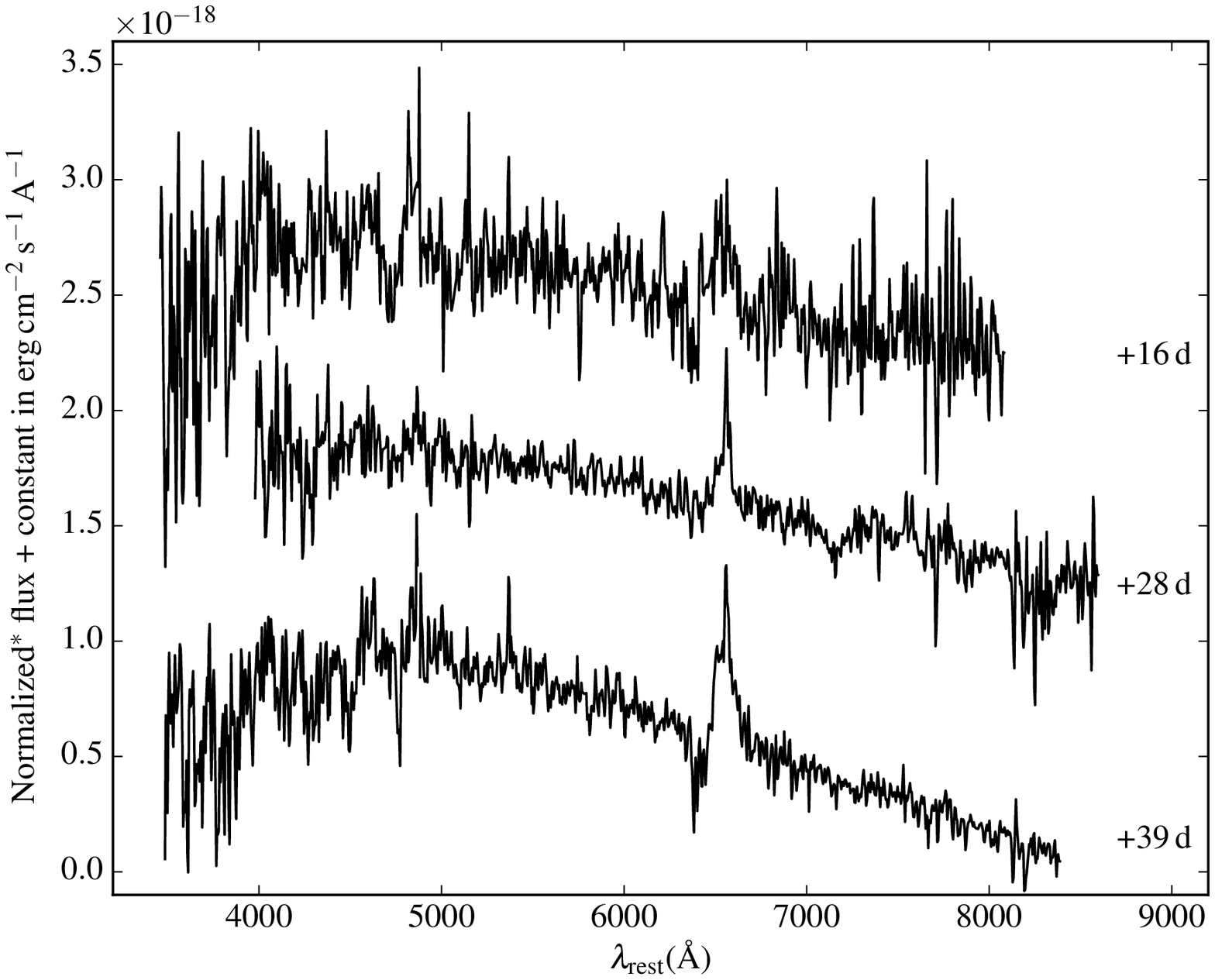}
\end{subfigure}%
   \caption{PS1-13baf. Left panel: $grizy_\mathrm{PS1}$ light curves of PS1-13baf. The vertical ticks on the top mark the epochs of the observed spectra. Right panel: PS1-13baf spectroscopy. $^*$Flux normalized to the maximum H$\alpha$ flux for better visibility of the features. The exact normalizations are: 
flux/4.1 for the +16\,d spectrum;
flux/3.2 for +28\,d;
flux/2.9 for +39\,d.
}
   \label{figure:PS1-13baf}
\end{figure*}

\begin{table*}
  \small
  \caption{PS1-13baf: Photometric observations}
  \label{table:PS1-13baf_photometry}
  \centering
  \begin{tabular}{c   c                      r            c                            c                        c           }
  \hline               
  \multirow{2}{*}{Date} & \multirow{2}{*}{MJD} & \multicolumn{1}{c}{Rest frame} & \multirow{2}{*}{Magnitude} & \multirow{2}{*}{Filter} & \multirow{2}{*}{Telescope} \\ 
                      &                      & \multicolumn{1}{c}{epoch$^*$}  &                            &                         &              \\
  \hline 
2013-05-04 & 56416.54 &   +6.16 & 22.92\,$\pm$\,0.19 & $g_\mathrm{PS1}$ & PS1 \\
2013-05-09 & 56421.55 &  +10.54 & 22.95\,$\pm$\,0.32 & $g_\mathrm{PS1}$ & PS1 \\
2013-05-12 & 56424.55 &  +13.16 & 23.08\,$\pm$\,0.37 & $g_\mathrm{PS1}$ & PS1 \\
2013-05-15 & 56427.45 &  +15.70 & 23.49\,$\pm$\,0.81 & $g_\mathrm{PS1}$ & PS1 \\
2013-05-31 & 56443.46 &  +29.69 & 23.41\,$\pm$\,0.63 & $g_\mathrm{PS1}$ & PS1 \\
2013-07-02 & 56475.43 &  +57.63 & 24.19\,$\pm$\,0.41 & $g_\mathrm{PS1}$ & PS1 \\
2013-07-05 & 56478.34 &  +60.18 & 23.51\,$\pm$\,0.30 & $g_\mathrm{PS1}$ & PS1 \\
2013-07-07 & 56480.41 &  +61.99 & 23.42\,$\pm$\,0.34 & $g_\mathrm{PS1}$ & PS1 \\
2013-07-13 & 56486.33 &  +67.17 & 23.57\,$\pm$\,0.27 & $g_\mathrm{PS1}$ & PS1 \\
2013-07-28 & 56501.40 &  +80.34 & 23.06\,$\pm$\,0.38 & $g_\mathrm{PS1}$ & PS1 \\
2013-08-08 & 56512.27 &  +89.84 & 23.37\,$\pm$\,0.33 & $g_\mathrm{PS1}$ & PS1 \\
\hline  
2013-05-04 & 56416.55 &   +6.17 & 22.64\,$\pm$\,0.12 & $r_\mathrm{PS1}$ & PS1 \\
2013-05-08 & 56420.57 &   +9.68 & 22.90\,$\pm$\,0.21 & $r_\mathrm{PS1}$ & PS1 \\
2013-05-12 & 56424.56 &  +13.17 & 22.38\,$\pm$\,0.22 & $r_\mathrm{PS1}$ & PS1 \\
2013-05-15 & 56427.46 &  +15.71 & 22.51\,$\pm$\,0.28 & $r_\mathrm{PS1}$ & PS1 \\
2013-05-31 & 56443.48 &  +29.70 & 23.52\,$\pm$\,0.90 & $r_\mathrm{PS1}$ & PS1 \\
2013-07-02 & 56475.44 &  +57.65 & 23.23\,$\pm$\,0.19 & $r_\mathrm{PS1}$ & PS1 \\
2013-07-05 & 56478.35 &  +60.19 & 24.01\,$\pm$\,0.51 & $r_\mathrm{PS1}$ & PS1 \\
2013-07-06 & 56479.37 &  +61.08 & 22.89\,$\pm$\,0.14 & $r_\mathrm{PS1}$ & PS1 \\
2013-07-13 & 56486.34 &  +67.18 & 23.28\,$\pm$\,0.23 & $r_\mathrm{PS1}$ & PS1 \\
2013-07-28 & 56501.41 &  +80.35 & 22.70\,$\pm$\,0.36 & $r_\mathrm{PS1}$ & PS1 \\
2013-08-03 & 56507.31 &  +85.50 & 23.34\,$\pm$\,0.40 & $r_\mathrm{PS1}$ & PS1 \\
2013-08-04 & 56508.31 &  +86.38 & 23.84\,$\pm$\,0.50 & $r_\mathrm{PS1}$ & PS1 \\
2013-08-08 & 56512.28 &  +89.85 & 22.76\,$\pm$\,0.26 & $r_\mathrm{PS1}$ & PS1 \\
2013-08-14 & 56518.29 &  +95.10 & 23.64\,$\pm$\,0.78 & $r_\mathrm{PS1}$ & PS1 \\
  \hline  
2013-04-29 & 56411.52 &   +1.77 & 23.44\,$\pm$\,0.48 & $i_\mathrm{PS1}$ & PS1 \\
2013-05-02 & 56414.52 &   +4.39 & 22.46\,$\pm$\,0.12 & $i_\mathrm{PS1}$ & PS1 \\
2013-05-05 & 56417.58 &   +7.07 & 22.01\,$\pm$\,0.27 & $i_\mathrm{PS1}$ & PS1 \\
2013-05-08 & 56420.55 &   +9.66 & 22.76\,$\pm$\,0.14 & $i_\mathrm{PS1}$ & PS1 \\
2013-05-10 & 56422.54 &  +11.40 & 22.31\,$\pm$\,0.19 & $i_\mathrm{PS1}$ & PS1 \\
2013-05-16 & 56428.54 &  +16.64 & 22.43\,$\pm$\,0.19 & $i_\mathrm{PS1}$ & PS1 \\
2013-05-20 & 56432.50 &  +20.11 & 22.47\,$\pm$\,0.15 & $i_\mathrm{PS1}$ & PS1 \\
2013-06-04 & 56447.48 &  +33.21 & 22.92\,$\pm$\,0.43 & $i_\mathrm{PS1}$ & PS1 \\
2013-06-12 & 56455.35 &  +40.09 & 22.36\,$\pm$\,0.07 & $i_\mathrm{PS1}$ & PS1 \\
2013-06-19 & 56462.44 &  +46.28 & 23.37\,$\pm$\,0.31 & $i_\mathrm{PS1}$ & PS1 \\
2013-06-27 & 56470.41 &  +53.25 & 22.84\,$\pm$\,0.27 & $i_\mathrm{PS1}$ & PS1 \\
2013-06-30 & 56473.44 &  +55.90 & 22.59\,$\pm$\,0.31 & $i_\mathrm{PS1}$ & PS1 \\
2013-07-03 & 56476.42 &  +58.50 & 22.68\,$\pm$\,0.25 & $i_\mathrm{PS1}$ & PS1 \\
2013-07-06 & 56479.35 &  +61.06 & 22.33\,$\pm$\,0.08 & $i_\mathrm{PS1}$ & PS1 \\
2013-07-08 & 56481.34 &  +62.81 & 22.61\,$\pm$\,0.25 & $i_\mathrm{PS1}$ & PS1 \\
2013-07-14 & 56487.35 &  +68.05 & 22.75\,$\pm$\,0.16 & $i_\mathrm{PS1}$ & PS1 \\
2013-07-18 & 56491.32 &  +71.53 & 23.02\,$\pm$\,0.15 & $i_\mathrm{PS1}$ & PS1 \\
2013-08-04 & 56508.33 &  +86.40 & 22.81\,$\pm$\,0.16 & $i_\mathrm{PS1}$ & PS1 \\
2013-08-09 & 56513.28 &  +90.72 & 23.55\,$\pm$\,0.45 & $i_\mathrm{PS1}$ & PS1 \\
2013-08-12 & 56516.30 &  +93.36 & 22.92\,$\pm$\,0.51 & $i_\mathrm{PS1}$ & PS1 \\
2013-09-03 & 56538.28 & +112.57 & 23.39\,$\pm$\,0.31 & $i_\mathrm{PS1}$ & PS1 \\
  \hline  
2013-04-28 & 56410.54 &   +0.91 & 22.28\,$\pm$\,0.47 & $z_\mathrm{PS1}$ & PS1 \\
2013-04-30 & 56412.48 &   +2.61 & 22.56\,$\pm$\,0.38 & $z_\mathrm{PS1}$ & PS1 \\
2013-05-03 & 56415.51 &   +5.26 & 22.21\,$\pm$\,0.26 & $z_\mathrm{PS1}$ & PS1 \\
2013-05-17 & 56429.43 &  +17.43 & 22.49\,$\pm$\,0.70 & $z_\mathrm{PS1}$ & PS1 \\
2013-05-30 & 56442.49 &  +28.84 & 22.47\,$\pm$\,0.20 & $z_\mathrm{PS1}$ & PS1 \\
2013-06-10 & 56453.50 &  +38.46 & 22.26\,$\pm$\,0.19 & $z_\mathrm{PS1}$ & PS1 \\
2013-06-16 & 56459.35 &  +43.58 & 22.46\,$\pm$\,0.25 & $z_\mathrm{PS1}$ & PS1 \\
2013-06-18 & 56461.47 &  +45.43 & 22.50\,$\pm$\,0.29 & $z_\mathrm{PS1}$ & PS1 \\
2013-06-20 & 56463.33 &  +47.06 & 22.61\,$\pm$\,0.16 & $z_\mathrm{PS1}$ & PS1 \\
2013-06-26 & 56469.46 &  +52.42 & 22.37\,$\pm$\,0.34 & $z_\mathrm{PS1}$ & PS1 \\
2013-07-01 & 56474.44 &  +56.77 & 22.48\,$\pm$\,0.27 & $z_\mathrm{PS1}$ & PS1 \\
2013-07-04 & 56477.31 &  +59.28 & 22.89\,$\pm$\,0.22 & $z_\mathrm{PS1}$ & PS1 \\
2013-07-07 & 56480.41 &  +61.99 & 22.68\,$\pm$\,0.27 & $z_\mathrm{PS1}$ & PS1 \\
2013-07-09 & 56482.35 &  +63.68 & 21.67\,$\pm$\,0.22 & $z_\mathrm{PS1}$ & PS1 \\
2013-07-15 & 56488.31 &  +68.90 & 22.70\,$\pm$\,0.16 & $z_\mathrm{PS1}$ & PS1 \\
2013-07-17 & 56490.33 &  +70.66 & 22.88\,$\pm$\,0.33 & $z_\mathrm{PS1}$ & PS1 \\
2013-07-19 & 56492.36 &  +72.43 & 22.48\,$\pm$\,0.24 & $z_\mathrm{PS1}$ & PS1 \\
2013-07-25 & 56498.36 &  +77.68 & 22.41\,$\pm$\,0.39 & $z_\mathrm{PS1}$ & PS1 \\
2013-08-02 & 56506.30 &  +84.62 & 22.58\,$\pm$\,0.24 & $z_\mathrm{PS1}$ & PS1 \\
2013-08-07 & 56511.31 &  +89.00 & 22.80\,$\pm$\,0.40 & $z_\mathrm{PS1}$ & PS1 \\
\hline  
  \end{tabular}
\end{table*}

\begin{table*}
  \small
  \caption*{Table \ref{table:PS1-13baf_photometry}: (continued)}
  \centering
  \begin{tabular}{c   c                      r            c                            c                        c           }
  \hline               
  \multirow{2}{*}{Date} & \multirow{2}{*}{MJD} & \multicolumn{1}{c}{Rest frame} & \multirow{2}{*}{Magnitude} & \multirow{2}{*}{Filter} & \multirow{2}{*}{Telescope} \\ 
                      &                      & \multicolumn{1}{c}{epoch$^*$}  &                            &                         &              \\
  \hline
2013-08-13 & 56517.30 &  +94.23 & 22.90\,$\pm$\,0.26 & $z_\mathrm{PS1}$ & PS1 \\
2013-08-15 & 56519.27 &  +95.96 & 23.34\,$\pm$\,0.48 & $z_\mathrm{PS1}$ & PS1 \\
2013-08-21 & 56525.29 & +101.22 & 23.70\,$\pm$\,0.54 & $z_\mathrm{PS1}$ & PS1 \\
  \hline  
2013-04-27 & 56409.49 &   +0.00 & 23.14\,$\pm$\,0.99 & $y_\mathrm{PS1}$ & PS1 \\
2013-04-30 & 56412.51 &   +2.63 & 21.88\,$\pm$\,0.59 & $y_\mathrm{PS1}$ & PS1 \\
2013-05-05 & 56417.54 &   +7.04 & 22.34\,$\pm$\,0.72 & $y_\mathrm{PS1}$ & PS1 \\
2013-05-23 & 56435.49 &  +22.73 & 22.61\,$\pm$\,0.75 & $y_\mathrm{PS1}$ & PS1 \\
2013-06-25 & 56468.40 &  +51.49 & 22.42\,$\pm$\,0.67 & $y_\mathrm{PS1}$ & PS1 \\
2013-08-18 & 56522.29 &  +98.60 & 22.81\,$\pm$\,0.56 & $y_\mathrm{PS1}$ & PS1 \\
2013-08-21 & 56525.32 & +101.24 & 22.61\,$\pm$\,0.60 & $y_\mathrm{PS1}$ & PS1 \\
  \hline  
  \end{tabular}
  \\[1.5ex]
  \flushleft
  The tabulated magnitudes are given ``as observed'', i.e. neither corrected for dust extinction nor $K$-corrected. $^*$Rest frame epochs (assuming a redshift of 0.144) with respect to the first detection on 56409.49	 (MJD). 
PS1 = Panoramic Survey Telescope \& Rapid Response System 1.
\end{table*}

Table \ref{table:PS1-13baf_journal_spectra} gives the journal of spectroscopic observations for PS1-13baf. The calibrated spectra of PS1-13baf are presented in the right panel of Figure \ref{figure:PS1-13baf}. They are corrected for galactic reddening and redshift ($z$ = 0.144). 
The spectra of PS1-13baf show strong H$\alpha$ and H$\beta$ lines with narrow components from the host galaxy overlayed in the +28 and +39\,d spectra. Iron lines cannot be identified, due to the poor signal-to-noise ratio of the spectra. 

\begin{table*}
  \caption{PS1-13baf: Journal of spectroscopic observations}
  \label{table:PS1-13baf_journal_spectra}
  \centering
  \begin{tabular}{c c     c            c              c            c                               }
  \hline 
\multirow{2}{*}{Date} & \multirow{2}{*}{MJD} & Epoch$^*$  & Wavelength   & Resolution & \multirow{2}{*}{Telescope+Instrument} \\
             &           & rest frame & range in \AA &  \AA       &                               \\
\hline 
  2013-05-15 & 56428.20  & +16.35 & 3750 - 9245 &  16.9      & GTC+OSIRIS+R300B              \\
  2013-05-28 & 56441.18  & +27.70 & 4500 - 9835 &  16.6      & GTC+OSIRIS+R300B              \\
  2013-06-10 & 56454.09  & +38.98 & 3800 - 9600 &  17.1      & GTC+OSIRIS+R300B              \\
  \hline  
  \end{tabular}
  \\[1.5ex]
  \flushleft
  $^*$Rest frame epochs (assuming a redshift of 0.144) with respect to the first discovery on 56409.49 (MJD). The resolution was determined from the FWHM of the O\,{\sc i} $\lambda$5577.34 sky line.
\end{table*}

\subsection{PS1-13bmf}
\label{appendix:section:PS1_13bmf_observations}

PS1-13bmf was discovered on 2013 May 8 by the PS1 MD survey. $g_{\mathrm{PS1}}r_{\mathrm{PS1}}i_{\mathrm{PS1}}z_{\mathrm{PS1}}y_{\mathrm{PS1}}$ photometry was obtained in the course of the PS1 MD survey up to 52\,days after the first detection. We  obtained four GTC spectra ranging from +5 to +47\,days after discovery. 

PS1-13bmf lies in the vicinity of the galaxy SDSS J122311.46+474426.7 for which the SDSS reports a spectroscopic redshift of 0.163\,$\pm$\,0.001. This redshift was adopted also for PS1-13bmf.

No Na\,{\sc i} doublet from interstellar gas either in the Milky Way or the host galaxy is apparent in the spectra of PS1-13bmf. The very low signal-to-noise ratio of the spectra prevented us from measuring a meaningful upper limit for the host extinction. 

The light curves of PS1-13bmf are presented in the left panel of Figure \ref{figure:PS1-13bmf}. Table \ref{table:PS1-13bmf_photometry} shows the log of imaging observations and the calibrated magnitudes.
PS1-13bmf was likely discovered very close to explosion, as can be inferred by the observed initial rise in the $g_{\mathrm{PS1}}$, $r_{\mathrm{PS1}}$, and $z_{\mathrm{PS1}}$ light curves and in particular the very steep rise in the $i_{\mathrm{PS1}}$ band. A fit to the early photometry results in an estimate for the explosion epoch of MJD 56420.0\,$\pm$\,0.1. The $g_{\mathrm{PS1}}$, $r_{\mathrm{PS1}}$, $i_{\mathrm{PS1}}$, and $z_{\mathrm{PS1}}$ light curves decline linearly, with the $r_{\mathrm{PS1}}$ and $i_{\mathrm{PS1}}$ decline rates being 1.05\,$\pm$\,0.11\,mag/50\,d and 0.76\,$\pm$\,0.11\,mag/50\,d, respectively. We therefore classify PS1-13bmf as a Type II-L SN following \citet{Li2011a}.

\begin{figure*}[t!]
   \centering
   \begin{subfigure}[t]{0.49\textwidth}
      \includegraphics[width=\columnwidth]{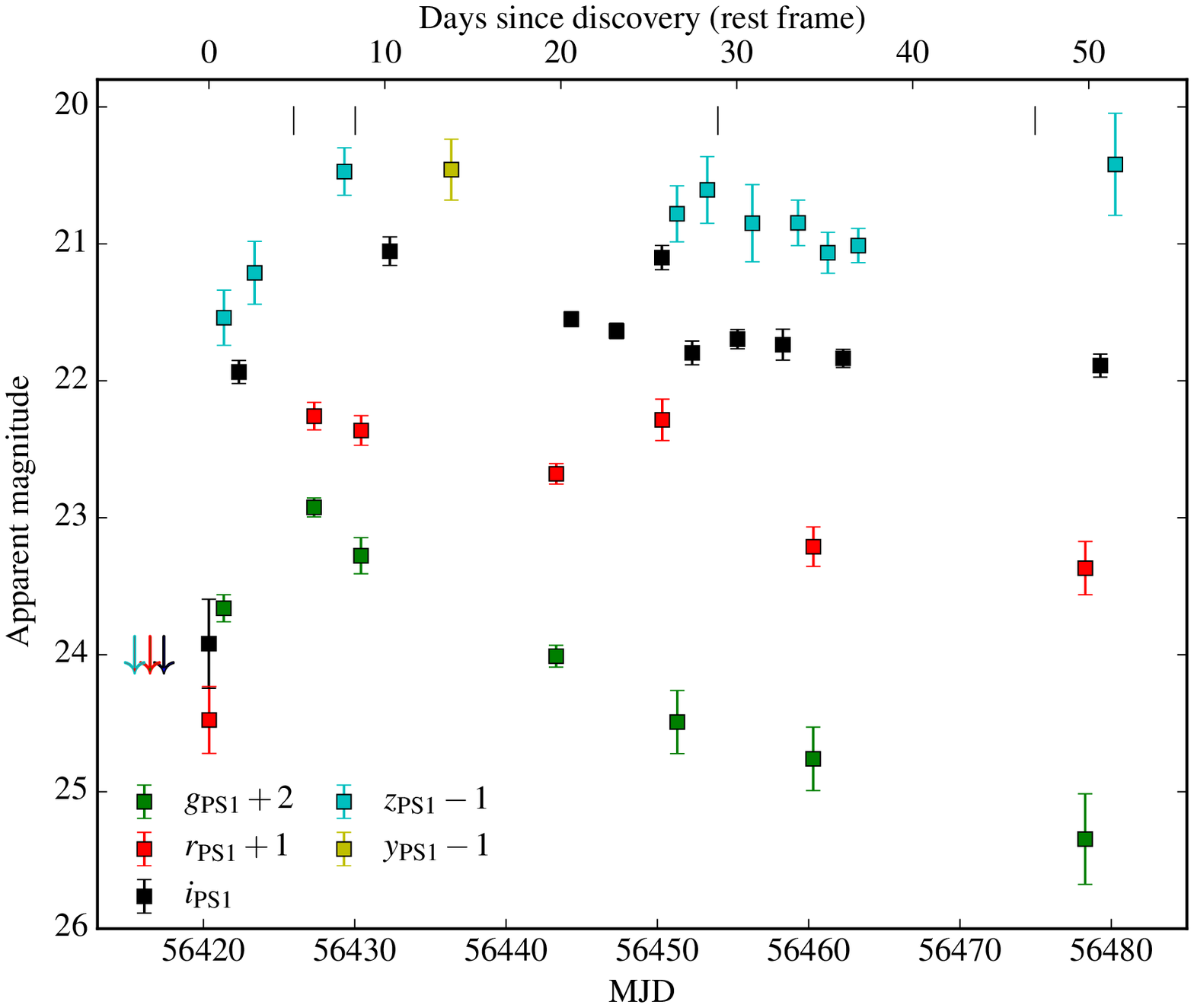}
   \end{subfigure}%
   \begin{subfigure}[t]{0.49\textwidth}
      \includegraphics[width=\columnwidth]{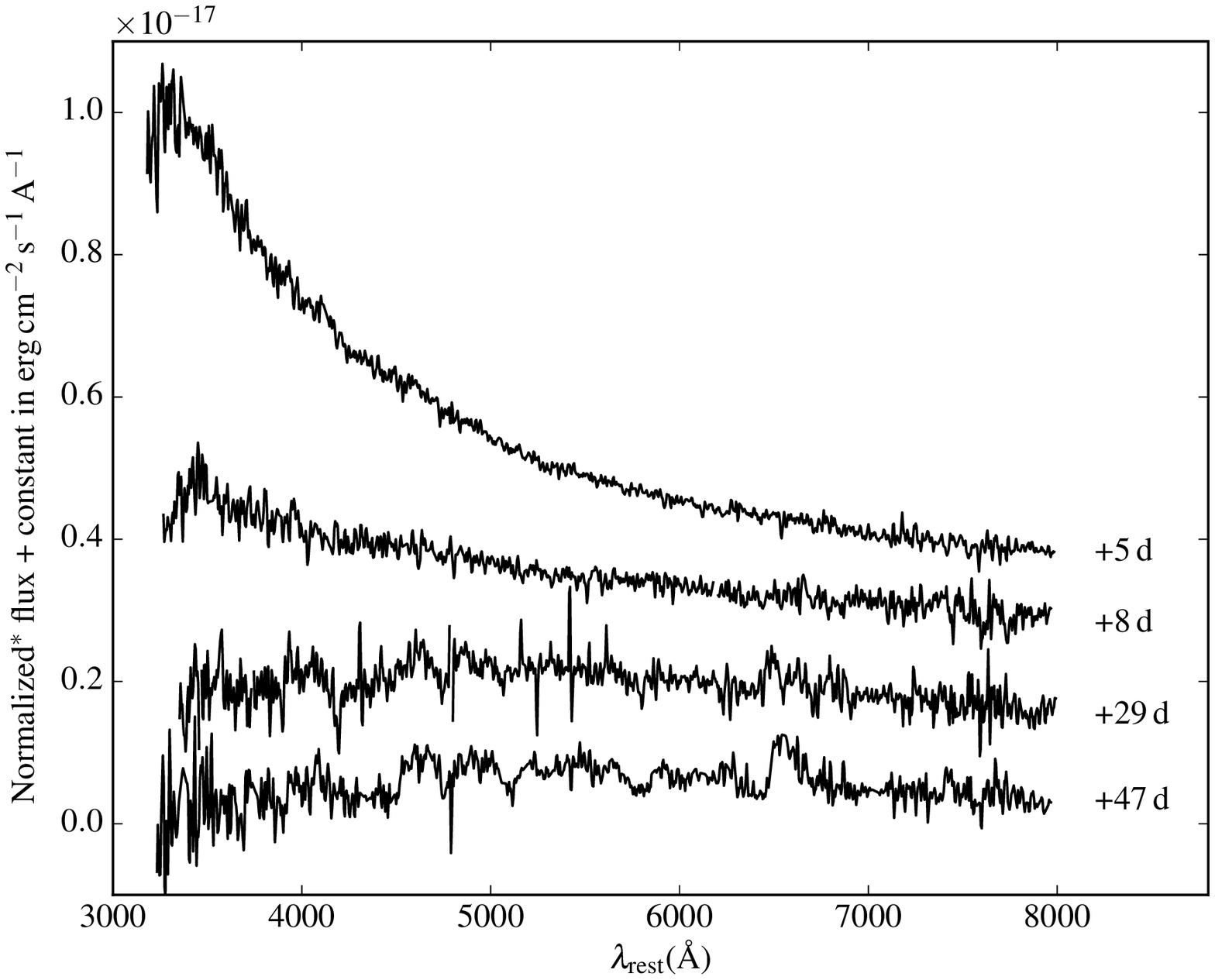}
\end{subfigure}%
   \caption{PS1-13bmf. Left panel: $grizy_\mathrm{PS1}$ light curves of PS1-13bmf. The vertical ticks on the top mark the epochs of the observed spectra. Right panel: PS1-13bmf spectroscopy. $^*$Flux normalized to the maximum H$\alpha$ flux for better visibility of the features. The exact normalizations are: 
flux/5.7 for the +5\,d spectrum;
flux/9.5 for +8\,d;
flux/8.8 for +29\,d;
flux/5.1 for +47\,d.
}
   \label{figure:PS1-13bmf}
\end{figure*}

\begin{table*}
  \small
  \caption{PS1-13bmf: Photometric observations}
  \label{table:PS1-13bmf_photometry}
  \centering
  \begin{tabular}{c   c                      r            c                            c                        c           }
  \hline               
  \multirow{2}{*}{Date} & \multirow{2}{*}{MJD} & \multicolumn{1}{c}{Rest frame} & \multirow{2}{*}{Magnitude} & \multirow{2}{*}{Filter} & \multirow{2}{*}{Telescope} \\ 
                      &                      & \multicolumn{1}{c}{epoch$^*$}  &                            &                         &              \\
  \hline 
2013-05-09 & 56421.34 &  +0.84 & 21.66\,$\pm$\,0.10 & $g_\mathrm{PS1}$ & PS1 \\
2013-05-15 & 56427.31 &  +5.97 & 20.92\,$\pm$\,0.07 & $g_\mathrm{PS1}$ & PS1 \\
2013-05-18 & 56430.42 &  +8.64 & 21.28\,$\pm$\,0.13 & $g_\mathrm{PS1}$ & PS1 \\
2013-05-31 & 56443.31 & +19.73 & 22.01\,$\pm$\,0.08 & $g_\mathrm{PS1}$ & PS1 \\
2013-06-08 & 56451.32 & +26.62 & 22.49\,$\pm$\,0.23 & $g_\mathrm{PS1}$ & PS1 \\
2013-06-17 & 56460.30 & +34.34 & 22.76\,$\pm$\,0.23 & $g_\mathrm{PS1}$ & PS1 \\
2013-07-05 & 56478.28 & +49.80 & 23.35\,$\pm$\,0.33 & $g_\mathrm{PS1}$ & PS1 \\
  \hline  
2013-05-08 & 56420.38 &  +0.02 & 23.48\,$\pm$\,0.24 & $r_\mathrm{PS1}$ & PS1 \\
2013-05-15 & 56427.32 &  +5.98 & 21.26\,$\pm$\,0.10 & $r_\mathrm{PS1}$ & PS1 \\
2013-05-18 & 56430.43 &  +8.65 & 21.36\,$\pm$\,0.11 & $r_\mathrm{PS1}$ & PS1 \\
2013-05-31 & 56443.32 & +19.74 & 21.68\,$\pm$\,0.07 & $r_\mathrm{PS1}$ & PS1 \\
2013-06-07 & 56450.33 & +25.77 & 21.29\,$\pm$\,0.15 & $r_\mathrm{PS1}$ & PS1 \\
2013-06-17 & 56460.31 & +34.35 & 22.21\,$\pm$\,0.14 & $r_\mathrm{PS1}$ & PS1 \\
2013-07-05 & 56478.29 & +49.81 & 22.37\,$\pm$\,0.19 & $r_\mathrm{PS1}$ & PS1 \\
\hline  
2013-05-08 & 56420.36 &  +0.00 & 23.92\,$\pm$\,0.32 & $i_\mathrm{PS1}$ & PS1 \\
2013-05-10 & 56422.34 &  +1.70 & 21.94\,$\pm$\,0.08 & $i_\mathrm{PS1}$ & PS1 \\
2013-05-20 & 56432.33 & +10.29 & 21.05\,$\pm$\,0.10 & $i_\mathrm{PS1}$ & PS1 \\
2013-06-01 & 56444.32 & +20.60 & 21.55\,$\pm$\,0.05 & $i_\mathrm{PS1}$ & PS1 \\
2013-06-04 & 56447.31 & +23.17 & 21.64\,$\pm$\,0.06 & $i_\mathrm{PS1}$ & PS1 \\
2013-06-07 & 56450.31 & +25.75 & 21.10\,$\pm$\,0.09 & $i_\mathrm{PS1}$ & PS1 \\
2013-06-09 & 56452.30 & +27.46 & 21.80\,$\pm$\,0.09 & $i_\mathrm{PS1}$ & PS1 \\
2013-06-12 & 56455.31 & +30.04 & 21.70\,$\pm$\,0.07 & $i_\mathrm{PS1}$ & PS1 \\
2013-06-15 & 56458.30 & +32.62 & 21.74\,$\pm$\,0.11 & $i_\mathrm{PS1}$ & PS1 \\
2013-06-19 & 56462.28 & +36.04 & 21.84\,$\pm$\,0.07 & $i_\mathrm{PS1}$ & PS1 \\
2013-07-06 & 56479.28 & +50.66 & 21.89\,$\pm$\,0.08 & $i_\mathrm{PS1}$ & PS1 \\
\hline  
2013-05-09 & 56421.36 &  +0.86 & 22.54\,$\pm$\,0.20 & $z_\mathrm{PS1}$ & PS1 \\
2013-05-11 & 56423.39 &  +2.60 & 22.21\,$\pm$\,0.23 & $z_\mathrm{PS1}$ & PS1 \\
2013-05-17 & 56429.32 &  +7.70 & 21.47\,$\pm$\,0.17 & $z_\mathrm{PS1}$ & PS1 \\
2013-06-08 & 56451.30 & +26.60 & 21.78\,$\pm$\,0.20 & $z_\mathrm{PS1}$ & PS1 \\
2013-06-10 & 56453.30 & +28.32 & 21.61\,$\pm$\,0.24 & $z_\mathrm{PS1}$ & PS1 \\
2013-06-13 & 56456.28 & +30.88 & 21.85\,$\pm$\,0.28 & $z_\mathrm{PS1}$ & PS1 \\
2013-06-16 & 56459.30 & +33.48 & 21.85\,$\pm$\,0.17 & $z_\mathrm{PS1}$ & PS1 \\
2013-06-18 & 56461.27 & +35.18 & 22.07\,$\pm$\,0.15 & $z_\mathrm{PS1}$ & PS1 \\
2013-06-20 & 56463.28 & +36.90 & 22.01\,$\pm$\,0.12 & $z_\mathrm{PS1}$ & PS1 \\
2013-07-07 & 56480.27 & +51.51 & 21.42\,$\pm$\,0.37 & $z_\mathrm{PS1}$ & PS1 \\
  \hline  
2013-05-24 & 56436.39 & +13.78 & 21.46\,$\pm$\,0.22 & $y_\mathrm{PS1}$ & PS1 \\
  \hline  
  \end{tabular}
  \\[1.5ex]
  \flushleft
  The tabulated magnitudes are given ``as observed'', i.e. neither corrected for dust extinction nor $K$-corrected. $^*$Rest frame epochs (assuming a redshift of 0.163) with respect to the first detection on 56420.36 (MJD). 
PS1 = Panoramic Survey Telescope \& Rapid Response System 1.
\end{table*}

Table \ref{table:PS1-13bmf_journal_spectra} gives the journal of spectroscopic observations for PS1-13bmf. The calibrated spectra of PS1-13bmf are presented in the right panel of Figure \ref{figure:PS1-13bmf}. They are corrected for galactic reddening and redshift ($z$ = 0.163). 
The +5\,d spectrum of PS1-13abg is a featureless blue continuum and displays no visible features. The +8\,d is similar with very weak H$\alpha$ and H$\beta$ visible. The last two spectra display more prominent lines of H$\alpha$ and H$\beta$ as well as weak lines of iron, in particular Fe\,{\sc ii}\,$\lambda$5169.

\begin{table*}
  \caption{PS1-13bmf: Journal of spectroscopic observations}
  \label{table:PS1-13bmf_journal_spectra}
  \centering
  \begin{tabular}{c c     c            c              c            c                               }
  \hline 
\multirow{2}{*}{Date} & \multirow{2}{*}{MJD} & Epoch$^*$  & Wavelength   & Resolution & \multirow{2}{*}{Telescope+Instrument} \\
             &           & rest frame & range in \AA &  \AA       &                               \\
\hline 
  2013-05-13 & 56425.96  &  +4.82 & 3700 - 9285 &  16.7      & GTC+OSIRIS+R300B              \\
  2013-05-17 & 56430.03  &  +8.31 & 3800 - 9265 &  21.5      & GTC+OSIRIS+R300B              \\
  2013-06-10 & 56454.00  & +28.92 & 3900 - 9295 &  16.5      & GTC+OSIRIS+R300B              \\
  2013-07-01 & 56474.97  & +46.96 & 3760 - 9265 &  17.0      & GTC+OSIRIS+R300B              \\
  \hline  
  \end{tabular}
  \\[1.5ex]
  \flushleft
  $^*$Rest frame epochs (assuming a redshift of 0.163) with respect to the first detection on 56420.36 (MJD). The resolution was determined from the FWHM of the O\,{\sc i} $\lambda$5577.34 sky line.
\end{table*}

\subsection{PS1-13atm}
PS1-13atm was discovered on 2013 April 14 by the PS1 MD survey and $g_{\mathrm{PS1}}r_{\mathrm{PS1}}i_{\mathrm{PS1}}z_{\mathrm{PS1}}y_{\mathrm{PS1}}$ photometry was obtained up to 22\,days after the first detection. We obtained three GTC spectra ranging from +16 to +44\,days after discovery. 

No spectroscopic redshifts of the host galaxy of PS1-13atm is available. The redshift of PS1-13atm was therefore determined by measuring the narrow H$\alpha$ component (presumably from the host galaxy) in all spectra and averaging the respective results yields $z = 0.220\,\pm\,0.001$. 

No features attributable to the Na\,{\sc i} doublet from interstellar gas either in the Milky Way or the host galaxy, are apparent in the spectra of PS1-13atm. Due to the low signal-to-noise ratio of the spectra. we were unable to determine a reliable value for the extinction.

The light curves of PS1-13atm are presented in the left panel of Figure \ref{figure:PS1-13atm}. Table \ref{table:PS1-13atm_photometry} shows the log of imaging observations and the calibrated magnitudes.
PS1-13atm was discovered only one day after the last non-detection in the $r_\mathrm{PS1}$-band on MJD 56395.29 which provides an explosion epoch of MJD\,$56395.8\,\pm\,0.5$. The photometric coverage is sparse -- in particular, the $g_{\mathrm{PS1}}$- and $r_{\mathrm{PS1}}$-band have only one data point -- rendering a distinction between Type II-P or Type II-L rather difficult.

\begin{figure*}[t!]
   \centering
   \begin{subfigure}[t]{0.49\textwidth}
      \includegraphics[width=\columnwidth]{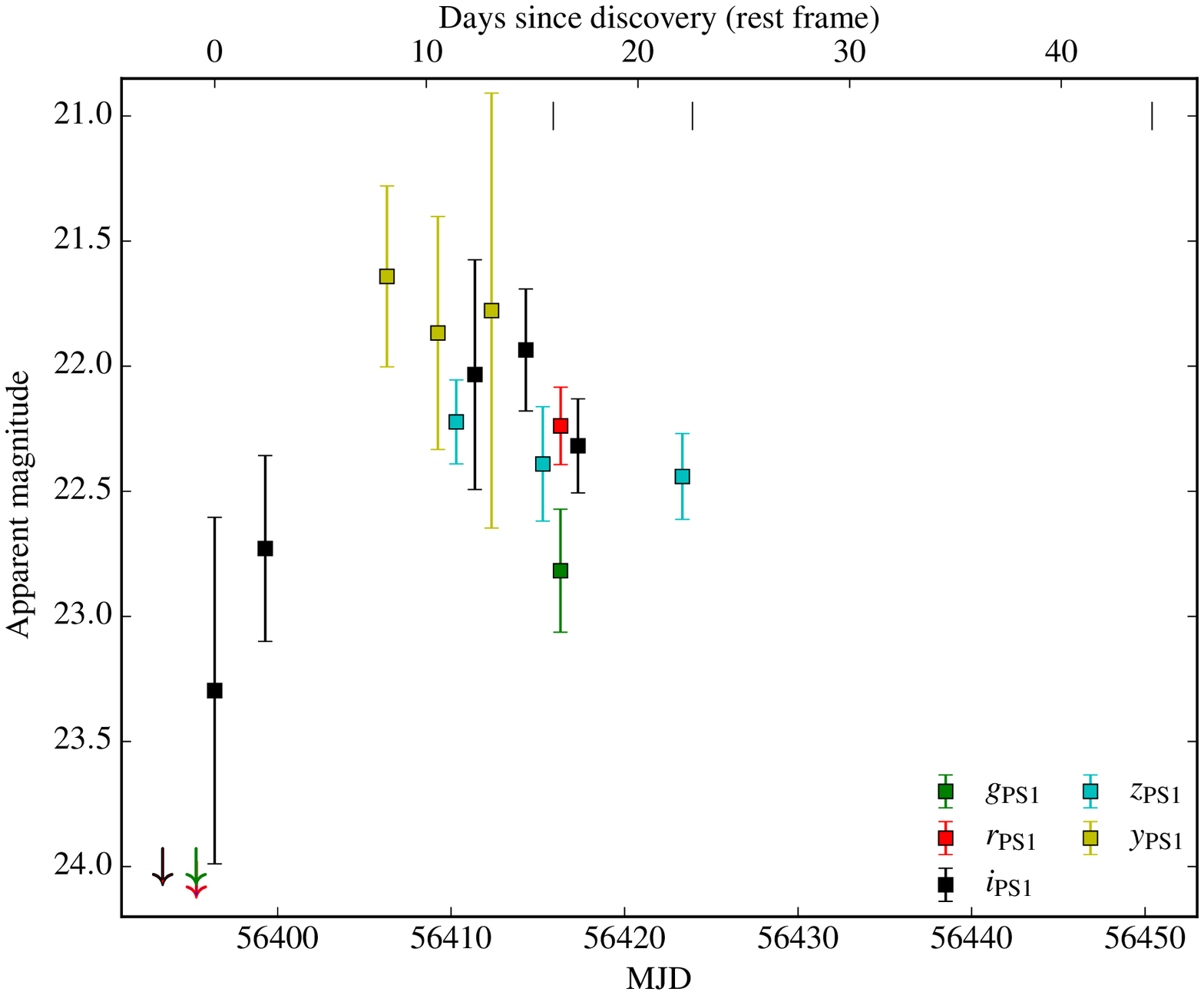}
   \end{subfigure}%
   \begin{subfigure}[t]{0.49\textwidth}
      \includegraphics[width=\columnwidth]{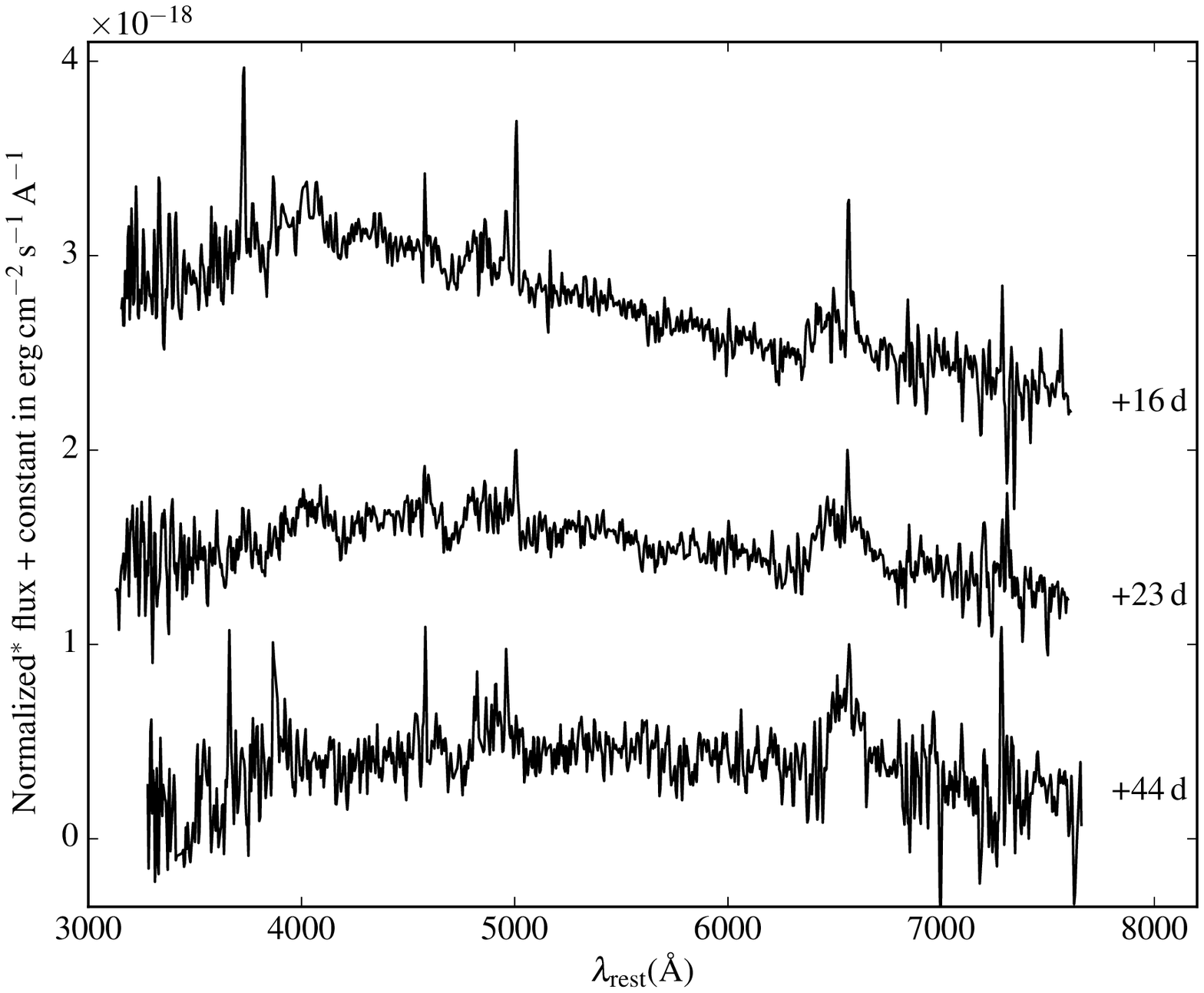}
\end{subfigure}%
   \caption{PS1-13atm. Left panel: $grizy_\mathrm{PS1}$ light curves of PS1-13atm. The vertical ticks on the top mark the epochs of the observed spectra. Right panel: PS1-13atm spectroscopy. $^*$Flux normalized to the maximum H$\alpha$ flux for better visibility of the features. The exact normalizations are: 
flux/4.3 for the +16\,d spectrum;
flux/6.1 for +23\,d;
flux/7.8 for +44\,d.
}
   \label{figure:PS1-13atm}
\end{figure*}

\begin{table*}
  \small
  \caption{PS1-13atm: Photometric observations}
  \label{table:PS1-13atm_photometry}
  \centering
  \begin{tabular}{c   c                      r            c                            c                        c           }
  \hline               
  \multirow{2}{*}{Date} & \multirow{2}{*}{MJD} & \multicolumn{1}{c}{Rest frame} & \multirow{2}{*}{Magnitude} & \multirow{2}{*}{Filter} & \multirow{2}{*}{Telescope} \\ 
                      &                      & \multicolumn{1}{c}{epoch$^*$}  &                            &                         &              \\
  \hline 
2013-05-04 & 56416.30 & +16.34 & 22.82\,$\pm$\,0.25 & $g_\mathrm{PS1}$ & PS1 \\
\hline
2013-05-04 & 56416.31 & +16.35 & 22.24\,$\pm$\,0.15 & $r_\mathrm{PS1}$ & PS1 \\
\hline
2013-04-14 & 56396.37 &  +0.00 & 23.30\,$\pm$\,0.69 & $i_\mathrm{PS1}$ & PS1 \\
2013-04-17 & 56399.28 &  +2.39 & 22.73\,$\pm$\,0.37 & $i_\mathrm{PS1}$ & PS1 \\
2013-04-29 & 56411.37 & +12.29 & 22.03\,$\pm$\,0.46 & $i_\mathrm{PS1}$ & PS1 \\
2013-05-02 & 56414.31 & +14.70 & 21.93\,$\pm$\,0.24 & $i_\mathrm{PS1}$ & PS1 \\
2013-05-05 & 56417.32 & +17.17 & 22.32\,$\pm$\,0.19 & $i_\mathrm{PS1}$ & PS1 \\
\hline
2013-04-28 & 56410.30 & +11.42 & 22.22\,$\pm$\,0.17 & $z_\mathrm{PS1}$ & PS1 \\
2013-05-03 & 56415.28 & +15.50 & 22.39\,$\pm$\,0.23 & $z_\mathrm{PS1}$ & PS1 \\
2013-05-11 & 56423.33 & +22.10 & 22.44\,$\pm$\,0.17 & $z_\mathrm{PS1}$ & PS1 \\
\hline
2013-04-24 & 56406.31 &  +8.14 & 21.64\,$\pm$\,0.36 & $y_\mathrm{PS1}$ & PS1 \\
2013-04-27 & 56409.24 & +10.55 & 21.87\,$\pm$\,0.47 & $y_\mathrm{PS1}$ & PS1 \\
2013-04-30 & 56412.33 & +13.08 & 21.78\,$\pm$\,0.87 & $y_\mathrm{PS1}$ & PS1 \\
  \hline  
  \end{tabular}
  \\[1.5ex]
  \flushleft
  The tabulated magnitudes are given ``as observed'', i.e. neither corrected for dust extinction nor $K$-corrected. $^*$Rest frame epochs (assuming a redshift of 0.220) with respect to the first detection on 56396.37 (MJD). 
PS1 = Panoramic Survey Telescope \& Rapid Response System 1.
\end{table*}

Table \ref{table:PS1-13atm_journal_spectra} gives the journal of spectroscopic observations for PS1-13atm. The calibrated spectra of PS1-13atm are presented in the right panel of Figure \ref{figure:PS1-13atm}. They are corrected for galactic reddening and redshift ($z$ = 0.220). 
The spectra of PS1-13atm clearly show H$\alpha$ and H$\beta$. The H$\alpha$ profile sports a very weak absorption and is contaminated by narrow components presumably from the host galaxy. No lines of iron can be discerned in any of the spectra. 

\begin{table*}
  \caption{PS1-13atm: Journal of spectroscopic observations}
  \label{table:PS1-13atm_journal_spectra}
  \centering
  \begin{tabular}{c c     c            c              c            c                               }
  \hline 
\multirow{2}{*}{Date} & \multirow{2}{*}{MJD} & Epoch$^*$  & Wavelength   & Resolution & \multirow{2}{*}{Telescope+Instrument} \\
             &           & rest frame & range in \AA &  \AA       &                               \\
\hline 
  2013-05-03 & 56415.89  & +16.00 & 3850 - 9280 &  17.0      & GTC+OSIRIS+R300B              \\
  2013-05-11 & 56423.91  & +22.58 & 3820 - 9265 &  16.8      & GTC+OSIRIS+R300B              \\
  2013-06-07 & 56450.40  & +44.29 & 4000 - 9415 &  16.8      & GTC+OSIRIS+R300B              \\
  \hline  
  \end{tabular}
  \\[1.5ex]
  \flushleft
  $^*$Rest frame epochs (assuming a redshift of 0.220) with respect to the first detection on 56396.37 (MJD). The resolution was determined from the FWHM of the O\,{\sc i} $\lambda$5577.34 sky line.
\end{table*}

\subsection{PS1-13bni}
PS1-13bni was discovered on 2013 April 27 by the PS1 MD survey and $g_{\mathrm{PS1}}r_{\mathrm{PS1}}i_{\mathrm{PS1}}z_{\mathrm{PS1}}y_{\mathrm{PS1}}$ photometry was obtained up to 99\,days after the first detection. We obtained five GTC spectra ranging from +14 to +51\,days after discovery. 

We used SNID \citep{Blondin2007} to determine a redshift of PS1-13bni. The results span from z=0.323 to z=0.344 with a median of z=0.335. We therefore adopt this values as the redshift of the SN, and give the range in redshifts mentioned above as a conservative estimate of its uncertainty: $z = 0.335^{+0.009}_{-0.012}$. 

Due to the low signal-to-noise ratio of the spectra we were unable to determine an upper limit for the host galaxy extinction.

The light curves of PS1-13bni are presented in the left panel of Figure~\ref{figure:PS1-13bni}. Table~\ref{table:PS1-13bni_photometry} shows the log of imaging observations and the calibrated magnitudes.
The light curves of PS1-13bni decline at rates of 
0.64\,$\pm$\,0.40\,mag/50\,d in the $r_{\mathrm{PS1}}$-band, 
0.29\,$\pm$\,0.50\,mag/50\,d in the $i_{\mathrm{PS1}}$-band, and 
0.28\,$\pm$\,0.35\,mag/50\,d in the $z_{\mathrm{PS1}}$-band. 
While PS1-13bni is likely a Type II-P SN, the relatively large uncertainties in the decline rates, stemming from similarly uncertain photometry make it difficult to clearly classify PS1-13bni as either a Type II-P or II-L SN. 

\begin{figure*}[t!]
   \centering
   \begin{subfigure}[t]{0.49\textwidth}
      \includegraphics[width=\columnwidth]{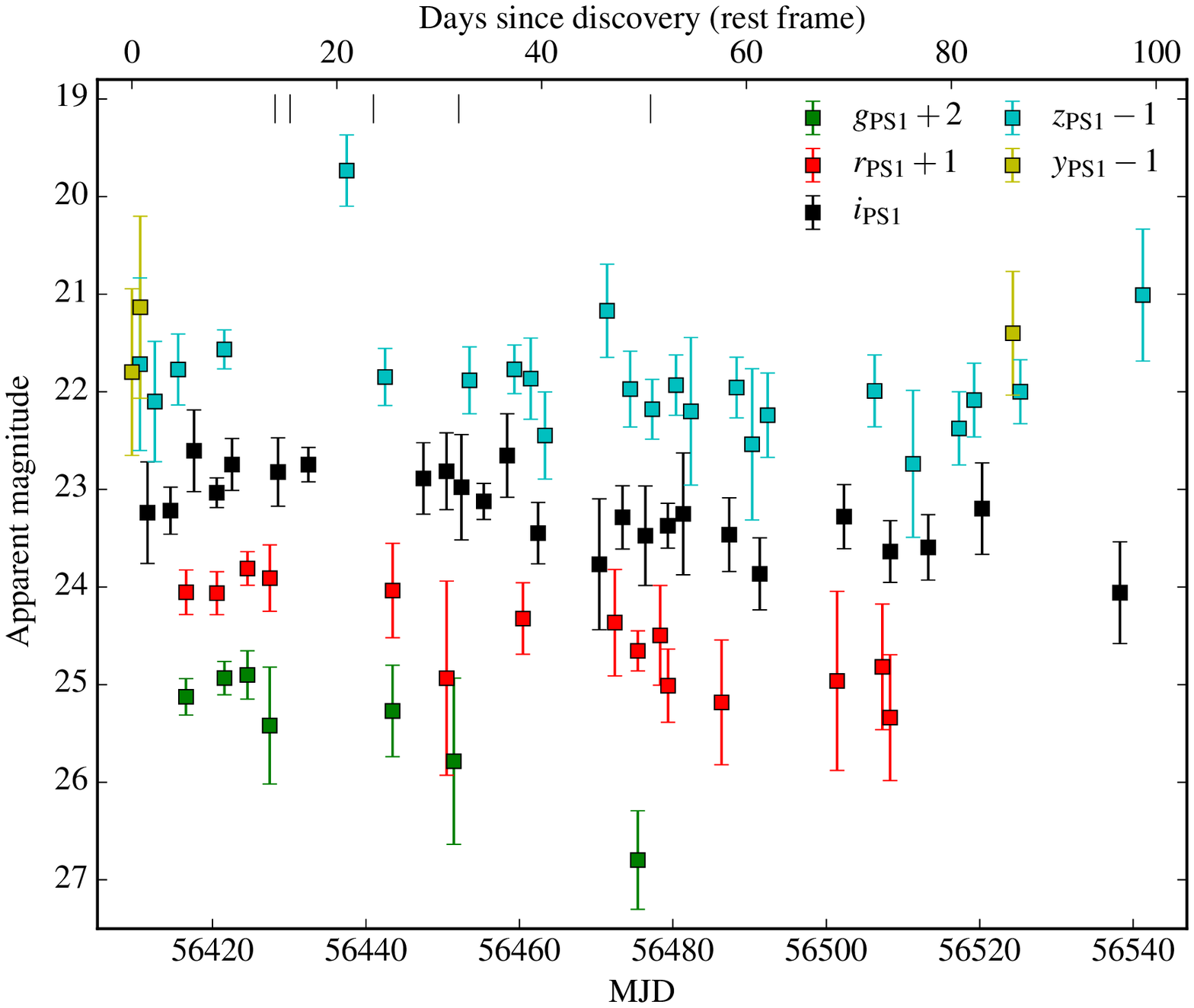}
   \end{subfigure}%
   \begin{subfigure}[t]{0.49\textwidth}
      \includegraphics[width=\columnwidth]{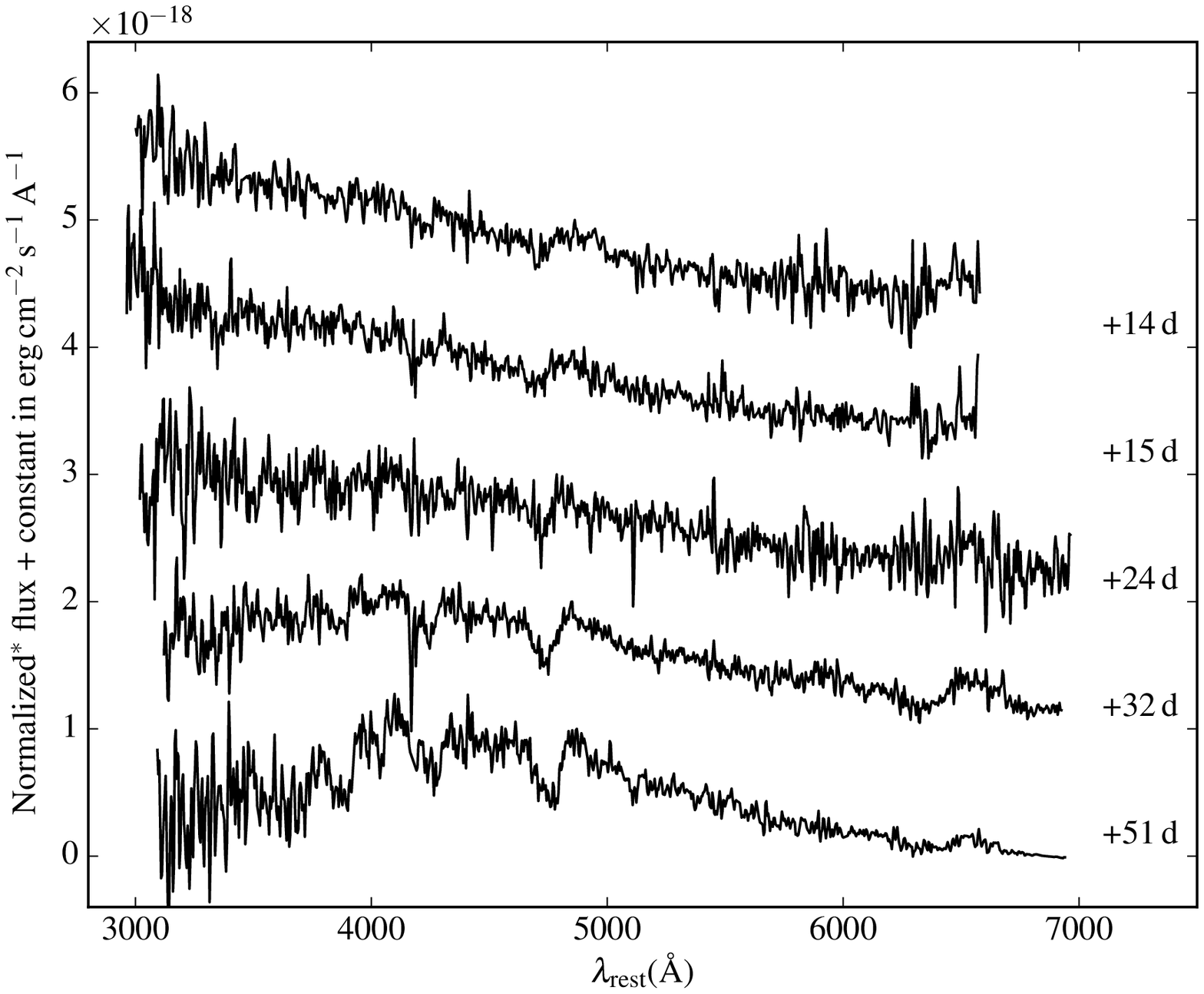}
\end{subfigure}%
   \caption{PS1-13bni. Left panel: $grizy_\mathrm{PS1}$ light curves of PS1-13bni. The vertical ticks on the top mark the epochs of the observed spectra. Right panel: PS1-13bni spectroscopy. $^*$Flux normalized to the maximum H$\alpha$ flux for better visibility of the features. The exact normalizations are: 
flux/2.5 for the +14\,d spectrum;
flux/2.5 for +15\,d;
flux/2.5 for +24\,d;
flux/2.2 for +32\,d;
flux/2.1 for +51\,d.
}
   \label{figure:PS1-13bni}
\end{figure*}

\begin{table*}
  \small
  \caption{PS1-13bni: Photometric observations}
  \label{table:PS1-13bni_photometry}
  \centering
  \begin{tabular}{c   c                      r            c                            c                        c           }
  \hline               
  \multirow{2}{*}{Date} & \multirow{2}{*}{MJD} & \multicolumn{1}{c}{Rest frame} & \multirow{2}{*}{Magnitude} & \multirow{2}{*}{Filter} & \multirow{2}{*}{Telescope} \\ 
                      &                      & \multicolumn{1}{c}{epoch$^*$}  &                            &                         &              \\
  \hline 
2013-05-04 & 56416.54 &  +5.28 & 23.12\,$\pm$\,0.19 & $g_\mathrm{PS1}$ & PS1 \\
2013-05-09 & 56421.55 &  +9.03 & 22.93\,$\pm$\,0.17 & $g_\mathrm{PS1}$ & PS1 \\
2013-05-12 & 56424.55 & +11.28 & 22.90\,$\pm$\,0.25 & $g_\mathrm{PS1}$ & PS1 \\
2013-05-15 & 56427.45 & +13.45 & 23.42\,$\pm$\,0.60 & $g_\mathrm{PS1}$ & PS1 \\
2013-05-31 & 56443.46 & +25.44 & 23.27\,$\pm$\,0.47 & $g_\mathrm{PS1}$ & PS1 \\
2013-06-08 & 56451.45 & +31.43 & 23.78\,$\pm$\,0.85 & $g_\mathrm{PS1}$ & PS1 \\
2013-07-02 & 56475.43 & +49.39 & 24.80\,$\pm$\,0.50 & $g_\mathrm{PS1}$ & PS1 \\
  \hline  
2013-05-04 & 56416.55 &  +5.29 & 23.05\,$\pm$\,0.23 & $r_\mathrm{PS1}$ & PS1 \\
2013-05-08 & 56420.57 &  +8.29 & 23.06\,$\pm$\,0.22 & $r_\mathrm{PS1}$ & PS1 \\
2013-05-12 & 56424.56 & +11.29 & 22.81\,$\pm$\,0.17 & $r_\mathrm{PS1}$ & PS1 \\
2013-05-15 & 56427.46 & +13.46 & 22.91\,$\pm$\,0.34 & $r_\mathrm{PS1}$ & PS1 \\
2013-05-31 & 56443.48 & +25.45 & 23.04\,$\pm$\,0.48 & $r_\mathrm{PS1}$ & PS1 \\
2013-06-07 & 56450.52 & +30.73 & 23.93\,$\pm$\,0.99 & $r_\mathrm{PS1}$ & PS1 \\
2013-06-17 & 56460.46 & +38.17 & 23.32\,$\pm$\,0.37 & $r_\mathrm{PS1}$ & PS1 \\
2013-06-29 & 56472.43 & +47.15 & 23.36\,$\pm$\,0.54 & $r_\mathrm{PS1}$ & PS1 \\
2013-07-02 & 56475.44 & +49.40 & 23.65\,$\pm$\,0.21 & $r_\mathrm{PS1}$ & PS1 \\
2013-07-05 & 56478.35 & +51.58 & 23.50\,$\pm$\,0.51 & $r_\mathrm{PS1}$ & PS1 \\
2013-07-06 & 56479.37 & +52.34 & 24.01\,$\pm$\,0.38 & $r_\mathrm{PS1}$ & PS1 \\
2013-07-13 & 56486.34 & +57.57 & 24.18\,$\pm$\,0.64 & $r_\mathrm{PS1}$ & PS1 \\
2013-07-28 & 56501.41 & +68.85 & 23.96\,$\pm$\,0.92 & $r_\mathrm{PS1}$ & PS1 \\
2013-08-03 & 56507.31 & +73.27 & 23.82\,$\pm$\,0.64 & $r_\mathrm{PS1}$ & PS1 \\
2013-08-04 & 56508.31 & +74.02 & 24.34\,$\pm$\,0.64 & $r_\mathrm{PS1}$ & PS1 \\
  \hline  
2013-04-29 & 56411.52 &  +1.52 & 23.24\,$\pm$\,0.52 & $i_\mathrm{PS1}$ & PS1 \\
2013-05-02 & 56414.52 &  +3.77 & 23.22\,$\pm$\,0.24 & $i_\mathrm{PS1}$ & PS1 \\
2013-05-05 & 56417.58 &  +6.06 & 22.60\,$\pm$\,0.42 & $i_\mathrm{PS1}$ & PS1 \\
2013-05-08 & 56420.55 &  +8.28 & 23.03\,$\pm$\,0.15 & $i_\mathrm{PS1}$ & PS1 \\
2013-05-10 & 56422.54 &  +9.77 & 22.74\,$\pm$\,0.27 & $i_\mathrm{PS1}$ & PS1 \\
2013-05-16 & 56428.54 & +14.26 & 22.82\,$\pm$\,0.35 & $i_\mathrm{PS1}$ & PS1 \\
2013-05-20 & 56432.50 & +17.23 & 22.75\,$\pm$\,0.18 & $i_\mathrm{PS1}$ & PS1 \\
2013-06-04 & 56447.48 & +28.45 & 22.89\,$\pm$\,0.37 & $i_\mathrm{PS1}$ & PS1 \\
2013-06-07 & 56450.50 & +30.72 & 22.81\,$\pm$\,0.39 & $i_\mathrm{PS1}$ & PS1 \\
2013-06-09 & 56452.45 & +32.18 & 22.98\,$\pm$\,0.54 & $i_\mathrm{PS1}$ & PS1 \\
2013-06-12 & 56455.35 & +34.35 & 23.12\,$\pm$\,0.18 & $i_\mathrm{PS1}$ & PS1 \\
2013-06-15 & 56458.41 & +36.64 & 22.65\,$\pm$\,0.43 & $i_\mathrm{PS1}$ & PS1 \\
2013-06-19 & 56462.44 & +39.66 & 23.45\,$\pm$\,0.31 & $i_\mathrm{PS1}$ & PS1 \\
2013-06-27 & 56470.41 & +45.63 & 23.77\,$\pm$\,0.67 & $i_\mathrm{PS1}$ & PS1 \\
2013-06-30 & 56473.44 & +47.90 & 23.29\,$\pm$\,0.32 & $i_\mathrm{PS1}$ & PS1 \\
2013-07-03 & 56476.42 & +50.13 & 23.47\,$\pm$\,0.51 & $i_\mathrm{PS1}$ & PS1 \\
2013-07-06 & 56479.35 & +52.32 & 23.37\,$\pm$\,0.23 & $i_\mathrm{PS1}$ & PS1 \\
2013-07-08 & 56481.34 & +53.82 & 23.25\,$\pm$\,0.62 & $i_\mathrm{PS1}$ & PS1 \\
2013-07-14 & 56487.35 & +58.32 & 23.46\,$\pm$\,0.38 & $i_\mathrm{PS1}$ & PS1 \\
2013-07-18 & 56491.32 & +61.29 & 23.86\,$\pm$\,0.37 & $i_\mathrm{PS1}$ & PS1 \\
2013-07-29 & 56502.32 & +69.53 & 23.28\,$\pm$\,0.33 & $i_\mathrm{PS1}$ & PS1 \\
2013-08-04 & 56508.33 & +74.04 & 23.64\,$\pm$\,0.32 & $i_\mathrm{PS1}$ & PS1 \\
2013-08-09 & 56513.28 & +77.74 & 23.59\,$\pm$\,0.34 & $i_\mathrm{PS1}$ & PS1 \\
2013-08-16 & 56520.31 & +83.01 & 23.20\,$\pm$\,0.47 & $i_\mathrm{PS1}$ & PS1 \\
2013-09-03 & 56538.28 & +96.47 & 24.06\,$\pm$\,0.52 & $i_\mathrm{PS1}$ & PS1 \\
  \hline  
2013-04-28 & 56410.54 &  +0.78 & 22.72\,$\pm$\,0.88 & $z_\mathrm{PS1}$ & PS1 \\
2013-04-30 & 56412.48 &  +2.23 & 23.10\,$\pm$\,0.62 & $z_\mathrm{PS1}$ & PS1 \\
2013-05-03 & 56415.51 &  +4.51 & 22.77\,$\pm$\,0.36 & $z_\mathrm{PS1}$ & PS1 \\
2013-05-09 & 56421.53 &  +9.02 & 22.57\,$\pm$\,0.20 & $z_\mathrm{PS1}$ & PS1 \\
2013-05-25 & 56437.49 & +20.97 & 20.73\,$\pm$\,0.36 & $z_\mathrm{PS1}$ & PS1 \\
2013-05-30 & 56442.49 & +24.71 & 22.85\,$\pm$\,0.29 & $z_\mathrm{PS1}$ & PS1 \\
2013-06-10 & 56453.50 & +32.96 & 22.88\,$\pm$\,0.34 & $z_\mathrm{PS1}$ & PS1 \\
2013-06-16 & 56459.35 & +37.34 & 22.77\,$\pm$\,0.25 & $z_\mathrm{PS1}$ & PS1 \\
2013-06-18 & 56461.47 & +38.93 & 22.86\,$\pm$\,0.42 & $z_\mathrm{PS1}$ & PS1 \\
2013-06-20 & 56463.33 & +40.33 & 23.45\,$\pm$\,0.45 & $z_\mathrm{PS1}$ & PS1 \\
2013-06-28 & 56471.43 & +46.39 & 22.17\,$\pm$\,0.48 & $z_\mathrm{PS1}$ & PS1 \\
2013-07-01 & 56474.44 & +48.65 & 22.97\,$\pm$\,0.39 & $z_\mathrm{PS1}$ & PS1 \\
2013-07-04 & 56477.31 & +50.80 & 23.18\,$\pm$\,0.31 & $z_\mathrm{PS1}$ & PS1 \\
2013-07-07 & 56480.41 & +53.12 & 22.93\,$\pm$\,0.31 & $z_\mathrm{PS1}$ & PS1 \\
2013-07-09 & 56482.35 & +54.57 & 23.20\,$\pm$\,0.76 & $z_\mathrm{PS1}$ & PS1 \\
2013-07-15 & 56488.31 & +59.04 & 22.96\,$\pm$\,0.31 & $z_\mathrm{PS1}$ & PS1 \\
2013-07-17 & 56490.33 & +60.55 & 23.54\,$\pm$\,0.77 & $z_\mathrm{PS1}$ & PS1 \\
2013-07-19 & 56492.36 & +62.07 & 23.24\,$\pm$\,0.43 & $z_\mathrm{PS1}$ & PS1 \\
2013-08-02 & 56506.30 & +72.51 & 22.99\,$\pm$\,0.37 & $z_\mathrm{PS1}$ & PS1 \\
  \hline  
  \end{tabular}
\end{table*}

\begin{table*}
  \small
  \caption*{Table \ref{table:PS1-13bni_photometry}: (continued)}
  \centering
  \begin{tabular}{c   c                      r            c                            c                        c           }
  \hline               
  \multirow{2}{*}{Date} & \multirow{2}{*}{MJD} & \multicolumn{1}{c}{Rest frame} & \multirow{2}{*}{Magnitude} & \multirow{2}{*}{Filter} & \multirow{2}{*}{Telescope} \\ 
                      &                      & \multicolumn{1}{c}{epoch$^*$}  &                            &                         &              \\
  \hline
2013-08-07 & 56511.31 & +76.26 & 23.74\,$\pm$\,0.75 & $z_\mathrm{PS1}$ & PS1 \\
2013-08-13 & 56517.30 & +80.75 & 23.38\,$\pm$\,0.38 & $z_\mathrm{PS1}$ & PS1 \\
2013-08-15 & 56519.27 & +82.23 & 23.09\,$\pm$\,0.38 & $z_\mathrm{PS1}$ & PS1 \\
2013-08-21 & 56525.29 & +86.74 & 23.00\,$\pm$\,0.33 & $z_\mathrm{PS1}$ & PS1 \\
2013-09-06 & 56541.27 & +98.71 & 22.01\,$\pm$\,0.68 & $z_\mathrm{PS1}$ & PS1 \\
  \hline  
2013-04-27 & 56409.49 &  +0.00 & 22.80\,$\pm$\,0.85 & $y_\mathrm{PS1}$ & PS1 \\
2013-04-28 & 56410.58 &  +0.81 & 22.13\,$\pm$\,0.93 & $y_\mathrm{PS1}$ & PS1 \\
2013-08-20 & 56524.31 & +86.01 & 22.40\,$\pm$\,0.63 & $y_\mathrm{PS1}$ & PS1 \\
  \hline  
  \end{tabular}
  \\[1.5ex]
  \flushleft
  The tabulated magnitudes are given ``as observed'', i.e. neither corrected for dust extinction nor $K$-corrected. $^*$Rest frame epochs (assuming a redshift of 0.335) with respect to the date of discovery on 56409.49 (MJD). 
PS1 = Panoramic Survey Telescope \& Rapid Response System 1.
\end{table*}

Table~\ref{table:PS1-13bni_journal_spectra} gives the journal of spectroscopic observations for PS1-13bni. The calibrated spectra of PS1-13bni are presented in the right panel of Figure \ref{figure:PS1-13bni}. They are corrected for galactic reddening and redshift ($z$ = 0.335). 
H$\beta$ becomes more prominent as time elapses. Due to the high redshift of $z$ = 0.335 of PS1-13bni H$\alpha$ falls near the end of the spectrum with poor sensitivity of the detector. No iron lines can be discerned in any of the spectra.

\begin{table*}
  \caption{PS1-13bni: Journal of spectroscopic observations}
  \label{table:PS1-13bni_journal_spectra}
  \centering
  \begin{tabular}{c c     c            c              c            c                               }
  \hline 
\multirow{2}{*}{Date} & \multirow{2}{*}{MJD} & Epoch$^*$  & Wavelength   & Resolution & \multirow{2}{*}{Telescope+Instrument} \\
             &           & rest frame & range in \AA &  \AA       &                               \\
\hline 
  2013-05-15 & 56428.16  & +13.98 & 3800 - 9245 &  21.3      & GTC+OSIRIS+R300B              \\
  2013-05-17 & 56430.10  & +15.43 & 3800 - 9245 &  21.4      & GTC+OSIRIS+R300B              \\
  2013-05-28 & 56440.97  & +23.58 & 4030 - 9295 &  21.3      & GTC+OSIRIS+R300B              \\
  2013-06-08 & 56452.09  & +31.90 & 3750 - 9245 &  21.6      & GTC+OSIRIS+R300B              \\
  2013-07-03 & 56477.09  & +50.63 & 3870 - 9265 &  21.5      & GTC+OSIRIS+R300B              \\
  \hline  
  \end{tabular}
  \\[1.5ex]
  \flushleft
  $^*$Rest frame epochs (assuming a redshift of 0.335) with respect to the first detection on 56409.49 (MJD). The resolution was determined from the FWHM of the O\,{\sc i} $\lambda$5577.34 sky line.
\end{table*}

\section{Additional Tables} 
\label{section:appendix:Additional_Tables}

\subsection{Temperature evolution} 
\label{section:appendix:Temperature_evolution}

Table~\ref{table:EPM_photometry} shows the rest frame photometry, interpolated to the times of spectral observations, and the derived temperature evolution for each SN (see Section \ref{Section:Temperature_evolution}). 

\begin{table*}
  \caption{Interpolated rest frame photometry and temperature evolution}
  \label{table:EPM_photometry}
  \centering
  \begin{tabular}{c    c         c          r@{.}l           c          c                 c           r@{\,$\pm$\,}l          }
  \hline  
  \multirow{2}{*}{SN}  & \multirow{2}{*}{Date} & \multirow{2}{*}{MJD} & \multicolumn{2}{c}{Epoch$^*$}      & $B^{**}$ & $V^{**}$ & $I^{**}$ & \multicolumn{2}{c}{$T_{\star,BVI}$}  \\
                       &                       &                      & \multicolumn{2}{c}{rest frame (d)} & mag      & mag       & mag      & \multicolumn{2}{c}{K} \\
\hline 
\multirow{4}{*}{SN 2013ca} & 2013-05-10 & 56422.97  & +29&63 & 19.27\,$\pm$\,0.11 & 18.88\,$\pm$\,0.06 & 18.41\,$\pm$\,0.15 & 7279&557  \\
                           & 2013-05-15 & 56428.02  & +34&46 & 19.43\,$\pm$\,0.11 & 18.92\,$\pm$\,0.06 & 18.43\,$\pm$\,0.15 & 6744&454  \\
                           & 2013-05-27 & 56439.99  & +45&92 & 19.77\,$\pm$\,0.12 & 19.05\,$\pm$\,0.07 & 18.58\,$\pm$\,0.14 & 6108&331  \\
                           & 2013-06-11 & 56454.91  & +60&20 & 20.26\,$\pm$\,0.14 & 19.19\,$\pm$\,0.08 & 18.62\,$\pm$\,0.13 & 5286&207  \\
\hline
\multirow{5}{*}{LSQ13cuw} & 2013-11-24 & 56620.08 & +25&49 & 19.49\,$\pm$\,0.12 & 19.09\,$\pm$\,0.09 & 18.55\,$\pm$\,0.09 & 7067&524 \\
                          & 2013-11-26 & 56622.21 & +27&53 & 19.62\,$\pm$\,0.12 & 19.27\,$\pm$\,0.09 & 18.68\,$\pm$\,0.11 & 7195&584 \\
                          & 2013-12-01 & 56627.22 & +32&32 & 19.96\,$\pm$\,0.12 & 19.60\,$\pm$\,0.09 & 19.04\,$\pm$\,0.09 & 7245&575 \\
                          & 2013-12-22 & 56648.15 & +52&33 & 21.08\,$\pm$\,0.12 & 20.62\,$\pm$\,0.09 & 20.04\,$\pm$\,0.09 & 6734&448 \\
                          & 2014-01-24 & 56681.13 & +83&86 & 22.29\,$\pm$\,0.12 & 21.30\,$\pm$\,0.09 & 21.00\,$\pm$\,0.09 & 5751&205 \\
\hline
\multirow{4}{*}{PS1-13wr} & 2013-03-11 & 56362.26  & +11&77 &  21.59\,$\pm$\,0.25 & 20.90\,$\pm$\,0.16 & 20.23\,$\pm$\,0.12 &  5856&573 \\
                          & 2013-03-17 & 56369.12  & +18&14 &  21.69\,$\pm$\,0.25 & 20.94\,$\pm$\,0.16 & 20.24\,$\pm$\,0.12 &  5653&501 \\
                          & 2013-03-25 & 56376.23  & +24&74 &  21.78\,$\pm$\,0.25 & 20.96\,$\pm$\,0.16 & 20.24\,$\pm$\,0.12 &  5477&440 \\
                          & 2013-04-11 & 56394.15  & +41&40 &  22.00\,$\pm$\,0.25 & 21.03\,$\pm$\,0.16 & 20.23\,$\pm$\,0.12 &  5111&338 \\
\hline  
\multirow{6}{*}{PS1-14vk} & 2014-03-26 & 56743.15 &  +2&44 & 20.32\,$\pm$\,0.05 & 20.03\,$\pm$\,0.12 & 19.96\,$\pm$\,0.15 & 8865&732 \\
                          & 2014-04-04 & 56752.03 & +10&66 & 20.60\,$\pm$\,0.05 & 20.31\,$\pm$\,0.12 & 20.20\,$\pm$\,0.15 & 8787&718 \\
                          & 2014-04-08 & 56755.91 & +14&26 & 20.72\,$\pm$\,0.05 & 20.40\,$\pm$\,0.12 & 20.32\,$\pm$\,0.15 & 8645&681 \\
                          & 2014-04-21 & 56769.13 & +26&50 & 21.68\,$\pm$\,0.21 & 20.84\,$\pm$\,0.12 & 20.56\,$\pm$\,0.15 & 6090&398 \\
                          & 2014-04-28 & 56775.93 & +32&79 & 22.09\,$\pm$\,0.21 & 21.03\,$\pm$\,0.12 & 20.59\,$\pm$\,0.15 & 5472&315 \\
                          & 2014-05-16 & 56793.93 & +49&46 & 22.77\,$\pm$\,0.21 & 21.43\,$\pm$\,0.12 & 20.99\,$\pm$\,0.15 & 5106&222 \\
\hline
\multirow{4}{*}{PS1-12bku} & 2012-09-16 & 56186.01  & +22&66 & 20.33\,$\pm$\,0.09 & 20.32\,$\pm$\,0.07 & 20.01\,$\pm$\,0.06 & 11479&1556 \\ 
                           & 2012-09-20 & 56190.90  & +27&15 & 20.50\,$\pm$\,0.09 & 20.37\,$\pm$\,0.07 & 20.06\,$\pm$\,0.06 &  9668&3485 \\ 
                           & 2012-10-05 & 56205.94  & +40&98 & 21.00\,$\pm$\,0.09 & 20.51\,$\pm$\,0.07 & 20.10\,$\pm$\,0.06 &  6990&342  \\ 
                           & 2012-11-17 & 56249.92  & +81&40 & 21.99\,$\pm$\,0.09 & 20.58\,$\pm$\,0.09 & 20.39\,$\pm$\,0.07 &  5288&95   \\
\hline  
\multirow{3}{*}{PS1-13abg} & 2013-04-05 & 56388.06  &  +6&85 & 21.89\,$\pm$\,0.05 & 21.85\,$\pm$\,0.07 & 21.25\,$\pm$\,0.09 & 9503&2939 \\
                           & 2013-05-03 & 56415.93  & +31&66 & 22.86\,$\pm$\,0.05 & 22.10\,$\pm$\,0.07 & 21.18\,$\pm$\,0.09 & 5265&150  \\
                           & 2013-05-17 & 56429.93  & +44&13 & 23.35\,$\pm$\,0.05 & 22.21\,$\pm$\,0.07 & 21.10\,$\pm$\,0.09 & 4481&108  \\
\hline
\multirow{3}{*}{PS1-13baf} & 2013-05-15 & 56428.20  & +16&35 & 23.15\,$\pm$\,0.21 & 22.84\,$\pm$\,0.22 & 22.12\,$\pm$\,0.12 & 7073&2439 \\
                           & 2013-05-28 & 56441.18  & +27&70 & 23.26\,$\pm$\,0.21 & 22.91\,$\pm$\,0.22 & 22.20\,$\pm$\,0.12 & 6931&2072 \\
                           & 2013-06-10 & 56454.09  & +38&98 & 23.36\,$\pm$\,0.21 & 22.98\,$\pm$\,0.22 & 22.29\,$\pm$\,0.12 & 6801&1981 \\
\hline
\multirow{4}{*}{PS1-13bmf} & 2013-05-13 & 56425.96 &  +4&82 & 21.21\,$\pm$\,0.04 & 21.60\,$\pm$\,0.06 & 21.66\,$\pm$\,0.11 & 108139&31530 \\
                           & 2013-05-17 & 56430.03 &  +8&31 & 21.44\,$\pm$\,0.04 & 21.48\,$\pm$\,0.06 & 21.22\,$\pm$\,0.11 &  12607&1221  \\
                           & 2013-06-10 & 56454.00 & +28&92 & 22.54\,$\pm$\,0.04 & 22.14\,$\pm$\,0.06 & 21.68\,$\pm$\,0.11 &   7280&347   \\
                           & 2013-07-01 & 56474.97 & +46&96 & 23.20\,$\pm$\,0.04 & 22.45\,$\pm$\,0.06 & 21.79\,$\pm$\,0.11 &   5727&157   \\
\hline
\multirow{5}{*}{PS1-13bni} & 2013-05-15 & 56428.16 & +13&98 & 23.30\,$\pm$\,0.20 & 23.17\,$\pm$\,0.25 & 22.93\,$\pm$\,0.20 & 10036&2359 \\
                           & 2013-05-17 & 56430.10 & +15&43 & 23.34\,$\pm$\,0.20 & 23.17\,$\pm$\,0.25 & 22.96\,$\pm$\,0.20 &  9655&2595 \\
                           & 2013-05-28 & 56440.97 & +23&58 & 23.55\,$\pm$\,0.20 & 23.23\,$\pm$\,0.25 & 23.08\,$\pm$\,0.20 &  8459&2302 \\
                           & 2013-06-08 & 56452.09 & +31&90 & 23.76\,$\pm$\,0.20 & 23.32\,$\pm$\,0.25 & 23.15\,$\pm$\,0.20 &  7690&1515 \\
                           & 2013-07-03 & 56477.09 & +50&63 & 24.20\,$\pm$\,0.20 & 23.41\,$\pm$\,0.25 & 23.13\,$\pm$\,0.19 &  6208&602  \\
\hline
\end{tabular}
  \\[1.5ex]
  \flushleft
  $^*$Epoch relative to discovery in the SN rest frame. $^{**}$Magnitudes in the Johnson-Cousins Filter System, $K$-corrected and corrected for dust extinction.
\end{table*}

\subsection{Fit parameters for Equation \ref{equation:velocity_evolution_fit}} 
\label{section:appendix:velocity_fit_parameters}

Table \ref{table:velocity_fit_parameters} gives the fit parameters that represent the velocity evolution of the SNe 1999em, 1999gi, 2004et, 2005cs and 2006bp following Equation \ref{equation:velocity_evolution_fit} for the H$\alpha$, H$\beta$ and Fe\,{\sc ii}\,$\lambda$5169 lines.

\begin{table*}
  \small
  \caption{Fit parameters for the H$\alpha$, H$\beta$ and Fe\,{\sc ii}\,$\lambda$5169 velocity evolutions }
  \label{table:velocity_fit_parameters}
  \centering
  \begin{tabular}{c   c                      c      c       c       c           }
  \hline               
  \multirow{2}{*}{SN} & \multirow{2}{*}{Line} & $a$ & $b$ & $c$ & RMS  \\ 
                      &        & 1000\,km\,s$^{-1}$  & 0.01\,d$^{-1}$ & 1000\,km\,s$^{-1}$ & km\,s$^{-1}$\\
  \hline 
\multirow{3}{*}{SN 1999em} & H$\alpha$                    & 11.25 & 1.67 &  0.41 & 290  \\
                           & H$\beta$                     & 10.34 & 2.97 &  1.45 & 222  \\
                           & Fe\,{\sc ii}\,$\lambda$5169  &  7.40 & 3.27 &  2.29 & 355  \\
\hline 
\multirow{3}{*}{SN 1999gi} & H$\alpha$                    & 13.82 & 2.08 &  1.17 & 199  \\
                           & H$\beta$                     & 11.68 & 2.52 &  1.39 &  56  \\
                           & Fe\,{\sc ii}\,$\lambda$5169  &  8.48 & 3.43 &  2.92 &  64  \\
\hline
\multirow{3}{*}{SN 2004et} & H$\alpha$                    &  6.61 & 1.18 &  3.59 & 139  \\
                           & H$\beta$                     &  7.47 & 2.12 &  3.54 & 128  \\
                           & Fe\,{\sc ii}\,$\lambda$5169  &  8.46 & 3.32 &  2.50 & 126  \\
\hline 
\multirow{3}{*}{SN 2005cs} & H$\alpha$                    &  9.36 & 1.78 & -1.58 & 354  \\
                           & H$\beta$                     &  8.22 & 3.10 & -0.46 & 182  \\
                           & Fe\,{\sc ii}\,$\lambda$5169  &  6.69 & 6.62 &  1.54 & 266  \\
\hline
\multirow{3}{*}{SN 2006bp} & H$\alpha$                    &  8.18 & 4.70 &  8.00 & 270  \\
                           & H$\beta$                     & 10.15 & 4.71 &  6.03 & 320  \\
                           & Fe\,{\sc ii}\,$\lambda$5169  & 13.21 & 5.80 &  4.76 & 107  \\
\hline
  \end{tabular}
  \\[1.5ex]
  \flushleft
  Fit parameters for the H$\alpha$, H$\beta$ and Fe\,{\sc ii}\,$\lambda$5169 velocity evolutions. RMS = root mean square error.
\end{table*}

\subsection{EPM quantities} 
\label{section:appendix:EPM_fit_plots}

Table~\ref{table:EPM_quantities} presents the derived EPM quantities for each SN used for the distance fits (see Section \ref{section:EPM_distances} and Table \ref{table:EPM_distances}).

\begin{table*}
  \caption{EPM quantities for the SNe in our sample}
  \label{table:EPM_quantities}
  \centering
  \begin{tabular}{c          c                 c           c         c                 c                  c   c c }
  \hline 
  \multirow{2}{*}{SN} & Epoch$^*$      & Velocity & Estimate of & \multirow{2}{*}{$\theta^\dag_B\,\times\,10^{12}$} & \multirow{2}{*}{$\theta^\dag_V\,\times\,10^{12}$} & \multirow{2}{*}{$\theta^\dag_I\,\times\,10^{12}$} & \multicolumn{2}{c}{Dilution factor} \\
     & rest frame (d) & km\,s$^{-1}$ & velocity via &  &  &   & $\zeta_{BVI}$ & reference \\
\hline 
\multirow{8}{*}{SN 2013ca} 
 & \multirow{2}{*}{+29.63} & \multirow{2}{*}{6245\,$\pm$\,312} & \multirow{8}{*}{Fe\,{\sc ii}\,$\lambda$5169} & 6.4\,$\pm$\,1.2 & 6.4\,$\pm$\,0.9 & 6.3\,$\pm$\,0.8 & 0.49 & H01 \\
 &                         &                                   &                                              & 5.5\,$\pm$\,1.0 & 5.6\,$\pm$\,0.8 & 5.5\,$\pm$\,0.7 & 0.64 & D05 \\
 & \multirow{2}{*}{+34.46} & \multirow{2}{*}{5885\,$\pm$\,294} &                                              & 6.8\,$\pm$\,1.2 & 7.0\,$\pm$\,1.0 & 6.7\,$\pm$\,0.8 & 0.53 & H01 \\
 &                         &                                   &                                              & 6.0\,$\pm$\,1.0 & 6.1\,$\pm$\,0.8 & 5.8\,$\pm$\,0.7 & 0.69 & D05 \\
 & \multirow{2}{*}{+45.92} & \multirow{2}{*}{5167\,$\pm$\,258} &                                              & 7.0\,$\pm$\,1.1 & 7.6\,$\pm$\,0.9 & 6.8\,$\pm$\,0.7 & 0.60 & H01 \\
 &                         &                                   &                                              & 6.2\,$\pm$\,1.0 & 6.7\,$\pm$\,0.8 & 6.0\,$\pm$\,0.6 & 0.77 & D05 \\
 & \multirow{2}{*}{+60.20} & \multirow{2}{*}{4314\,$\pm$\,216} &                                              & 7.6\,$\pm$\,1.0 & 8.9\,$\pm$\,0.9 & 7.5\,$\pm$\,0.7 & 0.76 & H01 \\
 &                         &                                   &                                              & 6.9\,$\pm$\,0.9 & 8.0\,$\pm$\,0.8 & 6.8\,$\pm$\,0.6 & 0.92 & D05 \\
\hline 
\multirow{2}{*}{LSQ13cuw} 
 & \multirow{2}{*}{+32.34} & \multirow{2}{*}{6873\,$\pm$\,674} & \multirow{2}{*}{H$\beta$}  & 4.7\,$\pm$\,0.9 & 4.6\,$\pm$\,0.7 & 4.7\,$\pm$\,0.5 & 0.49 & H01 \\
 &                         &                                   &                            & 4.1\,$\pm$\,0.8 & 4.0\,$\pm$\,0.6 & 4.1\,$\pm$\,0.5 & 0.64 & D05 \\
\hline 
\multirow{8}{*}{PS1-13wr} 
 & \multirow{2}{*}{+11.77} & \multirow{2}{*}{5786\,$\pm$\,289} & \multirow{8}{*}{Fe\,{\sc ii}\,$\lambda$5169} & 3.3\,$\pm$\,1.0 & 3.4\,$\pm$\,0.8 & 3.3\,$\pm$\,0.5 & 0.64 & H01 \\
 &                         &                                   &                                              & 2.9\,$\pm$\,0.9 & 3.1\,$\pm$\,0.7 & 2.9\,$\pm$\,0.5 & 0.81 & D05 \\
 & \multirow{2}{*}{+18.14} & \multirow{2}{*}{5528\,$\pm$\,276} &                                              & 3.4\,$\pm$\,1.0 & 3.6\,$\pm$\,0.8 & 3.4\,$\pm$\,0.5 & 0.68 & H01 \\
 &                         &                                   &                                              & 3.0\,$\pm$\,0.9 & 3.2\,$\pm$\,0.7 & 3.0\,$\pm$\,0.5 & 0.84 & D05 \\
 & \multirow{2}{*}{+24.74} & \multirow{2}{*}{4800\,$\pm$\,240} &                                              & 3.5\,$\pm$\,0.9 & 3.7\,$\pm$\,0.8 & 3.4\,$\pm$\,0.5 & 0.71 & H01 \\
 &                         &                                   &                                              & 3.1\,$\pm$\,0.8 & 3.3\,$\pm$\,0.7 & 3.1\,$\pm$\,0.5 & 0.88 & D05 \\
 & \multirow{2}{*}{+41.40} & \multirow{2}{*}{3450\,$\pm$\,173} &                                              & 3.7\,$\pm$\,0.9 & 4.0\,$\pm$\,0.7 & 3.7\,$\pm$\,0.5 & 0.81 & H01 \\
 &                         &                                   &                                              & 3.3\,$\pm$\,0.8 & 3.6\,$\pm$\,0.7 & 3.4\,$\pm$\,0.4 & 0.97 & D05 \\
\hline 
\multirow{8}{*}{PS1-14vk} 
 & \multirow{2}{*}{+10.66} & \multirow{2}{*}{6417\,$\pm$\,321} & \multirow{8}{*}{Fe\,{\sc ii}\,$\lambda$5169} & 2.5\,$\pm$\,0.4 & 2.6\,$\pm$\,0.4 & 2.3\,$\pm$\,0.3 & 0.43 & H01 \\
 &                         &                                   &                                              & 2.2\,$\pm$\,0.3 & 2.2\,$\pm$\,0.3 & 2.0\,$\pm$\,0.2 & 0.56 & D05 \\
 & \multirow{2}{*}{+26.50} & \multirow{2}{*}{5348\,$\pm$\,267} &                                              & 2.9\,$\pm$\,0.6 & 3.3\,$\pm$\,0.5 & 2.7\,$\pm$\,0.3 & 0.60 & H01 \\
 &                         &                                   &                                              & 2.6\,$\pm$\,0.5 & 3.0\,$\pm$\,0.5 & 2.4\,$\pm$\,0.3 & 0.77 & D05 \\
 & \multirow{2}{*}{+32.79} & \multirow{2}{*}{4882\,$\pm$\,244} &                                              & 3.0\,$\pm$\,0.6 & 3.6\,$\pm$\,0.5 & 2.9\,$\pm$\,0.4 & 0.71 & H01 \\
 &                         &                                   &                                              & 2.7\,$\pm$\,0.5 & 3.2\,$\pm$\,0.5 & 2.6\,$\pm$\,0.3 & 0.88 & D05 \\
 & \multirow{2}{*}{+49.46} & \multirow{2}{*}{4507\,$\pm$\,225} &                                              & 2.6\,$\pm$\,0.4 & 3.3\,$\pm$\,0.4 & 2.6\,$\pm$\,0.3 & 0.81 & H01 \\
 &                         &                                   &                                              & 2.4\,$\pm$\,0.4 & 3.0\,$\pm$\,0.4 & 2.4\,$\pm$\,0.2 & 0.97 & D05 \\
\hline 
\multirow{6}{*}{PS1-12bku} 
 & \multirow{2}{*}{+22.66} & \multirow{2}{*}{6050\,$\pm$\,303} & \multirow{6}{*}{Fe\,{\sc ii}\,$\lambda$5169} & 1.8\,$\pm$\,0.4 & 1.8\,$\pm$\,0.3 & 1.9\,$\pm$\,0.3 & 0.41 & H01 \\ 
 &                         &                                   &                                               & 1.6\,$\pm$\,0.3 & 1.6\,$\pm$\,0.3 & 1.7\,$\pm$\,0.2 & 0.51 & D05 \\ 
 & \multirow{2}{*}{+27.15} & \multirow{2}{*}{5619\,$\pm$\,281} &                                               & 2.2\,$\pm$\,1.4 & 2.2\,$\pm$\,1.1 & 2.3\,$\pm$\,0.9 & 0.42 & H01 \\ 
 &                         &                                   &                                               & 1.9\,$\pm$\,1.2 & 1.9\,$\pm$\,1.0 & 2.0\,$\pm$\,0.8 & 0.54 & D05 \\ 
 & \multirow{2}{*}{+40.98} & \multirow{2}{*}{4700\,$\pm$\,235} &                                               & 3.1\,$\pm$\,0.4 & 3.2\,$\pm$\,0.3 & 3.0\,$\pm$\,0.2 & 0.51 & H01 \\ 
 &                         &                                   &                                               & 2.7\,$\pm$\,0.3 & 2.8\,$\pm$\,0.3 & 2.6\,$\pm$\,0.2 & 0.67 & D05 \\ 
\hline 
\multirow{2}{*}{PS1-13abg} 
 & \multirow{2}{*}{+44.13} & \multirow{2}{*}{4733\,$\pm$\,237} & \multirow{2}{*}{Fe\,{\sc ii}\,$\lambda$5169} & 2.7\,$\pm$\,0.2 & 2.9\,$\pm$\,0.2 & 2.8\,$\pm$\,0.2 & 1.05 & H01 \\
 &                         &                                   &                                              & 2.5\,$\pm$\,0.2 & 2.7\,$\pm$\,0.2 & 2.6\,$\pm$\,0.2 & 1.19 & D05 \\
\hline 
\multirow{4}{*}{PS1-13baf} 
 & \multirow{2}{*}{+16.35} & \multirow{2}{*}{6762\,$\pm$\,423} & \multirow{4}{*}{H$\beta$}  & 1.1\,$\pm$\,0.9 & 1.1\,$\pm$\,0.7 & 1.2\,$\pm$\,0.6 & 0.50 & H01 \\
 &                         &                                   &                            & 1.0\,$\pm$\,0.8 & 0.9\,$\pm$\,0.6 & 1.0\,$\pm$\,0.5 & 0.66 & D05 \\
 & \multirow{2}{*}{+27.70} & \multirow{2}{*}{5103\,$\pm$\,417} &                            & 1.1\,$\pm$\,0.8 & 1.1\,$\pm$\,0.6 & 1.1\,$\pm$\,0.5 & 0.51 & H01 \\
 &                         &                                   &                            & 1.0\,$\pm$\,0.7 & 0.9\,$\pm$\,0.5 & 1.0\,$\pm$\,0.4 & 0.67 & D05 \\
\hline 
\multirow{4}{*}{PS1-13bmf} 
 & \multirow{2}{*}{+28.92} & \multirow{2}{*}{6194\,$\pm$\,310} & \multirow{4}{*}{Fe\,{\sc ii}\,$\lambda$5169} & 1.4\,$\pm$\,0.2 & 1.4\,$\pm$\,0.1 & 1.4\,$\pm$\,0.1 & 0.49 & H01 \\
 &                         &                                   &                                              & 1.2\,$\pm$\,0.1 & 1.2\,$\pm$\,0.1 & 1.2\,$\pm$\,0.1 & 0.64 & D05 \\
 & \multirow{2}{*}{+46.96} & \multirow{2}{*}{3385\,$\pm$\,169} &                                              & 1.6\,$\pm$\,0.1 & 1.7\,$\pm$\,0.1 & 1.6\,$\pm$\,0.1 & 0.66 & H01 \\
 &                         &                                   &                                              & 1.5\,$\pm$\,0.1 & 1.6\,$\pm$\,0.1 & 1.4\,$\pm$\,0.1 & 0.83 & D05 \\
\hline 
\multirow{4}{*}{PS1-13bni} 
 & \multirow{2}{*}{+15.43} & \multirow{2}{*}{8795\,$\pm$\,562} & \multirow{4}{*}{H$\beta$}  & 0.6\,$\pm$\,0.3 & 0.6\,$\pm$\,0.2 & 0.6\,$\pm$\,0.2 & 0.42 & H01 \\
 &                         &                                   &                            & 0.5\,$\pm$\,0.3 & 0.5\,$\pm$\,0.2 & 0.5\,$\pm$\,0.2 & 0.54 & D05 \\
 & \multirow{2}{*}{+23.58} & \multirow{2}{*}{7146\,$\pm$\,604} &                            & 0.7\,$\pm$\,0.4 & 0.7\,$\pm$\,0.3 & 0.6\,$\pm$\,0.2 & 0.44 & H01 \\
 &                         &                                   &                            & 0.6\,$\pm$\,0.3 & 0.6\,$\pm$\,0.3 & 0.6\,$\pm$\,0.2 & 0.58 & D05 \\
\hline  
  \end{tabular}
  \\[1.5ex]
  \flushleft
  $^*$Epoch in SN rest frame. H01 = \citet{Hamuy2001}; D05 = \citet{Dessart2005}.
\end{table*}

\section{EPM fits of individual SNe} 
\label{section:appendix:EPM_fit_plots}

Figures \ref{figure:SN2013ca_EPM_distances}-\ref{figure:PS1-13bni_EPM_distances} show the fits through $\chi$ for the $B$, $V$, and $I$ filters to determine the distance and -- when unknown -- the time of explosion of SNe 2013ca, LSQ13cuw, PS1-13wr, PS1-12bku, PS1-13abg, PS1-13baf, PS1-13bmf, and PS1-13bni. The fit was performed twice for each SN using the dilution factors as given either by \citet{Hamuy2001} or \citet{Dessart2005}. The distance of SN 2013eq was adopted from \citet{Gall2016} and the corresponding fit is illustrated in their Figure 4. The fit for PS1-14vk is presented in Section \ref{section:EPM_distances}.

\begin{figure*}[t!]
   \centering
   \begin{subfigure}[t]{0.47\textwidth}
      \includegraphics[width=\columnwidth]{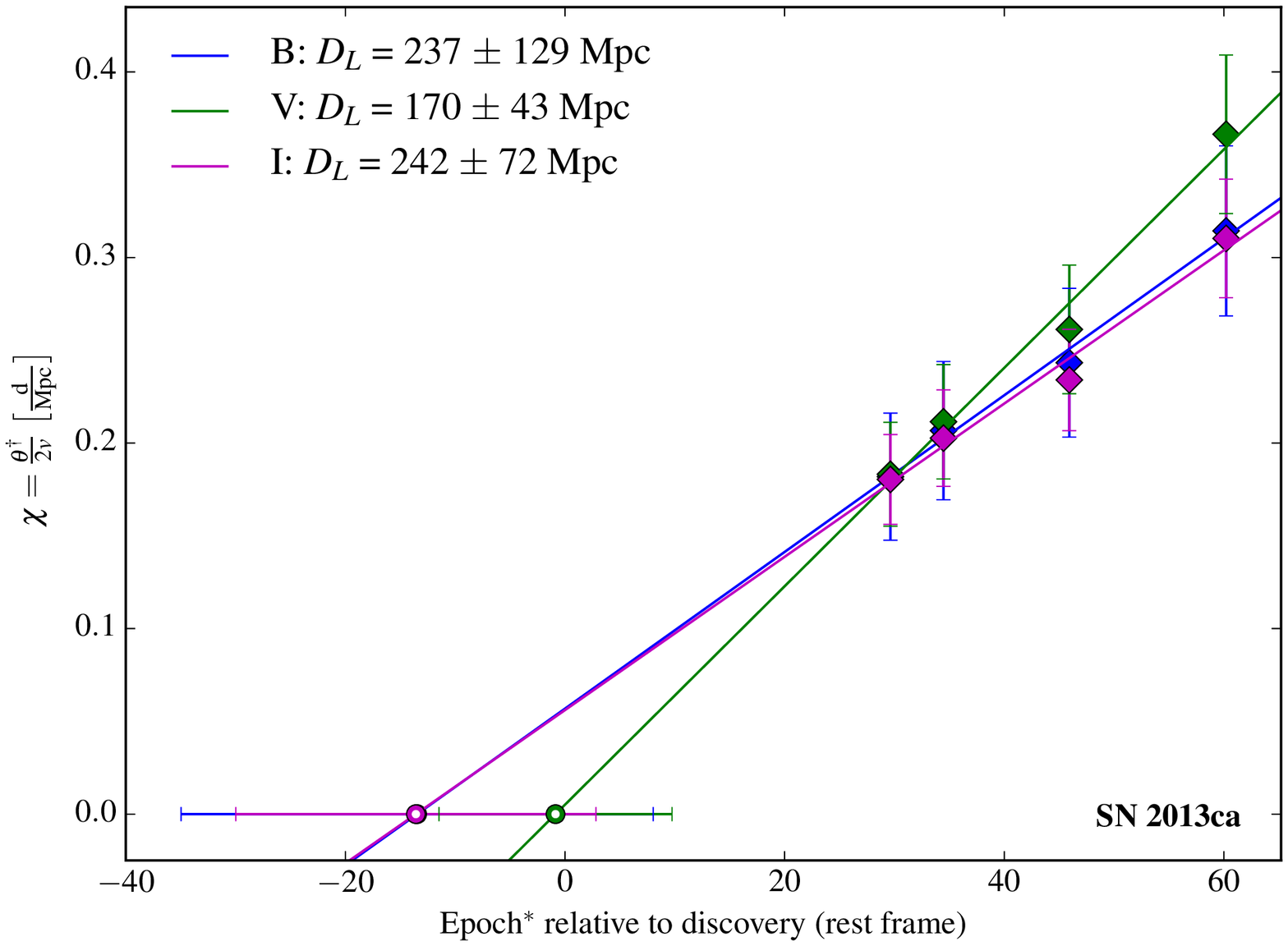}
   \end{subfigure}%
   \begin{subfigure}[t]{0.47\textwidth}
      \includegraphics[width=\columnwidth]{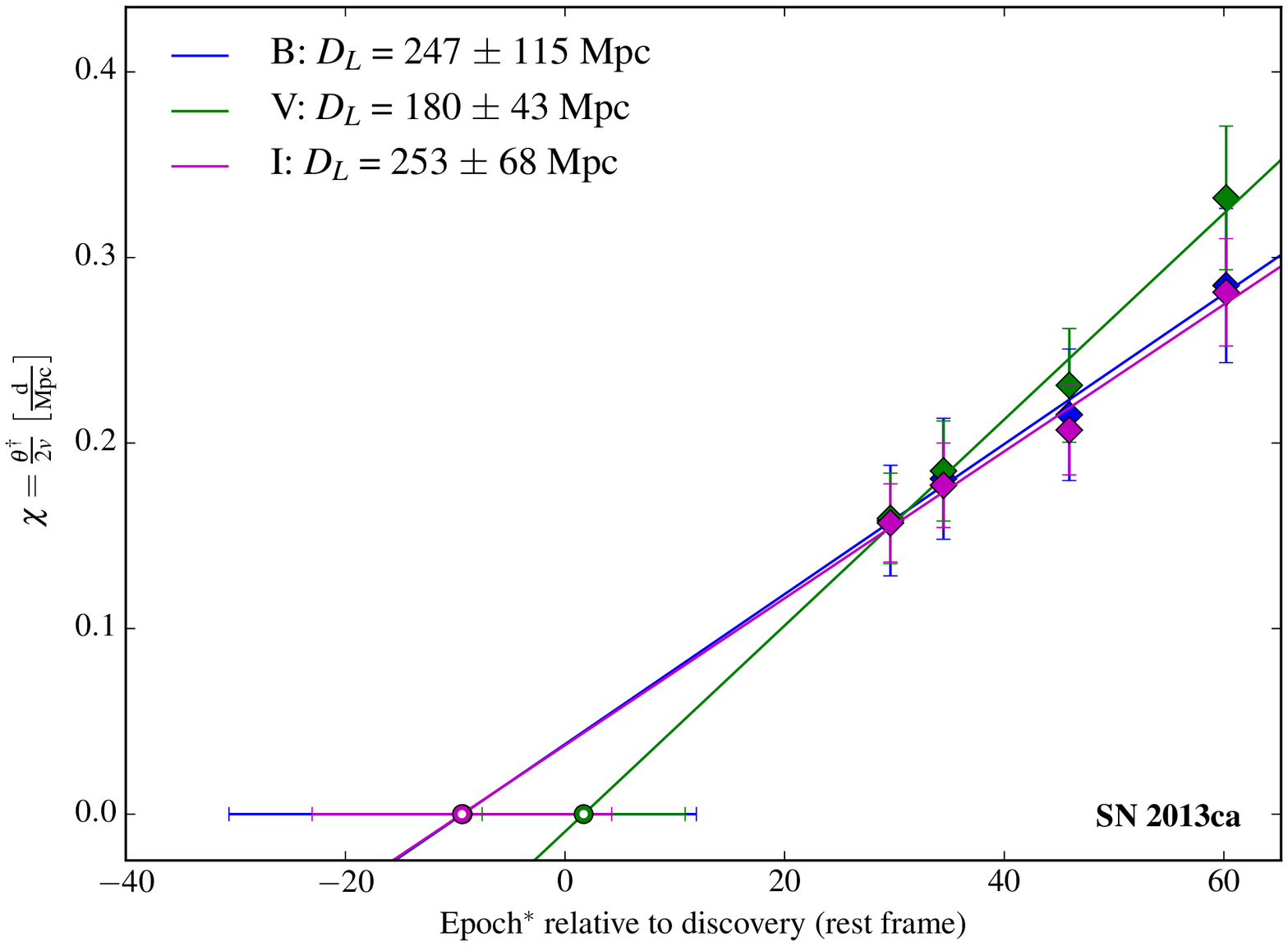}
   \end{subfigure}%
   \caption{Distance fit for SN 2013ca using $\zeta_{BVI}$ as given in \citet{Hamuy2001} (left panel) and \citet{Dessart2005} (right panel). The diamond markers denote values of $\chi$ through which the fit is made.
}
   \label{figure:SN2013ca_EPM_distances}
\end{figure*}

\begin{figure*}[t!]
   \centering
   \begin{subfigure}[t]{0.47\textwidth}
      \includegraphics[width=\columnwidth]{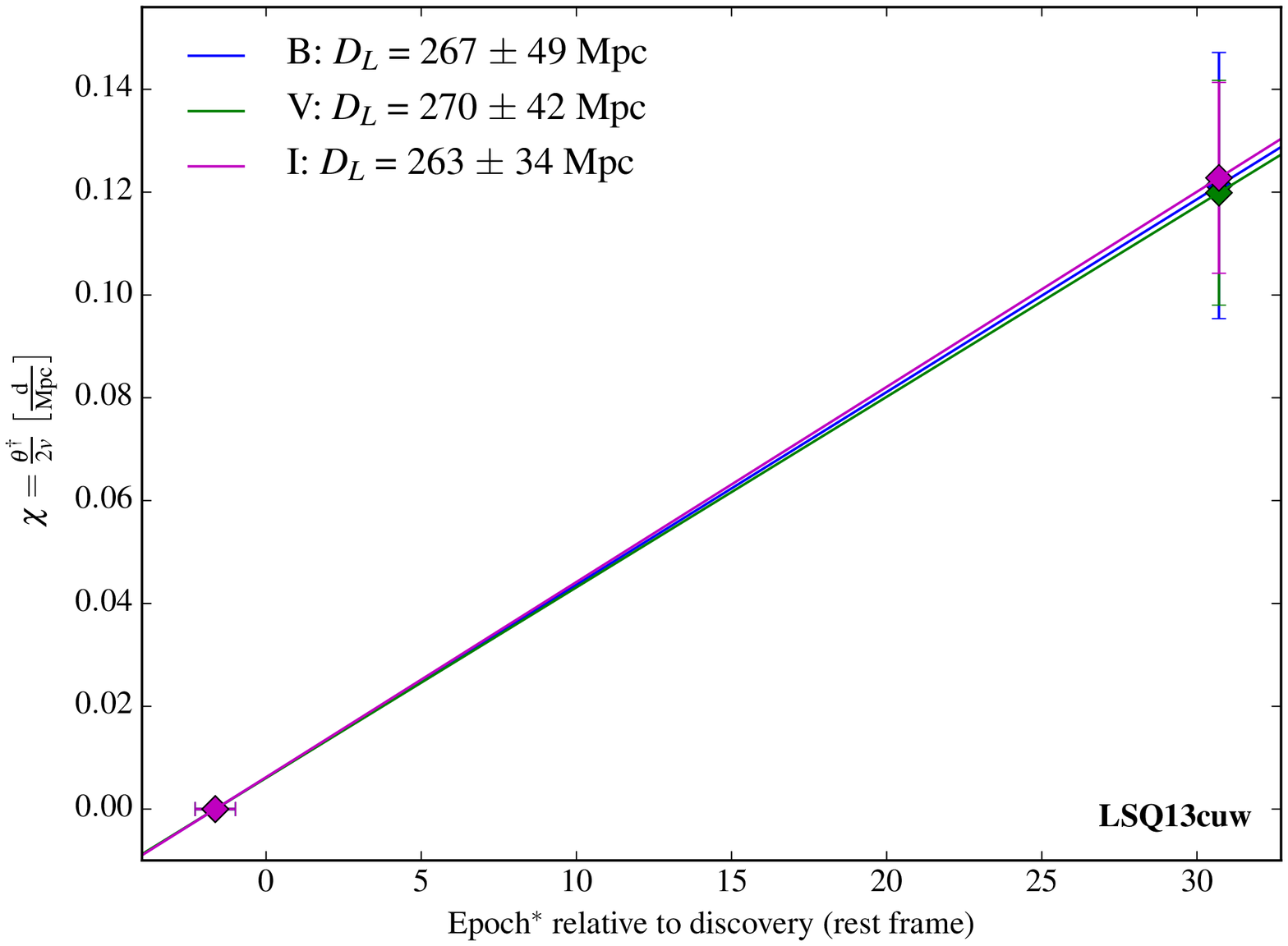}
   \end{subfigure}%
   \begin{subfigure}[t]{0.47\textwidth}
      \includegraphics[width=\columnwidth]{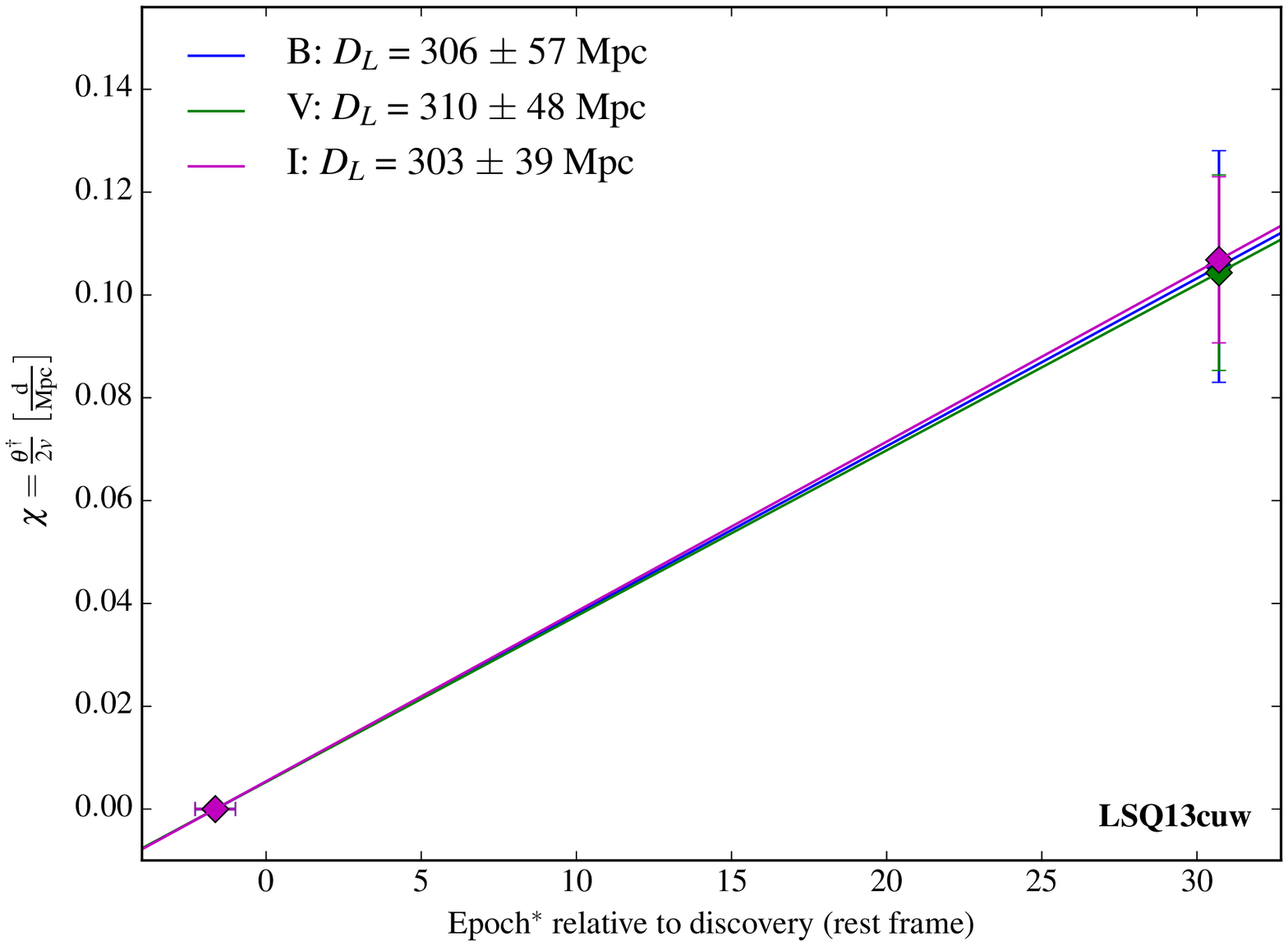}
   \end{subfigure}%
   \caption{Distance fit for LSQ13cuw using $\zeta_{BVI}$ as given in \citet{Hamuy2001} (left panel) and \citet{Dessart2005} (right panel). The diamond markers denote values of $\chi$ through which the fit is made.
}
   \label{figure:LSQ13cuw_EPM_distances}
\end{figure*}

\begin{figure*}[t!]
   \centering
   \begin{subfigure}[t]{0.47\textwidth}
      \includegraphics[width=\columnwidth]{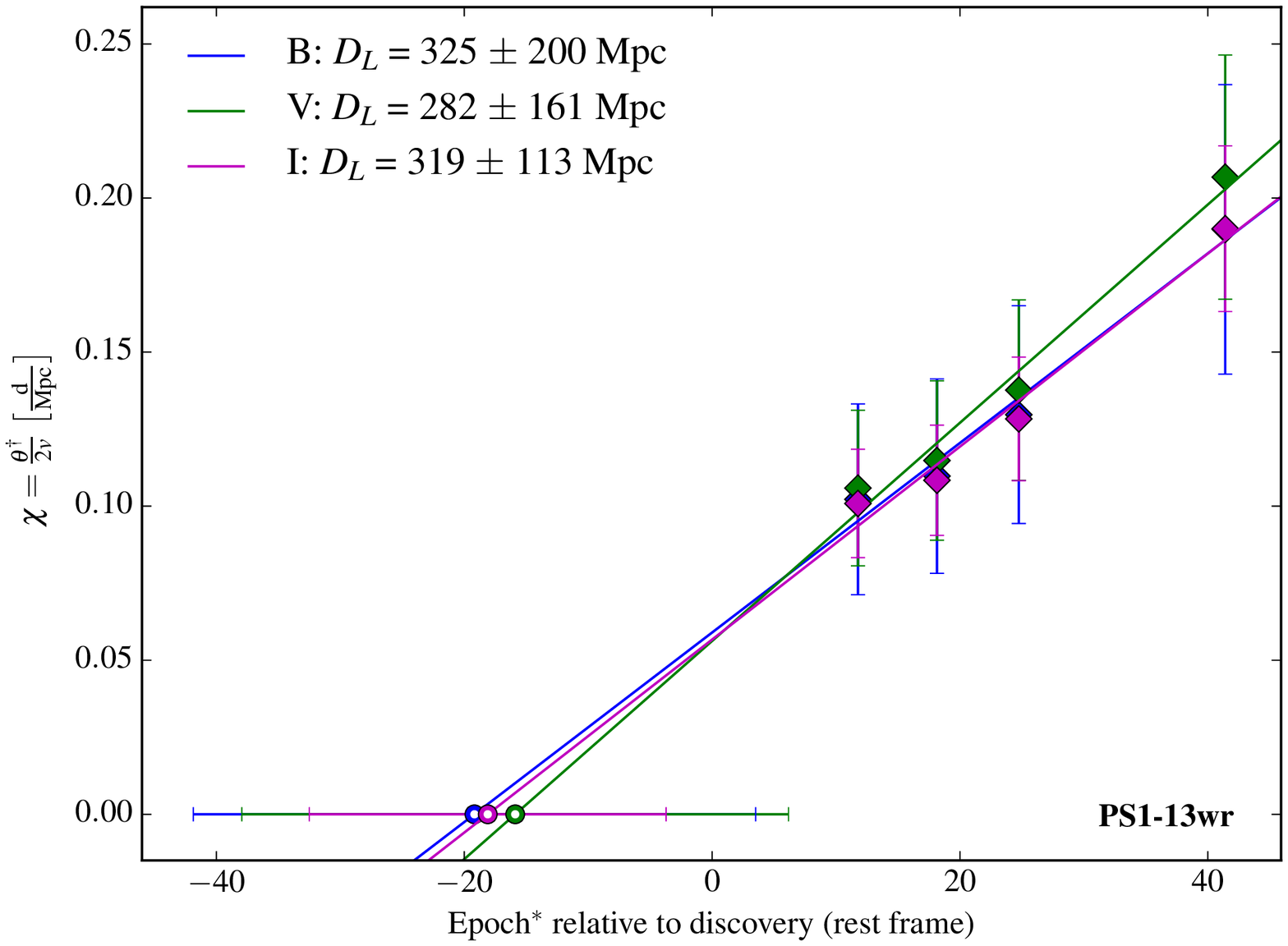}
   \end{subfigure}%
   \begin{subfigure}[t]{0.47\textwidth}
      \includegraphics[width=\columnwidth]{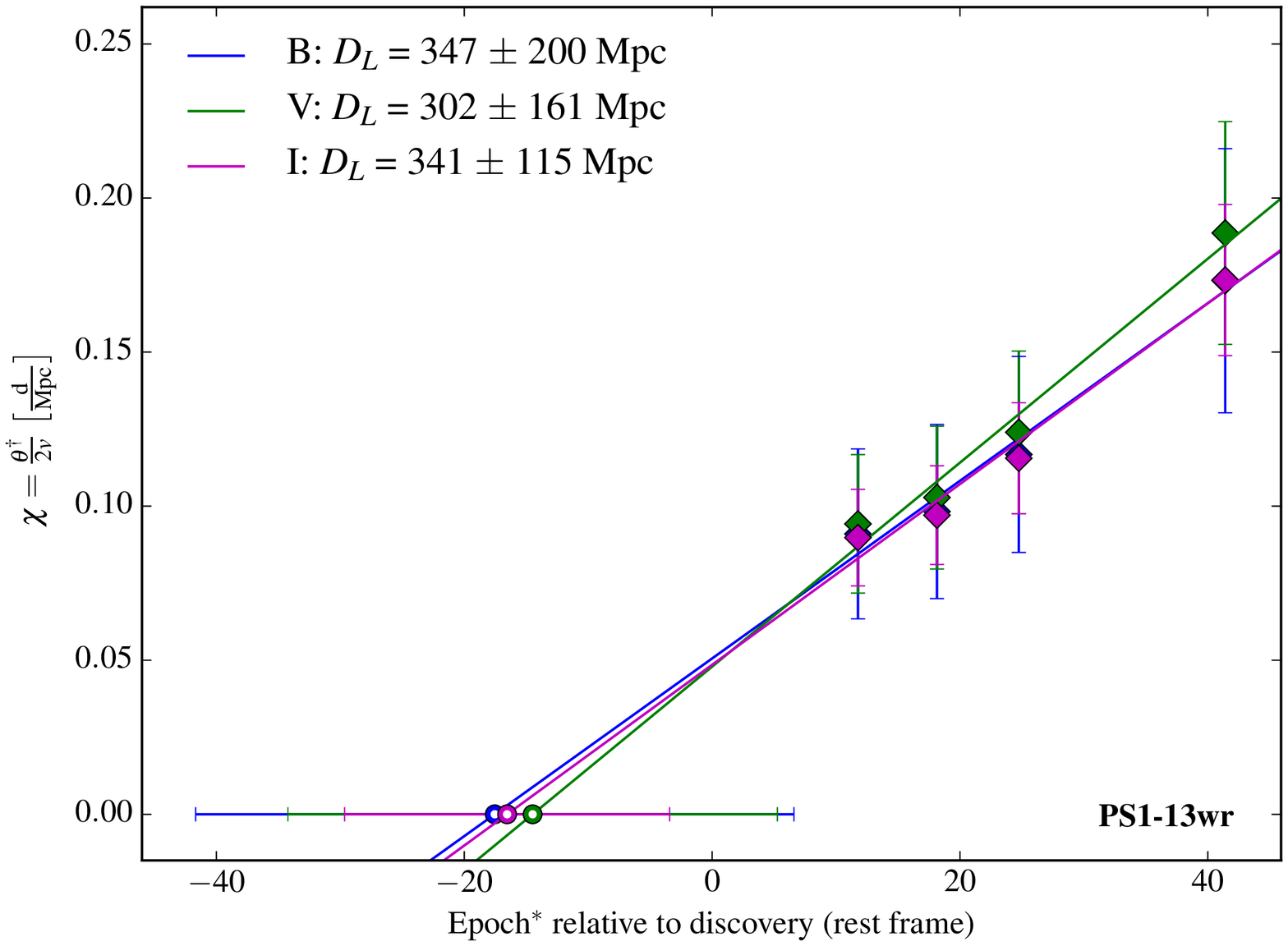}
   \end{subfigure}%
   \caption{Distance fit for PS1-13wr using $\zeta_{BVI}$ as given in \citet{Hamuy2001} (left panel) and \citet{Dessart2005} (right panel). The diamond markers denote values of $\chi$ through which the fit is made.
}
   \label{figure:PS1-13wr_EPM_distances}
\end{figure*}

\begin{figure*}[t!]
   \centering
   \begin{subfigure}[t]{0.47\textwidth}
      \includegraphics[width=\columnwidth]{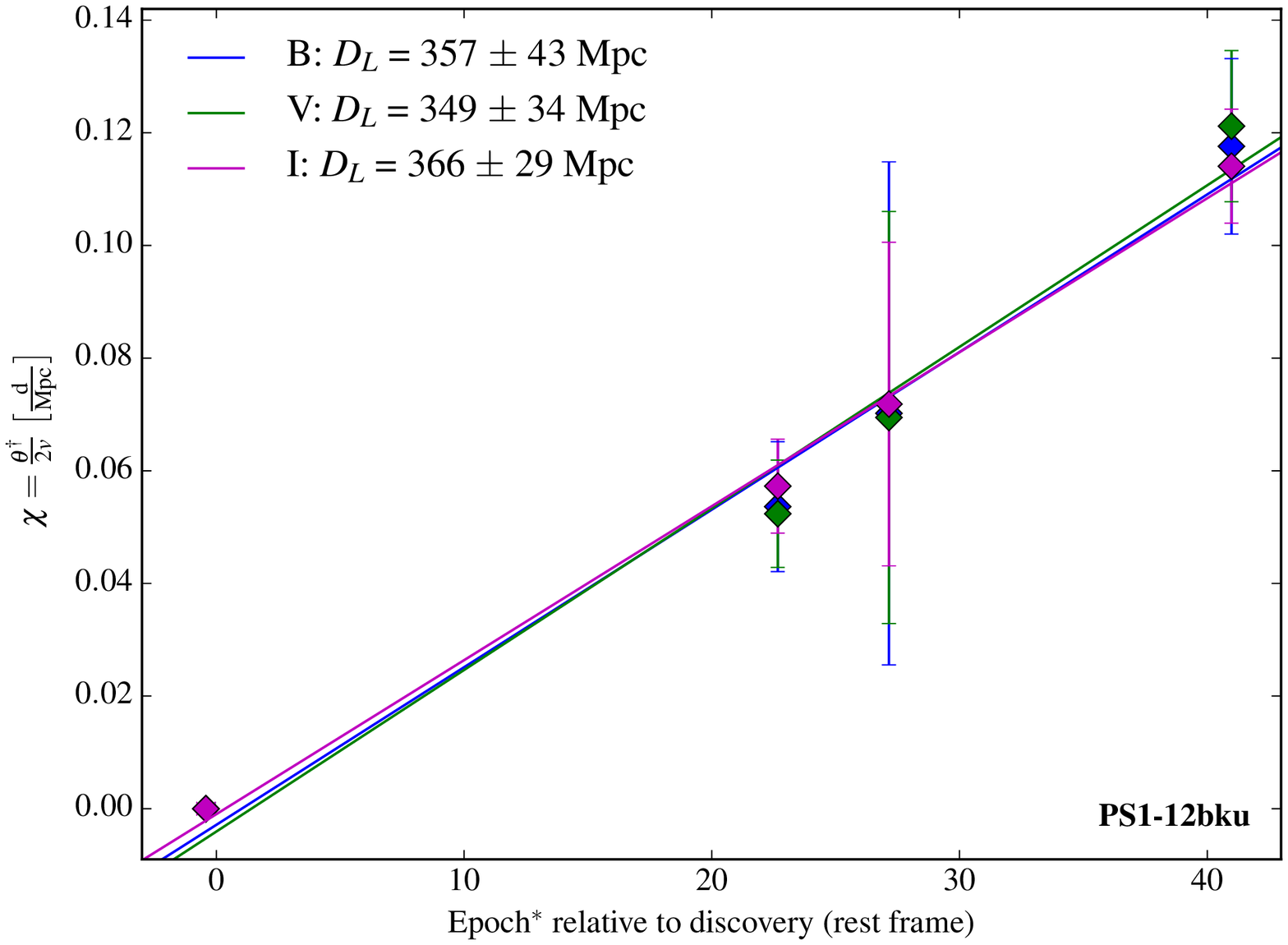}
   \end{subfigure}%
   \begin{subfigure}[t]{0.47\textwidth}
      \includegraphics[width=\columnwidth]{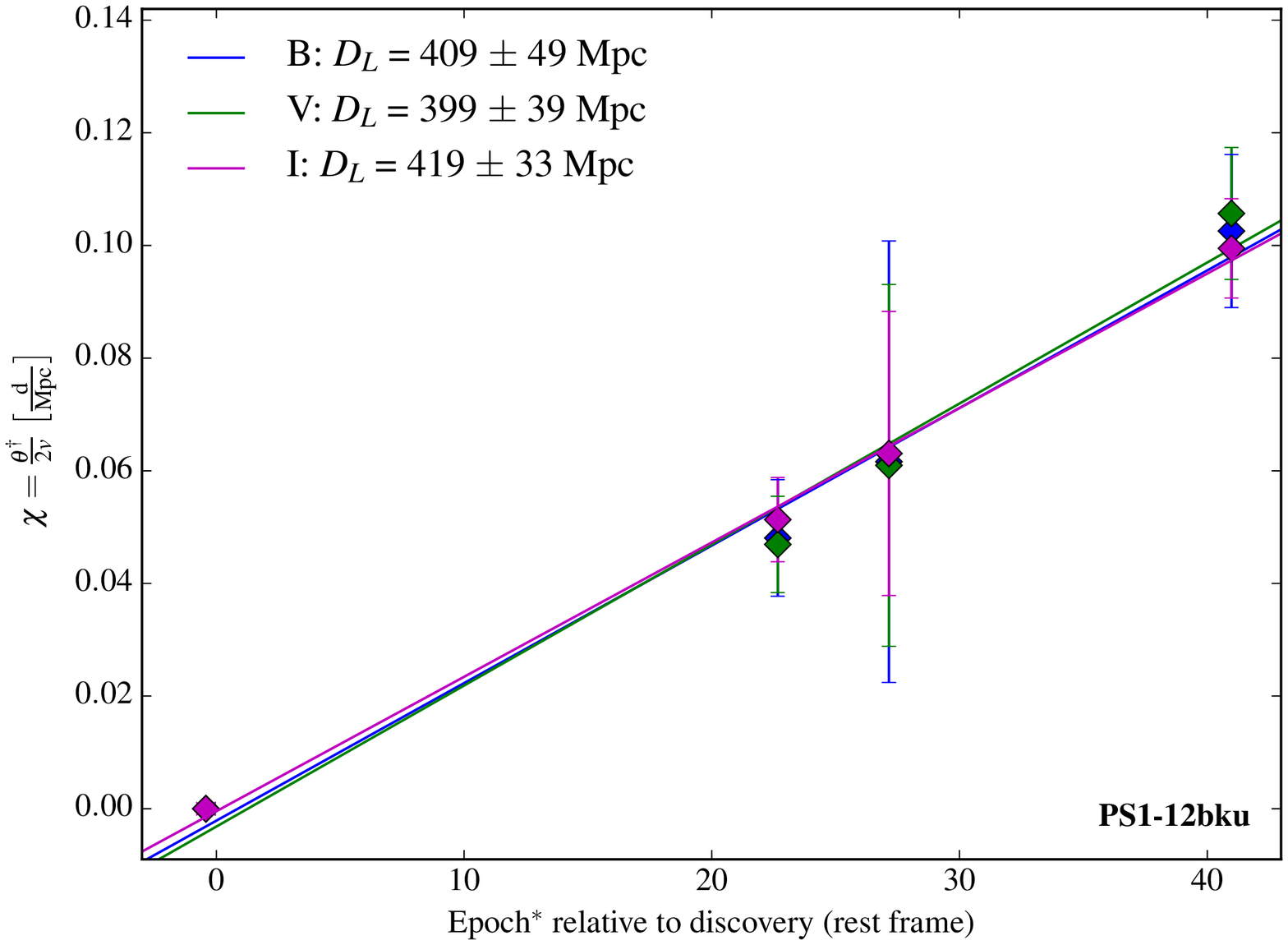}
   \end{subfigure}%
   \caption{Distance fit for PS1-12bku using $\zeta_{BVI}$ as given in \citet{Hamuy2001} (left panel) and \citet{Dessart2005} (right panel). The diamond markers denote values of $\chi$ through which the fit is made.
}
   \label{figure:PS1-12bku_EPM_distances}
\end{figure*}

\begin{figure*}[t!]
   \centering
   \begin{subfigure}[t]{0.47\textwidth}
      \includegraphics[width=\columnwidth]{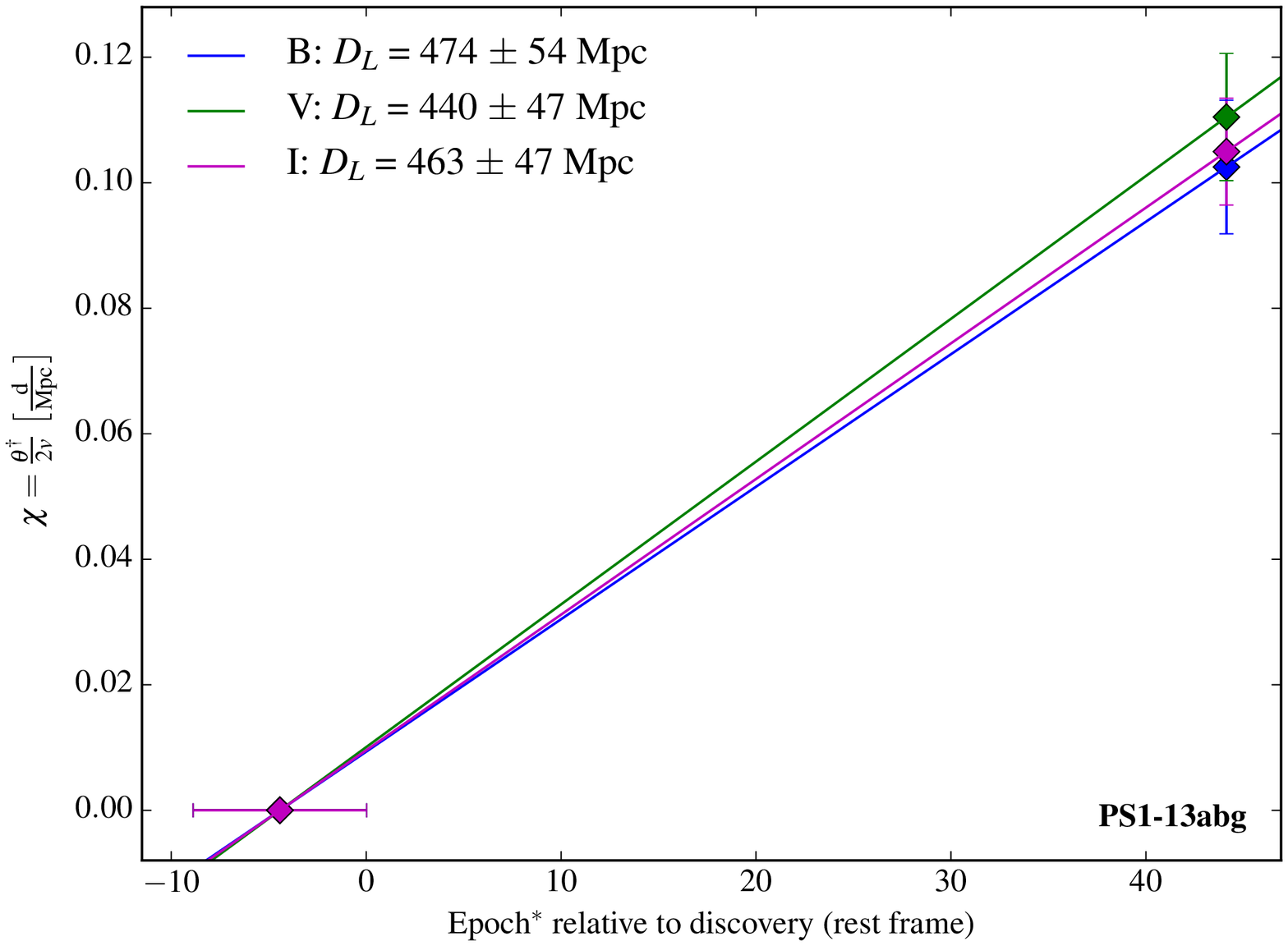}
   \end{subfigure}%
   \begin{subfigure}[t]{0.47\textwidth}
      \includegraphics[width=\columnwidth]{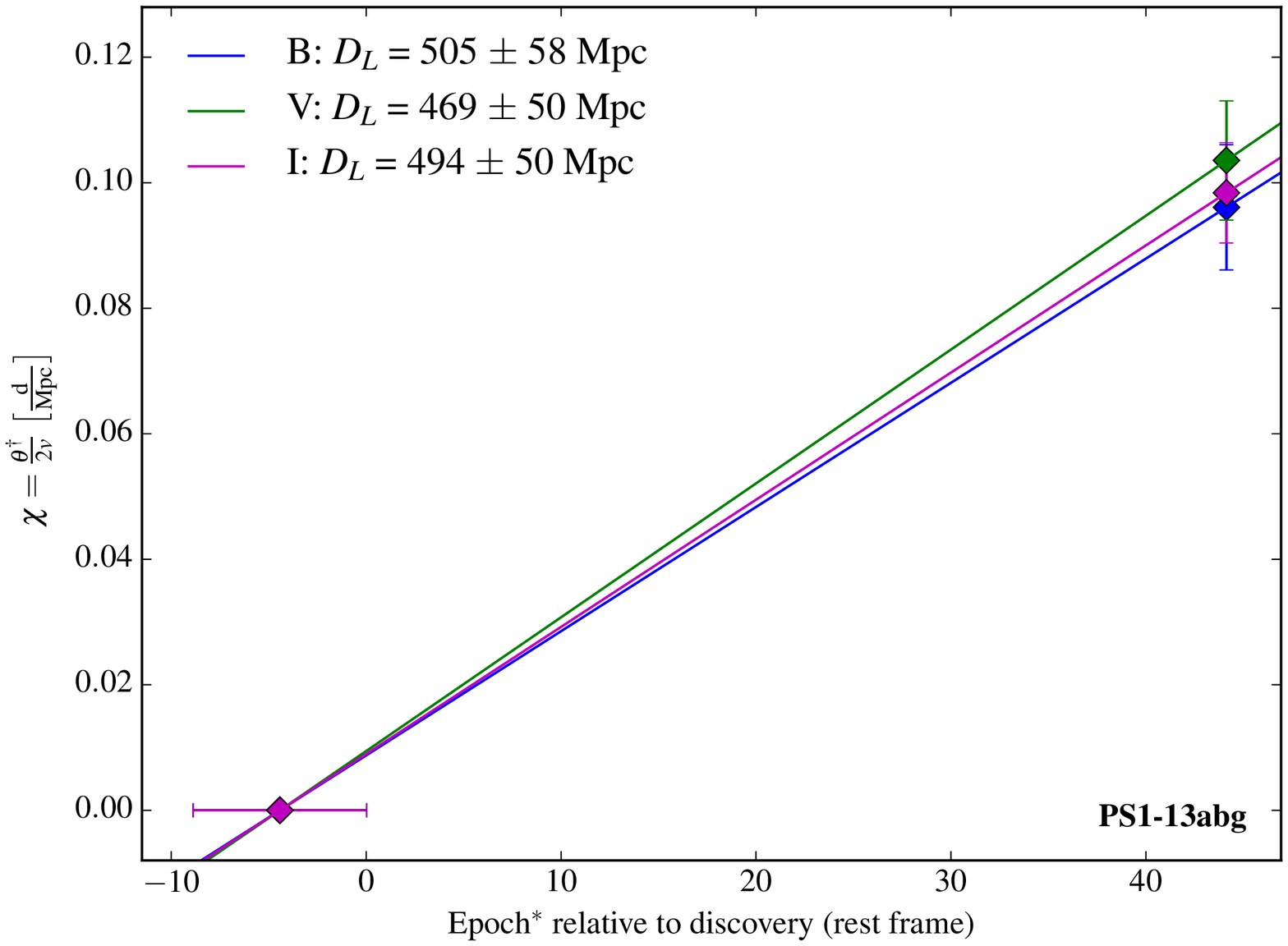}
   \end{subfigure}%
   \caption{Distance fit for PS1-13abg using $\zeta_{BVI}$ as given in \citet{Hamuy2001} (left panel) and \citet{Dessart2005} (right panel). The diamond markers denote values of $\chi$ through which the fit is made.
}
   \label{figure:PS1-13abg_EPM_distances}
\end{figure*}

\begin{figure*}[t!]
   \centering
   \begin{subfigure}[t]{0.47\textwidth}
      \includegraphics[width=\columnwidth]{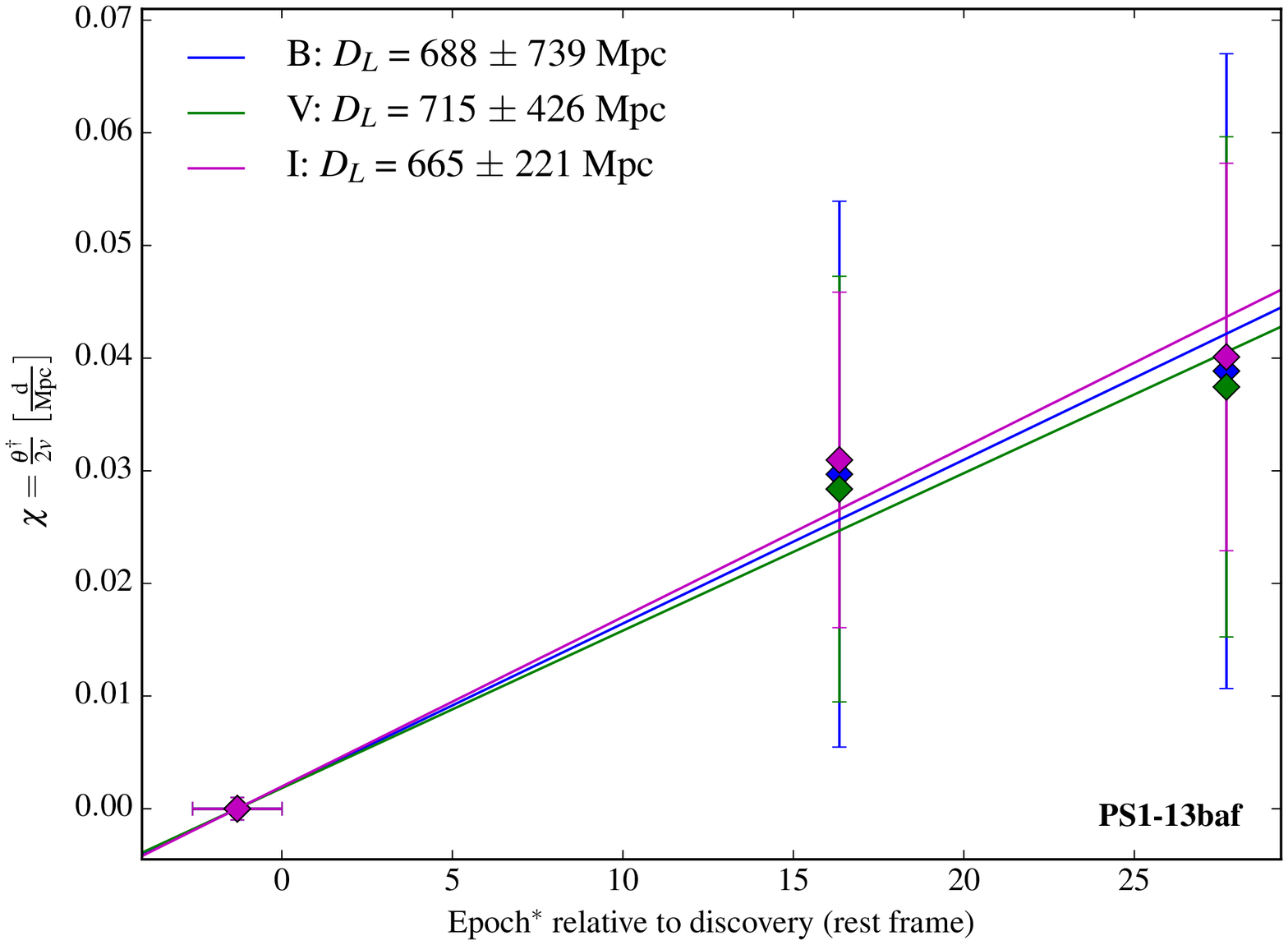}
   \end{subfigure}%
   \begin{subfigure}[t]{0.47\textwidth}
      \includegraphics[width=\columnwidth]{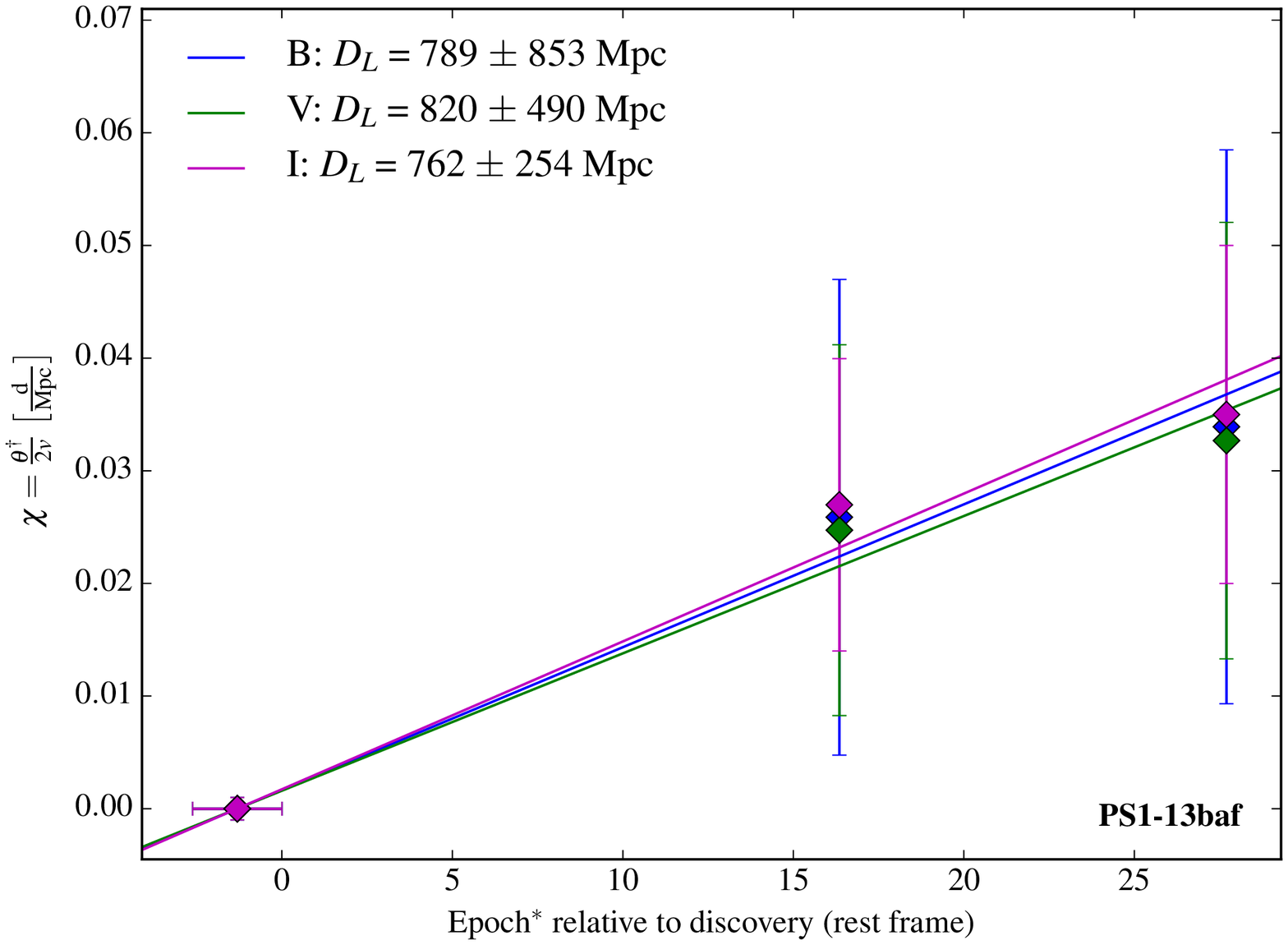}
   \end{subfigure}%
   \caption{Distance fit for PS1-13baf using $\zeta_{BVI}$ as given in \citet{Hamuy2001} (left panel) and \citet{Dessart2005} (right panel). The diamond markers denote values of $\chi$ through which the fit is made.
}
   \label{figure:PS1-13baf_EPM_distances}
\end{figure*}

\begin{figure*}[t!]
   \centering
   \begin{subfigure}[t]{0.47\textwidth}
      \includegraphics[width=\columnwidth]{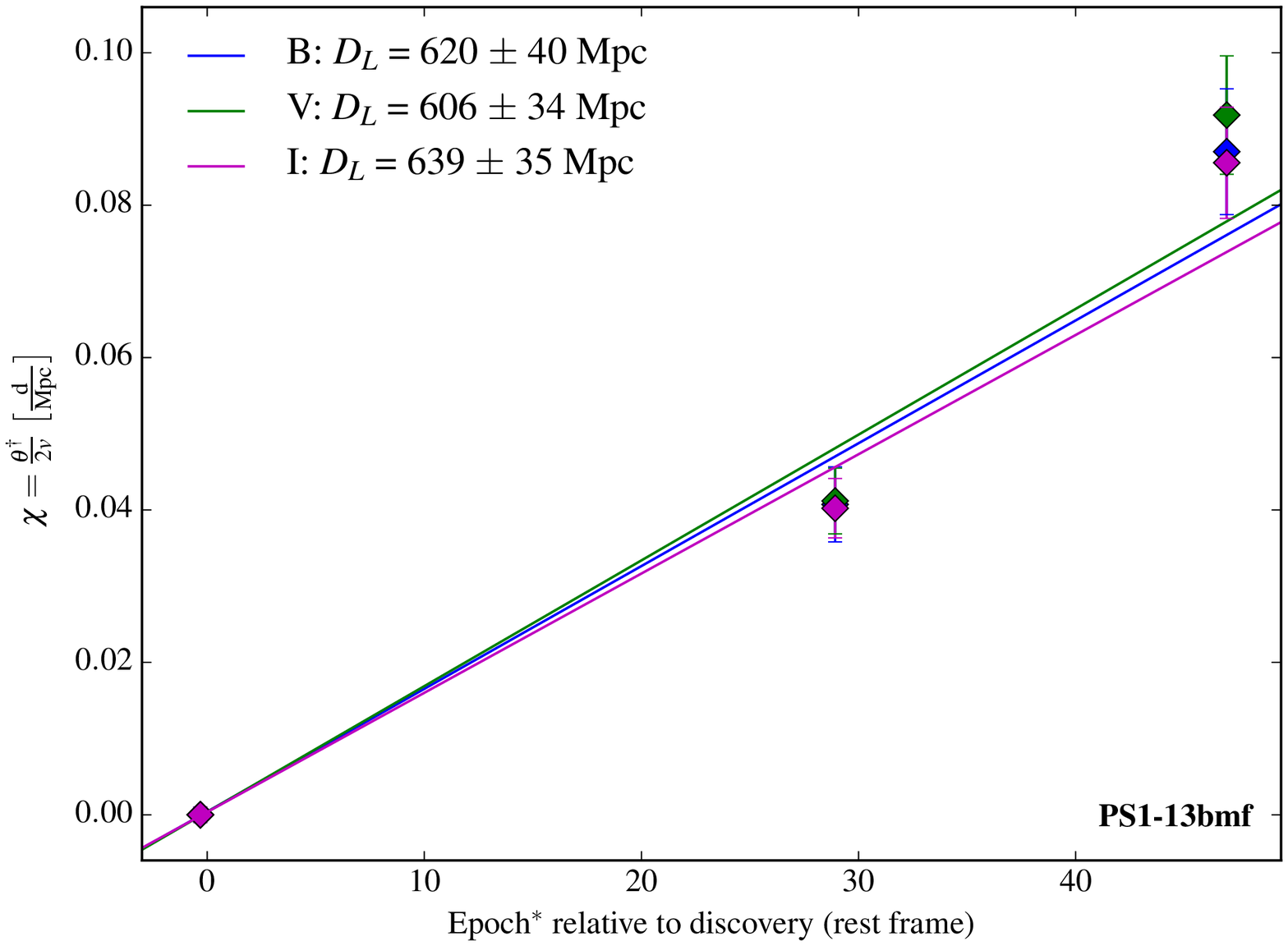}
   \end{subfigure}%
   \begin{subfigure}[t]{0.47\textwidth}
      \includegraphics[width=\columnwidth]{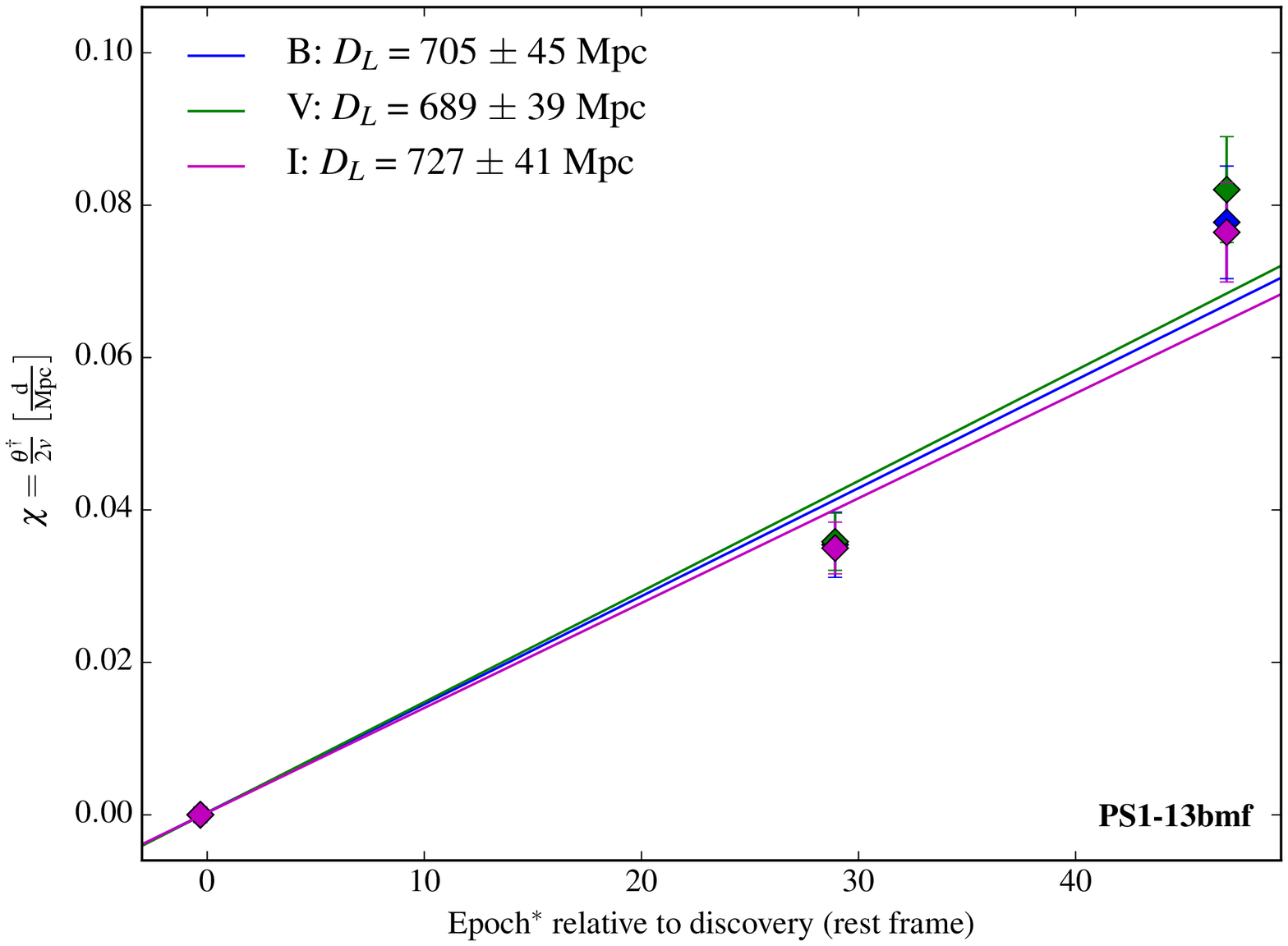}
   \end{subfigure}%
   \caption{Distance fit for PS1-13bmf using $\zeta_{BVI}$ as given in \citet{Hamuy2001} (left panel) and \citet{Dessart2005} (right panel). The diamond markers denote values of $\chi$ through which the fit is made.
}
   \label{figure:PS1-13bmf_EPM_distances}
\end{figure*}

\begin{figure*}[t!]
   \centering
   \begin{subfigure}[t]{0.47\textwidth}
      \includegraphics[width=\columnwidth]{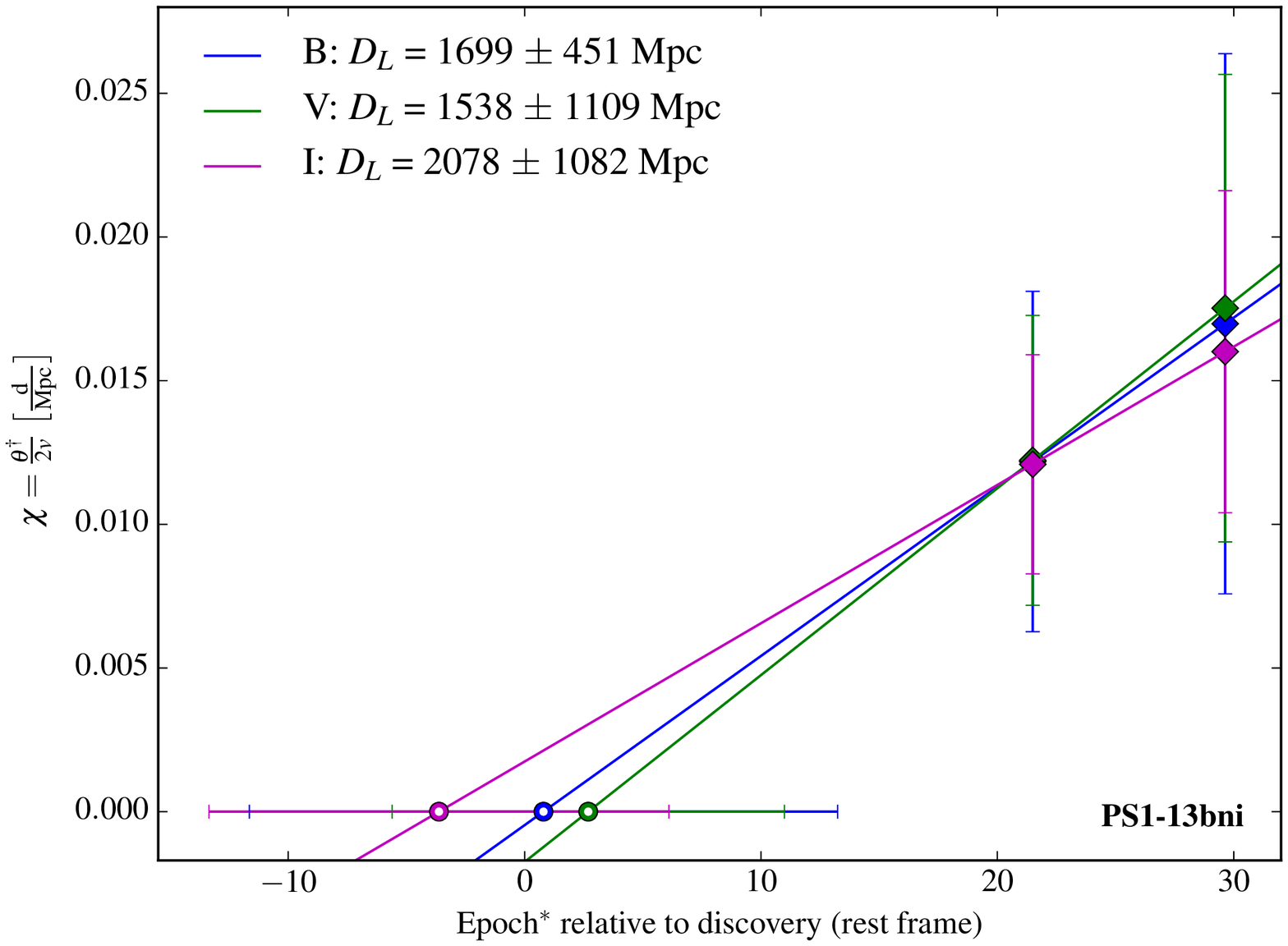}
   \end{subfigure}%
   \begin{subfigure}[t]{0.47\textwidth}
      \includegraphics[width=\columnwidth]{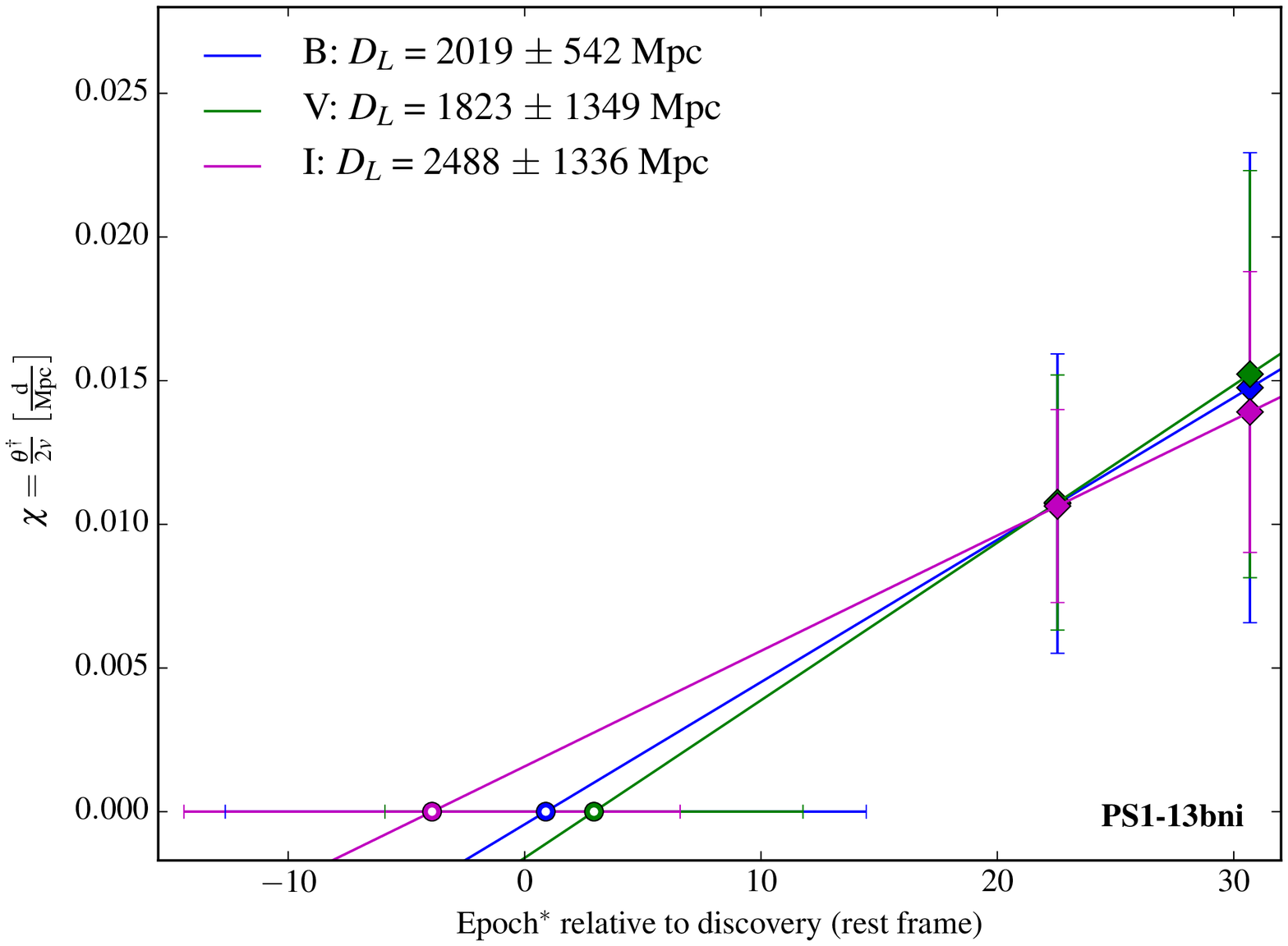}
   \end{subfigure}%
   \caption{Distance fit for PS1-13bni using $\zeta_{BVI}$ as given in \citet{Hamuy2001} (left panel) and \citet{Dessart2005} (right panel). The diamond markers denote values of $\chi$ through which the fit is made. Note that this figure shows the last step of the iteration as described in Section \ref{section:EPM:PS1_13bni}.
}
   \label{figure:PS1-13bni_EPM_distances}
\end{figure*}

\end{document}